\documentclass[11pt,a4paper]{article}
\usepackage{jheppub,multirow}
\usepackage{float}
\usepackage{graphicx}
\usepackage{url}
\usepackage{cancel}
\usepackage{slashed}
\usepackage{caption}
\usepackage{placeins}
\usepackage{feynman}
\usepackage{soul} 
\usepackage{bm}
\usepackage{amsmath}
\usepackage{amssymb}
\usepackage{listings}
\usepackage[toc, page]{appendix}
\usepackage{color}
\usepackage{booktabs,afterpage} 
\usepackage{pgf}
\usepackage{empheq}
\usepackage{tikz-feynman}
\usepackage{tikz}

\newcommand{\amc}{{\sc MadGraph5\_aMC@NLO}}

\def\G1{{\bf \gamma^{(1)}_N}}

\def\gsim{\gtrsim}
\def\lsim{\lesssim}

\newcommand{\g}{\gamma}

\definecolor{violet}{cmyk}{0,1,0,0.2}
\definecolor{red}{cmyk}{0,1,1,0}

\usepackage[colorlinks=true, linkcolor=black!50!blue, urlcolor=blue, citecolor=blue, anchorcolor=blue]{hyperref}
\usepackage[font=small,labelfont=bf,labelsep=period]{caption}
\usepackage{tabularx}
\newcolumntype{C}[1]{>{\centering\arraybackslash}p{#1}}

\newcommand{\be}{\begin{equation}}
\newcommand{\ee}{\end{equation}}
\newcommand{\bea}{\begin{eqnarray}}
\newcommand{\eea}{\end{eqnarray}}
\newcommand{\bi}{\begin{itemize}}
\newcommand{\ei}{\end{itemize}}
\newcommand{\ben}{\begin{enumerate}}
\newcommand{\een}{\end{enumerate}}

\newcommand{\lc}{\left[}
\newcommand{\rc}{\right]}
\newcommand{\lp}{\left(}
\newcommand{\rp}{\right)}

\numberwithin{equation}{section}
\numberwithin{figure}{section}
\numberwithin{table}{section}

\title{Parton distributions in the SMEFT from high-energy Drell-Yan tails}
\author[a,f]{Admir Greljo,}
\author[b]{Shayan Iranipour,}
\author[b]{Zahari Kassabov,}
\author[b]{Maeve Madigan,}
\author[b]{James Moore,}
\author[d,e]{Juan Rojo,}
\author[b]{Maria Ubiali,}
\author[c]{Cameron Voisey}
\affiliation[a]{Albert Einstein Center for Fundamental Physics, Institut f\"{u}r Theoretische Physik, Universit\"{a}t Bern, Sidlerstrasse 5, CH-3012 Bern, Switzerland}
\affiliation[b]{DAMTP, University of Cambridge, Wilberforce Road, Cambridge, CB3 0WA, United Kingdom}
\affiliation[c]{Cavendish Laboratory (HEP), JJ Thomson Avenue, Cambridge, CB3 0HE, United Kingdom}
\affiliation[d]{Department of Physics and Astronomy, Vrije Universiteit Amsterdam, NL-1081 HV Amsterdam, The Netherlands}
\affiliation[e]{ Nikhef Theory Group, Science Park 105, 1098 XG Amsterdam, The Netherlands}
\affiliation[f]{CERN, Theoretical Physics Department, CH-1211 Geneva 23, Switzerland}
\emailAdd{M.Ubiali@damtp.cam.ac.uk}

\abstract{
  The high-energy tails of charged- and neutral-current Drell-Yan  processes provide important constraints on the light quark and anti-quark parton distribution functions (PDFs) in the large-$x$ region.
  At the same time, short-distance new physics effects 
  such as those encoded by the Standard Model Effective Field Theory (SMEFT)
  would induce smooth distortions to the same high-energy Drell-Yan tails.
  In this work, we assess for the first time the interplay between PDFs and EFT effects for
  high-mass Drell-Yan processes at the LHC and quantify the impact that the consistent joint
  determination of PDFs and Wilson coefficients has on the bounds derived for the latter.
  We consider two well-motivated new physics scenarios: $1)$ electroweak oblique corrections $(\hat W, \hat Y)$ and $2)$ four-fermion interactions potentially related to the LHCb anomalies in $R(K^{(*)})$.
  We account for available Drell-Yan data, both from unfolded cross sections and from searches,
  and carry out dedicated projections for the High-Luminosity LHC.
  Our main finding is that, while the interplay between PDFs and EFT effects remains moderate
  for the current dataset, it will become a significant challenge
  for EFT analyses at the HL-LHC.
}

\arxivnumber{2104.02723}
\subheader{\small Cavendish-HEP-21/06, Nikhef-2020-040}

\begin{document}

\maketitle
\flushbottom

\section{Introduction}
\label{sec:introduction}

The Large Hadron Collider (LHC) at CERN is currently leading the exploration of the high-energy frontier.
Thanks to a wealth of precision experimental measurements, combined
with the staggering progress in precise theoretical calculations and simulations,
the Standard Model (SM) of particle physics is being stress-tested at high energies to
an unprecedented level.
The hunt for deviations from the SM predictions, which could signal the long-sought evidence
for beyond-the-SM (BSM) dynamics, is being pursued at the LHC with two
complementary strategies.
On the one hand, by means of direct searches for new particles that might be light enough to be
produced on shell.
On the other hand,  with indirect searches aiming to identify a consistent pattern of
deviations in the properties of known particles, such as their coupling strengths, which would be induced by
new particles and interactions  whose characteristic energy scale lies above
the direct reach of the LHC.

In the case of direct searches,
the target is  identifying
an abrupt deviation from the SM, for example in terms of a resonance peak on top of a
smoothly falling background, or with a
systematic mismatch in the number of measured events as compared to the SM
expectations for non-resonant searches.
Indirect searches instead often aim at pinning down subtle distortions with respect
to the SM predictions, such as a few-percent variation in the value of some production
cross sections or decay rates, induced by yet-unknown particles and interactions.
A powerful framework to identify, parametrise, and correlate
in a model-independent manner the possible deviations
from the SM predictions that arise
in such indirect searches
is provided by the effective field theory (EFT) framework, such
as the Standard Model EFT (SMEFT) (e.g.~\cite{Weinberg:1979sa,Buchmuller:1985jz,Georgi:1994qn,Giudice:2007fh,Grzadkowski:2010es,Alonso:2013hga,Jenkins:2013wua,Jenkins:2013zja,Henning:2014wua,Brivio:2017vri,Wells:2015uba,Englert:2019zmt,Fuentes-Martin:2020udw,Cohen:2020qvb})
or the Higgs EFT (HEFT) (e.g.~\cite{Feruglio:1992wf,Grinstein:2007iv,Contino:2010mh,Alonso:2012px,Alonso:2014wta,Buchalla:2015qju,Alonso:2015fsp,Cohen:2020xca}).

It is a common misconception that indirect BSM investigations based on EFTs
are relevant only for low-energy phenomena, while instead direct searches dominate the sensitivity
for high-$p_T$ observables.
On the contrary: accurate measurements of observables at high energies offer one of the most promising avenues towards a discovery of BSM physics at the LHC.
While the collider has now (almost) reached the design energy, its integrated luminosity continues to grow
steadily, thus greatly facilitating dedicated studies of the
(currently statistics-limited) high-energy tails of distributions.
With this motivation, several  high-$p_T$  LHC cross sections have been studied as indirect probes of new
physics in the EFT framework, from the (di)lepton distributions
in neutral- and charged-current Drell-Yan (DY) distributions~\cite{Cirigliano:2012ab, deBlas:2013qqa, Gonzalez-Alonso:2016etj, Faroughy:2016osc, Greljo:2017vvb, Cirigliano:2018dyk, Greljo:2018tzh,Bansal:2018eha,Angelescu:2020uug, Farina:2016rws, Alioli:2017nzr, Raj:2016aky, Schmaltz:2018nls, Brooijmans:2020yij,Ricci:2020xre,Fuentes-Martin:2020lea,Alioli:2017ces,Alioli:2017jdo,Alioli:2018ljm,Alioli:2020kez,Panico:2021vav,Sirunyan:2021khd,ATLAS:2021pvh,Marzocca:2020ueu,Afik:2019htr,Alves:2018krf} (the
main subject of this work) to inclusive jets and dijets~\cite{Krauss:2016ely,Alte:2017pme,Hirschi:2018etq,Goldouzian:2020wdq}, top quark pair
distributions~\cite{Buckley:2015lku,Englert:2016aei,Hartland:2019bjb,Brivio:2019ius,Durieux:2018tev,vanBeek:2019evb,Bissmann:2020mfi,Bruggisser:2021duo,Ellis:2020unq}, gauge boson pair
production~\cite{Falkowski:2015jaa,Falkowski:2016cxu,Baglio:2017bfe,Baglio:2020ibv,Ellis:2020unq,Panico:2017frx,Ethier:2021ydt}, and vector boson scattering~\cite{Ethier:2021ydt,Grojean:2018dqj,Gomez-Ambrosio:2018pnl,Dedes:2020xmo}, among several others.

Furthermore, many higher-dimensional EFT operators induce deviations from the SM which grow
with the energy of the partonic collision, enhancing the BSM sensitivity
of these high-energy tails.
For instance, in the Drell-Yan process,  the na\"{\i}ve  scaling of the $\psi \psi \to \psi \psi$ scattering
amplitude in an underlying EFT description leads to
$\mathcal{A} \propto E^2 / \Lambda^2$,  where $E$ is the energy of the process and $\Lambda$ is the new physics scale.
Thus, rather generically, a less precise measurement of a high-mass tail can compete with a low-energy precision measurement due to this energy enhancement, which can be traced back to the preservation of unitarity.
This property leads to the somehow unintuitive result that
the study of the high-energy lepton tails in the Drell-Yan process represents a competitive probe
of BSM dynamics compared to electroweak precision tests and low-energy
flavour physics test.

In order for the EFT interpretation of these high-$p_T$ cross sections
to convincingly uncover a BSM effect, it becomes crucial to ensure
full control over the SM inputs and their
uncertainties, such as those associated to the hard-scattering (partonic) cross sections or the
parton distribution functions (PDFs) of the nucleon~\cite{Gao:2017yyd}.
In this context, one of the key challenges hampering the applicability of the EFT programme
to the LHC high-energy tails is indeed related to the treatment of the PDFs.
These are determined from experimental data assuming  the validity of the SM, often
using exactly the same processes that enter EFT analyses.
This observation naturally prompts the question of how can one make sure that eventual BSM deviations arising in high-$p_T$ tails are not being
inadvertedly reabsorbed into the PDFs.

A first take on this challenge was presented in a proof-of-concept study~\cite{Carrazza:2019sec} where
the simultaneous determination of PDFs and EFT coefficients from deep inelastic scattering (DIS)
structure functions was demonstrated.
There, it was found that the EFT corrections can indeed be partially reabsorbed into the PDFs
but also that it is possible to robustly disentangle
QCD and BSM effects by exploiting their different energy scaling.
The main goal of the present work is to extend this approach to LHC processes, specifically with the
joint determination of PDFs and EFT coefficients from DIS and Drell-Yan data.
Drell-Yan processes in general, and high-mass measurements in particular,
provide information on the light quark and anti-quark PDFs in a broad region of $x$ representing
an important ingredient in modern global PDF fits~\cite{Ball:2014uwa,Ball:2017nwa,Hou:2019efy,Bailey:2020ooq}.
Furthermore, high-mass Drell-Yan data will be instrumental at the High-Luminosity LHC (HL-LHC)
to pin down the large-$x$ PDFs~\cite{Khalek:2018mdn}.
Considering that EFT signals can lead to significant deviations from the SM
in these same high-energy DY tails, one would like to assess to what extent they
can be reabsorbed into the PDFs and to define strategies to
separate QCD from BSM effects.

In this work, in order to interpret the Drell-Yan data in the EFT framework
we formulate simple, yet motivated, BSM benchmark scenarios, which are chosen to represent a wide
class of UV-complete theories.
In the first scenario~\cite{Farina:2016rws}, we consider the $\hat{W}$ and $\hat{Y}$ electroweak parameters generated in universal theories that modify the electroweak gauge boson propagators and lead to flavour-universal deviations which grow with the invariant mass.
In the second benchmark~\cite{Greljo:2017vvb}, we consider a flavour-specific scenario motivated by the evidence of lepton flavour universality (LFU) violation in $B$-meson decays recently reported by the LHCb collaboration~\cite{Aaij:2014ora, Aaij:2017vbb, Aaij:2019wad,Aaij:2021vac}.
Our analysis accounts for available unfolded high-mass Drell-Yan cross
section data, detector-level
searches  based on the full Run II luminosity, and dedicated HL-LHC projections.
We do not consider other data beyond the DIS and DY processes, in order to ensure a theoretically consistent
description of the parton distributions in the SMEFT.
The price to pay for this reduced dataset is the information loss on some flavour combinations,
specifically on the gluon PDF.

The structure of this paper is as follows.
First of all, in Sect.~\ref{sec:scenarios} we discuss the EFT benchmark scenarios that will be used
in the fits.
Then in Sect.~\ref{sec:data} we summarise the datasets used in our
analysis and the corresponding theoretical calculations, both in the
SM and in the SMEFT, and we discuss their impact on a PDF fit. We also
describe the methodology we use to simultaneously fit PDFs and SMEFT
coefficients, and how we deal with several sources of uncertainties
in the fits. 
In Sect.~\ref{sec:res1} we present the results for the simultaneous determination of the
SMEFT coefficients and the PDFs from the available high-mass DY data
from LHC Run I and Run II, in the two
scenarios presented in Sect.~\ref{sec:scenarios} and assess how they
modify the interpretation of BSM searches based on the SM PDFs.
In Sect.~\ref{sec:hllhc} we present a summary of the constraints we find on the two
scenarios we consider and we assess the outcome of a joint
PDF and EFT analysis at the HL-LHC.
Finally, we will outline our main conclusions and possible
future developments in Sect.~\ref{sec:conclusions}.

More technical discussions are collected in the appendices, and include
detailed comparisons of the SM PDF fits produced in this work
with previous NNPDF global fits
(App.~\ref{app:pdfs}), the quantitative assessment of the fit quality
to the various input datasets (App.~\ref{app:fit_quality}), a
benchmarking study for the calculation of EFT cross sections (App.~\ref{sec:benchmarking}),
and a study of the flavour dependence of the SMEFT PDFs (App.~\ref{add:smeftpdfs}).

\section{SMEFT benchmark scenarios}
\label{sec:scenarios}

In this section we present the two SMEFT benchmark scenarios that will be
used in this work to interpret the LHC Drell-Yan processes.
The first scenario belongs to the class of electroweak precision tests and is
sensitive to a broad range of UV-complete theories proposed in the literature.
The second benchmark represents a consistency check of the existing hints
of lepton universality violation in rare $B$-meson decays reported by the LHCb collaboration.
Both scenarios highlight the interplay between the PDFs and the EFT dynamics, illustrating
in particular how the former changes and how constraints to the latter are modified.

\subsection{Benchmark I: oblique corrections $\hat W$ and $\hat Y$ }
\label{sec:scenarioI}

The oblique corrections, as originally proposed in~\cite{Peskin:1991sw,Altarelli:1991fk}, play a key role in testing theories beyond the Standard Model.
They parametrise the self-energy $\Pi_V(q^2)$ of the electroweak gauge bosons $W^a_\mu$ and $B_\mu$,  where $V = W^3 W^3$, $B B$, $W^3 B$, and $W^+ W^-$.
Truncating the momentum expansion at order $q^4$, while imposing proper normalisation and symmetry constraints, one concludes that there are only four oblique
parameters which can be identified with dimension-six operators in the SMEFT.
These are the well-known  $\hat{S}$, $\hat{T}$, $\hat{W}$, and $\hat{Y}$
parameters~\cite{Barbieri:2004qk}.
The parameters $\hat{S}$ and $\hat{T}$ are well constrained from precision
LEP measurements~\cite{Barbieri:2004qk} and grow slowly with $q^2$,
while $\hat{W}$ and $\hat{Y}$ scale faster implying that their effects will be
enhanced for the high-energy dilepton tails at the LHC~\cite{Farina:2016rws}.
 While $\hat{T} = {\mathcal O}(q^0)$ and $\hat{S} = {\mathcal O}(q^2)$ ,
instead one has that $\hat{W},\hat{Y} = {\mathcal O}(q^4)$.
In the universal basis (see e.g.~\cite{Englert:2019zmt}), the $\hat{W}$ and $\hat{Y}$
parameters are the Wilson coefficients associated to the following two operators:
\begin{equation}
\label{eq:WY}
\mathcal{L}_{\rm SMEFT} \supset -\frac{\hat{W}}{4 m_W^2} (D_\rho W^a_{\mu\nu})^2 -\frac{\hat{Y}}{4 m_W^2} (\partial_\rho B_{\mu\nu})^2 ~,
\end{equation}
where $m_W$ indicates the $W$-boson mass, and $D_\rho$ is the covariant derivative.
The physical effects of the operators in Eq.~(\ref{eq:WY}) on the Drell-Yan process
arise from the difference in the propagators through the self-energy modifications, see Eq.~(1) of Ref.~\cite{Farina:2016rws}.

Alternatively, using the equations of motion, the same operators can be rotated to a basis in which the modifications to the Drell-Yan cross sections are instead captured by four-fermion contact interactions,
\begin{equation} \label{eq:JJ}
\mathcal{L}_{\rm SMEFT} \supset -\frac{g^2 \hat{W}}{2 m_W^2} \, J^{a}_{L \mu} J^{a \mu}_L -\frac{ g_Y^2 \hat{Y} }{2 m_W^2} \, J_{Y \mu} J^\mu_Y ~~,
\end{equation}
where $J_L$ and $J_Y$ are $SU(2)_L$ and $U(1)_Y$ conserved fermionic currents,
\begin{equation}
J^{a \mu}_L =  \frac{1}{2} \sum_{f = q, l}  \bar f \sigma^a \gamma^\mu f ~~,\quad ~~ J^\mu_Y= \sum_{f = q, l, u, d, e}  Y_f\,\bar f \gamma^\mu f~~.
\end{equation}
Here $q$, $l$ are the SM quark and lepton left-handed doublets, while $u$, $d$, $e$ are 
the right-handed singlets. Also, $g$ and $g_Y$ are the corresponding electroweak gauge couplings,
while $\sigma^a$ 
are the Pauli matrices, and the hypercharge $Y_f = 1/6$, $-1/2$, $2/3$, $-1/3$, and $-1$ for $q,l,u,d,e$, respectively. Summation over flavour indices is assumed, which implies
that in this scenario the fermionic currents respect the $U(3)^5$
global flavour symmetry.

Expanding Eq.~\eqref{eq:JJ}, one can relate the $\hat{W}$ and $\hat{Y}$ parameters
to the coefficients of dimension-six operators in the Warsaw basis~\cite{Grzadkowski:2010es}.
There, the operators  relevant to the description of the Drell-Yan process are given by
\begin{equation}
\begin{aligned}\label{eq:Warsaw}
\mathcal{O}_{ld} &= (\bar  l\g_\mu l)(\bar d \g^\mu d) \,, 
&
\mathcal{O}_{lu} &= (\bar  l\g_\mu l)(\bar u \g^\mu u) \,,
&
\mathcal{O}_{lq}^{(1)} &= (\bar l \g^\mu l) (\bar  q\g_\mu q)\,,   
\\
\mathcal{O}_{ed} &= (\bar  e\g_\mu e)(\bar d \g^\mu d) \,, 
&
\mathcal{O}_{eu} &= (\bar  e\g_\mu e)(\bar u \g^\mu u) \,,
&
\mathcal{O}_{qe} &=  (\bar  q\g_\mu q) (\bar e \g^\mu e)\,,
\\
\mathcal{O}_{lq}^{(3)} &=(\bar l \sigma^a \g^\mu l)  (\bar  q \sigma^a \g_\mu q) \, .
\end{aligned}
\end{equation}
Again, the flavour indices are contracted within the brackets, for example $(\bar  l\g_\mu l) \equiv (\bar  l^1 \g_\mu l^1 + \bar  l^2 \g_\mu l^2 + \bar  l^3 \g_\mu l^3)$. 
Taking into account this matching between the $\hat{W}$ and $\hat{Y}$ parameters
and the corresponding Wilson coefficients
in the Warsaw basis,
we can express the SMEFT Lagrangian in this scenario, Eq.~(\ref{eq:JJ}), as follows
\begin{align}\label{eq:WYWarsaw}
\mathcal{L}_{\rm SMEFT}=\mathcal{L}_{\rm SM}   &-\frac{g^2 \hat{W}}{4
                                                 m_W^2}
                                                 \mathcal{O}_{lq}^{(3)}
                                                 - \frac{ g_Y^2
                                                 \hat{Y} }{m_W^2}
                                                 \Big(Y_l Y_d \,
                                                 \mathcal{O}_{ld} +
                                                 Y_l Y_u \,
                                                 \mathcal{O}_{lu}
                                                 \nonumber \\&+Y_l Y_q
  \, \mathcal{O}_{lq}^{(1)} + Y_e Y_d \, \mathcal{O}_{ed} + Y_e Y_u \,
  \mathcal{O}_{eu} + Y_e Y_q \, \mathcal{O}_{q e} \Big) \,.
\end{align}
The parametrisation in Eq.~(\ref{eq:Warsaw}) has been implemented using the {\tt SMEFTsim} package~\cite{Brivio:2017btx} and cross-checked against the reweighting method used in Ref.~\cite{Greljo:2017vvb} (see also~\cite{Ricci:2020xre}), as will be discussed in Sect.~\ref{sec:theory} and in Appendix~\ref{sec:benchmarking}.

The analysis in the Ref.~\cite{Farina:2016rws} reports the following 95\% confidence level intervals on $\hat{W}$ assuming $\hat{Y}=0$,
\begin{equation}
\begin{split}
\hat W &\in \left[ - 3, 15 \right]\times 10^{-4}~\textrm{(ATLAS 8 TeV, 20.3 fb$^{-1}$~\cite{Aad:2016zzw})}~,\\
\hat W &\in \left[ - 5, 22 \right]\times 10^{-4}~\textrm{(CMS 8 TeV, 19.7 fb$^{-1}$~\cite{CMS:2014jea})}~,
\end{split}
\end{equation} 
as well as, the 95\% confidence level intervals for $\hat{Y}$ assuming $\hat{W}=0$,
\begin{equation}
\begin{split}
\hat Y &\in \left[ - 4, 24 \right]\times 10^{-4}~\textrm{(ATLAS 8 TeV, 20.3 fb$^{-1}$~\cite{Aad:2016zzw})}~,\\
\hat Y &\in \left[ - 7, 41 \right]\times 10^{-4}~\textrm{(CMS 8 TeV, 19.7 fb$^{-1}$~\cite{CMS:2014jea})}~.
\end{split}
\end{equation}
These bound have been computed by assuming SM PDFs. In our analysis,
for this benchmark scenario, we see how the limits based on SM PDFs
are modified once a consistent determination of the SMEFT PDFs is
done, requiring a simultaneous fit of the PDFs together with the
$\hat{W}$ and $\hat{Y}$ parameters from the high-mass Drell-Yan distributions.

\subsection{Benchmark II: left-handed muon-philic lepton-quark interactions} 
\label{sec:scenarioII}

Following Ref.~\cite{Greljo:2017vvb}, here we consider
gauge invariant four-fermion operators built from the SM quark and lepton $SU(2)_L$ doublets.
In the Warsaw basis, Eq.~(\ref{eq:Warsaw}), these correspond to the $\mathcal{O}_{lq}^{(3)}$ and $\mathcal{O}_{lq}^{(1)}$
operators.
Expanding the $SU(2)_L$ indices, we find that the SMEFT Lagrangian contains operators
of the form
\begin{equation}
  \mathcal L_{\rm SMEFT} \supset \frac{{\bf C}^{U\mu}_{ij}}{v^2} (\bar u^i_L \gamma_\mu u^j_L) (\bar \mu_L \gamma^\mu \mu_L)+ \frac{{\bf C}^{D\mu}_{ij}}{v^2} (\bar d^i_L \gamma_\mu d^j_L) (\bar \mu_L \gamma^\mu \mu_L)~,
  \label{eq:benchmark2}
\end{equation}
where $v \approx 246$~GeV is the Higgs vacuum expectation value and ${\bf C}^{U\mu}_{ij}$ and
${\bf C}^{D\mu}_{ij}$ represent matrices of Wilson coefficients.
In Eq.~(\ref{eq:benchmark2}), $i,j=1,2,3$ indicate quark flavour indices, and we have chosen
to focus on those operators that couple the quark fields exclusively to the second lepton family.

The operators highlighted in Eq.~(\ref{eq:benchmark2}) have received a lot of attention recently in the context of the LHCb anomalies reported in
rare $B$-meson decays~\cite{Aaij:2014ora, Aaij:2017vbb, Aaij:2019wad,Aaij:2021vac}. 
The reason is that the CKM-like flavour structure relates the $b \to s \mu^+ \mu^-$ decays to
the neutral-current Drell-Yan process at the LHC $p\, p \to \mu^+ \mu^-$~\cite{Greljo:2017vvb}.
The explicit models which successfully describe the LHCb anomalies, based on the $U(2)$ flavour symmetry and dominant dynamics with the third generation fermions~\cite{Barbieri:2011ci,Kagan:2009bn,Fuentes-Martin:2019mun,Faroughy:2020ina}, predict that the  flavour channel dominating EFT effects in the Drell-Yan production is $b \,\bar b \to  \mu^+ \mu^-$ (see~\cite{Greljo:2021xmg} for an explicit model example).
The direct $b \bar{s}$ production channel is suppressed by $V_{ts}$ and is therefore irrelevant. If the observed deviations in $R(K^{(*)})$ are due to new physics, in this class of models we generically expect $|{\bf C}^{D\mu}_{3 3}| \gtrsim 0.001$.

The ATLAS dimuon search reported in Ref.~\cite{ATLAS:2017wce} is
recast in Ref.~\cite{Greljo:2017vvb} to set the limit on this
scenario. 
In particular, the reported 95\% confidence level interval is
\begin{equation}
  \label{eq:scen2}
{\bf C}^{D\mu}_{33} \in \left[ -0.026, 0.021 \right]~\textrm{(ATLAS 13 TeV, 36.1 fb$^{-1}$~\cite{ATLAS:2017wce})}.
\end{equation} 

For this second benchmark scenario we will assume that, out of the operators listed in
Eq.~(\ref{eq:benchmark2}), only a single  Wilson coefficient is allowed to be non-zero.
Specifically, we allow ${\bf C}^{D\mu}_{3 3} \neq 0$, while setting to zero all the
coefficients of the other four-fermion operators.
This assumption implies that in this scenario the SMEFT Lagrangian is reduced to
\begin{equation}
  \label{eq:benchmark2simplified}
\mathcal L_{\rm SMEFT} = \mathcal L_{\rm SM} + \frac{{\bf C}^{D\mu}_{33}}{v^2} (\bar d^3_L \gamma_\mu d^3_L) (\bar \mu_L \gamma^\mu \mu_L)~.
\end{equation}
In contrast to the previous benchmark, now the electron channel is SM-like.
This feature provides a useful handle to separate PDF and EFT effects in the Drell-Yan process,
by using electron data to determine the former and muon data to constrain both.
Another difference with respect to the first benchmark
is that here the leading new physics effects arise at the dimension-six squared level,
since the interference of the operator in Eq.~(\ref{eq:benchmark2simplified}) with the SM is subleading \cite{Greljo:2017vvb}.
(Dijet production is another prominent process relevant for flavour physics~\cite{Bordone:2021cca} that enters into PDF fits. A detail study of the interplay is left for future work.)

To summarise, in this second benchmark scenario there is a single non-zero Wilson coefficient, ${\bf C}^{D\mu}_{33}$.
Therefore, the determination of the SMEFT PDFs from Drell-Yan data requires a simultaneous
fit of the PDFs together with the ${\bf C}^{D\mu}_{33}$ parameter.
One should also note that this
operator enters in the description of the DIS neutral-current structure functions via $\mu\,b $
scattering, though this contribution is highly suppressed due to the smallness
of the bottom PDFs and the low energy scale probed by DIS data.

\section{Experimental data, theory predictions, and fit settings}
\label{sec:data}

In Sect.~\ref{sec:dataset} we present the LHC experimental data  that will be used in the present analysis
for the simultaneous determination of the PDFs and
the EFT coefficients from high-mass Drell-Yan cross sections.
We then describe in Sect.~\ref{sec:theory} the corresponding theoretical calculations, both in the SM
and in the two SMEFT benchmark scenarios described in Sect.~\ref{sec:scenarios}.
In Sect.~\ref{sec:fitsettings} we  discuss the settings of the baseline SM PDF fit and assess the
specific impact of the Run I and Run II high-mass Drell-Yan data on PDFs.
Finally, in Sect.~\ref{eq:jointfits} we outline the fitting methodology adopted for
the determination of the PDFs in the SMEFT, along with their
simultaneous determination with the EFT Wilson coefficients.

\subsection{Experimental data}
\label{sec:dataset}

The present analysis is based on the DIS and DY measurements which were part of
the strangeness study of~\cite{Faura:2020oom}, which in turn was a variant of the
NNPDF3.1 global PDF determination~\cite{Ball:2017nwa}, extended with additional high-mass
DY cross sections.
The DIS structure functions include the same legacy HERA inclusive combination~\cite{Abramowicz:2015mha}
used in the DIS-only joint fit of PDF and EFT effects of~\cite{Carrazza:2019sec}.

No other datasets beyond DIS and DY are considered.
In particular, the inclusive jet and top quark production
measurements used in~\cite{Faura:2020oom} are excluded from the
present analysis.
%
The rationale behind this choice is the following.
SMEFT at dimension-6 level introduces 2499 independent parameters,
many of which contribute to the processes used to extract the parton
distribution functions. The full PDF fit in the SMEFT (with the
consistent power counting in the inverse powers of the new physics
scale) is the ultimate future goal of this line of research. Before
that, we are forced to make assumptions about the subset of operators
and processes involved. The restricted choice of DIS and DY is
motivated by the idea that other datasets, such as inclusive jet,
could potentially receive corrections from other SMEFT operators,
e.g. four-quark operators while being insensitive to the semi-leptonic
operators. Including all datasets to effectively determine PDF, while
considering one or two operators able to impact a subset of processes,
would misrepresent the realistic case.

For the purposes of our study, the DY data can be classified into low-mass,
on-shell, and high-mass datasets.
Table~\ref{tab:data-low-mass} summarises
the low-mass and on-shell datasets, where in 
each case we indicate the experiment, the centre-of-mass energy $\sqrt{s}$,
the publication reference, the physical observable, and the number of data points.
The only difference as compared to~\cite{Faura:2020oom} is the removal of the $W\to e\nu$
asymmetry measurements from D0~\cite{D0:2014kma}, which were found to be inconsistent
with the rest of the Drell-Yan data.

\begin{table}[t]
  \begin{center}
  \renewcommand{\arraystretch}{1.20}
\small
\begin{tabular}{ c c c c c }
 \toprule
 Exp.   & $\sqrt{s}$ (TeV) & Ref. & Observable & $n_{\rm dat}$ \\
 \midrule
 E886   & 0.8   & \cite{Towell:2001nh}            & $d\sigma^d_{\rm DY}/d\sigma^p_{\rm DY}$ & 15 \\
 E886   & 0.8   & \cite{Webb:2003ps, Webb:2003bj} & $d\sigma^p_{\rm DY}/(dy\,dm_{\ell\ell})$     & 89 \\
 \midrule
 E605   & 0.04  & \cite{Moreno:1990sf}            & $\sigma^p_{\rm DY}/(dx_F\,dm_{\ell\ell})$    & 85 \\
 \midrule
 CDF    & 1.96  & \cite{Aaltonen:2010zza}         & $d\sigma_Z/dy_Z$                     & 29 \\
 \midrule
 D0     & 1.96  & \cite{Abazov:2007jy}            & $d\sigma_Z/dy_Z$                     & 28 \\
 D0     & 1.96  & \cite{Abazov:2013rja}		  & $d\sigma_{W\to \mu \nu}/d\eta_\mu$ asy.  & 9 \\
 \midrule
 ATLAS  & 7     & \cite{Aad:2011dm}               & $d\sigma_W/d\eta_l,d\sigma_Z/dy_z$     & 30 \\
 ATLAS  & 7     & \cite{Aad:2014qja}              & $d\sigma_{Z\rightarrow e^+e^-}/dm_{e^+e^-}$  & 6 \\
 ATLAS  & 7     & \cite{Aaboud:2016btc}           & $d\sigma_W/d\eta_l,d\sigma_Z/dy_z$     & 61 \\
 ATLAS  & 7 	& \cite{Aad:2014xca}	          & $d\sigma_{W+c}/dy_c$	          & 22 \\
 ATLAS  & 8     & \cite{Aad:2015auj}              & $d\sigma_Z/dp_T$                       & 82 \\
 ATLAS  & 8 	& \cite{Aaboud:2017soa}		  & $d\sigma_{W+j}/dp_T$			  & 32 \\
 \midrule
 CMS 	& 7	& \cite{Chatrchyan:2012xt}	  & $d\sigma_{W\to l\nu}/d\eta_\ell$ asy.	          & 22 \\
 CMS 	& 7 	& \cite{Chatrchyan:2013uja}	  & $d\sigma_{W+c}/dy_c$		  & 5 \\
 CMS 	& 7 	& \cite{Chatrchyan:2013uja}	  & $d\sigma_{W^{+}+c}/d\sigma_{W^{-}+c}$	  & 5 \\
 CMS    & 8     & \cite{Khachatryan:2015oaa}      & $d\sigma_Z/dp_T$                       & 28 \\
 CMS    & 8 	& \cite{Khachatryan:2016pev}	  & $d\sigma_{W\rightarrow \mu\nu}/d\eta_\mu$  & 22 \\
 CMS 	& 13 	& \cite{Sirunyan:2018hde}	  & $d\sigma_{W+c}/dy_c$			              & 5 \\
 \midrule
 LHCb   & 7     & \cite{Aaij:2012vn}              & $d\sigma_{Z\rightarrow\mu^+\mu^-}/dy_{\mu^+\mu^-}$   & 9 \\
 LHCb   & 7     & \cite{Aaij:2015gna}             & $d\sigma_{W,Z}/d\eta$ & 29 \\
 LHCb   & 8     & \cite{Aaij:2012mda}             & $d\sigma_{Z \rightarrow e^+e^-}/dy_{e^+e^-}$         & 17 \\
 LHCb   & 8     & \cite{Aaij:2015zlq}             & $d\sigma_{W,Z}/d\eta$           & 30 \\
 \midrule
 {\bf Total}  &       &                                  &                                       & {\bf 659} \\
 \bottomrule
\end{tabular}
\end{center}
\caption{\small The low-mass and on-shell Drell-Yan datasets used in the present study.
  For each dataset we indicate the experiment, the centre-of-mass energy $\sqrt{s}$,
  the publication reference, the physical observable, and the number of data points
}
\label{tab:data-low-mass}
\end{table}


In Table~\ref{tab:data-high-mass} we provide the same
information as in Table~\ref{tab:data-low-mass} but for the neutral-current high-mass Drell-Yan datasets.
In Table~\ref{tab:data-high-mass} we also indicate the final state,
whether the distribution is 1D or 2D (thus differential only in the
lepton invariant mass or differential in the lepton invariant mass and
rapidity), the integrated luminosity $\mathcal{L}$,
and the values of the dilepton invariant mass $m_{\ell\ell}$ for the most energetic bin.
  We note that while the ATLAS and CMS measurements at $\sqrt{s}=7$ TeV~\cite{Aad:2013iua,Chatrchyan:2013tia}
  were already part of the  strangeness study of~\cite{Faura:2020oom},
  the corresponding 8 TeV and 13 TeV measurements from~\cite{Aad:2016zzw,CMS:2014jea,Sirunyan:2018owv}
were not and are being considered for the first time in this analysis.
For those datasets where data are available in terms of both Born
and dressed leptons, the ATLAS 7 TeV analysis being an example thereof, we use
the Born data so that it is not necessary to supplement our fixed-order predictions with
final-state QED radiation corrections. The CMS 13 TeV data on the
other hand are only provided in terms of dressed leptons. 
In total, there are either 270 or 313 data points in this high-mass category, depending on whether the
13 TeV CMS data are included in the combined channel or in the separate electron and muon channels.


\begin{table}[t]
  \renewcommand{\arraystretch}{1.20}
\begin{center}
\small
\begin{tabular}{cccccccc}
 \toprule
 Exp.   & $\sqrt{s}$ (TeV) & Ref. & $\mathcal{L}$ (fb$^{-1}$) & Channel & 1D/2D & $n_{\rm dat}$ & $m_{\ell\ell}^{\rm max}$ (TeV) \\
 \midrule
 ATLAS  & 7    & \cite{Aad:2013iua}         & 4.9   & $e^{-}e^{+}$									& 1D & 13 & [1.0, 1.5] \\
 ATLAS~{\bf (*)}  & 8    & \cite{Aad:2016zzw}         & 20.3  & $\ell^{-}\ell^{+}$									& 2D & 46 & [0.5, 1.5] \\
 \midrule
 CMS    & 7    & \cite{Chatrchyan:2013tia}  & 9.3   & $\mu^{-}\mu^{+}$								& 2D & 127 & [0.2, 1.5] \\
 CMS~{\bf (*)}    & 8    & \cite{CMS:2014jea}         & 19.7  & $\ell^{-}\ell^{+}$									& 1D & 41 & [1.5, 2.0] \\
  \midrule
 \multirow{2}{*}{ CMS~{\bf (*)}}    & \multirow{2}{*}{13}   & \multirow{2}{*}{\cite{Sirunyan:2018owv}}    & \multirow{2}{*}{5.1}   & $e^{-}e^{+}$, $\mu^{-}\mu^{+}$	& \multirow{2}{*}{1D} & 43, 43 & \multirow{2}{*}{[1.5, 3.0]} \\
 &  &   &    &   $\ell^{-}\ell^{+}$  &    & 43    &   \\
 \midrule
 {\bf Total}  &  &     &     &     &  & {\bf 270~(313)} \\
 \bottomrule
\end{tabular}
\end{center}
\caption{\small Same as Table~\ref{tab:data-low-mass} for the neutral-current high-mass Drell-Yan datasets
  considered in this work.
  We also indicate the final-state, whether the distribution is 1D (which are differential in 
the invariant mass, $m_{\ell\ell}$, of the final-state leptons) or 2D (which are differential in both the invariant mass of the 
leptons, $m_{\ell\ell}$, and in their rapidity, $y_{\ell\ell}$), and the values
  of $m_{\ell\ell}$ for the most energetic bin.
  Datasets indicated with {\bf (*)} are used for the first time in this analysis
  in comparison with~\cite{Faura:2020oom}.
}
\label{tab:data-high-mass}
\end{table}


From Table~\ref{tab:data-high-mass} one can observe that, with the exception of the CMS 13 TeV data,
only one specific leptonic final state is available to be used in the fit.
For the CMS 13 TeV measurement instead, one can select between the combined channel or the individual
electron and muon final states, which are statistically independent.
The separate use of the electron and muon channels is potentially beneficial when considering BSM effects that are
not lepton-flavour universal.
For example, in
benchmark scenario II described in Sect.~\ref{sec:scenarios},
the theoretical predictions for the DY electron data would be those of the SM while
those of the muon data should include EFT corrections.
On the other
hand in the (flavour-universal) $\hat{W}$ and $\hat{Y}$ scenario,
it is more convenient to include the data
from the combined channel, which displays reduced systematic
uncertainties.

\subsection{Theoretical predictions}
\label{sec:theory}

We now discuss the settings of the theoretical calculations, both in
the SM and in the SMEFT. 
Appendix~\ref{sec:benchmarking} contains further information regarding the computation and
benchmarking of the SMEFT corrections for both the DIS structure
functions, for which the effect in both scenarios is negligible, and
the DY cross sections, for which the impact of SMEFT corrections is
much more sizeable. 

\paragraph{SM cross sections.}
The SM cross sections are computed at next-to-next-to-leading order
(NNLO) in QCD and include next-to-leading order (NLO) EW corrections,
the latter being especially significant in the high-mass region relevant for this study.
In particular, the DIS reduced cross sections (combinations of structure functions)
are evaluated at NNLO in the FONLL-C general-mass 
variable flavour number scheme~\cite{Forte_2010} with {\tt APFEL}~\cite{Bertone_2014}
interfaced to {\tt APFELgrid}~\cite{Bertone:2016lga}.
The Drell-Yan differential distributions are computed using {\tt
MCFM}~\cite{Campbell:2019dru} and \amc~\cite{Frederix:2018nkq} interfaced to
{\tt APPLgrid}~\cite{Carli:2010rw} and {\tt APFELgrid} to generate fast NLO
interpolation tables which are then supplemented by bin-by-bin $K$-factors to
account for the NNLO QCD and NLO EW corrections.
These $K$-factors are defined as
\begin{equation}
 d\sigma_{pp}= \left(d\widehat\sigma_{ij}\big|_{\rm NLO\, QCD}\otimes {\cal L}^{\rm NNLO}_{ij}\right)\times K_{\rm QCD} \times   K_{\rm EW} \, ,
\end{equation}
where $\otimes$ represents the standard convolution product, $d\sigma_{pp}$ ($d\widehat\sigma_{ij}$) is the short-hand
notation for the bin-by-bin hadronic cross section (partonic cross
section for partons $i,j$) differential in $m_{\ell\ell}$ (in case of neutral-current (NC)
Drell-Yan) or $m_T$ (in case of charged-current (CC) Drell-Yan) and the partonic luminosities ${\cal L}_{ ij}$ are
defined as
\begin{equation}
\mathcal{L}_{ij}(\tau,m) = \int_{\tau}^1 \frac{d x}{x}~f_i (x,m) f_j (\tau/x,m) ~,
\end{equation}
where $m=m_{\ell\ell}$ in the NC case and $m=m_T$ in the CC case and
are evaluated at NNLO. The QCD and EW $K$-factors are defined as
\begin{align}
  K_{\rm QCD}&= \left( {\cal L}_{
             ij}^{\rm NNLO} \otimes d\widehat\sigma_{ij}\big|_{\rm NNLO\, QCD}\right)
             \big/\left({\cal L}_{
             ij}^{\rm NNLO} \otimes d\widehat\sigma_{ij}\big|_{\rm NLO\, QCD}\right) \, , \label{def:kqcd}\\
K_{\rm EW}&= \left({\cal L}_{ ij}^{\rm NNLO}\otimes
            d\widehat\sigma_{ij}\big|_{\rm NLO~QCD+EW}\right)
            \big/ \left({\cal L}_{ ij}^{\rm NNLO} \otimes d\widehat\sigma_{ij}\big|_{\rm NLO\, QCD} \right) \, , \label{def:kew}
\end{align}
The NNLO QCD $K$-factors have been computed using either {\tt
  MATRIX}~\cite{Grazzini_2018} or {\tt FEWZ}~\cite{Li:2012wna}
and cross-checked with the analytic computations of~\cite{Duhr:2020seh,Duhr:2020sdp}.
The NLO EW $K$-factors have been evaluated with \amc~\cite{Frederix:2018nkq}.
Eq.~(\ref{def:kew}) accounts also for photon-initiated contributions
(using the {\tt NNPDF3.1QED} PDF set~\cite{Bertone:2017bme})
and final-state radiation effects, except when the latter
has already been subtracted in the corresponding experimental analysis.

\begin{figure}
  \centering
  \includegraphics[width=0.49\textwidth]{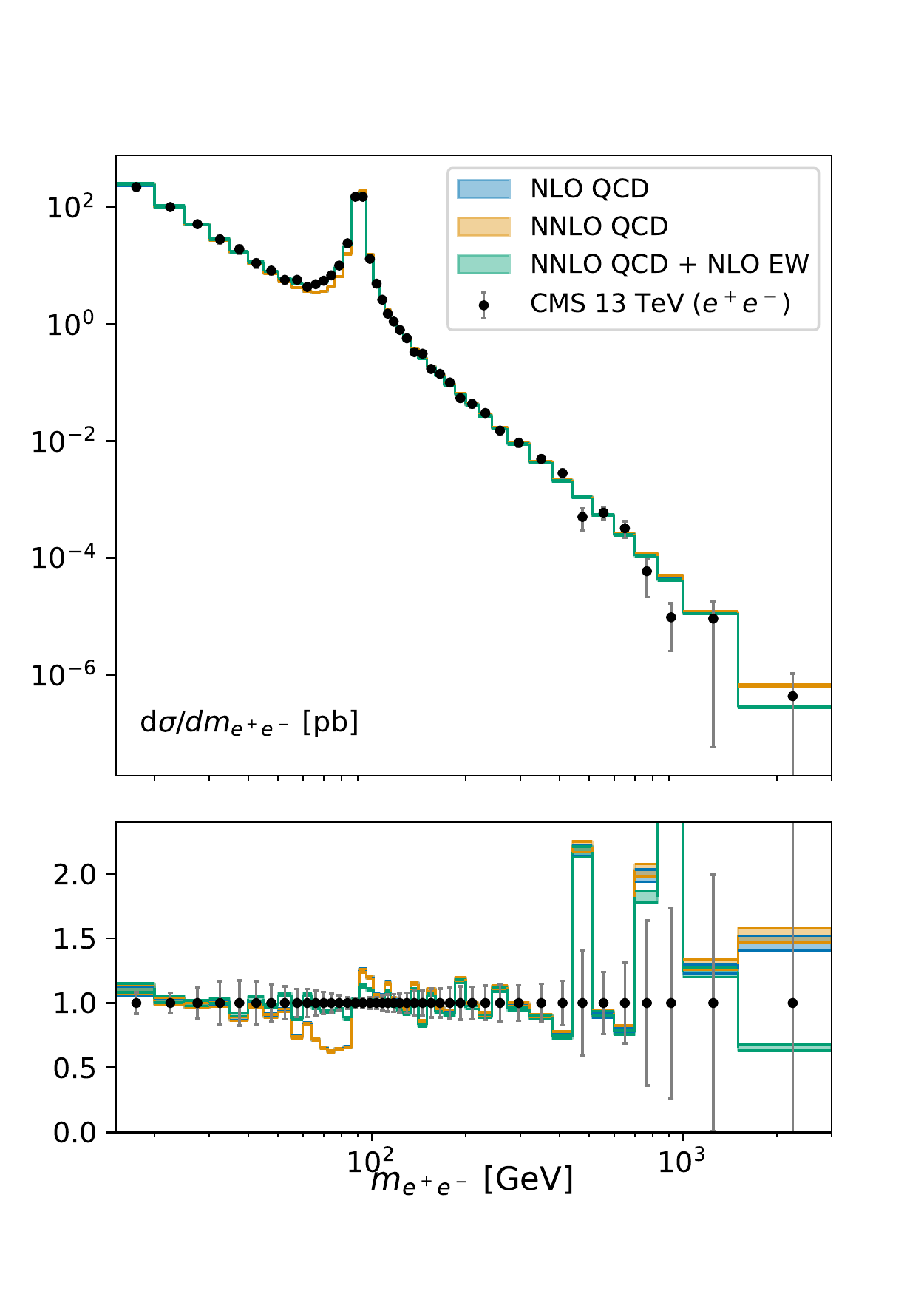}
\includegraphics[width=0.49\textwidth]{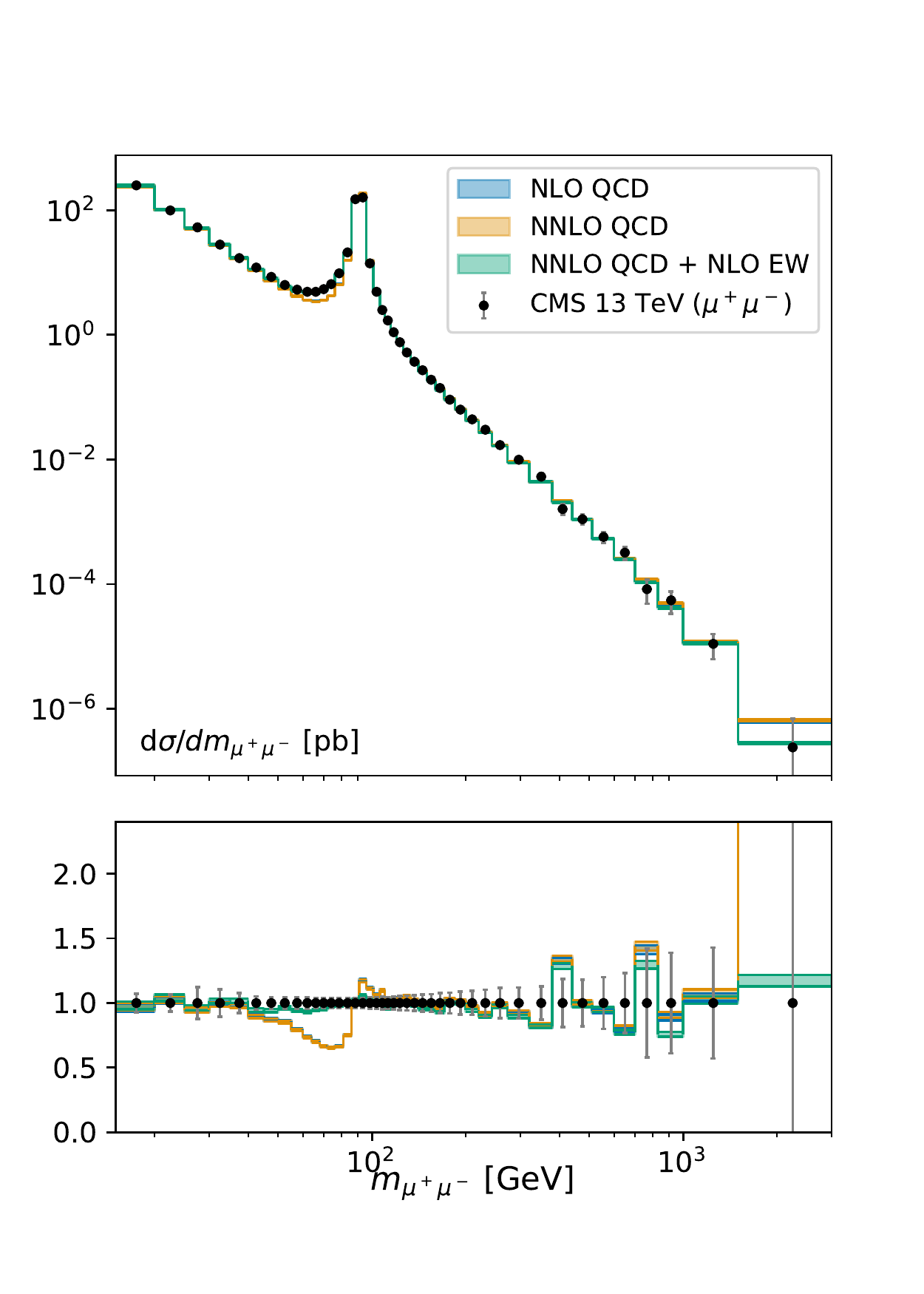}
\caption{\small Comparison of the CMS Drell-Yan
  13 TeV data with the corresponding theoretical calculations at different
  perturbative orders as a function of the
  dilepton invariant mass $m_{\ell\ell}$ in the dielectron (left)
  and dimuon (right panel) final states.
  The bottom panels display the ratio of the theory calculations
  to the central value of the experimental data.
  We display the sum in quadrature of the experimental uncertainties, and the error
  band in the theory predictions correspond to the one-sigma PDF uncertainties.
  \label{fig:dysm}}
\end{figure}

Fig.~\ref{fig:dysm} displays a comparison between the CMS Drell-Yan distributions
at 13 TeV and the corresponding theoretical predictions as a function of the
dilepton invariant mass $m_{\ell\ell}$, separately for the dielectron
and dimuon final states.
The theory calculations are presented at NLO QCD, NNLO QCD, and NNLO QCD combined with
NLO EW corrections, in all
cases with {\tt NNPDF3.1QED\_nnlo\_as\_0118} as input PDF set, to illustrate the
effect of the $K$-factors of Eq.~\eqref{def:kqcd} and
\eqref{def:kew}.
The CMS data are provided
in terms of dressed leptons, and hence final state radiation (FSR) QED effects must be included in the electroweak
corrections.
Accounting for these effects is essential to improve the agreement
between theory and data in the region below the $Z$-mass peak.
NLO electroweak corrections are also important in the high-energy tail in $m_{\ell\ell}$,
where they are driven by the interplay between (negative) virtual EW effects
and (positive) photon-initiated contributions.

A quantitative assessment of the agreement between theoretical predictions and experimental 
data for the high-mass DY datasets listed in Table~\ref{tab:data-high-mass}
is presented in Table~\ref{tab:dysm}, which collects the  values of the $\chi^2$ per data point
evaluated using the full information on correlated systematics
provided by the experimental covariance
matrix
\begin{equation}
  \label{eq:chi2}
\chi^2=\frac{1}{n_{\rm dat}}\sum_{i,j=1}^{n_{\rm
    dat}}(D_i-T_i)\,({\rm cov}^{-1})_{ij}\, (D_j-T_j),
\end{equation}
where $T_i$ are the theoretical predictions, $D_i$ the central value
of the experimental data and where the multiplicative uncertainties in
the experimental covariance matrix $({\rm cov}_{ij})$ are treated as explained in~\cite{Ball:2009qv,Ball:2012wy}.
One can observe how in general the NNLO QCD corrections are relatively
small and that the NLO electroweak
ones can be significant, especially for observables presented 
in terms of dressed leptons (such as the CMS 13 TeV ones) and are required to achieve a good description of the Drell-Yan
data in the whole kinematical range available.
Note that the input PDF sets used for these calculations include 
only a subset of these Drell-Yan measurements, in particular only the 7 TeV measurements, 
for which the data-theory agreement is comparable to the one observed in~\cite{Ball:2017nwa}.

The data-theory agreement before including the 
8 TeV and 13 TeV data in the PDF fit is generally good, once EW corrections 
are included, with the exception of the CMS 13 TeV data in the 
$e^+e^-$ channel, for which the $\chi^2$ per data point remains above 2.
As can be
observed in Fig.~\ref{fig:dysm}, the dielectron invariant mass distribution
in this channel presents dips at about
500 GeV and 900 GeV which are not present in the $\mu^+\mu^-$ channel.
These dips are the origin of this worse data-theory agreement,
which is partially reduced once the dataset is included in the fit
(see Sect.~\ref{sec:fitsettings}). 
We have verified that excluding this dataset from the fit does not 
change the results of the analysis, and therefore decided to keep it.
Further experimental analysis based on the full Runs II and III
datasets will tell whether the dips in the distributions in the
electron invariant mass will stay. 

\begin{table}[t]
  \renewcommand{\arraystretch}{1.40}
  \small
  \begin{center}
\begin{tabular}{c c c| c c c}
\toprule       
\multirow{3}{*}{Dataset} & \multirow{3}{*}{Final state} &  \multirow{3}{*}{$n_{\rm dat}$}    &  \multicolumn{3}{c}{$\chi^2/n_{\rm dat}$}   \\
      &  &      &  \multirow{2}{*}{NLO QCD}
           & \multirow{2}{*}{NNLO QCD}
& NNLO QCD \\
&  &      &  
           & 
& + NLO EW\\
\midrule
ATLAS 7 TeV  & $e^+e^-$            & 13  & 1.45  & 1.77  & 1.73 \\
ATLAS 8 TeV  & $\ell^+\ell^-$      & 46  & 1.67  & -     & 1.20 \\
\midrule
CMS 7 TeV    & $\mu^+\mu^-$        & 127 & 3.40  & 1.27  & 1.54 \\
CMS 8 TeV    & $\ell^+\ell^-$      & 41  & 2.22  & 2.21  & 0.70 \\
\midrule
CMS  13 TeV  & $\ell^+\ell^-$      & 43  & 18.7  & 19.7  & 1.91 \\
CMS  13 TeV  & $e^+e^-$            & 43  & 9.16  & 9.45  & 2.32 \\
CMS 13 TeV   & $\mu^+\mu^+$        & 43  & 15.7  & 15.8  & 0.81 \\
\bottomrule
\end{tabular}
\end{center}
  \caption{\label{tab:dysm}\small The values of the $\chi^2$ per data point
    evaluated for the  high-mass DY datasets listed in Table~\ref{tab:data-high-mass},
    using theoretical predictions computed at different perturbative accuracy.
    The PDF sets used here are {\tt NNPDF31\_nlo\_as\_0118}, {\tt
    NNPDF31\_nnlo\_as\_0118} and {\tt NNPDF31\_nnlo\_as\_0118\_luxqed} for
    the NLO QCD, NNLO QCD and NNLO QCD + NLO EW predictions respectively.
    For CMS 13 TeV, where different final states are available, we indicate the
    $\chi^2$ values for each of them.
    For the ATLAS 8 TeV data, we only evaluated the combined
    NNLO QCD + NLO EW correction, and hence the  pure NNLO QCD result is not given.
  }
\end{table}

\paragraph{SMEFT cross sections.}
In this work, we augment the SM calculations of the high-$Q^2$ DIS
reduced cross sections discussed in Ref.~\cite{Carrazza:2019sec} and
the high-mass Drell-Yan cross sections listed in
Table~\ref{tab:data-high-mass} with the effects of dimension-six SMEFT operators following the two benchmark scenarios presented
in Sect.~\ref{sec:scenarios}.
These corrections are negligible for dilepton invariant masses of $m_{\ell\ell}\le 200$
GeV and for DIS structure functions with $Q\le (120)$
GeV, and hence there can safely adopt the SM calculations.
In a similar manner as for higher-order QCD and EW corrections, we can
define correction factors that encapsulate  the linear and quadratic modifications induced
by the dimension-six SMEFT operators.
Adopting an operator normalisation such that
\be
\label{eq:lagrangian}
\mathcal{L}_{\rm SMEFT} = \mathcal{L}_{\rm SM} + \sum_{n=1}^{n_{\rm op}} \frac{c_n}{v^2} \,\mathcal{O}_n \, ,
\ee
with $n_{\rm op}$ indicating the number of operators that contribute to a given
benchmark scenario and $c_n$ being the (dimensionless) Wilson coefficient associated to $\mathcal{O}_n$,
the linear EFT corrections can be parametrised as
\begin{equation}
  \label{eq:mult_k_fac_app1}
   R_{\rm SMEFT}^{(n)} \equiv \displaystyle \left( {\cal L}_{ ij}^{\rm NNLO} \otimes d\widehat{\sigma}_{ij,{\rm SMEFT}}^{(n)}\right)
 \big/ \left( {\cal L}_{ ij}^{\rm NNLO} \otimes d\widehat{\sigma}_{ij,{\rm SM}} \right) \, , \quad
 n=1\,\ldots, n_{\rm op} \, ,
\end{equation}
with ${\cal L}_{ ij}^{\rm NNLO}$ being the usual partonic luminosity evaluated at NNLO QCD, 
$d\widehat{\sigma}_{ij,{\rm SM}}$
the bin-by-bin partonic SM cross section, and $d\widehat{\sigma}_{ij,{\rm SMEFT}}^{(n)}$
the corresponding partonic cross section associated to the interference between 
$\mathcal{O}_n$ and the SM amplitude $\mathcal{A}_{\rm SM}$ when setting $c_n = 1$.
Likewise, the ratio encapsulating the quadratic effects is defined as
\begin{equation}
  \label{eq:mult_k_fac_app2}
  R_{\rm SMEFT}^{(n,m)} \equiv \displaystyle \left( {\cal L}_{ ij}^{\rm NNLO} \otimes d\widehat{\sigma}_{ij,{\rm SMEFT}}^{(n,m)}\right)
  \big/ \left( {\cal L}_{ ij}^{\rm NNLO} \otimes d\widehat{\sigma}_{ij,{\rm SM}} \right) \, , \quad
 n,m=1\,\ldots, n_{\rm op} \, ,
\end{equation}
with the bin-by-bin partonic cross section
$d\widehat{\sigma}_{ij,{\rm SMEFT}}^{(n,m)}$ now being evaluated from the squared amplitude $\mathcal{A}_n\mathcal{A}_m$
associated to the operators $\mathcal{O}_n$ and $\mathcal{O}_m$ when $c_n = c_m = 1$.
The partonic cross sections in these ratios are computed at LO.
In terms of  Eqns.~(\ref{eq:mult_k_fac_app1}) and~(\ref{eq:mult_k_fac_app2}), we can define the EFT $K$-factors as
\bea \label{eq:theory_k_fac}
  K_{\rm EFT} &=& 1+\sum_{n=1}^{n_{\rm op}} c_n R_{\rm SMEFT}^{(n)}
  +\sum_{n,m=1}^{n_{\rm op}} c_n c_m R_{\rm SMEFT}^{(n,m)} \, ,
  \eea
  which allow us to express a general Drell-Yan or DIS cross sections accounting for
  the dimension-six operators in Eq.~(\ref{eq:lagrangian}) as
\bea
  \label{eq:theory_k_fac_app3}
  d\sigma_{\rm SMEFT} &=& d\sigma_{\rm SM}
  \times  K_{\rm EFT} \, 
\eea
where the $ d\sigma_{\rm SM}$ is the state-of-the-art SM prediction
including NNLO QCD and NLO EW corrections.
In this approach, the SMEFT predictions inherit factorisable higher-order radiative correction~\cite{Greljo:2017vvb,Ricci:2020xre}. 
The SMEFT $K$-factors in Eq.~(\ref{eq:theory_k_fac}) are
precomputed before the fit using a reference SM PDF set and then kept fixed.
The effect of varying the input NNLO PDF in
Eqns.~(\ref{eq:mult_k_fac_app1}) and~(\ref{eq:mult_k_fac_app2})
is quantitatively assessed in Appendix~\ref{sec:benchmarking} and it
is found to be at the permil level in Scenario I and slightly more significant but
still at most at the percent level in Scenario II.
As a result, this effect will be neglected in the following.
Further details about the implementation and validation of these $K$-factors
can be found in Appendix~\ref{sec:benchmarking}.

Fig.~\ref{fig:dysmeft} illustrates the size of the EFT corrections in
benchmark scenario I from Sect.~\ref{sec:scenarios} by comparing
$(K_{\rm EFT}$$-$$1)$
with the relative experimental
  uncertainties  for the ATLAS 7 TeV, CMS 8 TeV,
  and CMS 13 TeV Drell-Yan $m_{\ell\ell}$ distributions.
  We provide results for two representative points in the $(\hat{W}$, $\hat{Y})$
  parameter space, namely $(\hat{W}$, $\hat{Y})=(10^{-3},0)$ and $(0,10^{-3})$.
  One can observe how for these values of $(\hat{W}$, $\hat{Y})$, 
and particularly for the ATLAS 8 TeV data, the SMEFT corrections
  to the Drell-Yan cross sections become comparable with the experimental uncertainties, 
  increasing steadily with $m_{\ell\ell}$.
  %

\begin{figure}[t]
\centering
\includegraphics[width=0.49\textwidth]{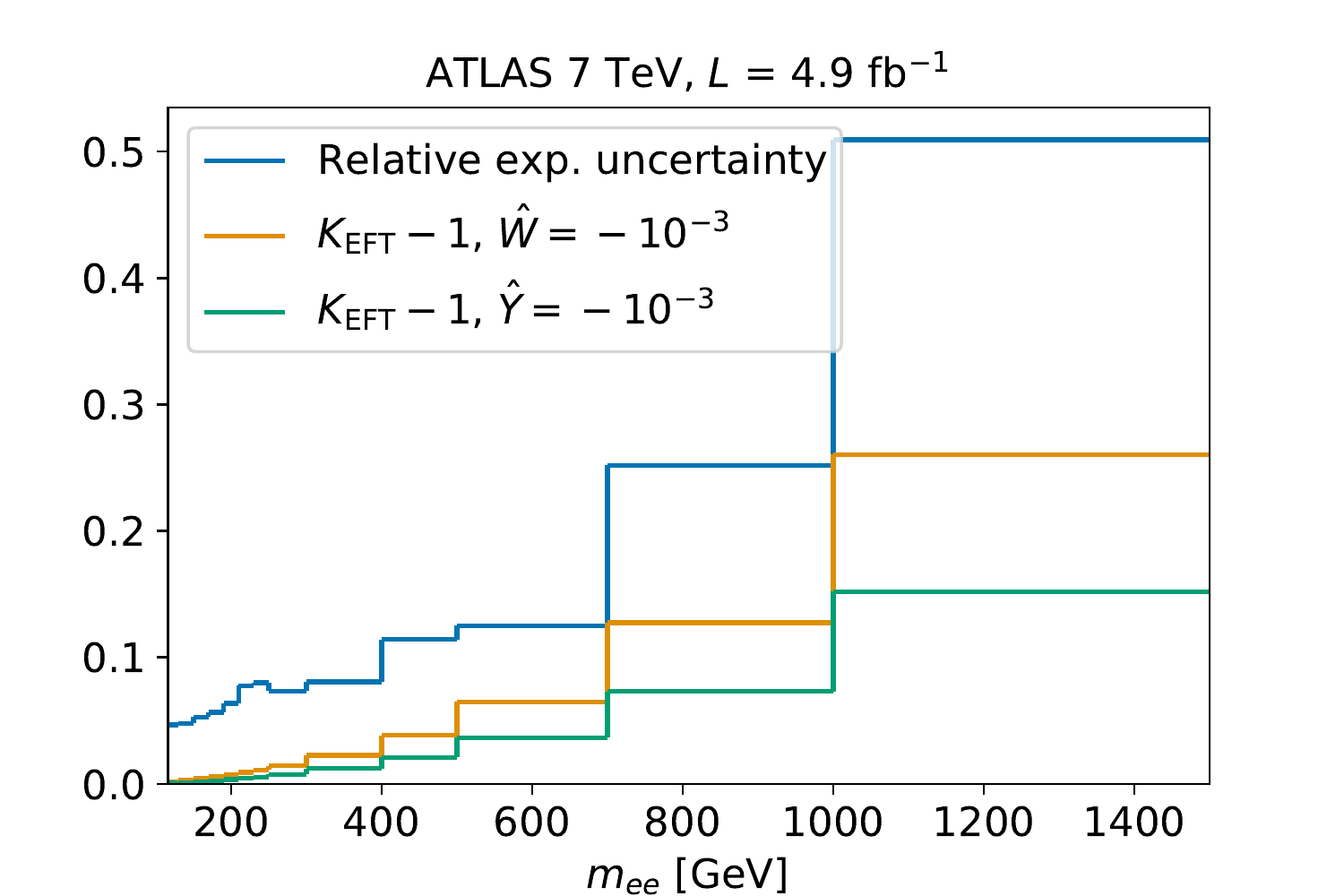}
\includegraphics[width=0.49\textwidth]{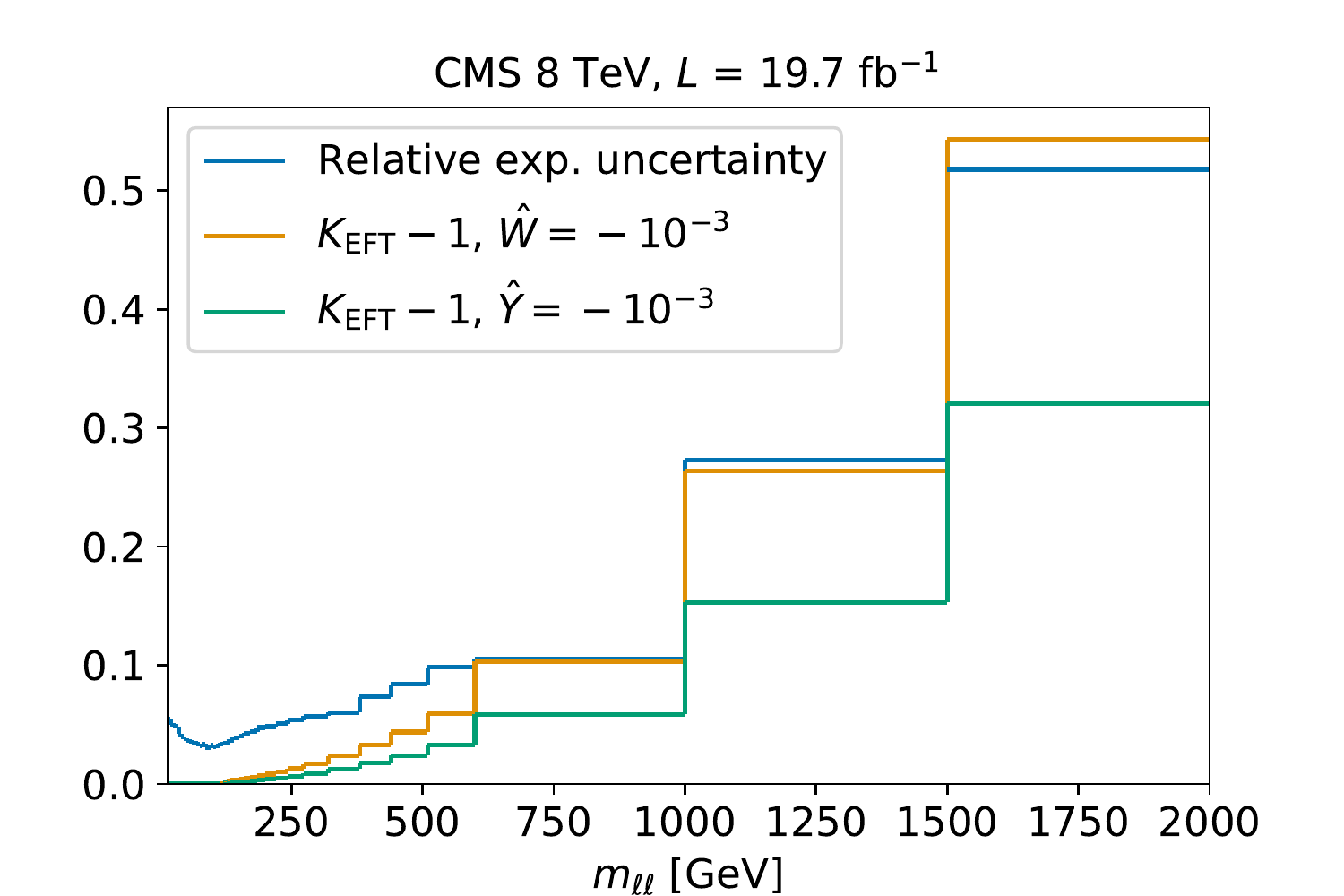}\\
\includegraphics[width=0.49\textwidth]{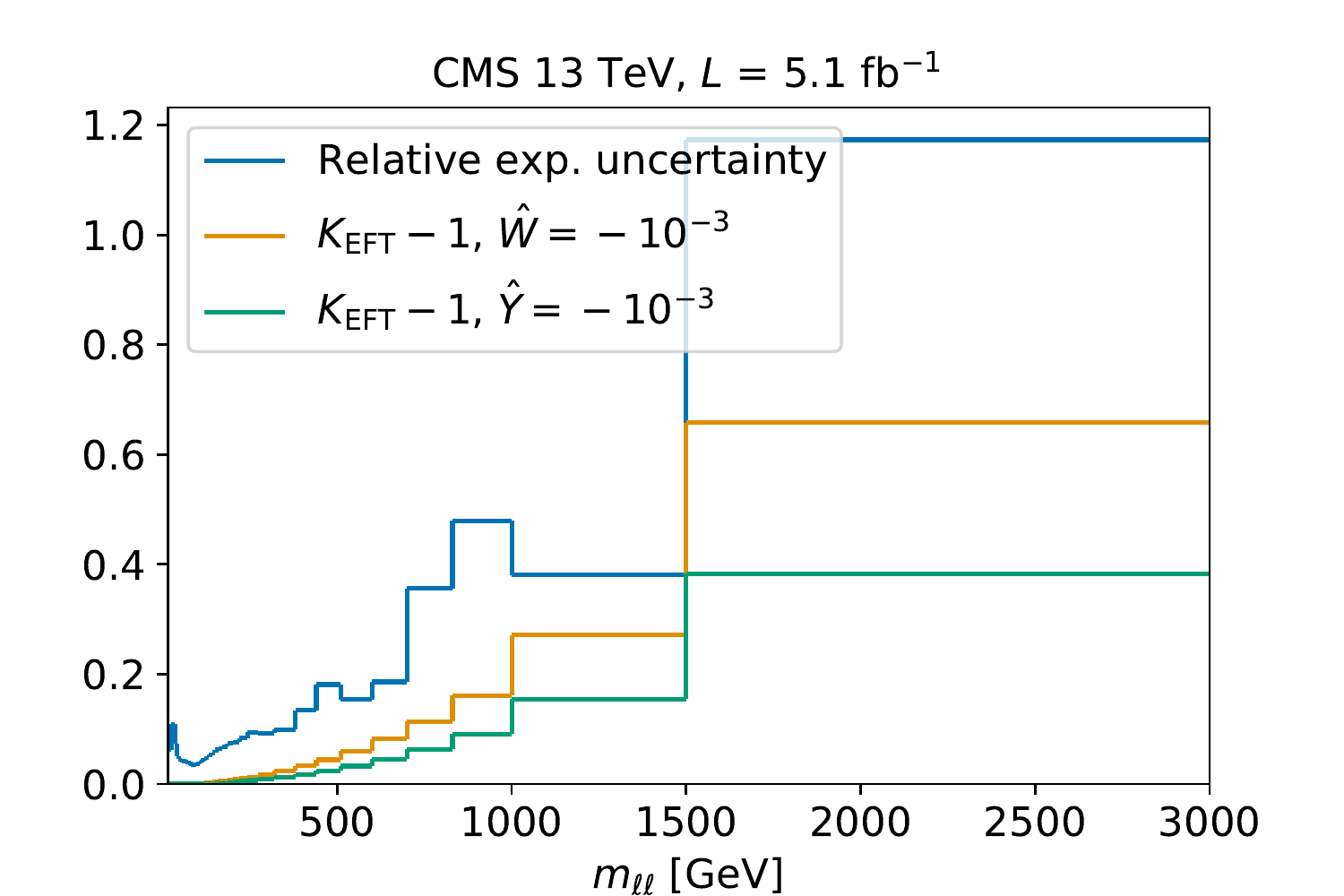}
\caption{\small Comparison between the (relative) experimental
  uncertainties and the corresponding EFT corrections,
$K_{\rm EFT}(\hat{W},\hat{Y})$$-$$1$ in Eq.~(\ref{eq:theory_k_fac}),
  for the ATLAS 7 TeV, CMS 8 TeV,
  and CMS 13 TeV Drell-Yan $m_{\ell\ell}$ distributions,
  for two representative values of  $\hat{W}$ and $\hat{Y}$.
  \label{fig:dysmeft}}
\end{figure}


\subsection{Baseline SM PDFs}
\label{sec:fitsettings}

We now describe the fit settings used to assemble the PDF set that
represents the baseline SM PDFs in this work.
In other words,
the results that we present in this section correspond to PDFs extracted from the experimental
data using the SM theoretical predictions, and then in the next section we will asses how these
PDFs are modified once EFT corrections to the DIS and DY cross
sections are accounted for in the fit of the PDFs.

The settings for this baseline SM PDF fit are the same as those used in the strangeness study
of~\cite{Faura:2020oom}, itself a variant of NNPDF3.1~\cite{Ball:2017nwa}.
As described in Sect.~\ref{sec:dataset}, in this work we consider only DIS
and Drell-Yan datasets, with the latter augmented as compared to~\cite{Faura:2020oom} with the new high-mass
measurements indicated in Table~\ref{tab:data-high-mass}.
A detailed comparison between this baseline SM PDF 
and the {\tt NNPDF3.1\_str} set from~\cite{Faura:2020oom} 
is provided in Appendix~\ref{app:pdfs},
while Table~\ref{tab:chi2-baseline} in App.~\ref{app:fit_quality} details the breakdown of the $\chi^2$
for all the datasets that enter the fit.
In general, the fit quality of the baseline SM PDF set
is similar to that of the global fit of~\cite{Faura:2020oom},
although the description of the CMS 13 TeV 
invariant mass distribution in the combined electron and muon 
channels remains sub-optimal.

\begin{figure}[t]
  \centering
  \includegraphics[width=0.32\textwidth]{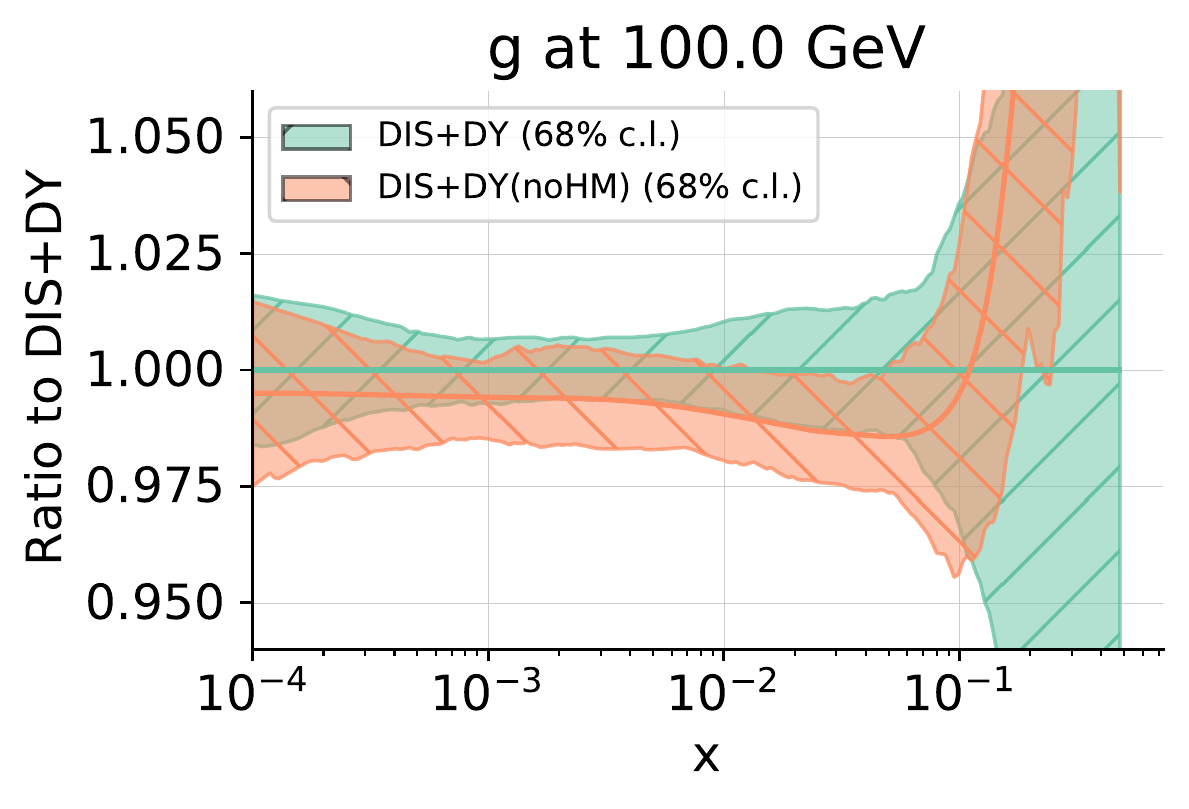}
  \includegraphics[width=0.32\textwidth]{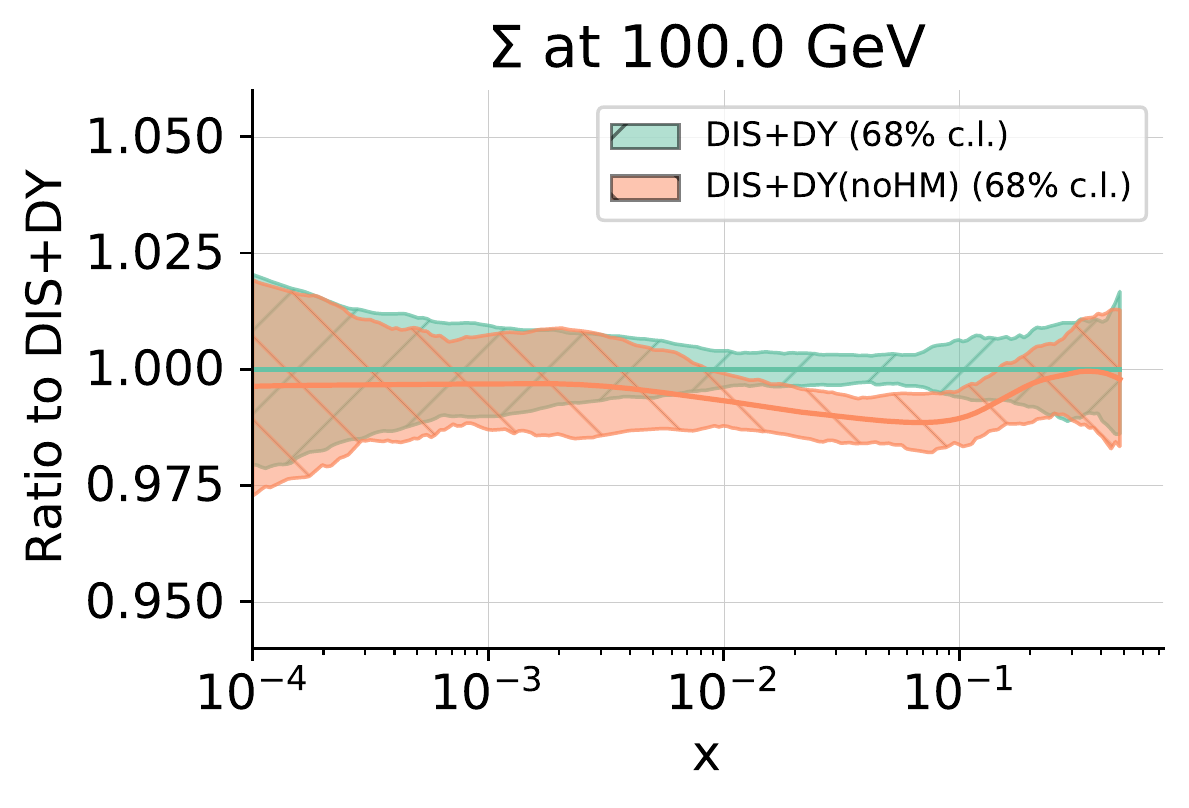}
  \includegraphics[width=0.32\textwidth]{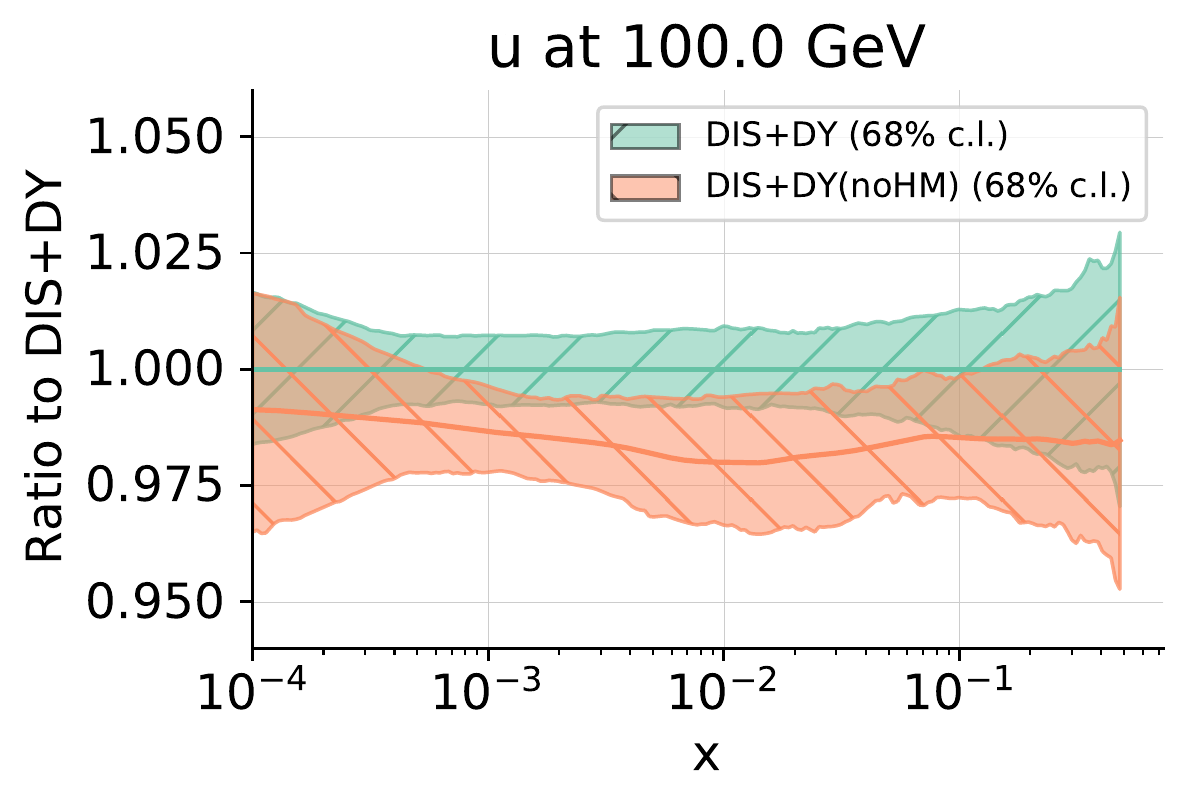}
  \includegraphics[width=0.32\textwidth]{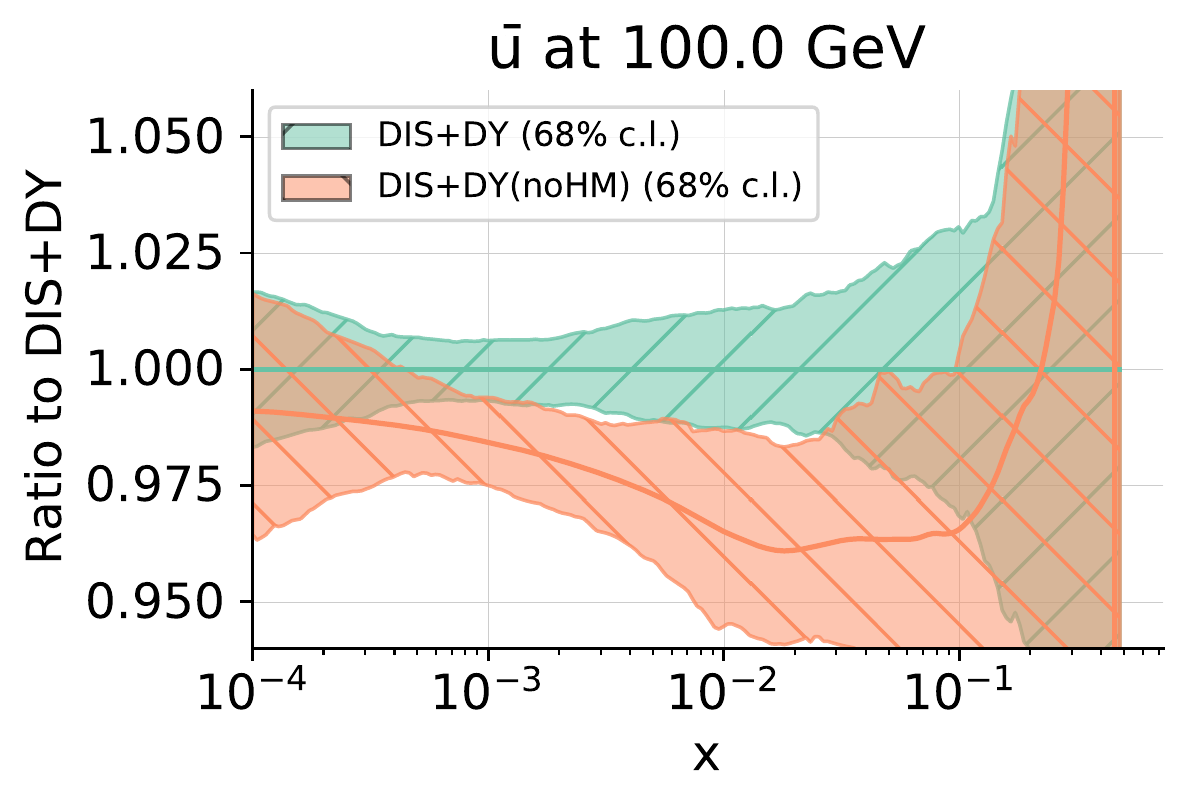}
  \includegraphics[width=0.32\textwidth]{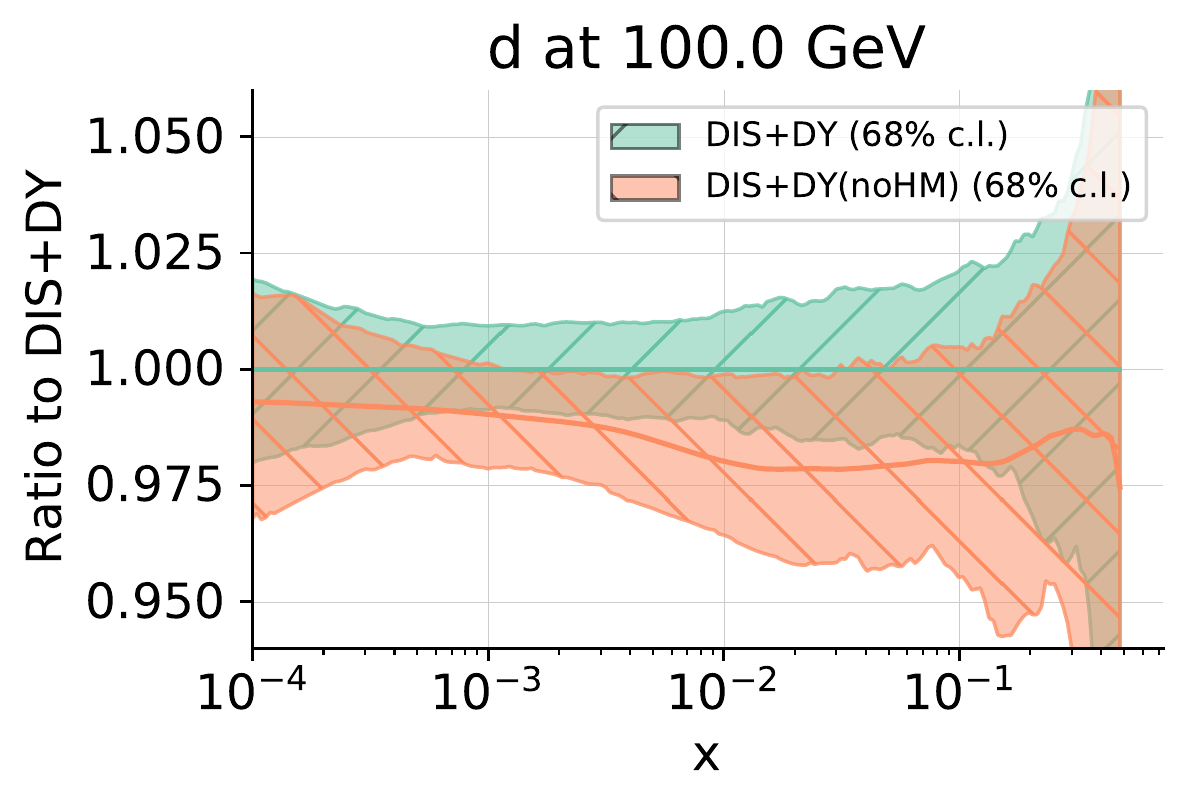}
  \includegraphics[width=0.32\textwidth]{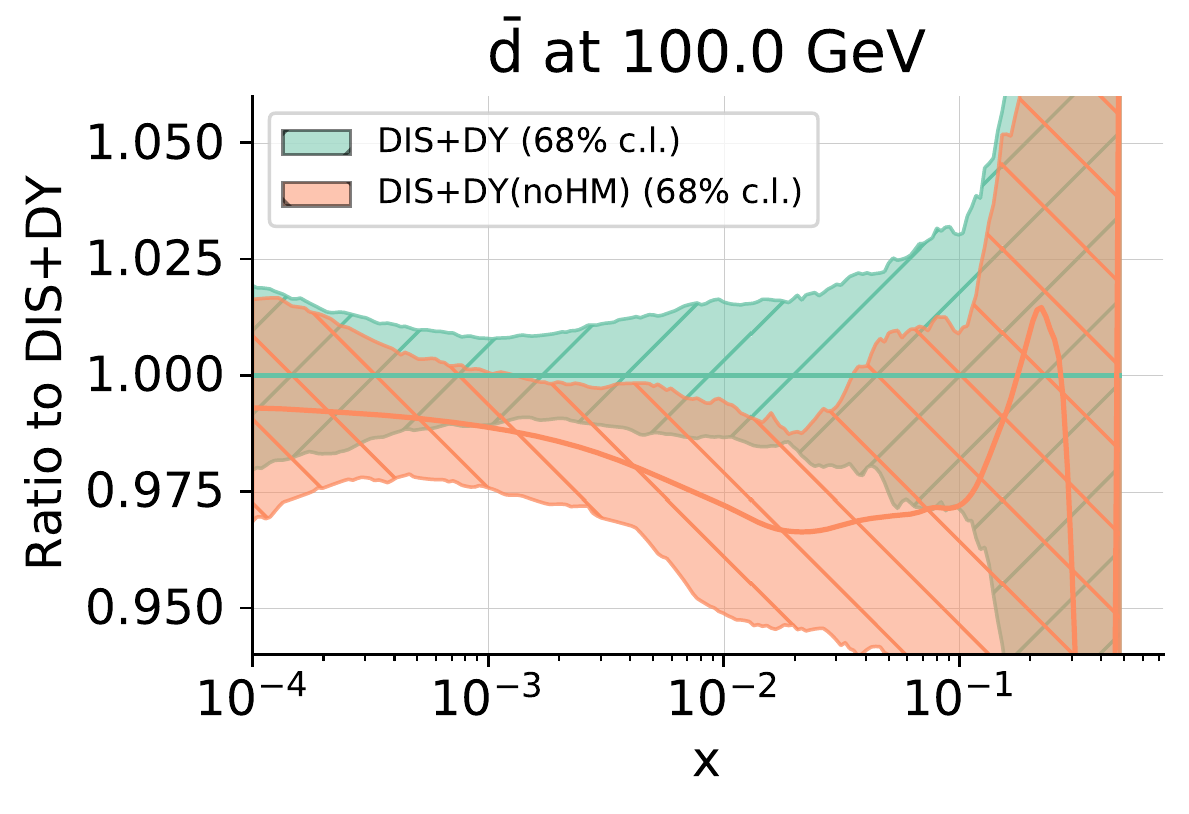}
  \includegraphics[width=0.32\textwidth]{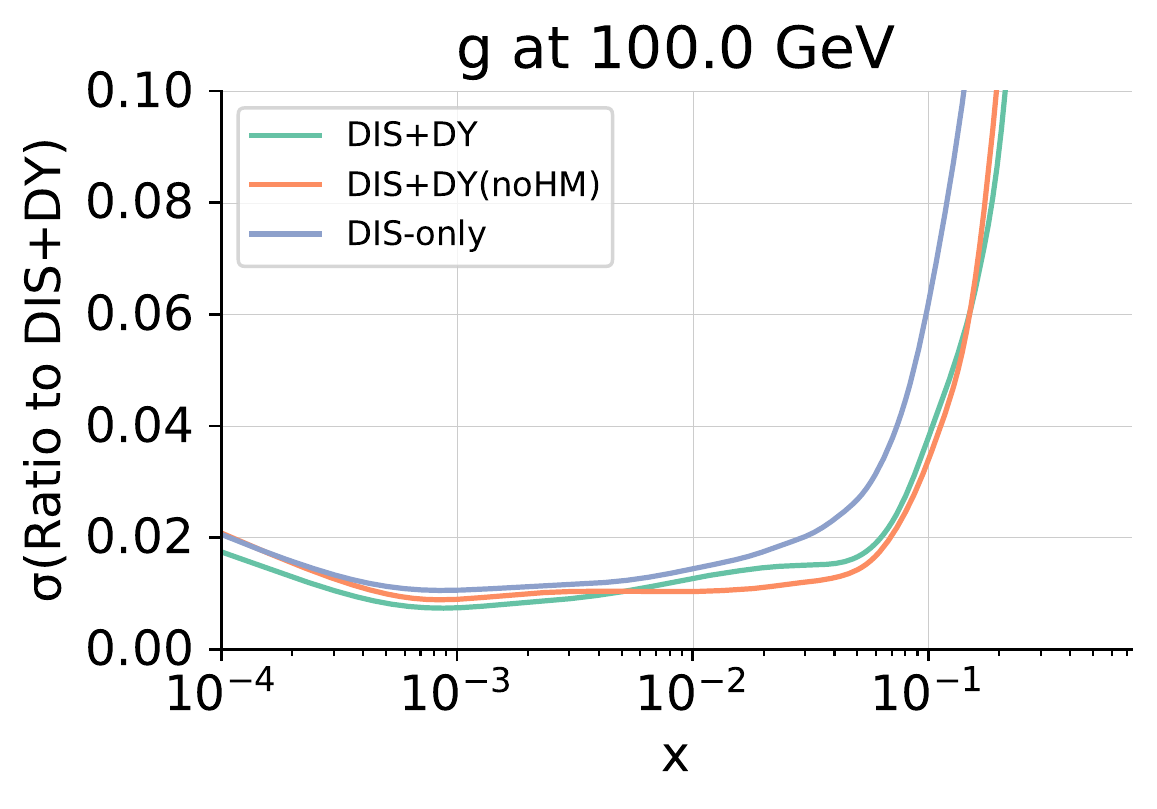}
  \includegraphics[width=0.32\textwidth]{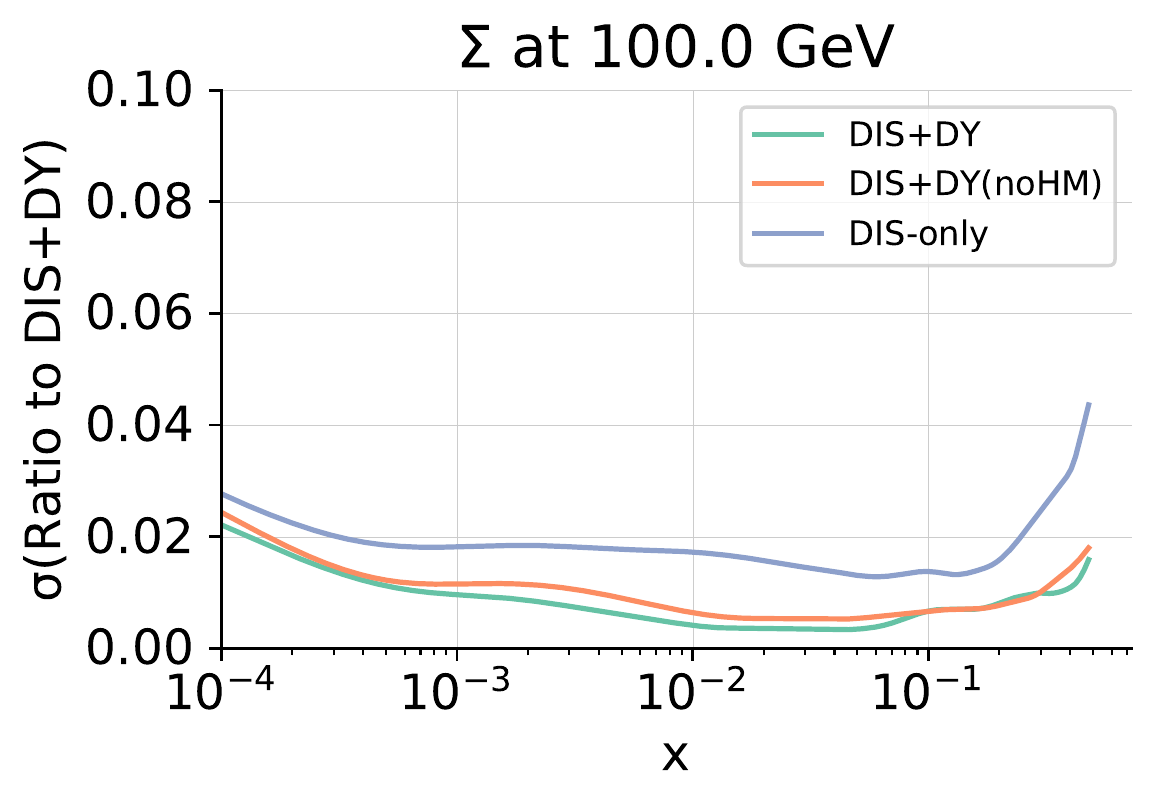}
  \includegraphics[width=0.32\textwidth]{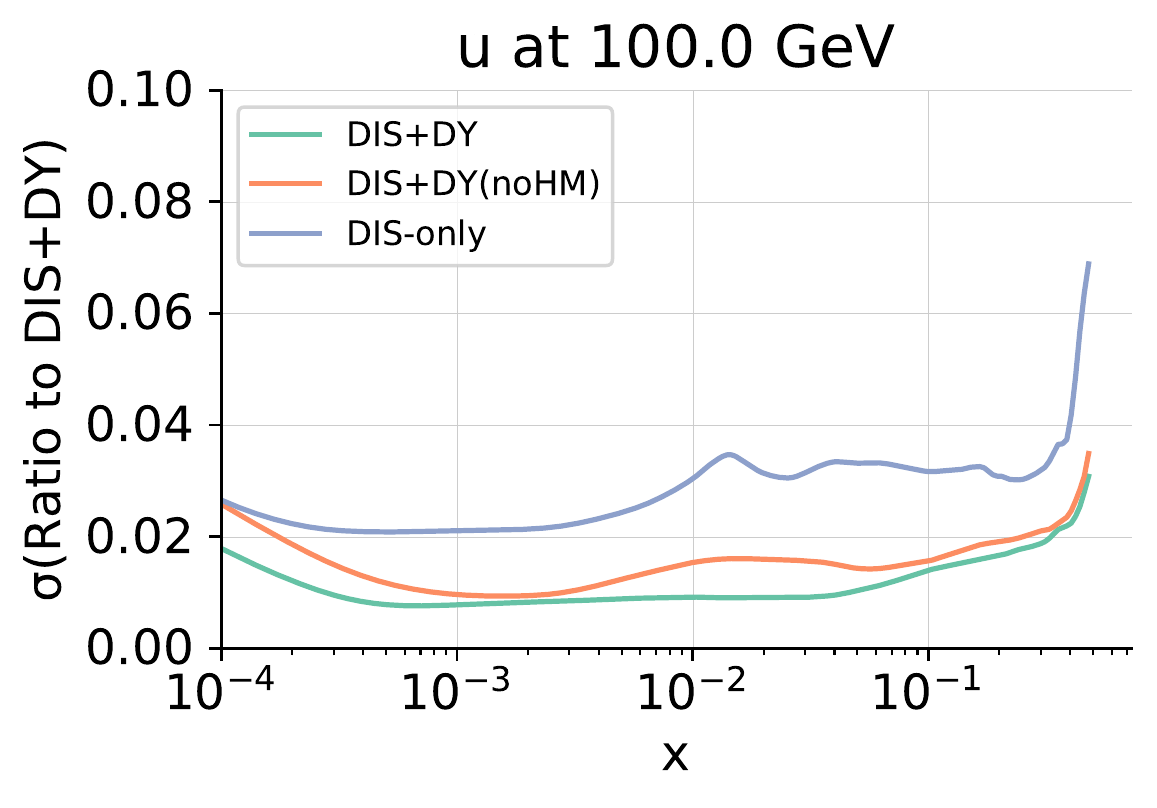}
  \includegraphics[width=0.32\textwidth]{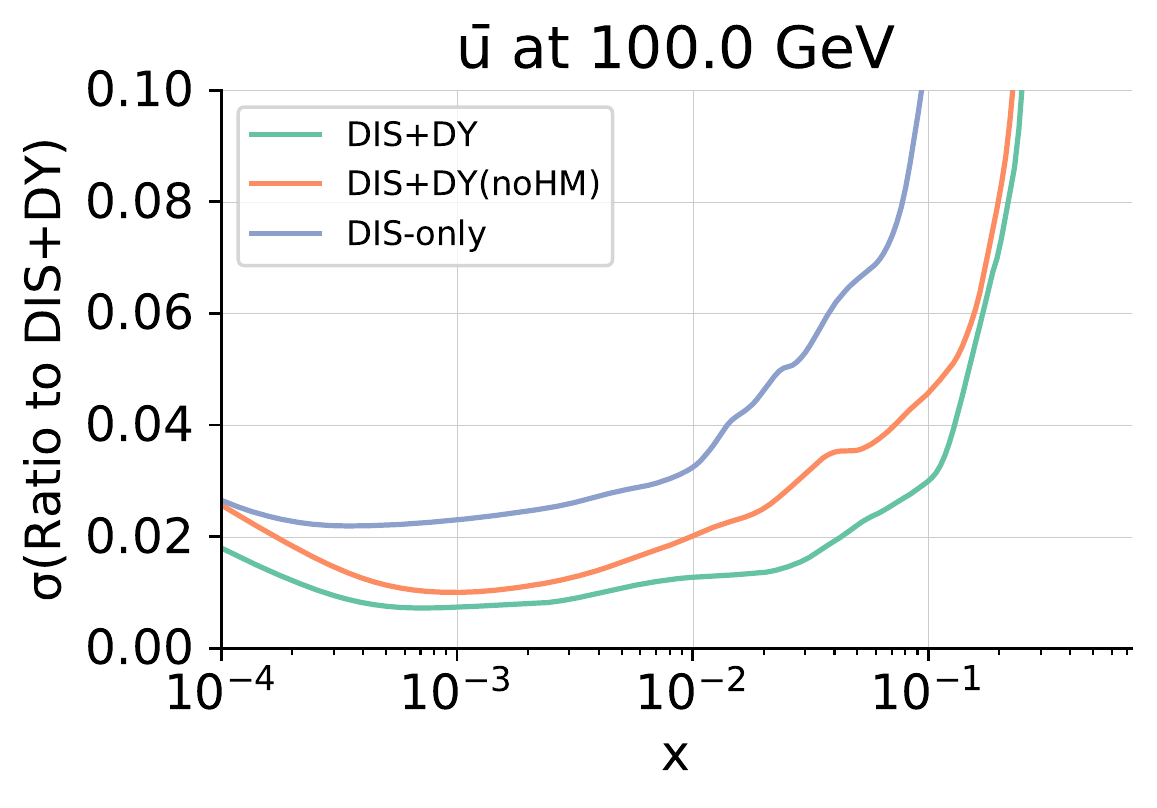}
  \includegraphics[width=0.32\textwidth]{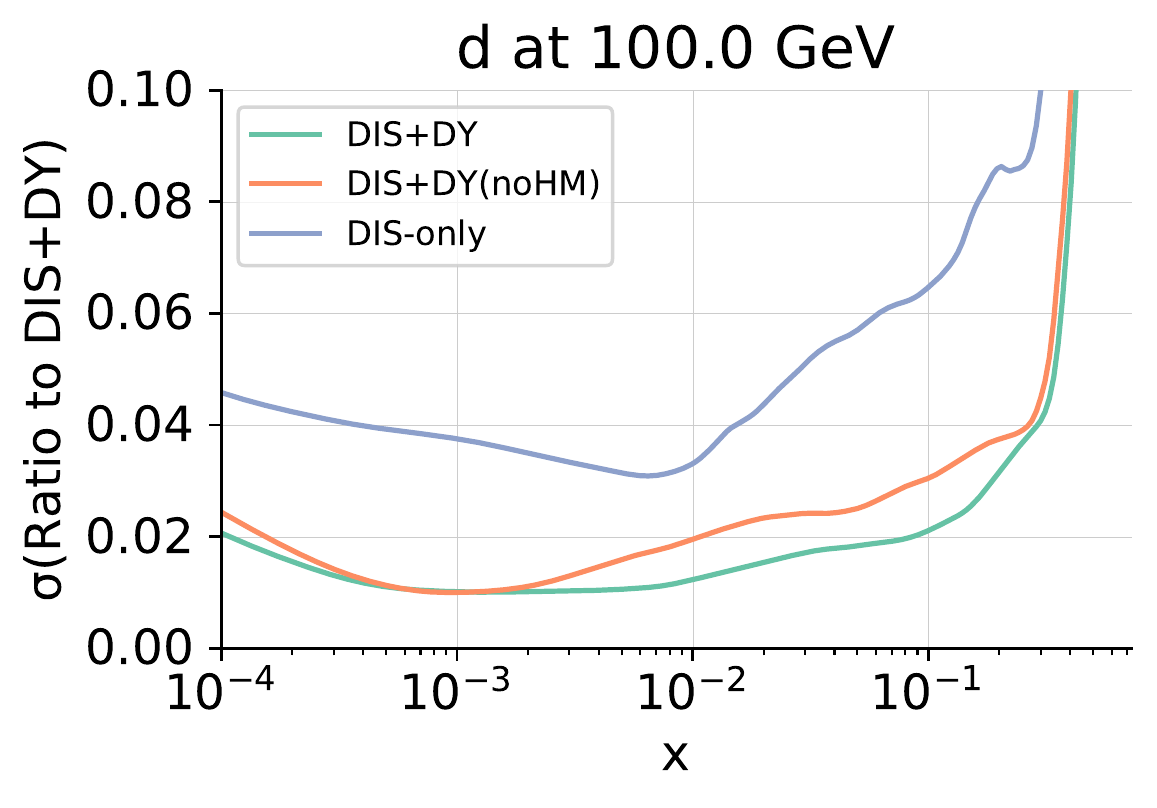}
  \includegraphics[width=0.32\textwidth]{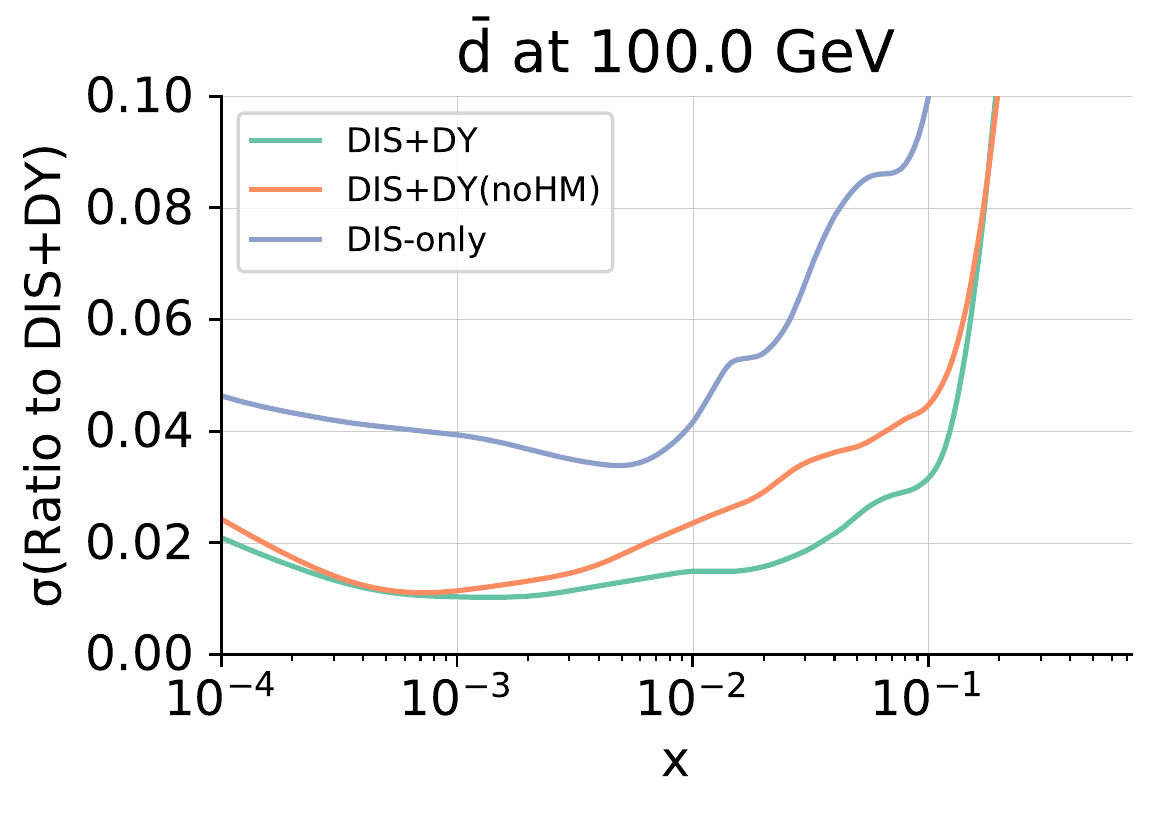}
  \caption{\small \small Comparison between the baseline SM PDF set of this work, labelled ``DIS+DY'', with
    the corresponding fit without high-mass DY data.
  We show results at $Q=100$ GeV for PDFs normalised to the central value of the baseline (upper) and
  for the relative PDF uncertainties (lower panels).
  In the latter case, we also display the PDF uncertainties from the DIS-only fit.
  \label{fig:pdfplot-impactHMDY}}
\end{figure}

Fig.~\ref{fig:pdfplot-impactHMDY} displays a comparison
between this baseline SM PDF set, labelled "DIS+DY'', with
the same fit but without any datapoints from the high-mass DY datasets
listed in Table~\ref{tab:data-high-mass}, labelled "DIS+DY (no HM)''.
 We show results at $Q=100$ GeV both for the PDFs normalised to the
 central value of the baseline and for the relative PDF
 uncertainties. In the latter case we also display the PDF uncertainties from a corresponding DIS-only fit.
The latter comparison shows that the DY cross sections significantly reduce
the PDF uncertainties of the DIS-only fit.
The addition of the high-mass DY data leads to a visible  
uncertainty reduction in the $0.005 \lsim x\lsim  0.3$ region
as compared to the "DIS+DY(noHM)'' reference as well an upwards shift
of the up and down quarks and antiquark PDF.

We therefore find that the available high-mass DY data can
have an appreciable impact on the light quark and antiquark PDFs,
despite the fact that in terms of Run II data our  analysis is restricted to a single low-luminosity 
 high-mass DY dataset.
Yet more stringent constraints on the PDFs are expected from the measurements based
on the full Runs II and III datasets,
as well as from those to be provided by the HL-LHC~\cite{Khalek:2018mdn}.
We study the anticipated impact of the HL-LHC measurements in
Sect.~\ref{sec:hllhc}.


\subsection{Methodology for the simultaneous PDF and EFT fits}
\label{eq:jointfits}

Let us denote by ${\mathbf{c}}=(c_1,c_2,\ldots,c_{N_{\rm op}})$ the array containing
the Wilson coefficients associated to the $N_{\rm op}$ dimension-six
operators contributing to a given SMEFT scenario, where $c_n$ are
defined as in Eq.~\eqref{eq:lagrangian}. 
For each point ${\mathbf{c}_i}$ in the scan of the EFT parameter space, we evaluate
the Drell-Yan and the DIS cross sections as described in Sect.~\ref{sec:theory}.
Subsequently, we determine the best-fit
PDFs associated to ${\mathbf{c}_i}$ by means of the standard NNPDF methodology, which
determines the minimum of the $\chi^2$ in the space of the PDF
parameters (subject to cross-validation, to avoid overlearning).
We note that this $\chi^2$, defined in
Eq.~\eqref{eq:chi2}, keeps fully into account the experimental
systematic correlations among all the
measurements $D_i$ included in the PDF analysis. 

This procedure results in a sampling of the $\chi^2$ values
in the EFT parameter space, which we denote by $\chi_{\rm eftp}^2({\mathbf{c}_i})$
(as in: EFT-PDFs).
Alternatively, one could also evaluate the same DIS and DY cross sections using instead
the baseline SM PDF set, ending up with  $\chi^2$ values which we denote by $\chi_{\rm smp}^2({\mathbf{c}_i})$
(as in: SM-PDFs).
The comparison between the resulting bounds on the EFT coefficients
obtained  from 
$\chi_{\rm eftp}^2({\mathbf{c}_i})$
and from $\chi_{\rm smp}^2({\mathbf{c}_i})$ quantifies the relevance of
producing consistent joint determinations of PDFs and Wilson coefficients
when studying EFTs in high-energy tails.
This strategy follows the one adopted in our proof-of-concept DIS-only
study~\cite{Carrazza:2019sec}, now extended to LHC processes.

Close enough to a local minimum $\chi_0^2=\chi^2\lp \mathbf{c}^{(0)}\rp$ associated with best-fit values $\mathbf{c}^{(0)}$,
the $\chi^2$ as a function of the EFT coefficients can be approximated by a quadratic form
\be
\label{functional-form}
\chi_i^2 \equiv \chi^2(\mathbf{c}_i) = \chi_0^2 +
\sum_{n,m=1}^{N_{\rm op}}\left( c_{n,i} - c^{(0)}_n\rp  H_{nm} 
    \lp c_{m,i} - c^{(0)}_m\rp\, ,
  \ee
with $H_{nm}$ being the usual Hessian matrix in the EFT parameter space.
Restricting the EFT calculations to their linear, $\mathcal{O}\lp \Lambda^{-2}\rp$, contributions, Eq.~(\ref{functional-form})
becomes exact
in the case of $\chi_{\rm smp}^2({\mathbf{c}_i})$ (where cross sections are
evaluated with SM PDFs).
The reason is that in this case
all dependence on the EFT coefficients is encoded in the partonic cross sections.

However, this is not true for $\chi_{\rm eftp}^2({\mathbf{c}_i})$, since now there will be a (non-linear) EFT back-reaction
onto the PDFs and hence Eq.~(\ref{functional-form}) is only valid up
to higher orders in the EFT expansion, even if the EFT cross sections themselves are evaluated in the linear
approximation.
Eq.~(\ref{functional-form}) can thus be only considered a reasonable approximation in the case that the SMEFT PDFs are not
too different from their SM counterparts.

Hence, if we work with linear EFT calculations,
provided the sampling in the EFT parameter space is sufficiently broad and fine-grained,
and that the EFT-induced distortion on the PDFs is moderate,
we can extract the parameters $\chi^2_0$ and
${\mathbf{c}}^{(0)}$ and the Hessian matrix $H$ using least-squares regression from Eq.~(\ref{functional-form}),
using $\chi^2_{\rm smp}$ for the SM PDFs and  $\chi^2_{\rm eftp}$ for the SMEFT PDFs.
The associated confidence level contours are determined by imposing
\begin{align}
  \label{eq:deltachi2def}
  \Delta\chi^2({\mathbf c}) \equiv \chi_i^2({\mathbf c})
  - \chi_0^2  = \sum_{n,m=1}^{N_{\rm op}}\left( c_{n} - c^{(0)}_n\rp  H_{nm} 
    \lp c_{m} - c^{(0)}_m\rp =\text{constant}\, ,
\end{align}
where this constant depends on the number of degrees of freedom.
For linear EFT two-parameter fits, such as those for
benchmark scenario I in the context of HL-LHC projections, 
imposing Eq.~(\ref{eq:deltachi2def}) leads to elliptic
contours in the $( \hat{W},\hat{Y})$ plane.

In the case of fits to $\chi^2$ profiles obtained from EFT calculations which include both linear, $\mathcal{O}\lp
\Lambda^{-2}\rp$, and quadratic, $\mathcal{O}\lp\Lambda^{-4}\rp$, contributions,
such as those arising in benchmark scenario II, rather than
working in the Hessian approximation 
we instead carry out a (one-dimensional) quartic fit of the form
\be
\label{eq:quarticfit}
 \chi^2\lp c\rp = \sum_{k=0}^{k_{\rm max}=4} a_k \lp c\rp^k
 \ee
 with the $\chi^2$ values being $\chi^2_{\rm smp}$~($\chi^2_{\rm eftp}$) for the SM (SMEFT) PDFs,
 and then determine confidence level intervals by imposing $\Delta\chi^2( c) = \chi_i^2( c)- \chi_0^2
 =\text{constant}$.
 We determine this constant
numerically by finding the likelihood contour,  $\mathcal{L} \propto \exp(-\chi^{2}/2)$, containing
95\% of the total probability (for the 95\% CL intervals).

To conclude this section, we give details on how we account for PDF
uncertainties and the statistical uncertainty associated to the finite
replica sample of the NNPDF Monte Carlo sets that we use here.

\paragraph{PDF uncertainty.} In
Sects.~\ref{sec:results_scenarioI},~\ref{sec:scenarioIIresults}
and~\ref{sec:hllhc_joint_fits} we
will present bounds on the EFT parameters  using the SM PDFs
with and without the PDF uncertainties being accounted for.
In order to estimate these, we follow the procedure detailed above to
determine the confidence level intervals for the EFT parameters but now
using the $k$th Monte Carlo replica of the PDF set, rather than the central replica $k=0$
as done when PDF uncertainties are neglected.
One ends up with  $N_{\rm rep}$ values of
the upper and lower bounds:
\be
\lc{\mathbf c}^{(k)}_{\rm min},{\mathbf c}^{(k)}_{\rm max}\rc \, ,\qquad k=1,\ldots,N_{\rm rep} \, ,
\ee
and then the outermost bounds in the \(68\%\) envelope are considered to be the bounds
on the EFT parameters ${\mathbf c}$, now including the 1$\sigma$-PDF uncertainty.
This is very important to account for, given that in the case of the
bounds determined using $\chi^2_{\rm eftp}$, the PDF
uncertainty is already included by
construction, given that the Wilson coefficients are determined from the
global set of PDFs, exactly as in the case of the $\alpha_s$
determination from a global set of PDFs of
\cite{Lionetti:2011pw,Ball:2011us}. A more sophisticated way to
extract parameters such as $\alpha_s$ of the Wilson coefficients from
a global fit of PDFs, that includes the correlations between these parameters
and the PDFs, is given by the correlated replica method proposed in the more recent $\alpha_s$ determination in
\cite{Ball:2018iqk}. The latter would allow better accounting of the correlations
between Wilson coefficients and PDFs. However we do not use it here due to the fact that these correlations of the
PDFs with the Wilson coefficients are much smaller than those with the strong
coupling constant and due to its large computational cost. We endeavour to address this issue in future work.

\paragraph{Methodological uncertainty.} In a simultaneous fit of PDFs and EFT coefficients,
for each set of Wilson coefficients \(\mathbf{c}_i\)
one has a PDF fit composed of \(N_\text{rep}\) Monte Carlo replicas.
The major methodological uncertainty is associated to  finite-\(N_\text{rep}\) effects
can be estimated by bootstrapping across the replicas, as explained in the $\alpha_s(m_Z)$
extraction of~\cite{Ball:2018iqk}.
Specifically, for each value of \(\mathbf{c}_i\)
we perform \(N_\text{res}\) re-samples of all \(N_\text{rep}\) replicas with replacement, and compute the theory predictions:
\begin{equation}
  \mathbf{T}^{\text{(res)}}_{i,lk} \, ,\quad
  \begin{matrix}
    l = 1,\dots,N_\text{res} \\
    k = 1,\dots,N_\text{rep}
  \end{matrix} \, ,
\end{equation}
such that there are \(N_\text{res}\) re-samples each composed of an \(N_\text{rep}\)-sized array
of theory predictions.
Since this re-sampling is done with replacement,
it differs from the original sample in that it contains  duplicates and missing values.
The average theory prediction is then obtained for each of these bootstrapped sets:
\begin{equation}
  \overline{\mathbf{T}}_{i,l} = \left<\mathbf{T}^\text{(res)}_{i,lk}\right>_\text{rep}\, ,
  \quad l = 1,\dots,N_{\text{res}} \, .
\end{equation}
These bootstrapped  theory predictions $\overline{\mathbf{T}}_{i,l}$ are used to evaluate the $\chi^2$ to data,
with the finite-size uncertainty given by the standard deviation across each bootstrap re-sample:
\begin{equation}
  \label{eq:finitesizechi2}
  \sigma_{\chi^2_i} = \text{std}\left(\chi^2_{i,l}\right)\Big|_\text{res}\, .
\end{equation}
A value of $N_\text{res}\simeq 10^4$ re-samples is found to be sufficient to achieve stable results
for the estimate of the finite-size uncertainties defined by Eq.~(\ref{eq:finitesizechi2}).


\section{Results}
\label{sec:res1}

In this section, we start by presenting results for the SMEFT PDFs extracted from DIS and Drell-Yan
data in benchmark scenario I. We compare them with their SM counterparts at the level
of partonic luminosities
and assess how the bounds obtained on the $\hat{W}$ and $\hat{Y}$
parameters in this simultaneous SMEFT and PDF fit compare to those based on assuming SM PDFs.
We then investigate the sensitivity of available high-mass Drell-Yan
data to benchmark scenario II, where only the dimuon final state is modified by EFT effects.
Finally, we quantify the impact that the consistent use of SMEFT PDFs has
on the reinterpretation of high-mass dilepton BSM searches.

\subsection{PDF and EFT interplay in current high-mass Drell-Yan data}
\label{sec:results_scenarioI}

By deploying the methodology described in Sect.~\ref{eq:jointfits},
we have extended the PDF analysis based on SM predictions presented in
Sect.~\ref{sec:fitsettings} to account for the effects of non-zero EFT coefficients within benchmark scenario I defined in Sect.~\ref{sec:scenarioI}.
 Here, we present results for one-dimensional fits where only one of the $\hat{W}$ or the $\hat{Y}$
parameter is allowed to be non-zero.
The reason for this choice is that, in a fit including only high-mass neutral-current
Drell-Yan processes, there exists a flat direction when $\hat{W}$ and $\hat{Y}$ are varied
simultaneously, since both operators scale as \(q^4\) and thus cannot both be constrained by a single 1D distribution.
This degeneracy can only be lifted once high-mass charged-current DY data is included in the fit.
As we  demonstrate in Sect.~\ref{sec:hllhc}, thanks to the HL-LHC it will be possible 
to carry out a simultaneous fit of the PDFs and the two EFT parameters $(\hat{W},\hat{Y}) $.

Taking into account the existing bounds reported in Sect.~2,
as well as the sensitivity of available high-mass Drell-Yan
data to the EFT coefficients illustrated by Fig.~\ref{fig:dysmeft}, here we have adopted
the following sampling ranges for the $\hat{W}$ and $\hat{Y}$ parameters:
\be
\label{eq:samplingrange}
\lp \hat{W}\times 10^4\rp \in \lc -22, 14 \rc \, , \qquad \lp \hat{Y}\times 10^4\rp \in \lc -20, 20 \rc \, .
\ee
We used 21 sampling values of $\hat{Y}_i$ equally spaced in this interval,
hence in steps of $\Delta \hat{Y}=2\times 10^{-4}$.
In the case of $\hat{W}_i$ it was found convenient to instead use
15 points equally spaced between $-14\times 10^{-4}$ and $14\times 10^{-4}$
in steps of $\Delta \hat{W}=2\times 10^{-4}$,
and then to add two more values at $\hat{W}_i=-18\times 10^{-4}$ and $-22\times 10^{-4}$.

\begin{figure}[t]
\begin{center}
  \includegraphics[width=0.49\textwidth]{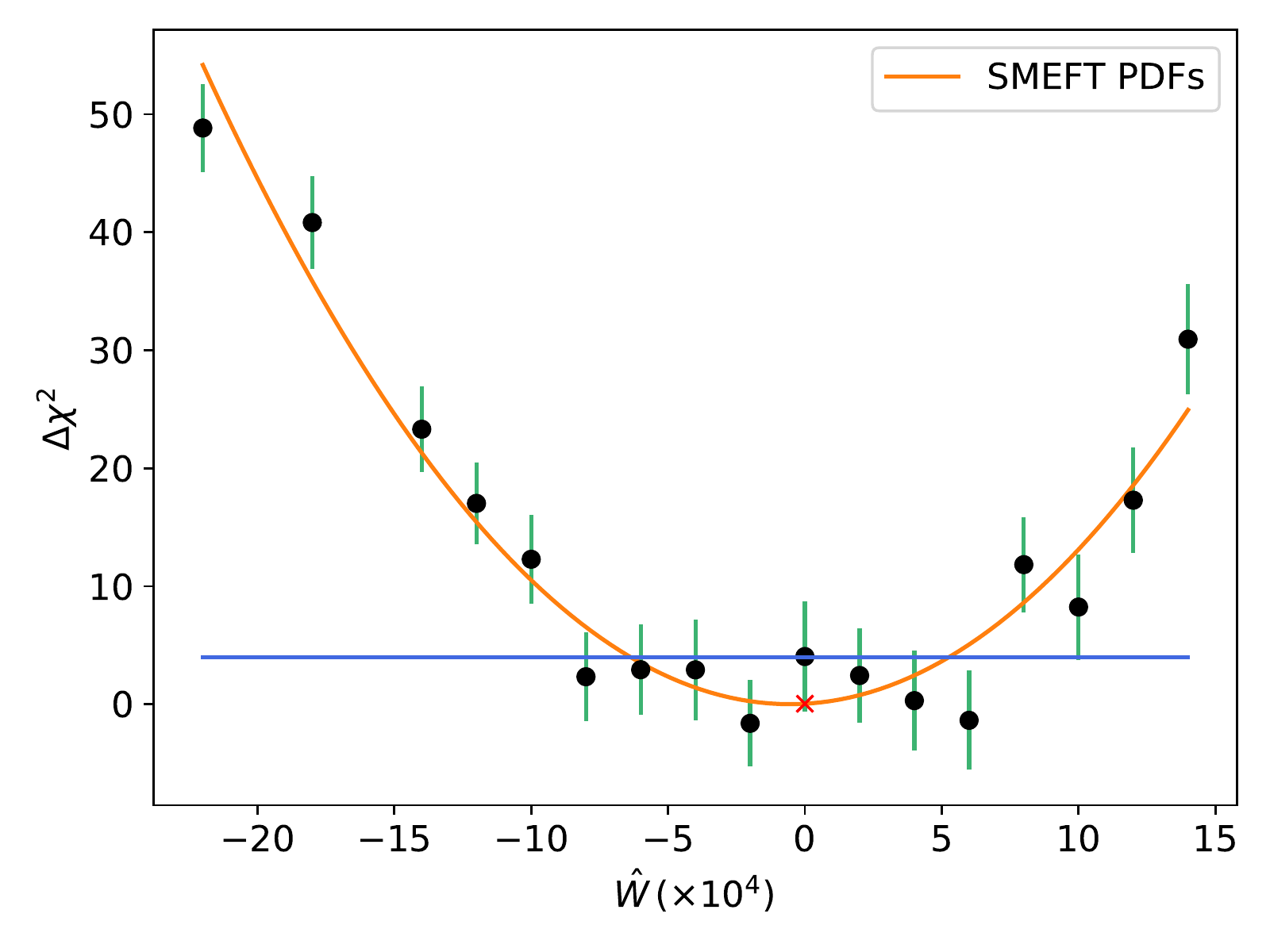}
  \includegraphics[width=0.49\textwidth]{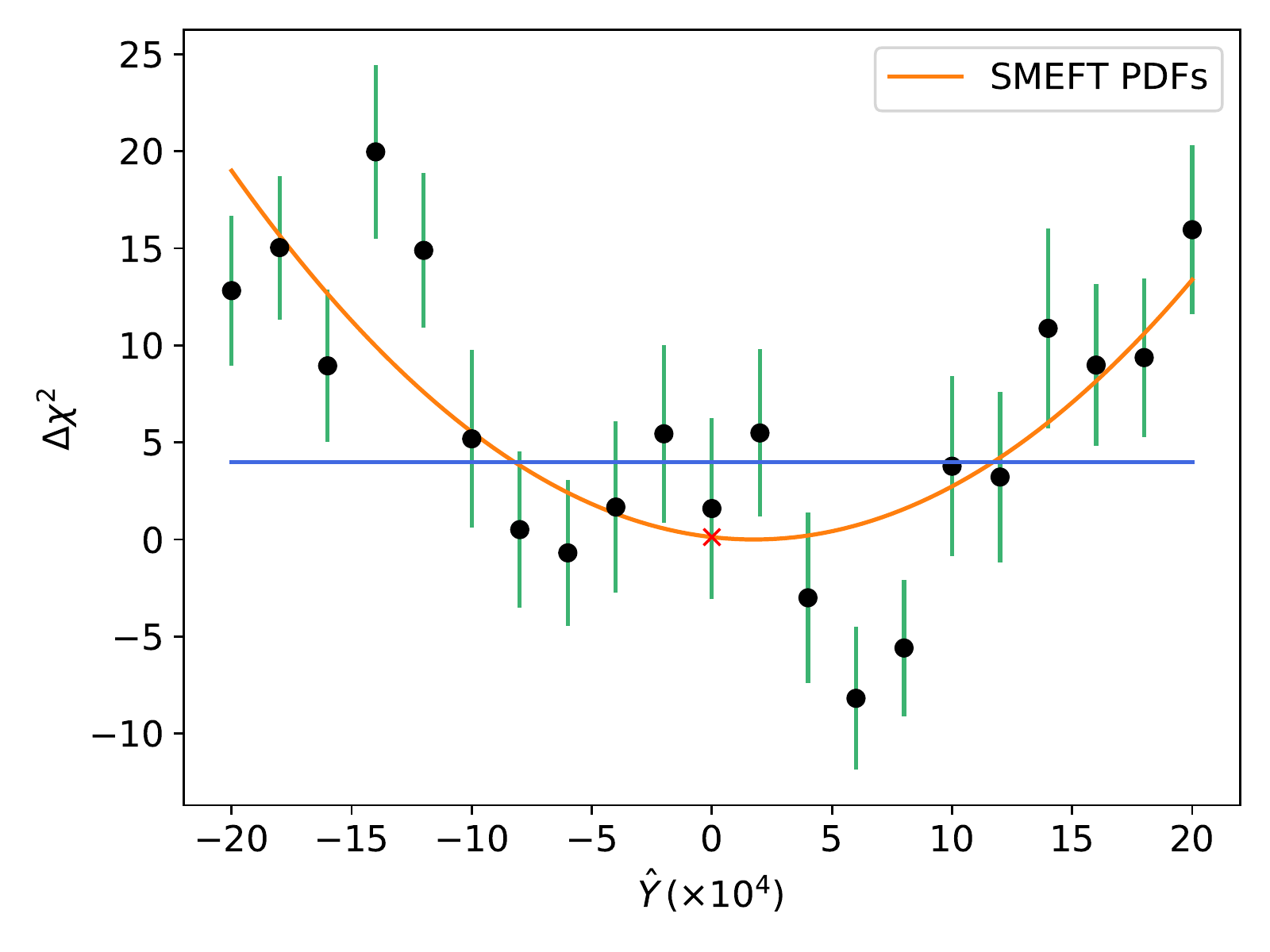}
    \caption{\label{fig:parabolas1} The values of $\Delta \chi^2$,
      Eq.~(\ref{eq:deltachi2def}), obtained for the SMEFT PDFs (thus using the
      $\chi_{\rm eftp}^2({\mathbf{c}_i})$ values)
  as a function of $\hat{W}_i$ (left)
      and $\hat{Y}_i$ (right panel) in the sampling ranges of
      Eq.~(\ref{eq:samplingrange})
      together with the corresponding parabolic fits.
      The error bars indicate the finite-size
      uncertainties and the horizontal line corresponds to the $\Delta \chi^2=4$ condition
      defining the 95\% CL intervals.
      The red cross indicates the SM expectation, $\hat{W}=\hat{Y}=0$.
      }
\end{center}
\end{figure}

Fig.~\ref{fig:parabolas1} displays the obtained values of $\Delta \chi^2$, Eq.~(\ref{eq:deltachi2def}),
as a function of $\hat{W}_i$ 
and $\hat{Y}_i$ in the case of the SMEFT PDFs. That is, using
the values of $\chi_{\rm eftp}^2({\mathbf{c}_i})$.
These $\chi^2$ values are  evaluated as a sum over
those datasets from Table~\ref{tab:data-low-mass} and
\ref{tab:data-high-mass} that receive
non-zero EFT corrections, namely the DIS datasets that
have a reach in  $Q^2$ above $(120)^2$ GeV$^2$ (namely HERA and NMC), and the ATLAS and CMS high-mass Drell-Yan measurements in Table~\ref{tab:data-high-mass}.
%
The use of such a partial $\chi^2$ rather than the global $\chi^2$ is
a necessary approximation due to the limitation of our current
methodology.  The statistical fluctuations of the global $\chi^2$ are
significantly larger than those of the partial $\chi^2$ and
can only be tamed by running a very large batch of replicas for each
benchmark point in $\hat{W}$ and $\hat{Y}$ and by increasing the density of benchmark points in
the region that is explored, as it was done for the scan of $\alpha_s$
in Ref.~\cite{Ball:2018iqk}. However, while for the scan of
$\alpha_s$ all processes contribute to the parabolic behaviour of the
$\Delta\chi^2$, in this case the dominant contributions to $\chi^2_{\rm eftp}$ come by far from the SMEFT
 corrections to the hard cross section of these processes, and
 from the changes in the PDFs induced by non-zero Wilson coefficients.
 The latter changes in PDFs are confined to the large-$x$ light quark
 and antiquark distributions, which affect the high-mass Drell-Yan
 data.  The analysis of the  $\chi^2_{\rm eftp}$ computed on the subset of
 data captures the dominant effects, while minimising the level of
 statistical fluctuations.\\
A further approximation is given by the fact that only linear EFT effects are included in the calculation of the DIS and DY 
cross sections, while the (subleading) quadratic corrections are neglected
in this scenario.
The error bars in the $\Delta\chi^2_i$ points of
Fig.~\ref{fig:parabolas1} indicate the
methodological finite-size
uncertainties evaluated with the bootstrapping method described in
Sect.~3.4  and the horizontal line corresponds to the $\Delta \chi^2=4$ condition associated
to a 95\% CL interval.
We also show in Fig.~\ref{fig:parabolas1} the results of the associated parabolic fits,
\be
\label{eq:parabolicfit}
\Delta \chi^2(\hat{W}) = \lp \hat{W} - \hat{W}^{(0)} \rp^2 / \lp \delta \hat{W} \rp^2 \, ,
\ee
and likewise for $\Delta \chi^2(\hat{Y})$.
From the results in Fig.~\ref{fig:parabolas1}, one observes that
both the $\hat{W}$ and $\hat{Y}$ parameters agree with the SM expectation
within uncertainties.

\begin{figure}[t]
\begin{center}
  \includegraphics[width=0.49\textwidth]{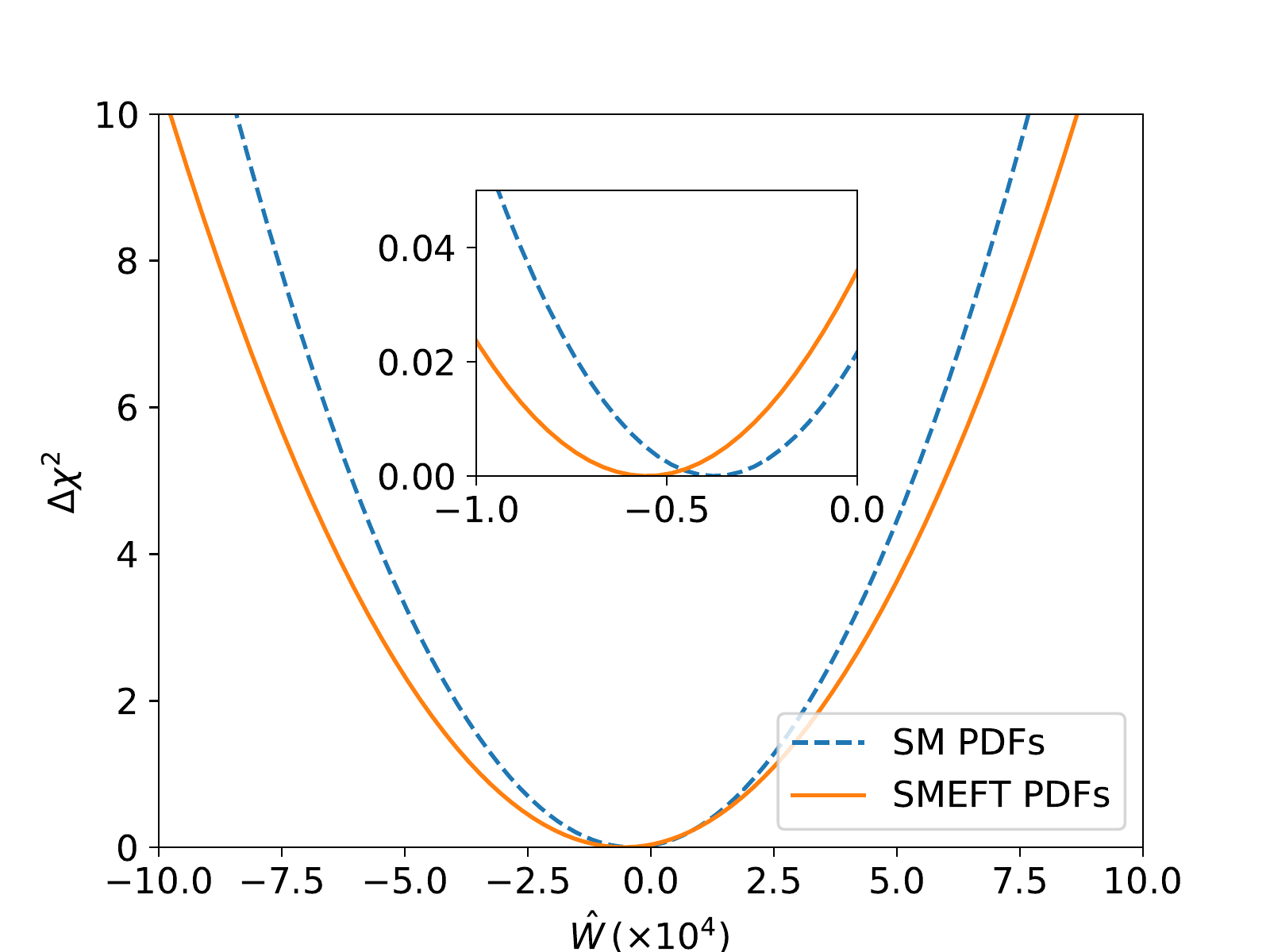}
    \includegraphics[width=0.49\textwidth]{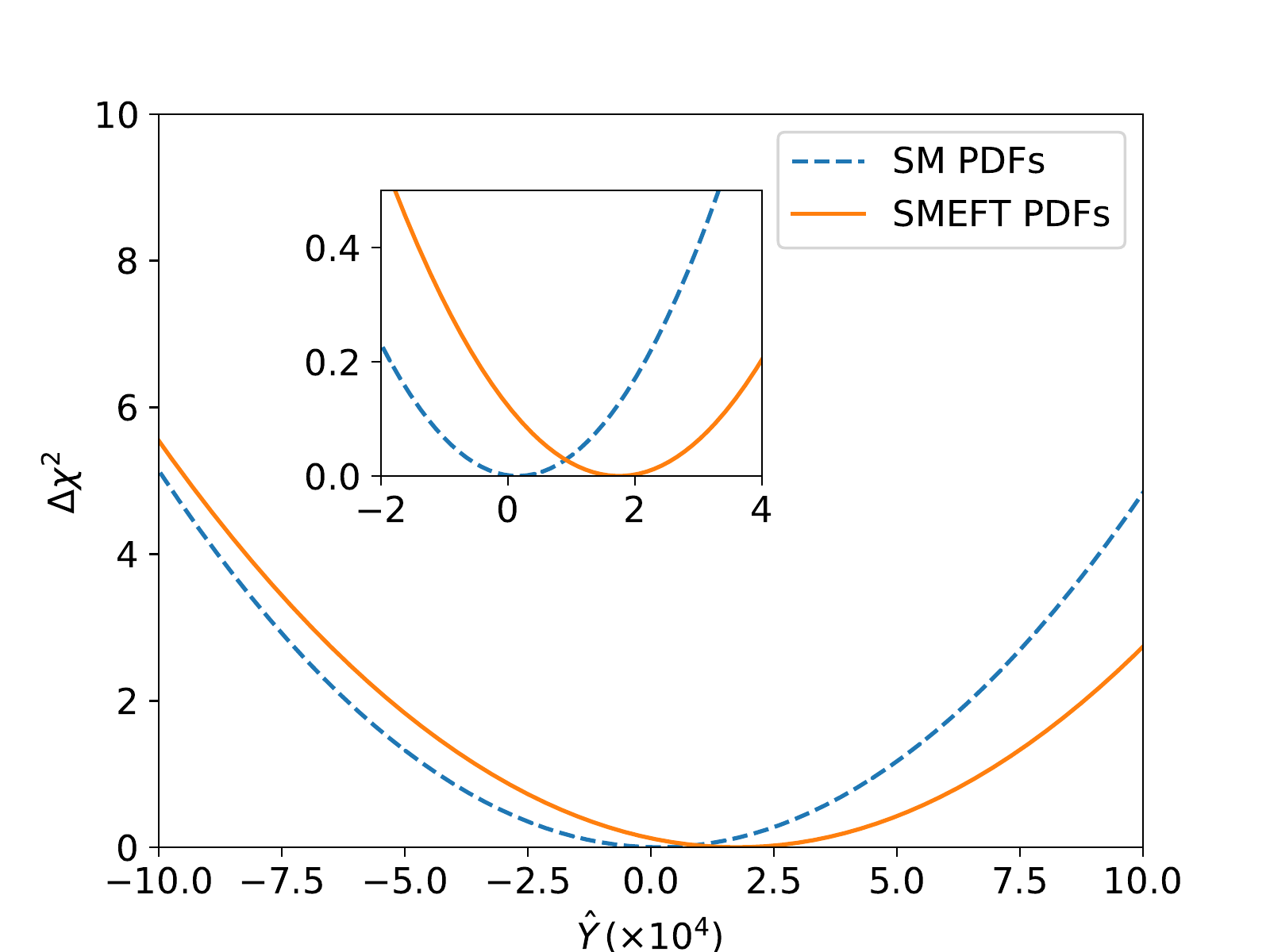}
    \caption{\label{fig:parabolas2} Comparison
      between the results of the parabolic fits to $\Delta\chi^2$, Eq.~(\ref{eq:parabolicfit}),
      for the $\hat{W}$ (left) and $\hat{Y}$ (right panel) parameters for either the SMEFT PDFs
      ($\chi^2_{\rm eftp}$, already displayed in Fig.~\ref{fig:parabolas1}) or
      the SM PDFs (hence with $\chi^2_{\rm smp}$).
      The insets zoom on the region close to $\Delta\chi^2\simeq 0$.
    }
\end{center}
\end{figure}

Fig.~\ref{fig:parabolas2} then compares the results
of the parabolic fits based on the SMEFT PDFs
as displayed in Fig.~\ref{fig:parabolas1} with their
counterparts
obtained in the case of the SM PDFs.
That is, in the latter case one carries out parabolic fits to
the $\chi^2_{{\rm smp}}$ values, as is customary
in the literature for the EFT analyses.
The insets highlight the region close to $\Delta\chi^2\simeq 0$.
For the $\hat{W}$ parameter, the consistent use of SMEFT PDFs leaves
the best-fit value essentially unchanged but increases the coefficient
uncertainty $\delta \hat{W}$, leading to a
broader parabola.
Similar observations can be derived for the $\hat{Y}$ parameter, though here
one also finds a upwards shift in the best-fit values by $\Delta\hat{Y}\simeq 2\times 10^{-4}$
in addition to a parabola broadening, when SMEFT PDFs are
consistently used.
We note that the SM PDF parabolas in Fig.~\ref{fig:parabolas2} are evaluated
using the central PDF replica and hence do not account for PDF uncertainties.

Table~\ref{tab:bound1w} summarises
the 68\% and 95\% CL bounds on the $\hat{W}$ and $\hat{Y}$
parameters obtained from the corresponding parabolic $\Delta\chi^2$ fits 
using either the SM or the SMEFT PDFs shown in Fig.~\ref{fig:parabolas2}.
The fourth and fifth column indicate the absolute shift in best-fit values
and the percentage broadening of the fit parameter uncertainties when the SMEFT PDFs
are consistently used instead of the SM PDFs (either without or with PDF uncertainties):
\be
\label{eq:shift}
{\rm best~fit~shift}\equiv \lp \hat{W}^{(0)}\Big|_{\rm SMEFT\,PDF}-\hat{W}^{(0)}\Big|_{\rm SM\,PDF}\rp \, ,
\ee
\be
\label{eq:broadening}
{\rm broadening}\equiv \lp \delta\hat{W}^{(0)}\Big|_{\rm SMEFT\,PDF}-\delta\hat{W}^{(0)}\Big|_{\rm SM\,PDF}\rp\bigg/\delta\hat{W}^{(0)}\Big|_{\rm SM\,PDF} \, ,
\ee
and likewise for the $\hat{Y}$ parameter.

\begin{table}[t]
  \renewcommand{\arraystretch}{1.40}
  \centering
  \begin{tabular}{l|c|c|c|c}
    & $\quad$ SM PDFs $\quad$  & SMEFT PDFs  & best-fit shift  & broadening  \\
    \toprule
    \multirow{2}{*}{$\hat{W}\times 10^4$ (68\% CL)} & $[-3.0, 2.2] $ & \multirow{2}{*}{$[-3.5, 2.4]$}  & $-0.2$    &+13\%\\
    & $[-4.3, 3.8] $ &   &  $-0.3$   & $-27\%$ \\
    \midrule
    \multirow{2}{*}{$\hat{W}\times 10^4$ (95\% CL)} & $[-5.5, 4.7] $ &  \multirow{2}{*}{ $[-6.4, 5.3] $} & $-0.2$  & +15\%\\
     &   $[-6.8, 6.3] $   &  & $-0.3$      &$-11\%$ \\
    \midrule
    \multirow{2}{*}{$\hat{Y}\times 10^4$ (68\% CL)} & $[-4.4, 4.7] $ &  \multirow{2}{*}{$[-3.4, 6.9]$} & $+1.6$ &  $+13\%$\\
       & $[-6.7, 7.5] $ &   & $+1.4$ & $-27\%$ \\
    \midrule
    \multirow{2}{*}{$\hat{Y}\times 10^4$ (95\% CL)} & $[-8.8, 9.2] $ &  \multirow{2}{*}{$[-8.3, 11.8]$} & $+1.6$ & +12\% \\
     & $[-11.1, 12.0] $ &   & $+1.3$ & $-13\%$ \\
    \bottomrule
  \end{tabular}
  \caption{\label{tab:bound1w} \small The 68\% CL and 95\% CL bounds on the $\hat{W}$ and $\hat{Y}$
    parameters obtained from the  corresponding parabolic fits to
    the $\Delta\chi^2$ values calculated from  either the SM or the the SMEFT PDFs.
    For the SM PDF results, we indicate the bounds obtained without (upper)
    and with (lower entry) PDF uncertainties accounted for; the SMEFT PDF
    bounds already include  PDF uncertainties by construction, while
    the methodological (finite-size) uncertainty is included according to the
    approached described in Sect.~3.4. 
    The fourth and fifth column indicate the absolute shift in best-fit values,
    Eq.~(\ref{eq:shift})
    and the percentage broadening of the EFT parameter uncertainties, Eq.~(\ref{eq:broadening}),
    when the SMEFT PDFs
    are consistently used instead of the SM PDFs.
}
\end{table}

In the specific case of the SM PDF results,
Table~\ref{tab:bound1w} indicates
the bounds obtained without (upper) and with (lower entry) PDF uncertainties accounted for;
recall that the SMEFT PDF bounds already include PDF uncertainties by construction
(see Sect.~\ref{eq:jointfits}). The methodological (finite-size)
uncertainty is included according to the approach described in
Sect.~3.4 and it amounts to $4.7\cdot 10^{-5}$ in the case of
$\hat{W}$ and $1.0 \cdot 10^{-4}$ in the case of $\hat{Y}$,
corresponding to 4\% and 5\% respectively of the 95\% C.L. bounds for the $\hat{W}$ and $\hat{Y}$ coefficients.

By comparing the bounds obtained when PDF uncertainties are accounted for to
those neglecting PDF uncertainty, one observes a systematic broadening of the
bounds from both the lower and upper limits, as was also reported in~\cite{Carrazza:2019sec}.

When PDF uncertainties are neglected (accounted for) when using the SM PDFs
to constrain the EFT parameters,
 the consistent use of the SMEFT PDFs leads to both a shift
in the best-fit values of magnitude $\Delta\hat{W}=-2\times 10^{-5}$
and $\Delta\hat{Y}=+1.6\times 10^{-4}$ 
as well as to an increase (decrease) of the fit parameter uncertainties, with $\delta \hat{W}$
and $\delta \hat{Y}$ growing by 15\% and 12\% (decreasing by
11\% and  13\%) respectively.
This result shows that, given available Drell-Yan data and once PDF uncertainties
are accounted for, the bounds on the EFT parameters are actually {\it improved}
once SMEFT PDFs are adopted.

All in all, the effect of the consistent treatment of the SMEFT PDFs
in the interpretation of high-mass DY cross sections
is moderate but not negligible, either loosening or tightening up the obtained
bounds on the EFT parameters (depending on whether or not PDF uncertainties
are accounted for to begin with) by up to 15\% and, in the case of
$\hat{Y}$ parameter, shifting its central value by  one-third
of the 68\% CL parameter uncertainty.
Such a relatively moderate effect can be partly understood from the limited availability
of high-mass DY measurements for
EFT interpretations, with a single dataset at 13 TeV, and even in this case, with it being restricted to
a small fraction of the Run II luminosity.
As we will demonstrate in Sect.~\ref{sec:hllhc}, the impact of SMEFT
PDFs  becomes much more significant once higher-statistics measurements of the NC and CC Drell-Yan tails
become available at the HL-LHC, loosening the bounds on $\hat{W}$ and $\hat{Y}$
by up to a factor 5.
  
Comparing the limits on the $\hat{W}$ and $\hat{Y}$ parameters from Table~\ref{tab:bound1w} with
those of Ref.~\cite{Farina:2016rws} and reported
in Sect.~2, we observe that our bounds are more stringent.
There are two main reasons that could explain this difference.
On the one hand, on top of the ATLAS and CMS 
high-mass DY cross sections at 8 TeV, we also include  the corresponding 7 and 13 TeV data that provide
additional weight to the  high invariant 
mass region of the spectrum in the fit.
On the other hand, in our analysis we fit the whole invariant mass spectrum and do not 
cut away the low $m_{\ell\ell}$ region below 120 GeV, thus we do not ignore the correlations  
between the low and high ends of the spectrum which are important even if the former is not affected by SMEFT 
corrections.

\begin{figure}[t]
\begin{center}
\includegraphics[width=0.32\textwidth]{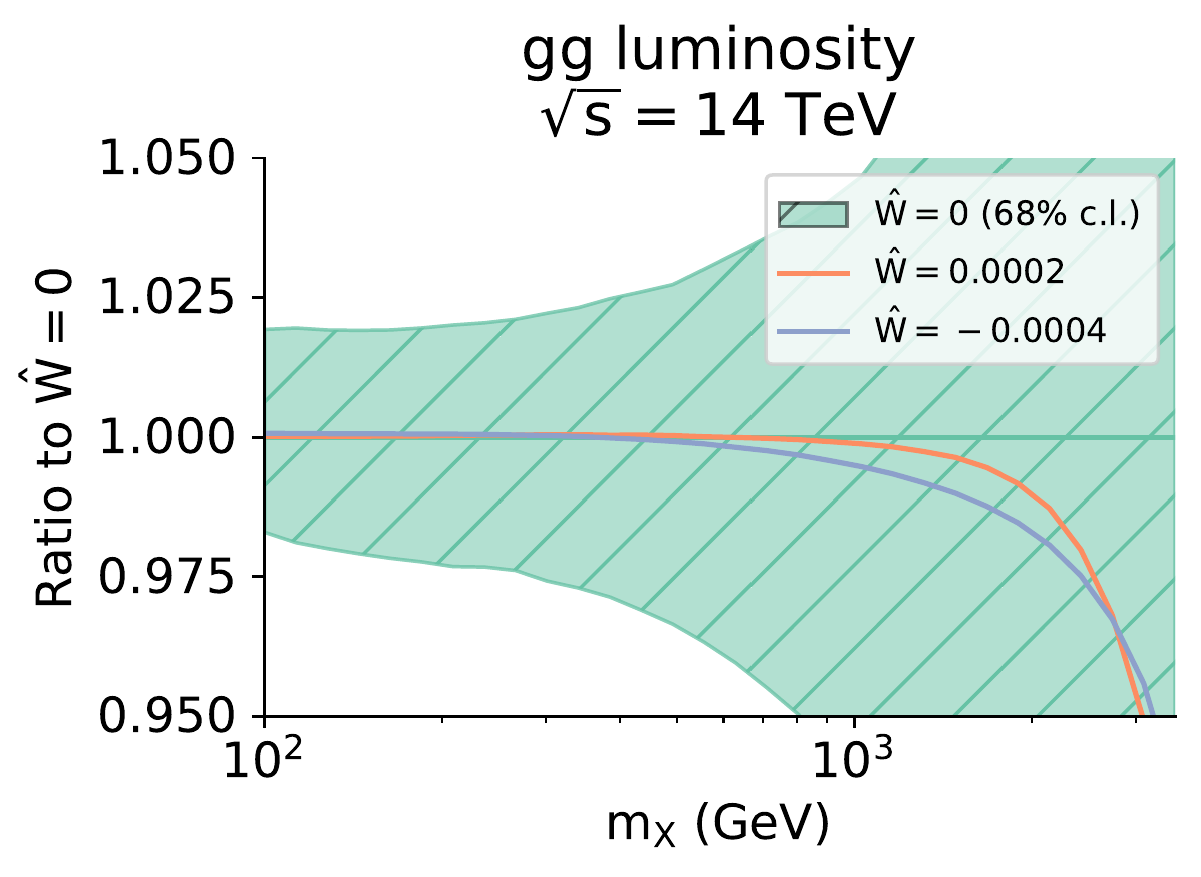}
\includegraphics[width=0.32\textwidth]{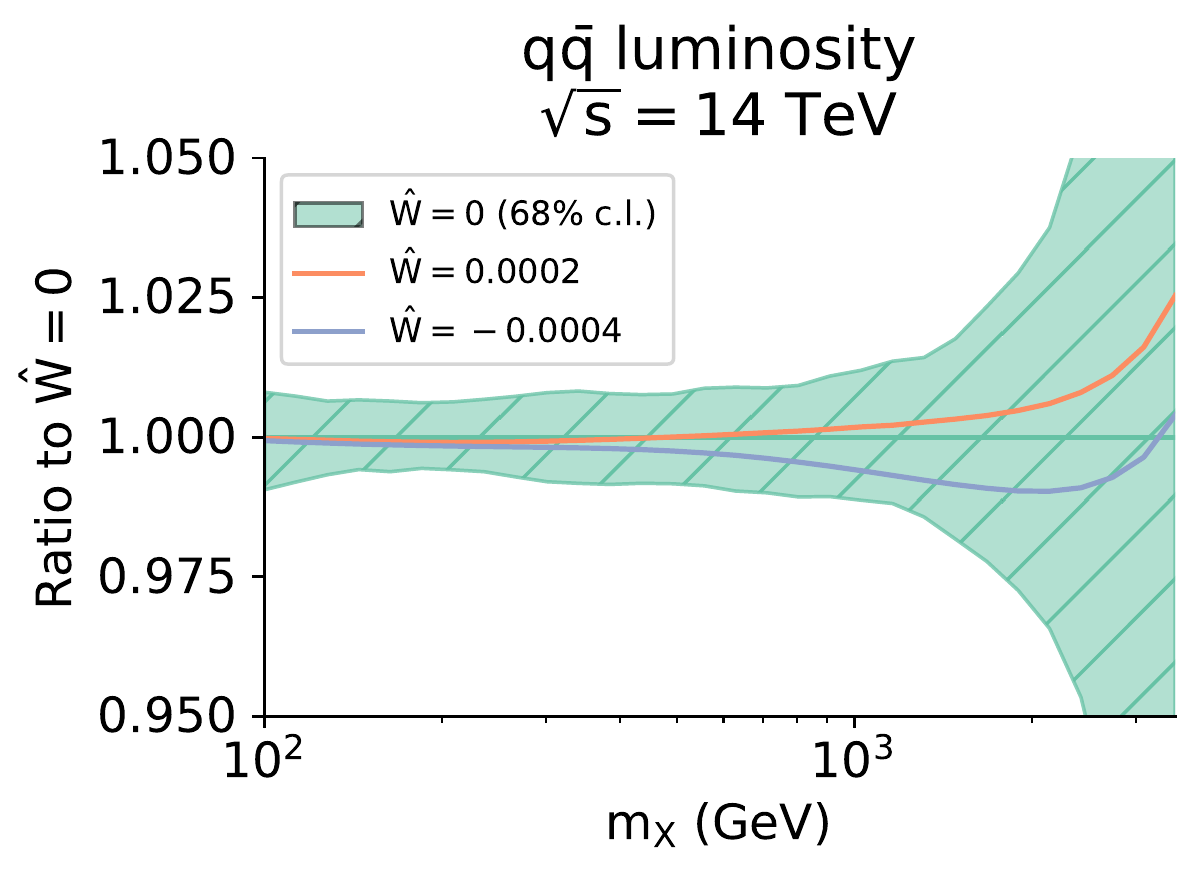}
\includegraphics[width=0.32\textwidth]{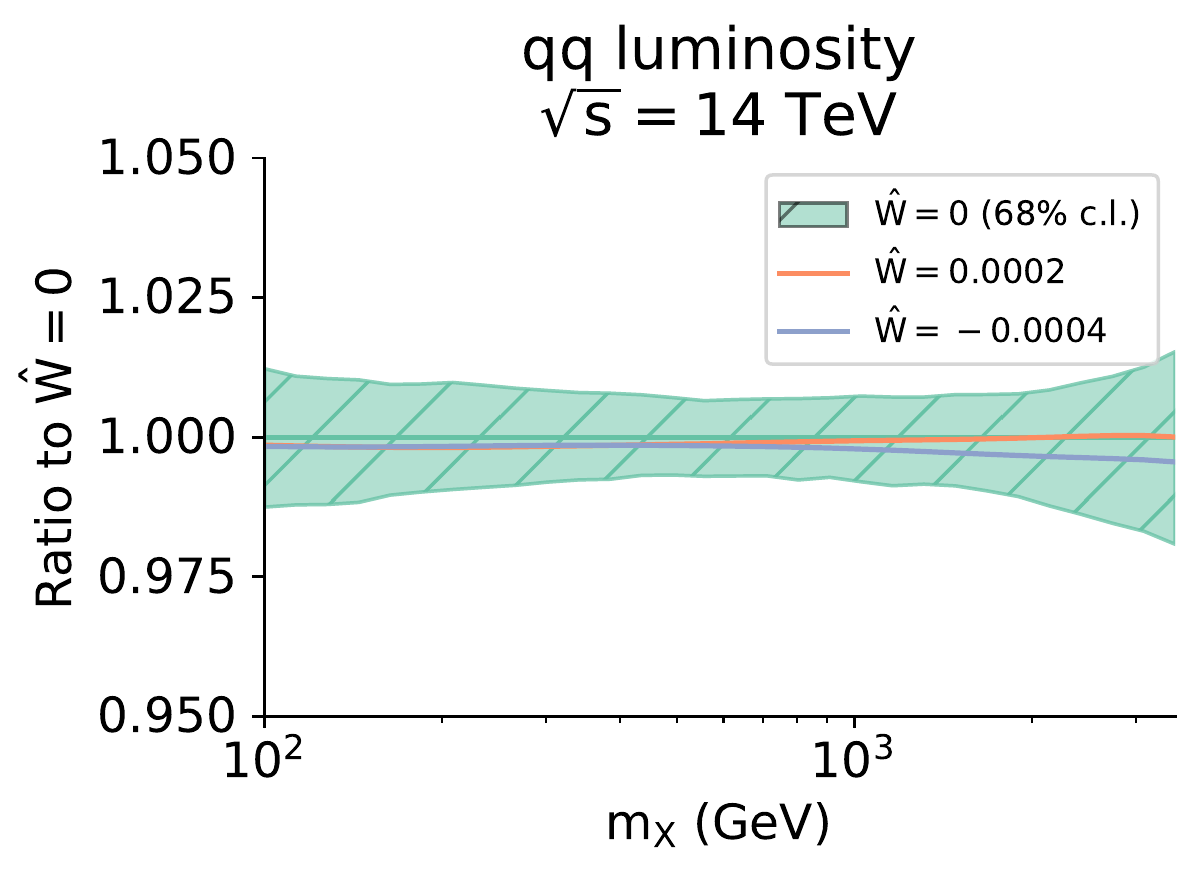}
\includegraphics[width=0.32\textwidth]{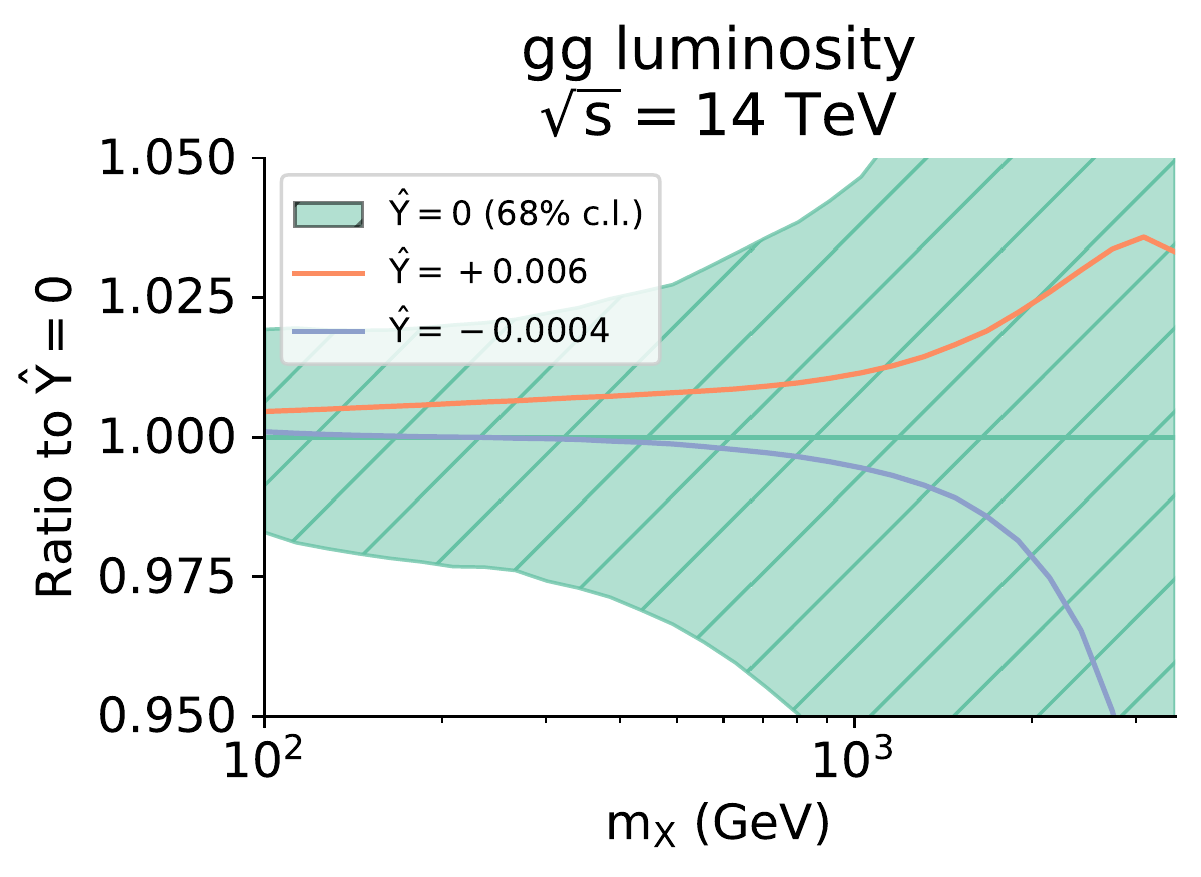}
\includegraphics[width=0.32\textwidth]{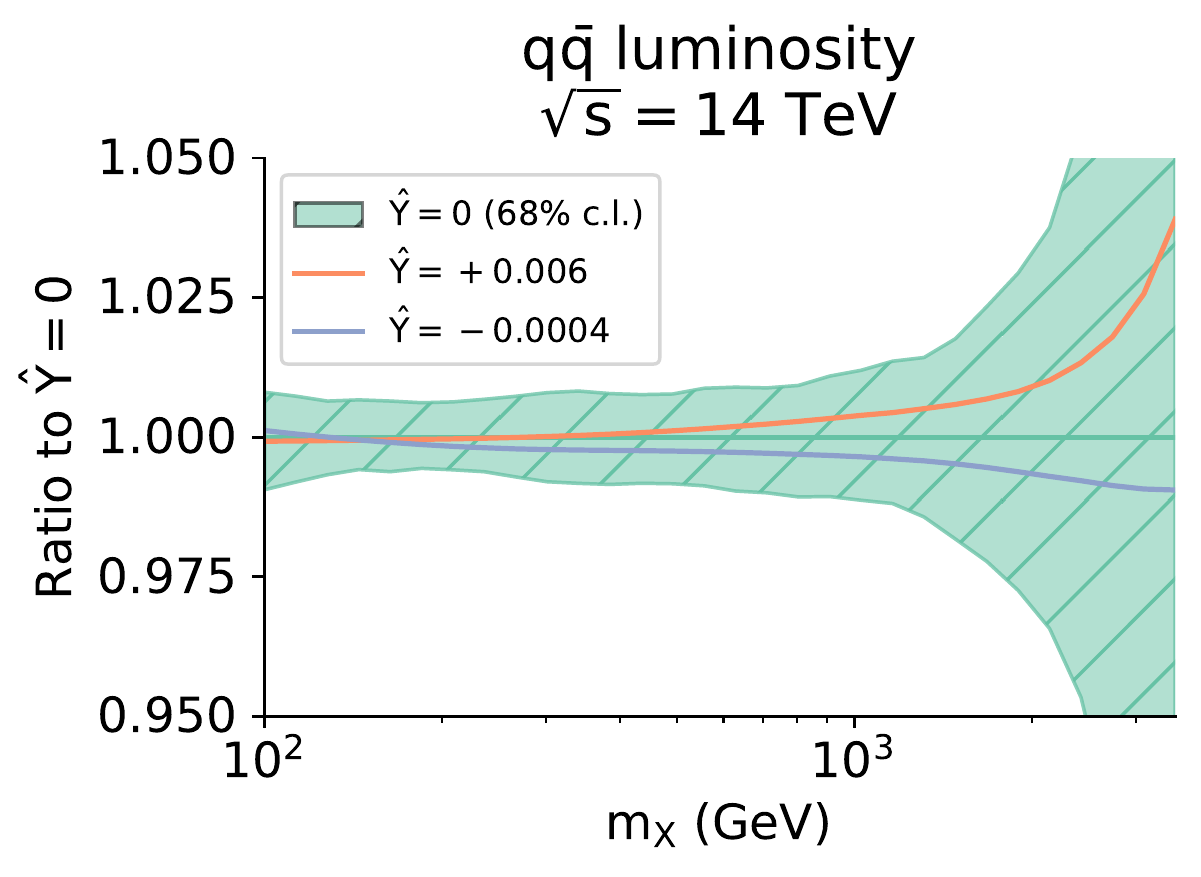}
\includegraphics[width=0.32\textwidth]{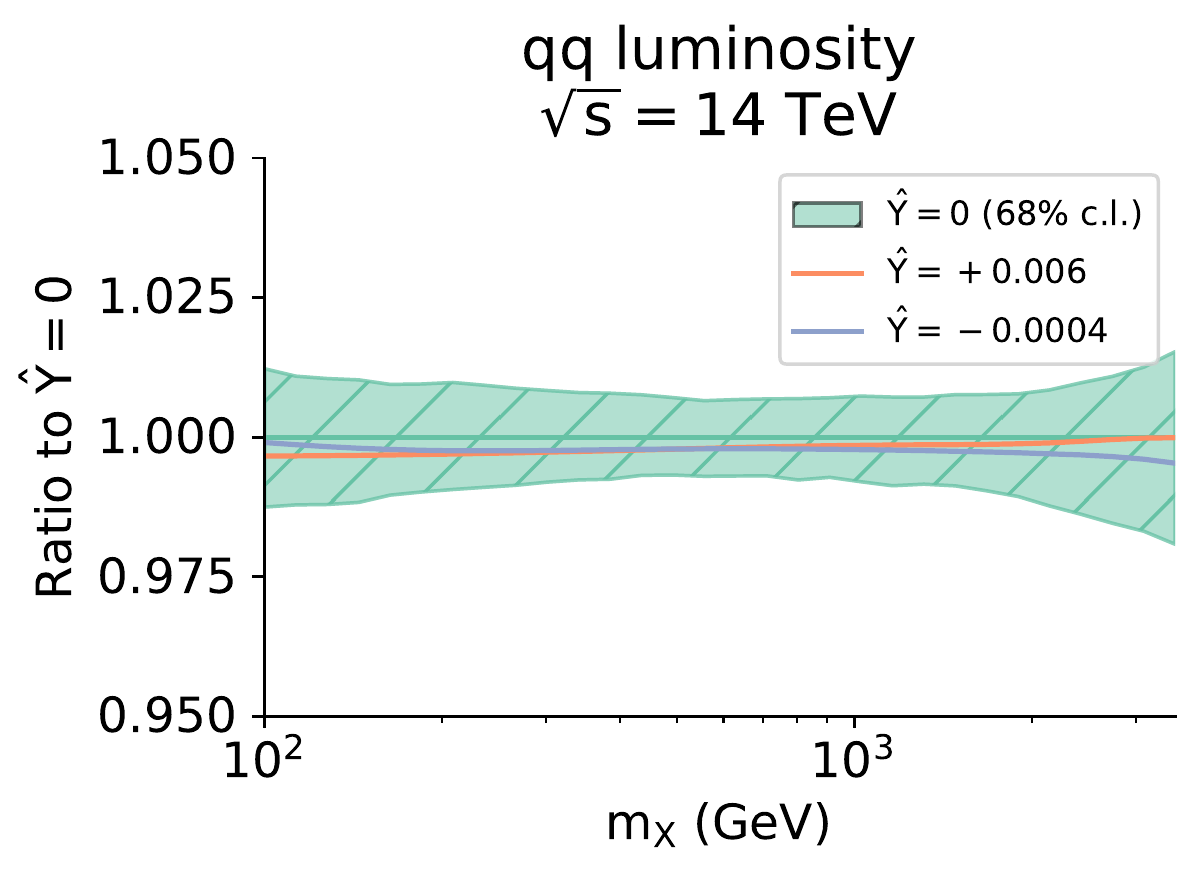}
\caption{\label{fig:SMEFT_lumis} 
Comparison between the SM PDF luminosities with their SMEFT counterparts, 
displayed as ratios to the central value of the SM luminosities, 
for representative values of the $\hat{W}$ (upper) and $\hat{Y}$ (lower panel)
parameters.
  The values of $\hat{W}$ and $\hat{Y}$ are chosen to be close 
 to the upper and lower limits of the 68\% CL intervals reported in Table~\ref{tab:bound1w}.}
\end{center}
\end{figure}

We now move to assess how the SMEFT PDFs relate to their SM counterparts,
and determine the extent to which it is possible to reabsorb EFT effects into the PDFs.
Fig.~\ref{fig:SMEFT_lumis} displays a
comparison between the SM and the SMEFT PDF luminosities 
for representative values of the $\hat{W}$ (upper) and $\hat{Y}$ (lower panel) parameters.
  The values of $\hat{W}$ and $\hat{Y}$ are chosen to be close 
  to the upper and lower limits of the 68\% CL intervals reported in Table~\ref{tab:bound1w}.
  The error band in the SM PDFs corresponds to the 68\% CL PDF uncertainty, while for the SMEFT
  PDFs only the central values are shown.

In all cases, one finds that the EFT-induced shifts on the luminosities are smaller
than their standard deviation.
The biggest differences, relative to uncertainties, are observed in the
quark-antiquark luminosities for $m_X \gsim$ 500 GeV.
This finding can be understood from the fact that the NC Drell-Yan cross section
is proportional to the $u\bar{u}$ and $d\bar{d}$ combinations at leading order,
but the up and down quark PDFs are already well constrained by lower-energy DIS measurements.
Furthermore, we have verified
that the size PDF uncertainties is unchanged in the SMEFT fits.
The results of  Fig.~\ref{fig:SMEFT_lumis} are consistent with those of 
Table~\ref{tab:bound1w} and demonstrate
that, with current data, the interplay between EFT effects and 
PDFs in the high-mass Drell-Yan tails
is appreciable but remains subdominant as compared to other sources of uncertainty.

One important question in this context concerns how one could disentangle the EFT-induced shifts in
the PDF luminosities   displayed in Fig.~\ref{fig:SMEFT_lumis} (see also
the corresponding PDF comparisons in Fig.~\ref{fig:SMEFT_PDFs}) from other possible sources
of deviations, such as internal inconsistencies in some datasets or missing higher
orders in the SM calculations.
An attractive strategy in this respect is based on exploiting the energy-growing
effects associated to the higher-dimensional EFT operators, which translate into an enhanced
sensitivity to the $\hat{W}$ and $\hat{Y}$ parameters for large values of the
 dilepton invariant mass $m_{\ell\ell}$.
To this purpose, it is useful to define the following ratio:
\be
\label{mllratio}
R_{\chi^2}\lp m_{\ell\ell}^{(\rm max)} ,\hat{W},\hat{Y}\rp\equiv \frac{\chi^2\lp m_{\ell\ell}^{(\rm max)}, \hat{W},\hat{Y}
  \rp}{ \chi^2\lp m_{\ell\ell}^{(\rm max)}=120~{\rm GeV} ,\hat{W},\hat{Y}\rp } \, ,
\ee
where  $m_{\ell\ell}^{(\rm max)}$ is the upper bound
on the value of the dilepton invariant mass bins that enter the $\chi^2$ calculation.
In Eq.~(\ref{mllratio}),
both the numerator and the denominator are evaluated
using either $\chi^2_{\rm smp}$ (for the SM PDFs) or $\chi^2_{\rm eftp}$ (for the SMEFT PDFs), and 
the denominator corresponds to the  $\chi^2$ value (per data point) in the kinematic
region for which EFT effects are negligible.\footnote{
  Note that Eq.~(\ref{mllratio}) is computed {\it a posteriori} using existing fits, and that the kinematical
cut in $m_{\ell\ell}^{(\rm max)}$ is absent from the actual fits and it is only evaluated as a diagnosis tool.}

The $R_{\chi^2}$ estimator defined in Eq.~(\ref{mllratio}) allows for the isolation of the contribution to
the total $\chi^2$ that arises from the high-$m_{\ell\ell}$ bins that dominate the overall sensitivity
to the  $\hat{W}$ and $\hat{Y}$ parameters.
For small values of $m_{\ell\ell}^{(\rm max)}$, say 200 GeV, one is cutting away all
$m_{\ell\ell}$ bins with EFT sensitivity and hence one expects $R_{\chi^2}\simeq 1$.
As  $m_{\ell\ell}^{(\rm max)}$ is increased, the $\chi^2$ will include
the contributions from the $m_{\ell\ell}$ bins more sensitive to  EFT effects,
and thus one expects to find a large deviation with respect to the $R_{\chi^2}\simeq 1$ reference value.
Furthermore, EFT effects should induce an approximately monotonic growth of $R_{\chi^2}$  with
 $m_{\ell\ell}^{(\rm max)}$,
which would instead be absent from other possible sources of PDF distortion and thus represent
a smoking gun for BSM physics in the high-energy DY tails.

\begin{figure}[t]
\begin{center}
\includegraphics[width=0.49\textwidth]{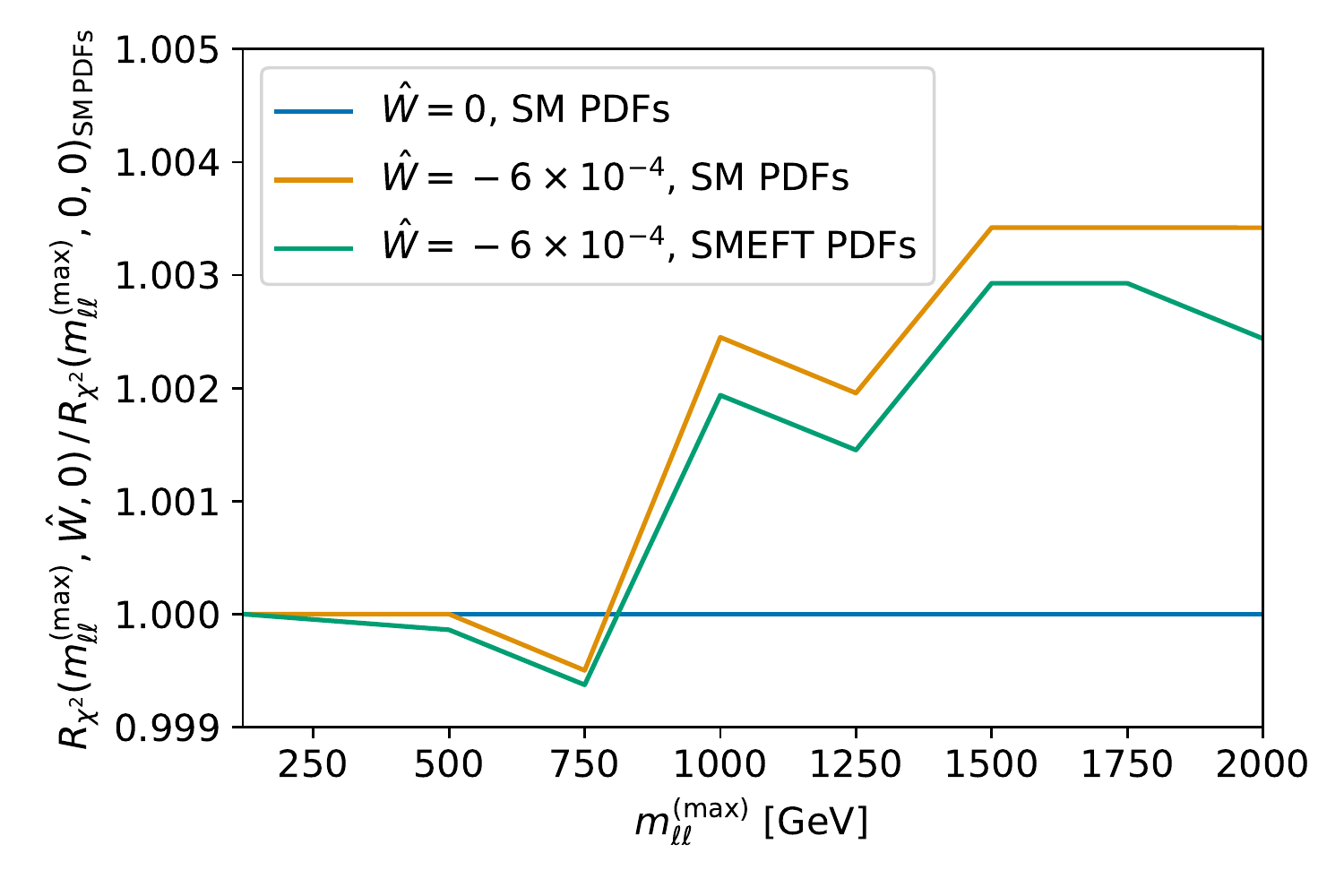}
\includegraphics[width=0.49\textwidth]{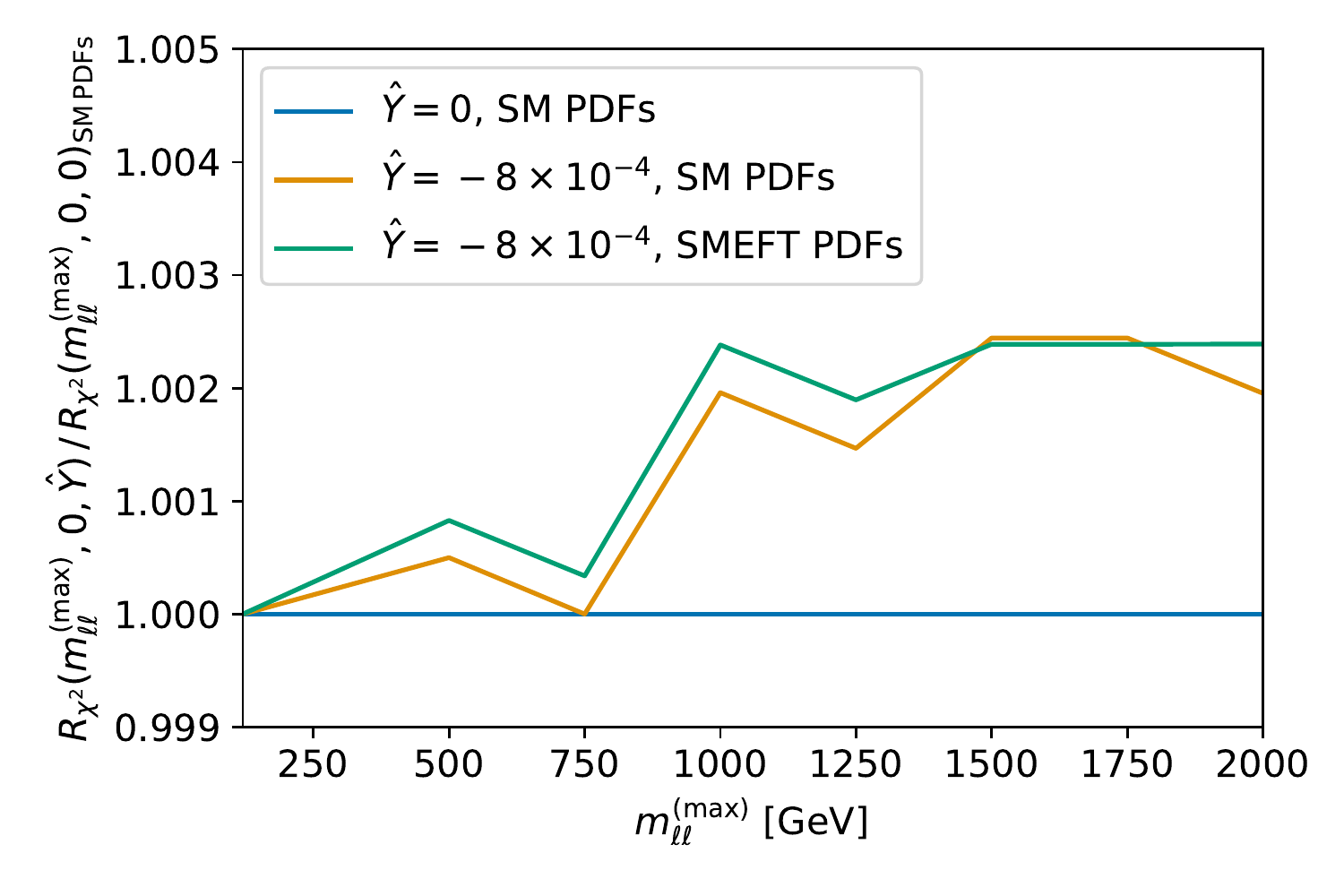}
\caption{\label{fig:cutmax} The $R_{\chi^2}$ estimator, Eq.~(\ref{mllratio}),
  normalised to its SM value,
  as a function for $m_{\ell\ell}^{(\rm max)}$ for representative values of $\hat{W}$
  (left) and $\hat{Y}$ (right panel).
  We display the results obtained both with SM and SMEFT PDFs,
  with the horizontal line indicating the reference SM value of $R_{\chi^2}$.
}
\end{center}
\end{figure}

These expectations are verified in Fig.~\ref{fig:cutmax}, which displays
the $R_{\chi^2}$ estimator
(normalised to its SM value)
  as a function for $m_{\ell\ell}^{(\rm max)}$ for representative values of the $\hat{W}$
  and $\hat{Y}$ parameters both for the SM and the SMEFT PDFs,
   where the horizontal line indicates its reference SM value.
  Indeed we observe an approximately monotonic growth of $R_{\chi^2}$ arising
  from the energy-growing effects in the EFT.
  Due to the limited experimental information the binning in $m_{\ell\ell}$
  is rather coarse, explaining the observed fluctuations.
  In the specific case of the $\hat{W}$ parameter, the SMEFT PDF curve
  lies slightly below the SM PDF one,
  highlighting how EFT effects
  are being partially (but not completely) reabsorbed into the PDFs.

\subsection{EFT constraints on scenario II from current high-mass
  Drell-Yan data}
\label{sec:scenarioIIresults}

In contrast to benchmark scenario I, which is flavour
universal, the second SMEFT scenario to be explored in this work and
described in Sect.~\ref{sec:scenarioII} contains a four-fermion
interaction involving muons but not electrons, which therefore modifies
the rates of the dilepton process $pp\to \mu^+\mu^-$ but not
those of $pp\to e^+e^-$.
This property implies that, without introducing further assumptions, the
Wilson coefficient ${\bf C}^{D\mu}_{33}$ can be only constrained from DY measurements carried out
in the dimuon (rather than in the dielectron or in the combined) final
state.
As indicated in Table~\ref{tab:data-high-mass}, only the
CMS data at 7 TeV and 13 TeV include DY distributions in
the dimuon final state.

Due to these restrictions in the input dataset, the
interplay between PDFs and  SMEFT effects is expected
to be milder as compared to the results presented in 
Sect.~\ref{sec:results_scenarioI}.
For this reason, here we do not attempt to perform a joint determination of
the PDFs and the ${\bf C}^{D\mu}_{33}$ coefficient, but rather restrict
ourselves to quantifying the information that available DY data in the dimuon
final state provide on this operator. We instead present a simultaneous
determination including projections for the HL-LHC in Sect.~\ref{sec:hllhc}.

\begin{figure}[t]
\begin{center}
  \includegraphics[width=0.8\textwidth]{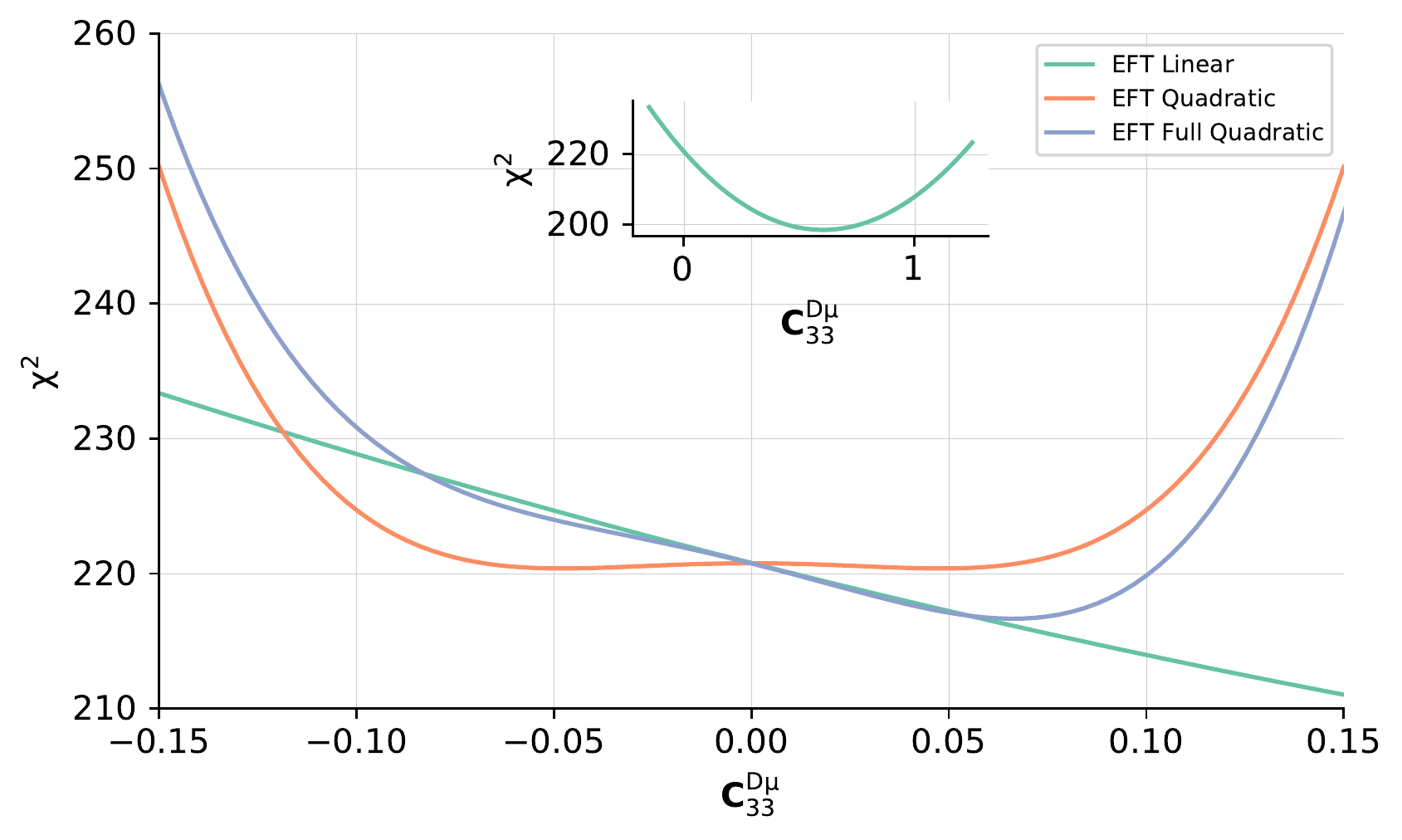}
  \caption{\label{fig:parabola_ii} The results of
    polynomial fits to  $\chi^2\lp {\bf C}^{D\mu}_{33}\rp$, Eq.~(\ref{eq:parabolicfit}),
    in scenario II.
    This $\chi^2$ includes only the contributions
    from the two DY measurements in the dimuon final state.
    We display results for fits based on cross sections that account only for the linear,
    only for the quadratic, and for both linear and quadratic terms in the EFT expansion,
    in all cases using the baseline SM PDF set.
    The inset displays the fit to the linear EFT values with an enlarged $x$-axis range.
     }
\end{center}
\end{figure}

Fig.~\ref{fig:parabola_ii} displays the results
of three quartic fits to the $\chi^2\lp {\bf C}^{D\mu}_{33}\rp$ profile in
benchmark scenario II, based on Eq.~(\ref{eq:quarticfit}),
 where here $\chi^2_{\rm smp}$ includes only the contributions
    from the two available DY measurements in the dimuon final state.
We present fits based on cross sections that account only for the linear,
only for the quadratic, and for both the linear and quadratic terms in the EFT expansion.
In all cases, these cross sections are computed using the baseline SM PDF set.
     The inset displays the outcome of the linear EFT fit with an enlarged $x$-axis range.

The results of Fig.~\ref{fig:parabola_ii} indicate that ${\bf C}^{D\mu}_{33}$ is essentially unconstrained
at the linear EFT level, and only once quadratic corrections ${\cal O}(\Lambda^{-4})$ are accounted for
is one able to obtain reasonable bounds on this coefficient.
The reason for this behaviour
is that for this operator the interference with the SM amplitude is suppressed,
and hence the leading EFT effects arise at the quadratic level from the square of the
EFT amplitude, thus being proportional to $\lp {\bf C}^{D\mu}_{33}\rp^2$ \cite{Greljo:2017vvb}.
In the case of the polynomial fit to the $\chi^2$ profile evaluated on the full quadratic
EFT cross sections, we find the following 95\% CL limits on this Wilson coefficient:
\begin{equation}
  \label{eq:cdmu33bound_1}
  \lp {\bf C}^{D\mu}_{33}\times 10^2\rp \in \left[-1.2, 10.7 \right] \, ,
\end{equation}
which can be compared with the  bounds on the same
operator obtained in~\cite{Greljo:2017vvb} from recasting
the ATLAS dilepton search data of~\cite{ATLAS:2017wce},
given by Eq.~\eqref{eq:scen2}.
The fact that our bound in Eq.~(\ref{eq:cdmu33bound_1})
is around a factor three looser than in Eq.~(\ref{eq:scen2})
is explained because the dilepton search data from~\cite{ATLAS:2017wce}
benefits from an extended coverage in $m_{\ell\ell}$ as compared
to the available unfolded DY cross sections.
The same result, this time for the
$\hat{W}$ and $\hat{Y}$ parameters,
 will be obtained in the next section
where we assess the impact of the SMEFT PDFs
in the EFT interpretation of the ATLAS dilepton search dataset.

\subsection{On the EFT interpretation of high-mass dilepton searches}
\label{sec:res2}

As mentioned above, a single high-mass DY cross section measurement
is available at 13~TeV, and even in this case it is only based on a small subset of the Run II
luminosity.
As a consequence, the highest energy bin of this dataset is rather wide, $m_{\ell\ell}\in \lc 1.5, 3.0\rc$ TeV.
This implies limited sensitivity to deviations in the tails of DY distributions, 
for which using a large number of narrow bins is most beneficial to constrain heavy
resonances, for instance.
Here we would like to quantify the interplay between PDF and EFT effects at the level
of a recent ATLAS 13~TeV search for $Z'$ bosons in the dilepton channel~\cite{Aad:2019fac}
based on the complete Run II luminosity of $\mathcal{L}=139$ fb$^{-1}$.
Since these are detector-level measurements, which cannot therefore be included
in a PDF analysis, our aim is to use the SMEFT PDFs to investigate
how the bounds on BSM physics are modified as compared to the standard
approach based on computing theory predictions using SM PDFs.

\begin{figure}[t]
  \centering
  \includegraphics[width=1.\textwidth]{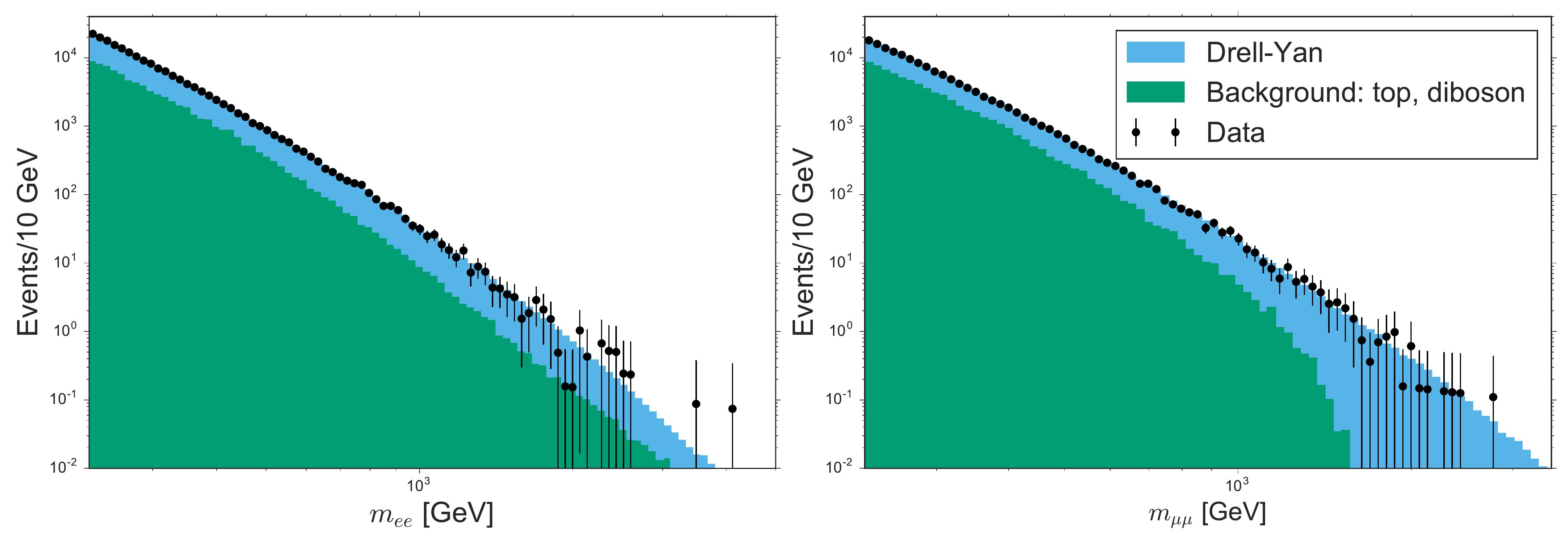}
  \caption{\label{fig:search} The data (number of events per 10 GeV bin)
    from the ATLAS $Z'$ search from~\cite{Aad:2019fac} 
  in the di-electron (left) and di-muon (right) channels.
  We also
  display the theoretical predictions associated to the contributions from  Drell-Yan
  and from the rest of the backgrounds, taken from the ATLAS publication.}
\end{figure}

The ATLAS data, displayed in Fig.~\ref{fig:search},
consist of event counts in 100 dilepton invariant mass, $m_{\ell\ell}$, bins in
both the dimuon and dielectron channels in the range
$m_{\ell\ell} \in (225,6000)$ GeV.  We take this data from
\texttt{HEPdata}~\cite{Maguire:2017ypu}
and denote the event count in the $i^{\rm th}$ bin by $n_{i}$.
The narrow binning and broad  $m_{\ell\ell}$ coverage allowed ATLAS to constrain $Z'$ masses to
$M_{Z'} \gsim 4$ TeV.  This is a much higher reach than the DY cross section
measurements used for the SMEFT PDF fits in
the previous subsections, and
should therefore provide  stronger constraints on the EFT benchmark scenarios described in
Sect.~\ref{sec:scenarios}.
By including this search data in our study,
we can investigate whether such strong constraints are sensitive to
the EFT-induced modifications in the PDF luminosities highlighted in Fig.~\ref{fig:SMEFT_lumis}.

In order to constrain the $\hat{W}$ and $\hat{Y}$ parameters in benchmark scenario I
from the ATLAS dilepton search data, 
for each bin we compute a theory prediction $y_{i} = y_{i}(\hat{W},\hat{Y})$ 
given by the sum of background $b_{i}$ (top, diboson) and signal $s_{i}(\hat{W},\hat{Y})$ 
(Drell-Yan) components. 
The ATLAS search provides an estimate of the total SM contribution (sum of top, diboson, and DY) 
without a breakdown into components.  This estimate is provided as a continuous function of $m_{\ell \ell}$. 
We thus estimate our background (top and diboson) 
by subtracting our own DY simulation from the estimated total SM event counts
found by evaluating this function at each bin centre.
We compute the DY signal in each bin as
\begin{equation}
  \label{eq:searchSi}
s_{i}(\hat{W},\hat{Y}) = s_{i, \textrm{SM}} \times K(\hat{W},\hat{Y})\, ,
\end{equation}
where $s_{i, \textrm{SM}}$ indicates the detector-level prediction
for the $i^{\rm th}$ bin of the $m_{\ell\ell}$ distribution
evaluated at NLO QCD using \amc~\cite{Alwall:2014hca}, {\tt Pythia}~\cite{Sjostrand:2014zea}
and {\tt Delphes}~\cite{deFavereau:2013fsa} and using the {\tt
  NNPDF31\_nnlo\_as\_0118} set as PDF input set.
In Eq.~(\ref{eq:searchSi}), 
$K(\hat{W},\hat{Y})$ is the $K$-factor calculated as the ratio of cross sections in each bin, accounting 
for the impact of non-zero  EFT corrections $\hat{W},\hat{Y}\ne 0$
both in the partonic cross section and in the PDFs 
\begin{equation}
\label{eq:pdfsmeft}
K_{\rm eftp}(\hat{W},\hat{Y}) \equiv \frac{\sum_{q} \displaystyle \int d \tau \mathcal{L}^{\rm SMEFT}_{q \bar{q}} (\tau, \mu_{F}, \hat{W},\hat{Y}) \hat{\sigma}(\tau s_{0}, \hat{W},\hat{Y} )}{\sum_{q} \displaystyle \int d \tau \mathcal{L}^{\rm SM}_{q \bar{q}} (\tau, \mu_{F}) \hat{\sigma}(\tau s_{0}, 0,0)},
\end{equation}
where the integration in $\tau$ goes from $\tau_{\rm min}$ to $\tau_{\rm max}$ in each bin.
For comparison, we will also present results where the $K$-factor is instead
evaluated as usual in terms of the SM PDFs,
\begin{equation}
\label{eq:pdfsm}
K_{\rm smp}(\hat{W},\hat{Y}) \equiv \frac{\sum_{q}\displaystyle \int d \tau \mathcal{L}^{\rm SM}_{q \bar{q}} (\tau, \mu_{F}) \hat{\sigma}(\tau s_{0}, \hat{W},\hat{Y})}{\sum_{q}\displaystyle \int d \tau  \mathcal{L}^{\rm SM}_{q \bar{q}} (\tau, \mu_{F}) \hat{\sigma}(\tau s_{0}, 0,0)} \, .
\end{equation}
The likelihood $\mathcal{L}$ is defined as the product of Poisson probabilities in each bin,
\begin{equation}
\mathcal{L}(n | \hat{W},\hat{Y}) = \prod_{i=1}^{m} \frac{y_{i}(\hat{W},\hat{Y})^{n_{i}}}{n_{i}!} e^{-y_{i}(\hat{W},\hat{Y})}\, ,
\end{equation}
and the best-fit values of $\hat{W}$, $\hat{Y}$ are determined as the maximum-likelihood estimates.  
The test statistics in the individual fits of $\hat{W}$ and $\hat{Y}$ are the profile likelihood ratios defined as
\begin{eqnarray}
\lambda(\hat{W}) &=& - 2 \ln \lp \frac{\mathcal{L}(n|\hat{W},0)}{\mathcal{L}(n|\hat{W}, 0)_{\rm max}}\rp\\
\lambda(\hat{Y}) &=& - 2 \ln \lp \frac{\mathcal{L}(n|0,\hat{Y})}{\mathcal{L}(n|0, \hat{Y})_{\rm max}}\rp \nonumber
\end{eqnarray}
and follow a $\chi^{2}$ distribution with $n_{\rm dof} = 1$. The $1\sigma$ and $2 \sigma$ bounds 
are found by solving the implicit equations $\lambda(\hat{W}) = 1,4$ respectively. 

PDF uncertainties can be included by taking the confidence level
intervals on the bounds given by the NNPDF Monte Carlo replicas, as it
is explained in Sect.~3.4. 

\begin{figure}[t]
  \centering
  \includegraphics[width=0.49\textwidth]{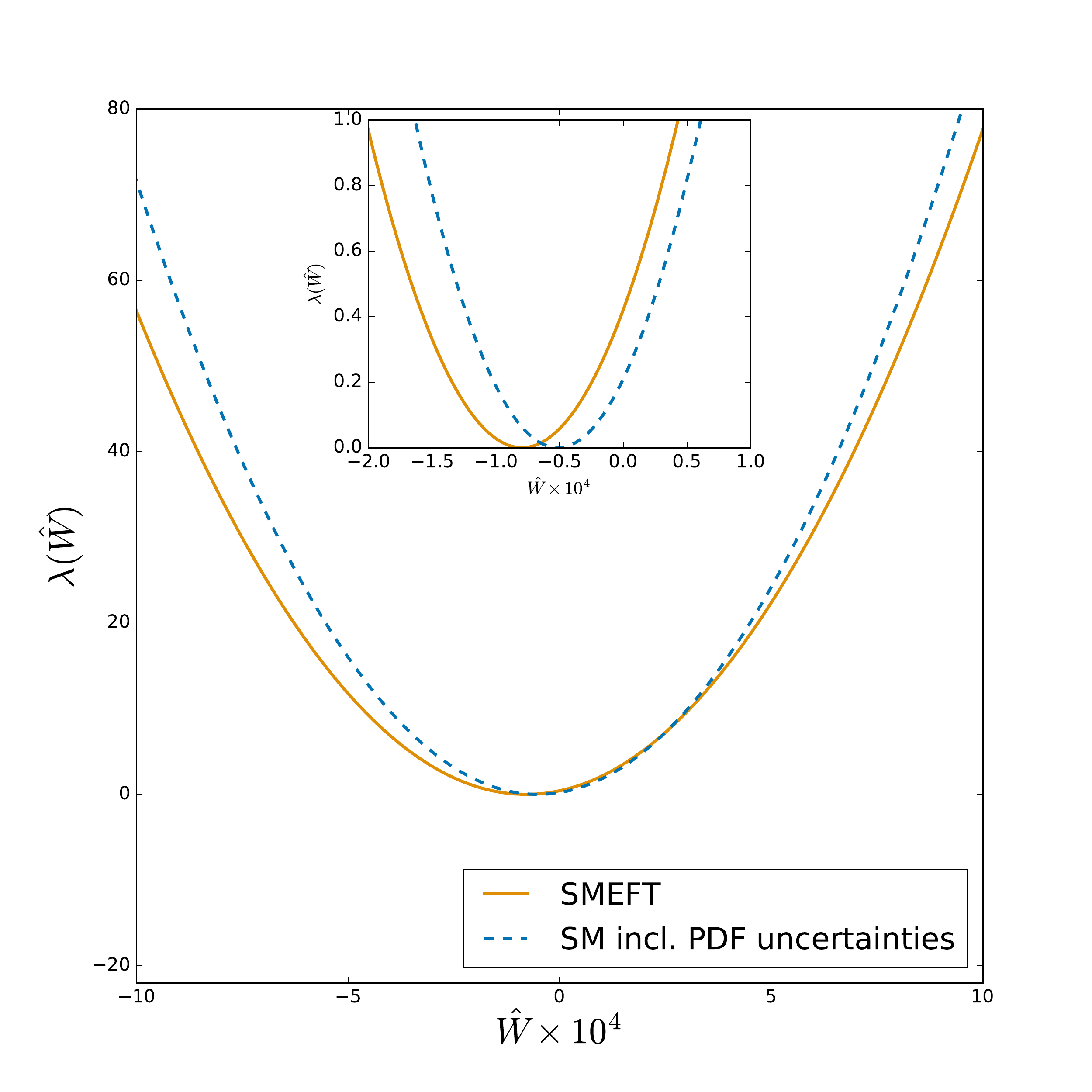}
  \includegraphics[width=0.49\textwidth]{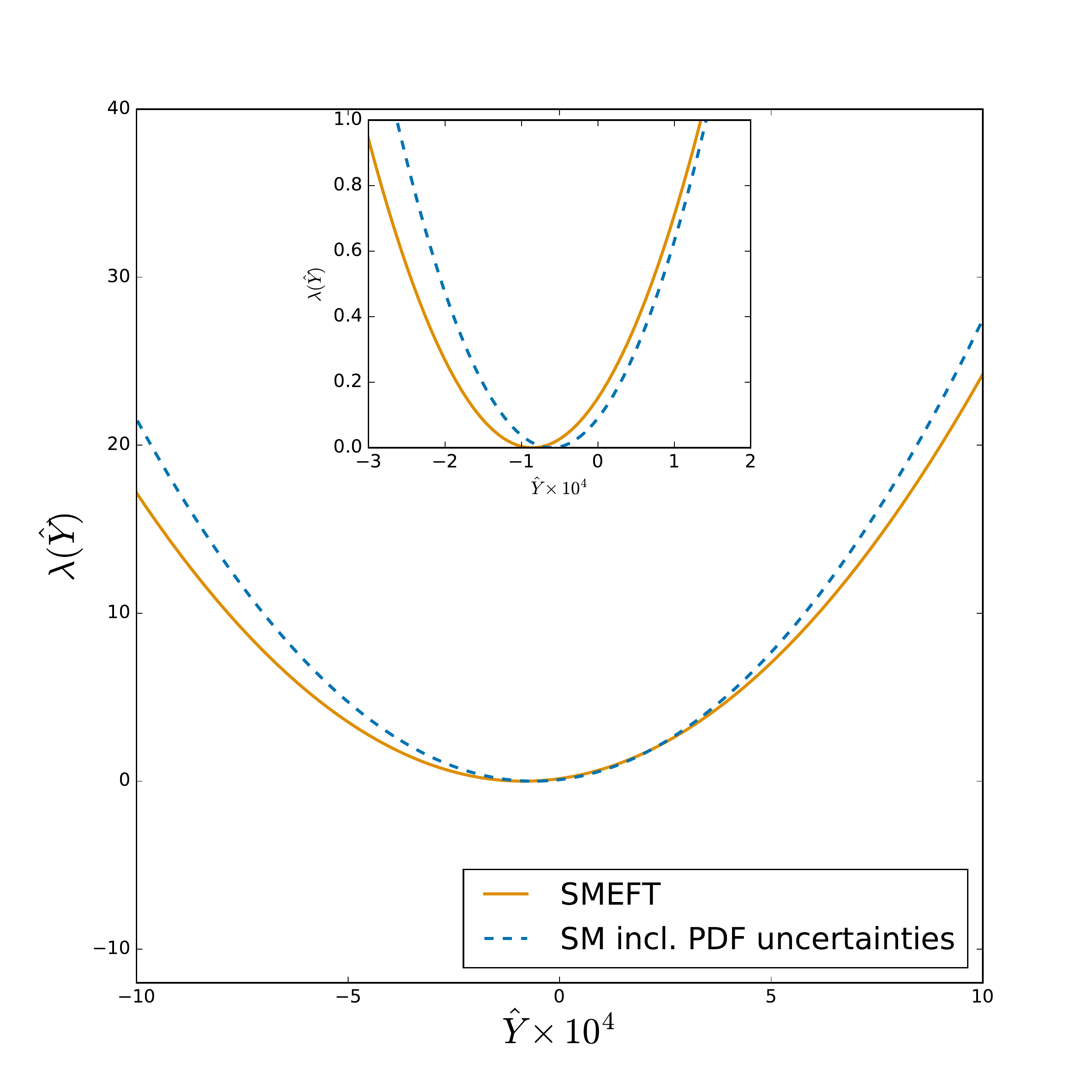}
  \caption{\label{fig:search1D}Comparison
      between the results of the parabolic fits to the ATLAS search data~\cite{Aad:2019fac}
      for $\hat{W}$ (left) and $\hat{Y}$ (right panel) when
      using the SMEFT PDFs ($\chi^2_{\rm eftp}$)
      as compared to the SM PDF baseline ($\chi^2_{\rm smp}$).
      The insets zoom on the region close to $\lambda(\hat{W}), \lambda(\hat{Y}) \simeq 0$.}
\end{figure}

 The results for the fits obtained by using SM PDFs in the $K$-factors, Eq.~\eqref{eq:pdfsm},
 compared to those obtained by using SMEFT PDFs, Eq.~\eqref{eq:pdfsmeft}, are displayed 
 in Fig.~\ref{fig:search1D}, with the corresponding bounds being provided in Table~\ref{tab:search1D}.
 We find that inclusion of PDF uncertainties has a 
 much smaller impact on the
 parabolic fit than in the previous analysis and thus only the
 parabola including PDF uncertainties is displayed. 
%
Secondly, we observe that the shift in the bounds 
that one has using SM versus SMEFT PDFs is not entirely negligible.
This finding indicates that it is important to use consistent PDFs determined 
with the same settings as the theoretical predictions in the partonic cross section.
We can similarly recast the ATLAS search data to constrain the scenario II Wilson coefficient ${\bf C}_{33}^{D \mu}$, finding
the following constraints at 95\% CL:
\begin{equation}
  \label{eq:cdmu33bound_search}
\lp {\bf C}^{D\mu}_{33}\times 10^2\rp \in \left[ -1.6, 2.4 \right] \, ,
\end{equation}
in good agreement with the results reported in Eq.~\eqref{eq:cdmu33bound_1}.

\begin{table}
 \renewcommand{\arraystretch}{1.40}
 \centering
   \begin{tabular}{l|c|c|c|c}
    & $\quad$ SM PDFs $\quad$  & SMEFT PDFs  & best-fit shift  & broadening  \\
    \toprule
    \multirow{2}{*}{$\hat{W}\times 10^4$ (68\% CL)} & $[-1.6, 0.6] $ & \multirow{2}{*}{$[-1.9, 0.5]$}  & $-0.2$    &+9\%\\
    & $[-1.6, 0.6] $ &   &  $-0.2$   & +9\% \\
    \midrule
    \multirow{2}{*}{$\hat{W}\times 10^4$ (95\% CL)} & $[-2.7, 1.7] $ &  \multirow{2}{*}{ $[-3.1, 1.6] $} & $-0.25$  & +7\%\\
     &   $[-2.7, 1.7] $   &  & $-0.25$      &+7\% \\
    \midrule
    \multirow{2}{*}{$\hat{Y}\times 10^4$ (68\% CL)} & $[-2.6, 1.5] $ &  \multirow{2}{*}{$[-3.1, 1.4]$} & $-0.3$ &  +10\%\\
       & $[-2.6, 1.6] $ &   & $-0.35$ & +7\% \\
    \midrule
    \multirow{2}{*}{$\hat{Y}\times 10^4$ (95\% CL)} & $[-4.7, 3.5] $ &  \multirow{2}{*}{$[-5.3, 3.6]$} & $-0.25$ & +9\% \\
     & $[-4.7, 3.6] $ &   & $-0.30$ & +7\% \\
    \bottomrule
   \end{tabular}
   
  \caption{ \label{tab:search1D}\small The 68\% and 95\% CL bounds on the $\hat{W}$ and $\hat{Y}$
    parameters obtained from the  corresponding parabolic fits to the 
    ATLAS search data of~\cite{Aad:2019fac} when using either the SMEFT PDFs  
    or their SM counterparts.
         For the SM PDF results, we indicate the bounds obtained without (upper)
    and with (lower entry) PDF uncertainties accounted for; the SMEFT PDF
    bounds already include  PDF uncertainties by construction.
    The fourth and fifth column indicate the absolute shift in best-fit values
    and the percentage broadening of the EFT parameter uncertainties when the SMEFT PDFs
    are consistently used instead of the SM PDFs.
}
\end{table}

\subsection{Overview of current constraints}
\begin{figure}[t]
\begin{center}
\includegraphics[width=0.99\textwidth]{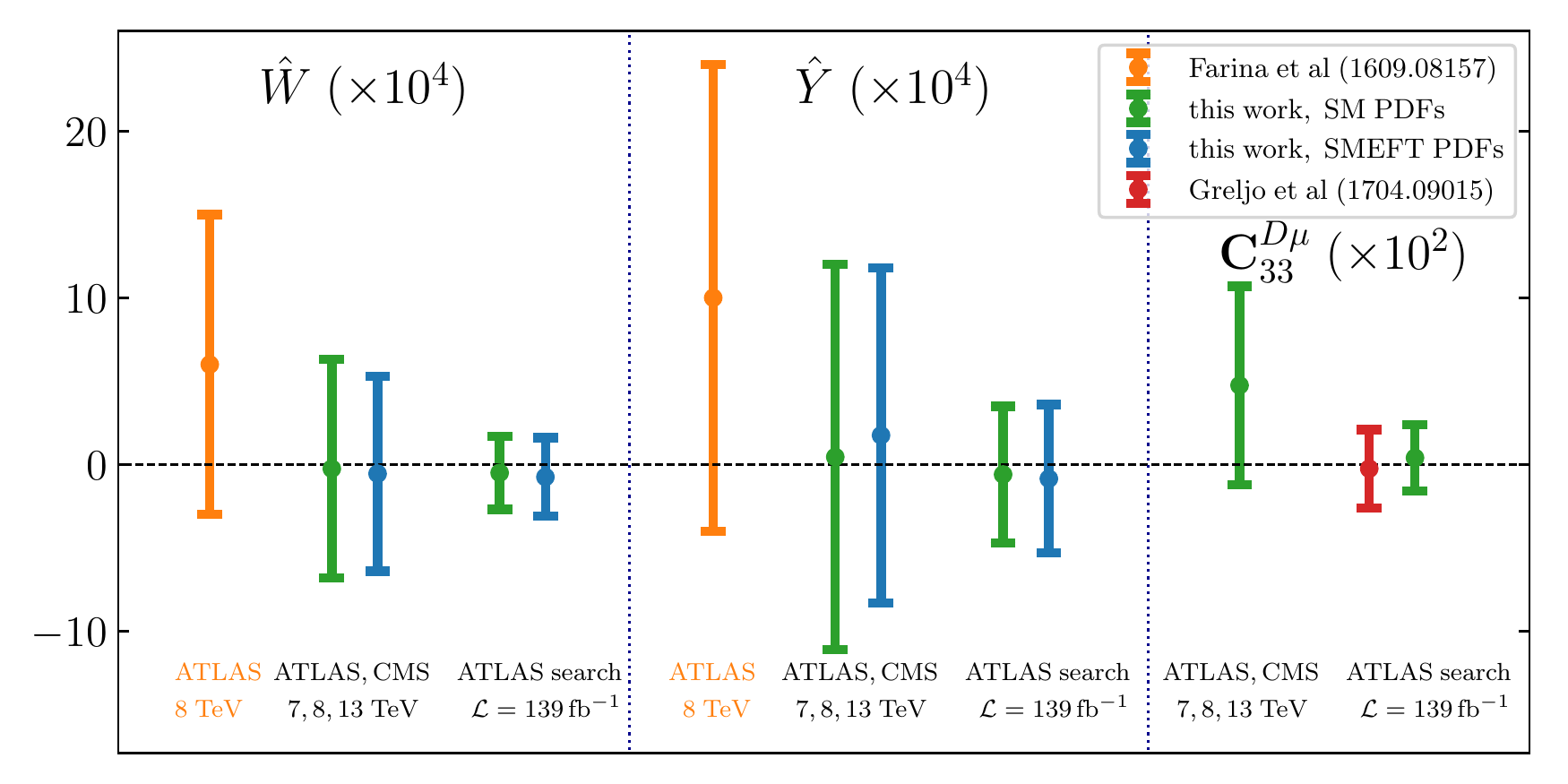}
\caption{\label{fig:plot_bounds.pdf} Overview of the results obtained in this
  section concerning the EFT parameters $\hat{W}$ and $\hat{Y}$ (in scenario I)
  and ${\bf C}^{D\mu}_{33}$ (in scenario II).
  We compare the 95\% CL bounds derived in~\cite{Farina:2016rws} with those
  obtained in this work from the high-mass DY cross section
  measurements (Table~\ref{tab:bound1w}) and from the ATLAS $Z'$ search data (Table~\ref{tab:search1D}),
  in both cases displaying the
  results obtained with either the SM or the SMEFT PDFs.
  In the former case, we indicate the results that account for PDF uncertainties; these
  are included by construction for the SMEFT PDFs.
}
\end{center}
\end{figure}

In order to summarise the results obtained in this section,
Fig.~\ref{fig:plot_bounds.pdf} displays the 95\% CL bounds
derived on the EFT parameters $\hat{W}$ and $\hat{Y}$ (in scenario I)
and  on ${\bf C}^{D\mu}_{33}$ (in scenario II),
both from the high-mass DY cross section
measurements (Table~\ref{tab:bound1w}) and from the ATLAS $Z'$ search data (Table~\ref{tab:search1D}).
These bounds are shown in the case of theoretical calculations
evaluated either with  SM PDFs or with SMEFT PDFs,
and in  the former case we indicate the results that account for PDF uncertainties (these
  are included by construction for the SMEFT PDFs).
To compare with previous works, we also display the bounds derived in~\cite{Farina:2016rws} for
the $\hat{W}$ and $\hat{Y}$ parameters
from the ATLAS 8 TeV data and in~\cite{Greljo:2017vvb} for the ${\bf C}^{D\mu}_{33}$ coefficient
from the same ATLAS $Z'$ search data.

As discussed above, our main findings are that the consistent simultaneous determination of the PDFs together
with the EFT parameters leads to a moderate increase in the uncertainties (in this case,
up to 10\%)
as well as to a small shift in their central values.
As we demonstrate in the next section, the interplay between PDFs and EFT coefficients becomes
much more marked in the case of the high-mass DY measurements that will become available
at the HL-LHC.

\section{Projections for the High-Luminosity LHC}
\label{sec:hllhc}
The results presented in the previous section indicate that,
given the available unfolded Drell-Yan measurements,
the impact of a  simultaneous determination of the PDFs together with the EFT parameters
remains moderate.
However, it is conceivable that this
interplay between PDFs and BSM effects in the high-energy tails of
Drell-Yan cross sections will become more significant once more data are accumulated.
With this motivation, we revisit the analysis of Sect.~\ref{sec:res1}
now accounting for the impact of projected High-Luminosity LHC pseudo-data generated for the present study.
                 We demonstrate that in the scenario under
                 consideration, in which no other data apart from the
                 high-mass Drell-Yan constrain the large-$x$ quark and
                 antiquark distributions, a consistent joint
                 determination of PDFs is crucial for EFT studies at
                 the HL-LHC. We will also discuss how the inclusion of
                 further LHC data, which can constrain the large-$x$
                 region without being affected by potential
                 energy-growing new physics effects,
                 can soften the interplay observed in this study and
                 disentangle new-physics effects.

\subsection{Generation of HL-LHC pseudo-data}

Following the strategy adopted in~\cite{Khalek:2018mdn} to estimate the ultimate PDF reach
of the HL-LHC measurements (see also~\cite{Azzi:2019yne,Cepeda:2019klc}), here
we generate HL-LHC pseudo-data for NC and CC high-mass Drell-Yan
cross sections at $\sqrt{s}=14$ TeV and for a total integrated luminosity of $\mathcal{L}=6$ ab$^{-1}$
(from the combination of ATLAS and CMS, which provide
$\mathcal{L}=3$ ab$^{-1}$ each).
For these projections, theoretical predictions are evaluated at NNLO
in QCD including NLO EW corrections, as is
explained in detail in Sect.~\ref{sec:theory}. The PDF set used as an input to
generate the theoretical prediction is the DIS+DY baseline that was presented
in Sect.~\ref{sec:fitsettings}.

For the generation of the NC pseudo-data, we adopt as reference the CMS  measurement
at 13 TeV~\cite{Sirunyan:2018owv} based on $\mathcal{L}=2.8$ fb$^{-1}$.
The dilepton invariant mass distribution $m_{\ell\ell}$ is evaluated using the same selection
and acceptance cuts of~\cite{Sirunyan:2018owv} but now with an extended
binning in the $m_{\ell\ell}$ to account for the increase in luminosity.
We assume equal cuts for electrons and muons and impose $|\eta_\ell|\le 2.4$,
$p_T^{\rm lead}\ge 20$ GeV, and $p_T^{\rm sublead}\ge 15$ GeV for the two
leading charged leptons of the event.
In the case of the  CC pseudo-data, the lack of unfolded measurements
of the $m_T$ distribution at 13 TeV to be used as reference forces
us to base our projections on the
ATLAS search for $W'$ bosons in the dilepton channel~\cite{Aad:2019wvl}.
As in the case of the NC projections, theory predictions for
the $m_T$ distribution at high-mass are generated
using the same selection and acceptance cuts as in~\cite{Aad:2019wvl}
but now using an extended coverage in $m_T$.

\begin{figure}[t]
\begin{center}
  \includegraphics[width=0.49\textwidth]{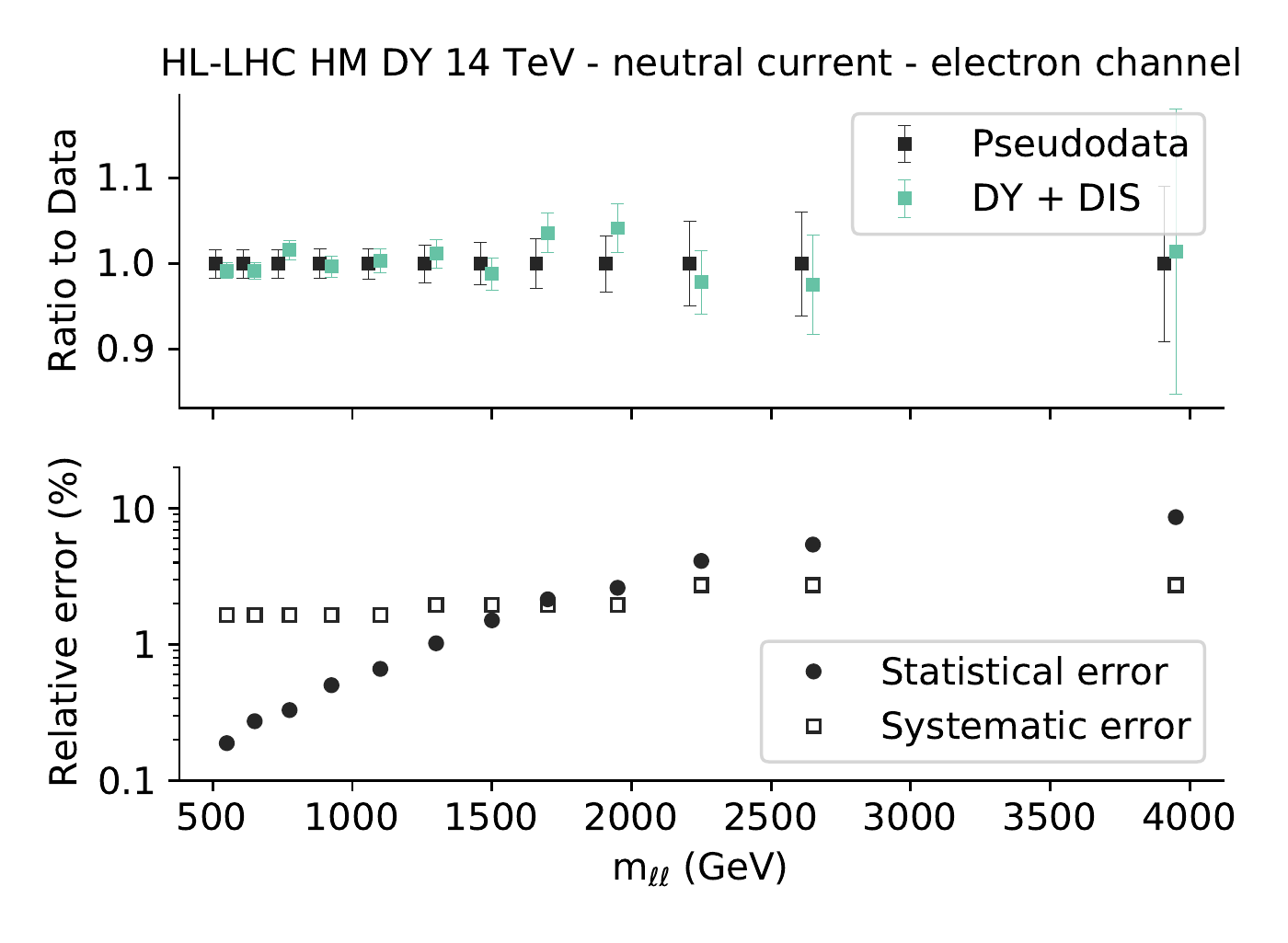}
  \includegraphics[width=0.49\textwidth]{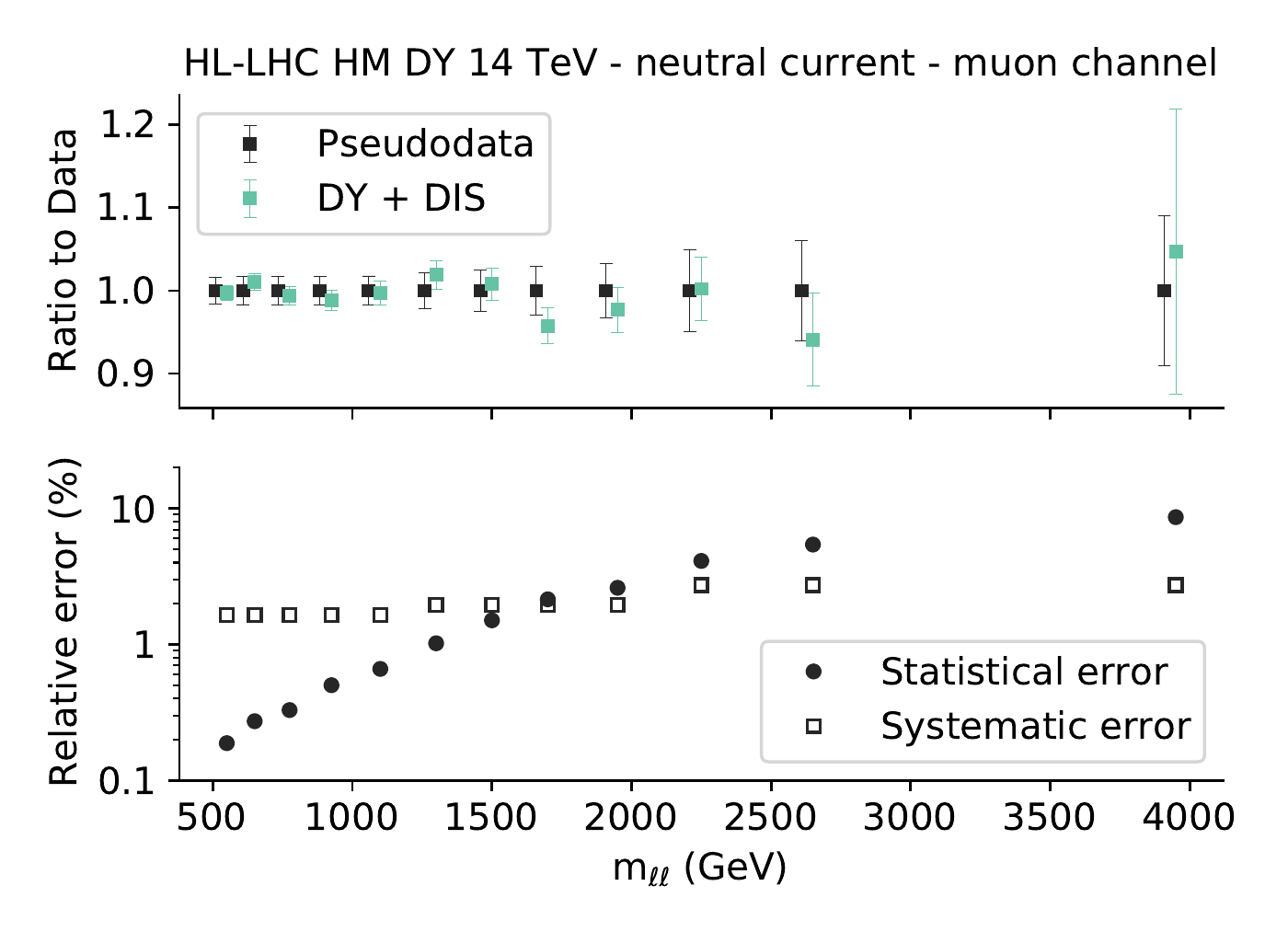}
  \caption{\label{fig:hllhc-nc} Top panels: comparison  of the
    projected HL-LHC pseudo-data for high-mass neutral-current Drell-Yan in the dielectron (left)
    and dimuon (right) final states
    as a function of $m_{\ell\ell}$ with
    the corresponding theory predictions obtained from the SM PDF
    baseline. The theoretical predictions, generated according to
    Eq.~\eqref{eq:hllhc}, are accompanied by their corresponding PDF
    uncertainties (green bars). 
    Lower panels: the percentage statistical and systematic uncertainty in each  $m_{\ell\ell}$ bin
    of the HL-LHC pseudo-data.
    }
\end{center}
\end{figure}

\begin{figure}[t]
\begin{center}
  \includegraphics[width=0.49\textwidth]{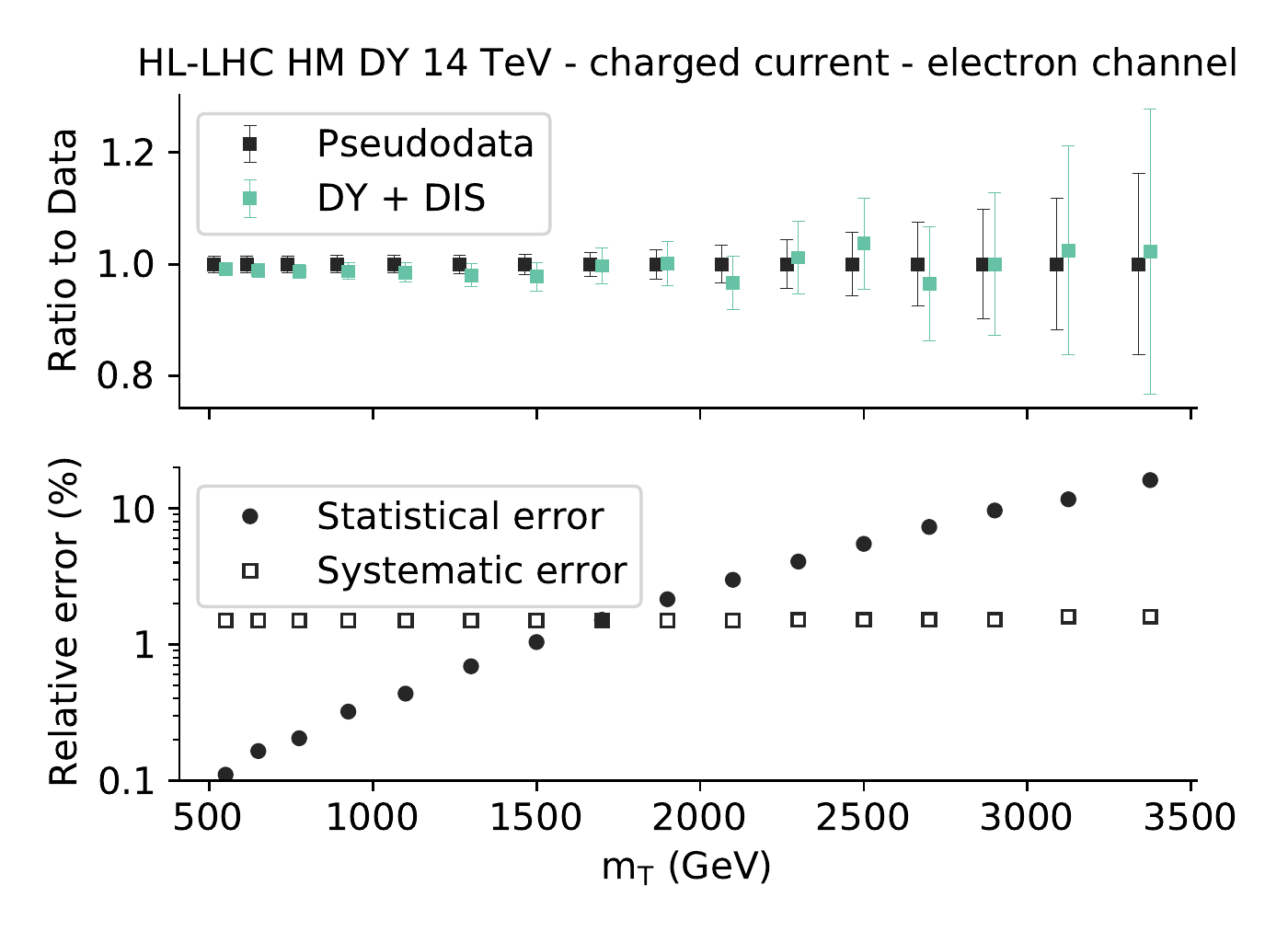}
  \includegraphics[width=0.49\textwidth]{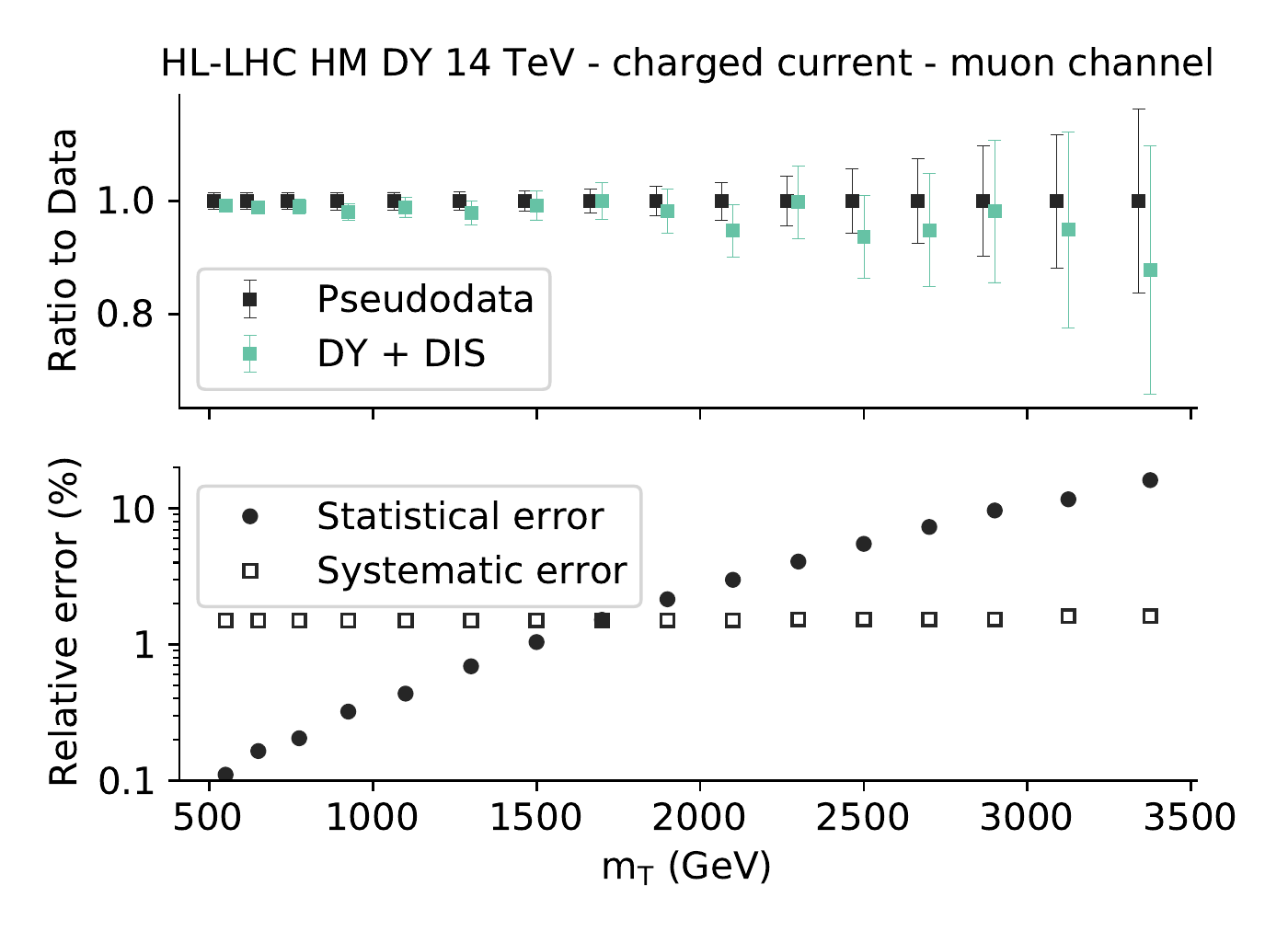}
    \caption{\label{fig:hllhc-cc} Same as Fig.~\ref{fig:hllhc-nc} for
      charged-current Drell-Yan in bins of the transverse mass $m_T$.
    }
\end{center}
\end{figure}

Both in the case of NC and CC Drell-Yan cross sections, we restrict ourselves
to  events with either $m_{\ell\ell}$ or $m_T$ greater than 500 GeV.
Otherwise, the total experimental uncertainty would be limited by our modelling
of the expected systematic errors and thus our projections could become unreliable.
Furthermore, we require that the expected number of events per bin
is bigger than 30 to ensure the applicability of Gaussian statistics.
Taking into account these considerations, our choice of binning
for the $m_{\ell\ell}$~($m_T$) distribution at the HL-LHC
is displayed in Fig.~\ref{fig:hllhc-nc}~(Fig.~\ref{fig:hllhc-cc}),
with the highest energy bins reaching $m_{\ell\ell}\simeq 4 $ TeV
($m_T\simeq 3.5$ TeV) for neutral-current (charged-current) scattering.

The percentage statistical and systematic uncertainties associated to the HL-LHC pseudo-data
are displayed in the lower panels of Figs.~\ref{fig:hllhc-nc} and~\ref{fig:hllhc-cc}
and have been estimated as follows.
Let us denote by  $\sigma^{\rm th}_{i}$ the theoretical prediction for the
DY cross section, including all relevant selection cuts
as well as the leptonic branching fractions.
The expected number of events in this bin and the associated (relative)
statistical uncertainty $\delta_i^{\rm stat}$ are given by 
\begin{equation}
  N_{i}^{\rm th}= \sigma^{\rm th}_{i}\times \mathcal{L} \,, \quad \delta_i^{\rm stat}\equiv
  \frac{\lp \delta N_{i}\rp_{\rm stat}}{N_{i}^{\rm th}} = \frac{1}{\sqrt{N_{i}^{\rm th}}} \, .
\end{equation}
Note that this bin-by-bin relative statistical uncertainty is the same both at the level
of number of events and at the level of fiducial cross sections.

The HL-LHC systematic uncertainties are also estimated from the same reference
measurements.
If $\delta_{i,j}^{\rm sys}$ denotes the $j^{\rm th}$ relative systematic uncertainty
associated to the $i^{\rm th}$ bin of the reference
measurement, and if this bin contains $N_i^{\rm th}$ events,
then for our projections we assume that the same systematic error
associated to a bin with a similar number of expected events will be given by 
$f_{{\rm red},j}\delta_{i,j}^{\rm sys}$, where $f_{{\rm red},j}$
is the expected reduction in systematic errors foreseen at the HL-LHC.\footnote{
  The binning of  the CC reference measurement, the ATLAS $W'$ search, is much finer
than for our  HL-LHC projections and hence we first match them by means
of a weighted average.
}
This assumption is justified since most systematic errors improve with the sample size
thanks to {\it e.g.} better calibration.

Adding in quadrature systematic uncertainties with the statistical error,
the total relative uncertainty for the $i$th bin of our HL-LHC projections
is
\begin{equation}
\delta_{{\rm tot},i}^{\rm exp} = \left( \left( \delta_i^{\rm stat}\right)^2 + \sum_{j=1}^{n_{\rm sys}}
\left( f_{{\rm red},j}\delta_{i,j}^{\rm sys} \right)^2\right)^{1/2} \, ,
\end{equation}
where $n_{\rm sys}$ indicates the number of systematic error sources.
The central values for the HL-LHC pseudo-data is then generated
by fluctuating the reference theory prediction by the expected total experimental
uncertainty, namely
\begin{equation}
  \label{eq:hllhc}
\sigma^{\rm hllhc}_{i} \equiv \sigma^{\rm th}_{i} \left( 1+ \lambda
  \delta_{\cal L}^{\rm exp} +  r_i\delta_{{\rm tot},i}^{\rm exp}   \right) \, , \qquad i=1,\ldots,n_{\rm bin} \, ,
\end{equation}
where $\lambda,r_i$ are univariate Gaussian random numbers, $\delta_{{\rm tot},i}^{\rm exp}$
is the total (relative) experimental uncertainty corresponding to this
specific bin
(excluding the luminosity and normalisation uncertainties), and $\delta_{\cal L}^{\rm exp}$
is the luminosity uncertainty, which is fully correlated amongst all
the pseudo-data bins of the same experiment. We take this luminosity uncertainty to be
$\delta_{\cal L}^{\rm exp}=1.5$\%  for both ATLAS and CMS, as done in Ref.~\cite{Khalek:2018mdn}.

Here we adopt the baseline SM PDF set described in Sect.~\ref{sec:res1},
which is denoted as ``DIS+DY'',
to evaluate the $ \sigma^{\rm th}_{i}$ cross sections entering Eq.~(\ref{eq:hllhc}).
We have verified that, both at the pre- and post-fit levels, the fit quality to the HL-LHC pseudo-data
satisfies $\chi^2/n_{\rm bin} \simeq 1$ in the case of the SM PDFs
as expected.
Furthermore,
we assume $f_{{\rm red},j}=0.2$ for all systematic sources, as done in the optimistic scenario of
Ref.~\cite{Khalek:2018mdn}.
We note that more conservative values for the reduction of systematic errors,
such as $f_{{\rm red},j}=0.5$, are not expected to qualitatively
modify our results.
The reason is that, as indicated by
the bottom panels of Figs.~\ref{fig:hllhc-nc} and~\ref{fig:hllhc-cc},
for the highest energy bins (which dominate
the EFT sensitivity), specifically above
$m_{\ell\ell}\approx$ 1.7 TeV and $m_T \approx$ 1.5 TeV, the measurement
will be limited by statistical uncertainties.

\subsection{Impact on PDF uncertainties}

From Figs.~\ref{fig:hllhc-nc} and \ref{fig:hllhc-cc},  one can observe that
the PDF uncertainties in the SM PDF baseline used to generate the pseudo-data
are either comparable or larger than the corresponding
projected experimental uncertainties at the  HL-LHC.
Specifically, for the highest $m_{\ell\ell}$ bin
of the NC distribution the PDF errors are twice the experimental ones,
while in the CC case the associated PDF errors become clearly larger
than the experimental ones starting at $m_T\simeq 2 $ TeV.
This comparison suggests that one should expect a significant 
uncertainty reduction once the HL-LHC pseudo-data is included
in the PDF fit.

\begin{figure}[t]
\begin{center}
  \includegraphics[width=0.49\textwidth]{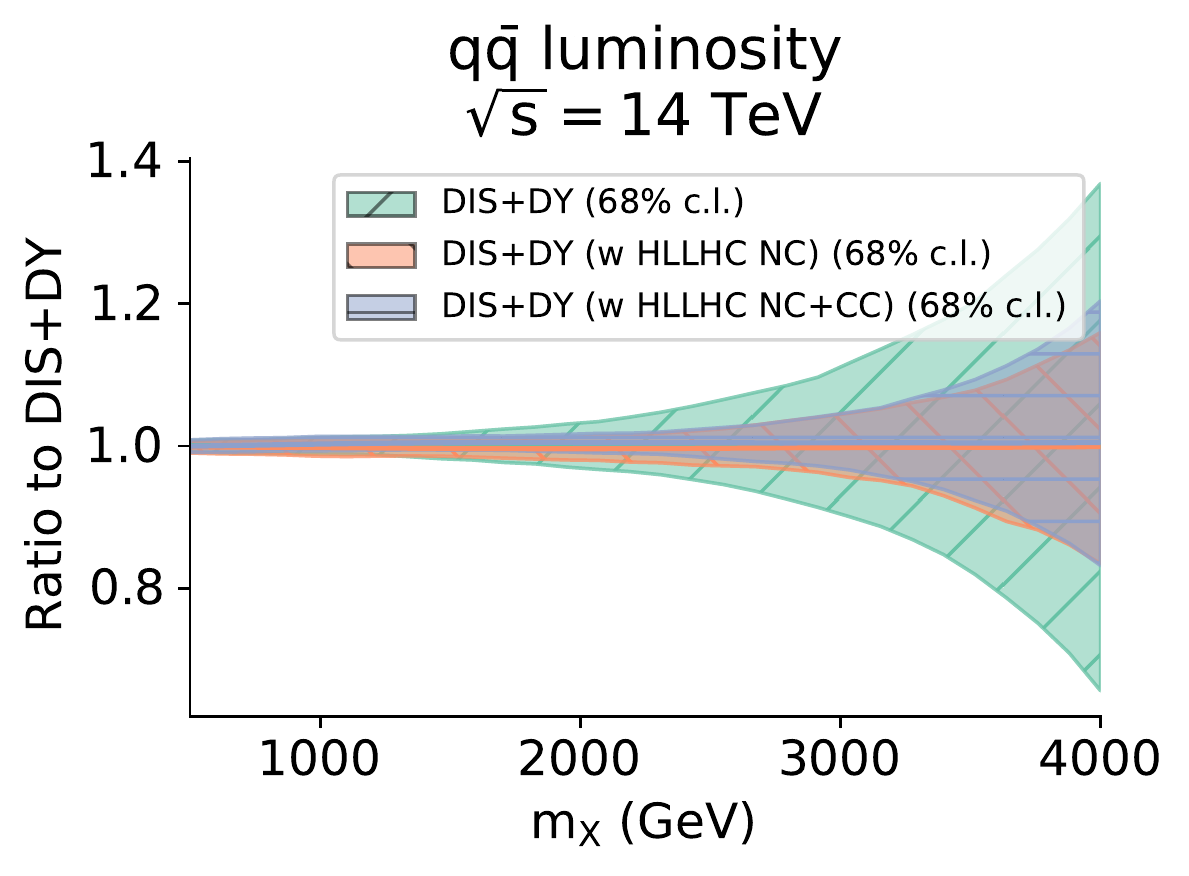}
  \includegraphics[width=0.49\textwidth]{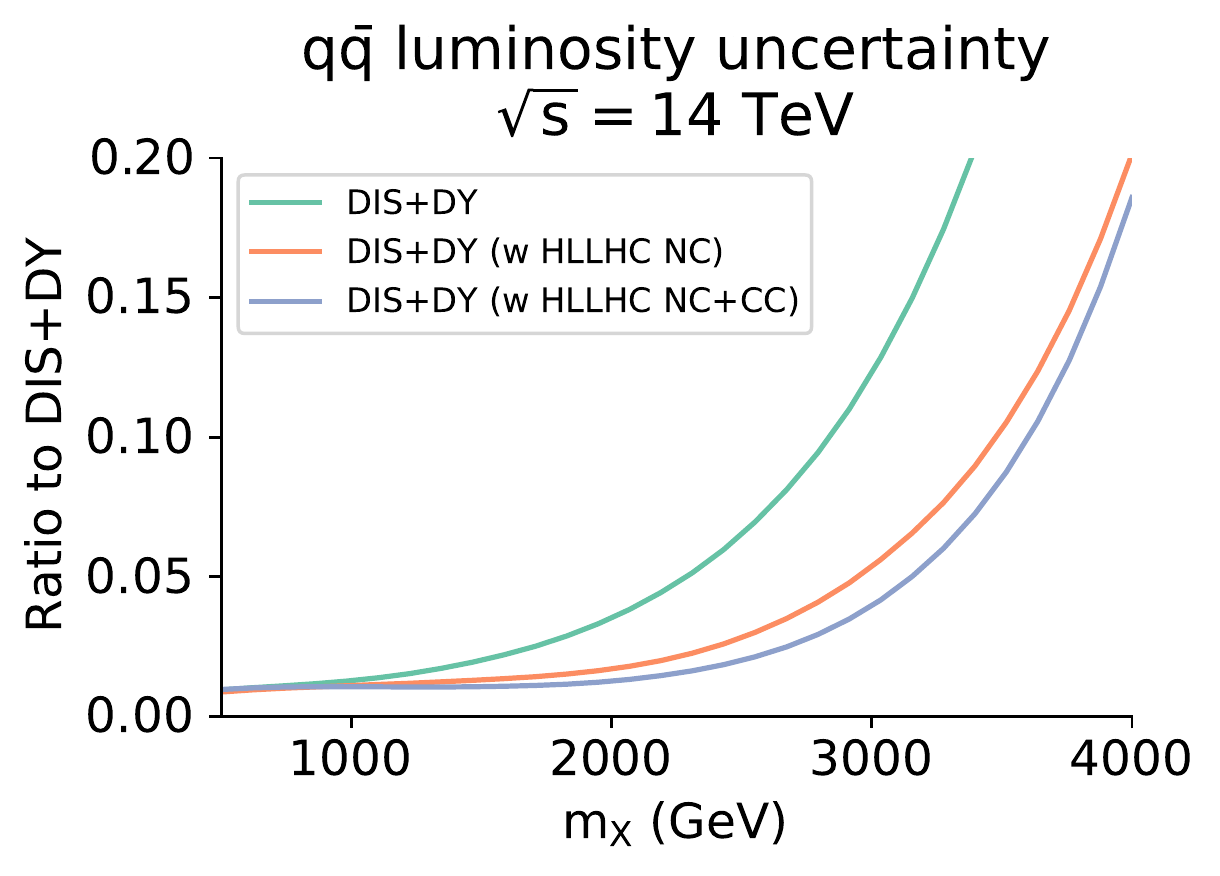}
    \caption{\label{fig:hllhc-lumi-qqb} Impact of the
      HL-LHC pseudo-data on the quark-antiquark luminosity $\mathcal{L}_{q\bar{q}}$
      of the SM PDF baseline fit as a function
      of  $m_X$.
      Left: the luminosities $\mathcal{L}_{q\bar{q}}$ for the DIS+DY baseline
      and the corresponding fits including the HL-LHC pseudo-data, either only NC
      or also with CC cross sections, presented as a ratio to the central value of the former.
      Right: the relative PDF uncertainty in $\mathcal{L}_{q\bar{q}}$ (with the central
      value of the DIS+DY baseline as reference) for the same fits.
    }
\end{center}
\end{figure}

To validate this expectation, Fig.~\ref{fig:hllhc-lumi-qqb} displays
the impact of the
HL-LHC pseudo-data on the quark-antiquark luminosity $\mathcal{L}_{q\bar{q}}$
as a function
of the final state invariant mass $m_X$ at $\sqrt{s}=14$ TeV.
We compare  $\mathcal{L}_{q\bar{q}}$ for the SM PDF baseline fit (DIS+DY)
with the same quantity from the corresponding fits including the HL-LHC pseudo-data, either only NC
or also with CC cross sections.
The right panel displays the associated relative PDF uncertainties.
We find a significant reduction of the PDF uncertainties
affecting the quark-antiquark luminosity (and hence the Drell-Yan
cross sections) in the high mass ($m_X\gsim$ 1 TeV) region
once the HL-LHC pseudo-data constraints are accounted for.
For instance, at  $m_X\gsim$ 2 TeV, PDF uncertainties on $\mathcal{L}_{q\bar{q}}$
decrease from $\simeq5\%$ in the baseline down to $\simeq 2.5\%$~($\simeq 1.5\%$)
once the NC~(NC+CC) HL-LHC pseudo-data is included in the fit. The
effect of the inclusion of HL-LHC projections becomes more dramatic as
$m_X$ increases. 
On the other hand, other partonic luminosities such as the quark-quark
and gluon-gluon ones are essentially unaffected by the HL-LHC constraints.
In terms of fit quality, the only noticeable effect is a mild
improvement in the $\chi^2$ of the high-mass DY datasets listed in
Table~\ref{tab:data-high-mass}. 

\subsection{PDF and EFT interplay at the HL-LHC}
\label{sec:hllhc_joint_fits}

The finding  that the projected HL-LHC pseudo-data has a significant
impact on the quark-antiquark PDF luminosity, summarised in  Fig.~\ref{fig:hllhc-lumi-qqb}, 
 suggests that
the interplay between PDFs and EFT effects in the high-energy
DY tails should become enhanced
as compared to the results reported in the previous section.
With this motivation, we first of all repeat the joint
determination of PDFs and the $\hat{W},\hat{Y}$ coefficients
from EFT  scenario I presented in Sect.~\ref{sec:results_scenarioI}
now accounting for the constraints of the HL-LHC pseudo-data.
An important difference in this case is that the
inclusion of CC data lifts the flat direction in the $(\hat{W},\hat{Y})$
plane, making a full two-dimensional fit possible.
Secondly, the availability of the HL-LHC pseudo-data allows
us to assess the interplay between the PDFs and the EFT
coefficient ${\bf C}^{D\mu}_{33}$  from benchmark scenario II,
whose analysis in Sect.~\ref{sec:scenarioIIresults} was restricted
to fixed SM PDFs.

\paragraph{Scenario I.}
%
For the simultaneous determination of PDFs and the $\hat{W},\hat{Y}$ coefficients
accounting for the constraints provided by the HL-LHC pseudo-data,
we use 35 sampling values of $(\hat{W}_i,\hat{Y}_i)$, 25 of which are equally spaced in
either $\hat{W}\in (-1.6,1.6)\times 10^{-5}$ or $\hat{Y}\in (-8,+8)\times 10^{-5}$
(hence in steps of $\Delta\hat{W}=0.8 \times 10^{-6}$ and
$\Delta\hat{Y}=4 \times 10^{-6}$ respectively), and then 10 additional points along the
diagonals.
In order to assess the robustness of the results, we added 12
more sampling values, 8 further away from the origin and 4 more along
the $\hat{W}=0$ and $\hat{Y}=0$ axes, and verified that the confidence
level countours are stable upon their addition. 

We find that the constraints on the $(\hat{W},\hat{Y})$ parameters are completely
dominated by the HL-LHC projections and that current data exhibit a much
smaller pull, consistent with the findings of previous
studies~\cite{Farina:2016rws,Ricci:2020xre}. Also, the $\chi^2_{\rm
  eftp}$ contour is more stable and requires less replicas if only the HL-LHC projections
are included in the computation of the $\chi^2$.
The corresponding 
marginalised bounds on $\hat{W}$ and $\hat{Y}$
are reported in Table~\ref{tab:hlbounds} using the same
format as in  Table~\ref{tab:bound1w}.

\begin{table}
 \renewcommand{\arraystretch}{1.40}
  \centering
  \begin{tabular}{l|c|c|c|c}
    & SM PDFs & SMEFT PDFs  & best-fit shift  & broadening  \\
    \hline
    \multirow{2}{*}{ $\hat{W}\times 10^5$ (68\% CL)} & $[-0.7, 0.5]$ & \multirow{2}{*}{$[-4.5, 6.9]$}  & 1.3 & 850\% \\
      & $[-1.0, 0.9]$ &  & 1.3  & 500\% \\
    \midrule
    \multirow{2}{*}{$\hat{W}\times 10^5$ (95\% CL)} & $[-1.0, 0.8]$ & \multirow{2}{*}{$[-8.1, 10.6]$} & 1.4 & 940\%\\
      & $[-1.4, 1.2]$ &  &  1.4 & 620\%  \\
   \hline
   \multirow{2}{*}{$\hat{Y}\times 10^5$ (68\% CL)} & $[-1.8, 3.2]$ & \multirow{2}{*}{$[-6.4, 8.0]$} & 0.1 & 190\% \\
       & $[-3.7, 4.7] $ &  & 0.3  & 70\% \\
   \midrule
   \multirow{2}{*}{$\hat{Y}\times 10^5$ (95\% CL)} & $[-3.4, 4.7] $ & \multirow{2}{*}{$[-11.1, 12.6]$} & 0.1 & 190\%\\
   & $[-5.3, 6.3] $ &  & 0.3  & 110\% \\
    \hline
  \end{tabular}
  \caption{ \label{tab:hlbounds}\small
    Same as Table~\ref{tab:bound1w} for the
 68\% CL and 95\% CL
    marginalised bounds on the $\hat{W}$ and $\hat{Y}$
    parameters obtained from the two-dimensional ($\hat{W}$,$\hat{Y}$)
    fits that include the HL-LHC pseudo-data for NC and CC
    Drell-Yan distributions.
    As in Table~\ref{tab:bound1w},
    for the SM PDFs we indicate the bounds obtained without (upper)
    and with (lower entry) PDF uncertainties accounted for.
}
\end{table}

From Table~\ref{tab:hlbounds},
one can observe how including high-mass data at the LHC both in a fit of PDFs and
in a fit of SMEFT coefficients and neglecting the interplay between
them could result in a significant underestimate
of the uncertainties associated to the EFT parameters.
Indeed, the marginalised 95\% CL bound on the $\hat{W}$~($\hat{Y}$) parameter becomes looser
once SMEFT PDFs are consistently used, with a broadening, defined in Eq.~(\ref{eq:broadening}),
of 500\%~(110\%), even once PDF uncertainties are fully accounted for. 
This effect would have been even more marked if PDF uncertainties had not
been accounted for in EFT fits based on SM PDFs,
where the same broadening factors would be 940\% and 190\%
respectively.

%
\begin{table}
 \renewcommand{\arraystretch}{1.40}
  \centering
  \begin{tabular}{l|c|c|c|c}
    & SM cons. PDFs & SMEFT PDFs  & best-fit shift  & broadening  \\
    \hline
    \multirow{2}{*}{ $\hat{W}\times 10^5$ (68\% CL)} & $[-1.0, 0.0]$ & \multirow{2}{*}{$[-4.5, 6.9]$}  & 1.7 & 1000\% \\
                                                                        & $[-4.0, 2.8]$ &  & 1.8  & 70\% \\
    \midrule
    \multirow{2}{*}{$\hat{W}\times 10^5$ (95\% CL)} & $[-1.4, 0.4]$ & \multirow{2}{*}{$[-8.1, 10.6]$} & 1.8 & 940\%\\
    & $[-4.3, 3.1]$ &  &  1.9 & 150\%  \\
    \hline
   \multirow{2}{*}{$\hat{Y}\times 10^5$ (68\% CL)} & $[2.1, 7.0]$ & \multirow{2}{*}{$[-6.4, 8.0]$} & -3.7 & 190\% \\
       & $[-3.4,11.2] $ &  & -3.6  & -1\% \\
   \midrule
   \multirow{2}{*}{$\hat{Y}\times 10^5$ (95\% CL)} & $[0.5, 8.5] $ & \multirow{2}{*}{$[-11.1, 12.6]$} & -3.7 & 200\%\\
   & $[-5.0, 13.7] $ &  & -3.6  & 30\% \\
    \hline
  \end{tabular}
  \caption{ \label{tab:hlboundscons}\small
    Same as Table~\ref{tab:hlbounds} for the 68\% and 95\% CL
    marginalised bounds on the $\hat{W}$ and $\hat{Y}$
    parameters obtained from the two-dimensional ($\hat{W}$,$\hat{Y}$)
    fits that include the HL-LHC pseudo-data for NC and CC
    Drell-Yan distributions. The input PDF set for the analysis done
    using fixed SM PDFs (corresponding to the results displayed in the
    column ``SM cons. PDFs'') is a conservative PDF set that does not
    include any of the high-mass
    distributions or the HL-LHC projections nor the Run I and
    Run II high-mass dataset listed in Table
    ~\ref{tab:data-high-mass}. The limits obtained from the simultaneous fit
    of PDFs and Wilson coefficients  (corresponding to the results displayed on the
    column ``SMEFT PDFs'') are the same as those in Table~\ref{tab:hlbounds}. 
}
\end{table}
%

A further important question is whether the bounds obtained with SM
PDFs appearing on the left column of Table~\ref{tab:hlbounds}  would become
more comparable to those obtained from the simultaneous fit of PDFs
and SMEFT coefficients, in case a
conservative set of PDF was used in the analysis based on SM PDFs. To
address this question, in Table~\ref{tab:hlboundscons} we display the
bounds that are obtained using a PDF set that does not include any of
the high-mass Drell-Yan sets (neither the HL-LHC projections nor the
current datasets listed in Table.~\ref{tab:data-high-mass}) and
compare the bounds obtained using this set of PDFs to those obtained
consistently using SMEFT PDFs.
We observe that, once this set of conservative PDF is used as an input
PDF set and the PDF uncertainty is included in the computation of the
bounds, the latter increases as compared to the bounds in
Table~\ref{tab:hlbounds}. As a result, the size of the bounds obtained
by keeping fixed SM PDFs is closer to the size obtained from the
simultaneous fits, although still slightly underestimated.
At the same time, the shift in the best-fit becomes more
marked.

\begin{figure}[t]
\begin{center}
  \includegraphics[width=0.9\textwidth]{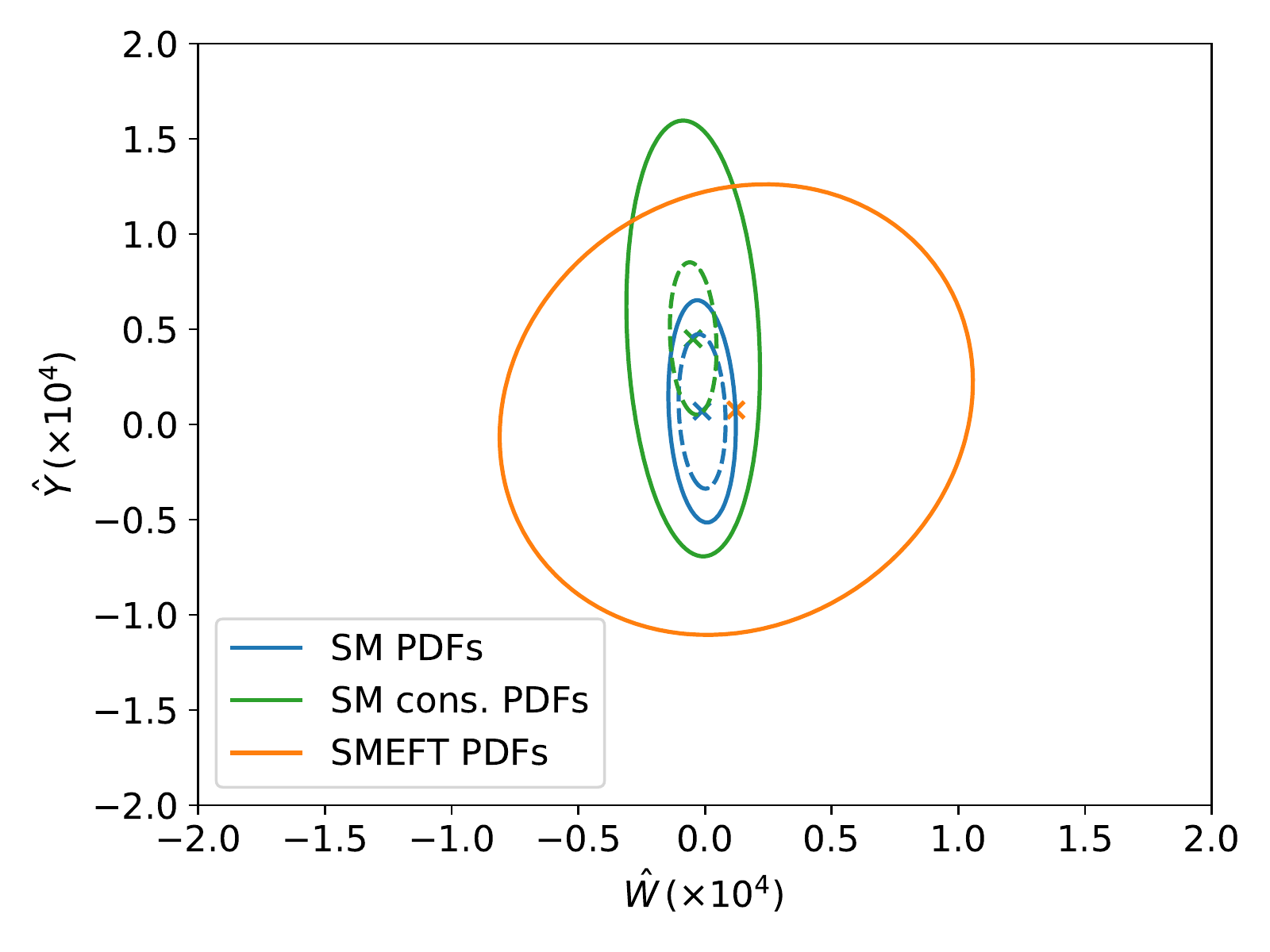}
  \caption{\label{fig:hllhc-ellipse} The 95\% confidence level contours
    in the ($\hat{W}$,$\hat{Y}$) plane obtained from the DIS+DY fits that
    include the high-mass Drell-Yan HL-LHC pseudo-data (both in the NC
    and CC channels) 
    when using either SM PDFs (blue) or conservative SM PDFs (green). In both cases
    the ellipses are obtained by performing a parabolic fit to
    $\chi^2_{\rm smp}$ with fixed PDFs. PDF
    uncertainties are included in the solid lines and not included in
    the dashed lines.  The results are compared to
    those obtained in a simultaneous fit, namely with SMEFT PDFs
    (orange). In this case, the parabolic fit is performed to $\chi^2_{\rm eftp}$ by varying
    simultaneously the Wilson Coefficients and the PDFs. 
    The crosses indicate the best fits in the three cases discussed in
    the text. }
\end{center}
\end{figure}
Results are graphically displayed in
Fig.~\ref{fig:hllhc-ellipse}, where the 95\% confidence level contours
    in the ($\hat{W}$,$\hat{Y}$) plane obtained from the DIS+DY fits that
    include the high-mass Drell-Yan HL-LHC pseudo-data  when using either SM PDFs, SM conservative
    PDFs or SMEFT PDFs are compared. All solid countours include PDF
    uncertainties, while the dashed contours that do not include PDF
    uncertainties are also indicated to visualise the impact of the
    inclusion of the PDF uncertainties.

  To conclude, we should also emphasise that, while in this work we use pseudo-data
and hence the best-fit values are by construction unchanged, this
would not necessarily be the case in the analysis of real data, where
improper treatment of PDFs could result in a spurious EFT `signal', or
even missing a signal which is indeed present in the data. A detailed
study aimed at a precise definition of `conservative' PDFs in a more
general scenario is beyond the scope of this paper and will be the
topic of future work; a thorough comparison of the consistent
simultaneous approach, versus the use of conservative PDF sets, will
be of particular interest in cases of EFT manifestations of new
physics.

\begin{figure}[t]
\begin{center}
  \includegraphics[width=0.49\textwidth]{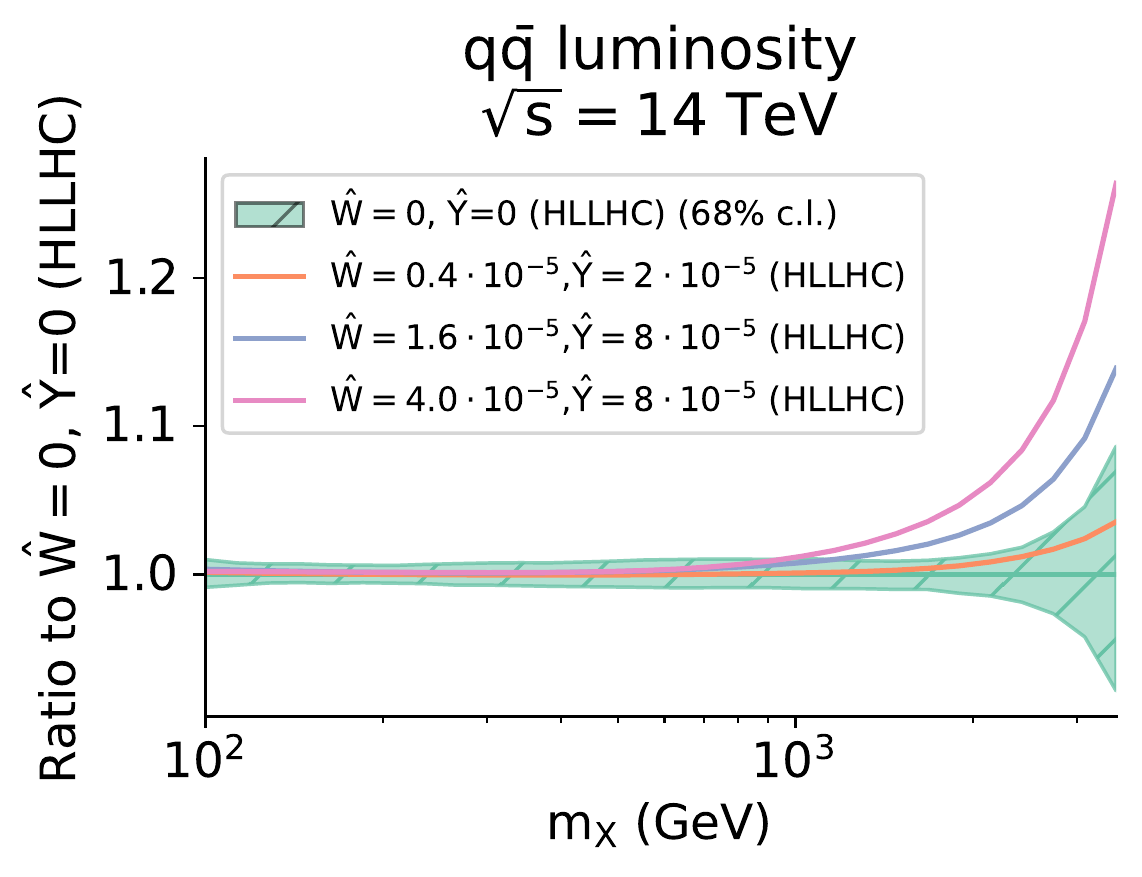}
  \includegraphics[width=0.49\textwidth]{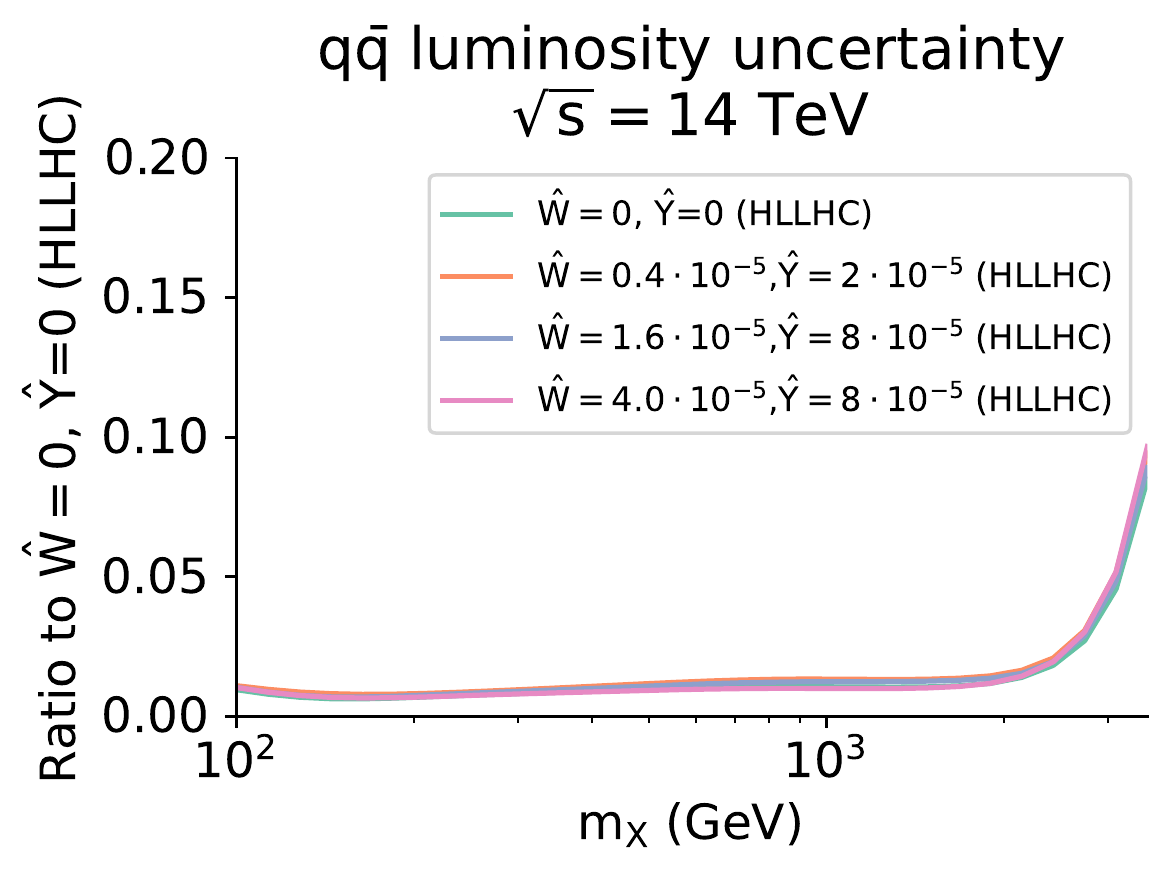}
  \caption{\label{fig:HLlumiWY} Same as Fig.~\ref{fig:hllhc-lumi-qqb}, now comparing
    the quark-antiquark SM PDF luminosity in the fits including the HL-LHC pseudo-data
    with those obtained in the SMEFT PDF fits for representative values
    of the $\hat{W}$ and $\hat{Y}$ parameters.
    The corresponding comparison in the case of fits to available Drell-Yan data
    was shown in Fig.~\ref{fig:SMEFT_lumis}.
  }
\end{center}
\end{figure}

The increased role that the interplay between PDFs and EFT coefficients
will play at the HL-LHC can also be illustrated by
comparing the expected behaviour of the quark-antiquark
luminosity, displayed in Fig.~\ref{fig:HLlumiWY},
for the SMEFT PDFs corresponding
to representative values of the $\hat{W}$ and $\hat{Y}$ parameters of benchmark scenario I
as compared to the  SM PDFs.
Note that the corresponding comparison for $\mathcal{L}_{q\bar{q}}$ 
in the fits to available Drell-Yan data
    was displayed in Fig.~\ref{fig:SMEFT_lumis}.
    Indeed, the central value of the quark-antiquark luminosity
    for SMEFT PDFs corresponding to values of $(\hat{W}_i,\hat{Y}_i)$ selected along the grid used
    to derive
Fig.~\ref{fig:HLlumiWY} changes greatly, well
outside the one-sigma error band of the SM PDFs, while the
PDF uncertainties themselves are unchanged. 
This change in central value of the large-$x$ PDFs partially reabsorbs the effects in the
partonic cross section induced by the SMEFT operators and leads to better
$\chi^2$ values as compared to those obtained with the SM PDFs.

Even neglecting SMEFT PDF effects, we note that our marginalised bounds on
the $\hat{W}$ and $\hat{Y}$ coefficients from HL-LHC pseudo-data using SM
PDFs turn out to be more stringent than those reported
in~\cite{Ricci:2020xre} by around a factor of 4 for $\hat{W}$ and a factor 2
for $(\hat{Y})$.
This is due to a combination of factors. First of all we use the 13
TeV measurements as reference to produce the HL
projections. Furthermore we assume a total integrated
luminosity of $\mathcal{L}=6$ fb$^{-1}$ (from the combination
of ATLAS and CMS) rather than 3 fb$^{-1}$ as well as a more optimistic
scenario concerning the reduction of the experimental systematic
uncertainties.

\begin{figure}[t]
\begin{center}
\includegraphics[width=0.49\textwidth]{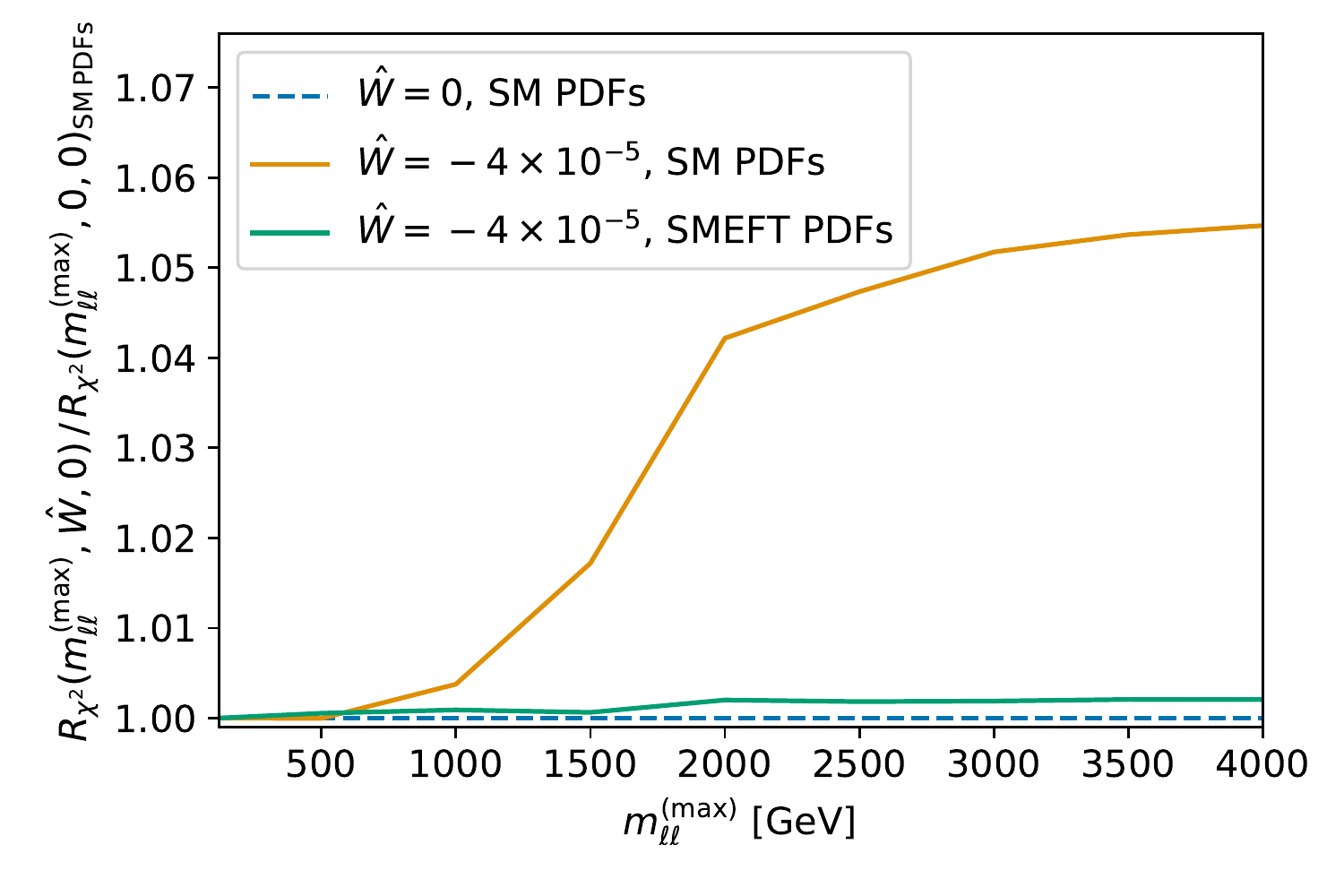}
\includegraphics[width=0.49\textwidth]{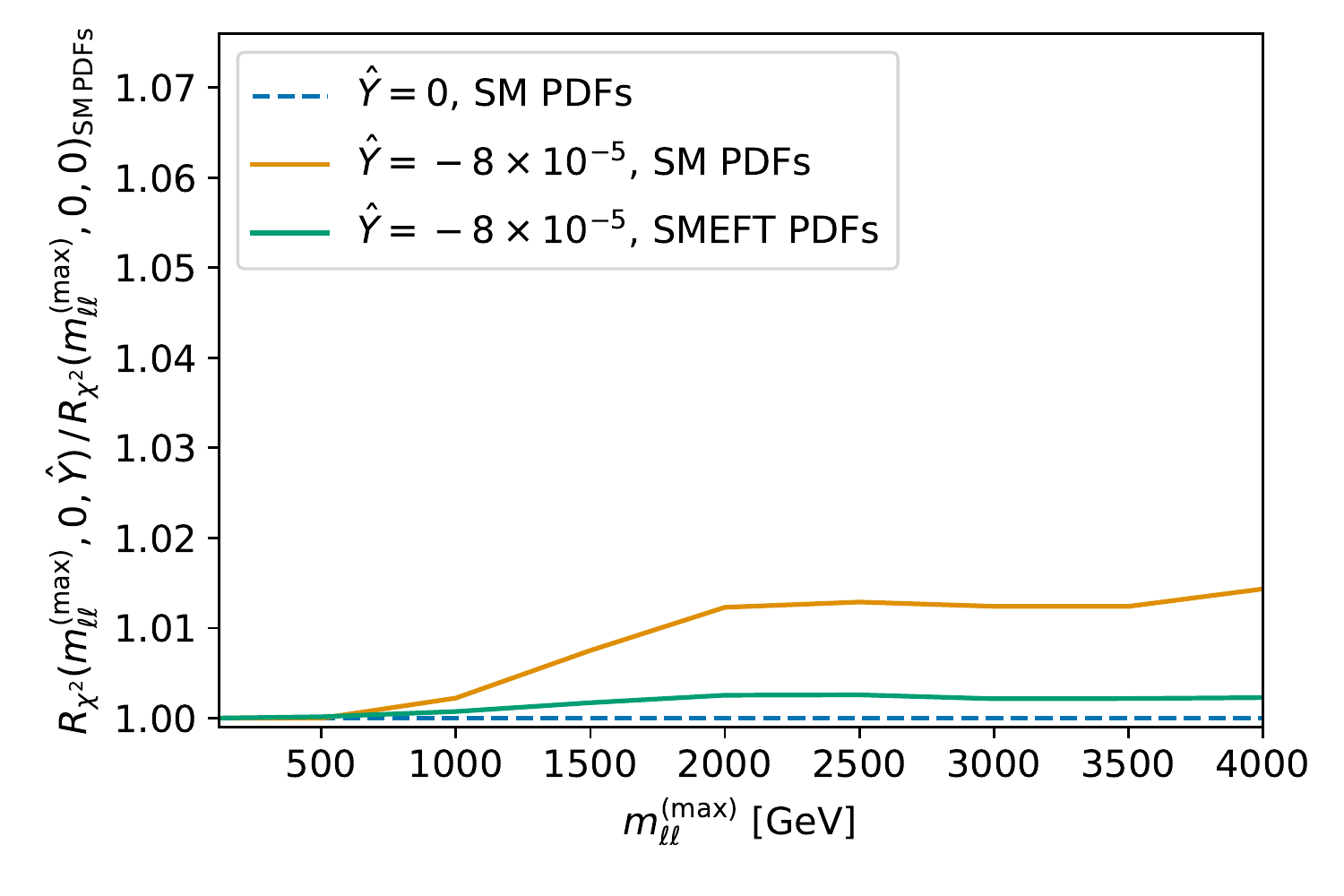}
\caption{\label{fig:cutmax_hl} Same as Fig.~\ref{fig:cutmax} now
  for the fits including the HL-LHC pseudo-data.
  Note that in this case the $m_{\ell\ell}^{(\rm max)}$ cut refers to
 the transverse mass $m_T$ for the charged-current distributions.
}
\end{center}
\end{figure}

Fig.~\ref{fig:cutmax_hl} then displays the $R_{\chi^2}$ estimator, defined in
Eq.~(\ref{mllratio}) and shown in Fig.~\ref{fig:cutmax} for the case
of available LHC data, now evaluated from
the fits including the HL-LHC pseudo-data.
In this case, the $m_{\ell\ell}^{(\rm max)}$ cut applies to $m_{\ell\ell}$
for the neutral-current distributions and to
 the transverse mass $m_T$ for the charged-current ones.
 As in the case of Fig.~\ref{fig:cutmax},
 we observe an approximately monotonic growth of $R_{\chi^2}$
for the SM PDFs arising
from the energy-growing EFT corrections that dominate the high-energy
DY tails, an effect which is now rather larger thanks to the presence
of the HL-LHC pseudo-data.
The most striking difference as compared to  Fig.~\ref{fig:cutmax}
is that now the  SMEFT PDF curve is much flatter,
indicating that EFT effects
are being almost totally reabsorbed into the PDFs.

The findings summarised by Fig.~\ref{fig:cutmax_hl} demonstrate that, at the HL-LHC, EFT-induced deviations
could be indeed inadvertently "fitted away'' into a PDF redefinition,
explaining the large broadenings reported in Fig.~\ref{fig:HLlumiWY},
and highlight the need to devise novel strategies to disentangle the effects
of PDFs and EFT contributions from the high-energy tails of LHC cross-sections.
Such strategies could exploit, for instance, the availability of measurements sensitive
to large-$x$ PDFs but not to high scales,
such as forward electroweak gauge boson production by LHCb~\cite{Khalek:2018mdn}.

\paragraph{Scenario II.}
%
We now turn to present the corresponding results of the simultaneous fits of the PDFs
and the EFT coefficients including the HL-LHC pseudo-data for the case
of benchmark scenario II.
As motivated in Sect.~\ref{sec:scenarioII}, a non-zero value of the ${\bf C}^{D\mu}_{33}$ coefficient
 affects only the NC and CC muon final states, while the electron ones remain described
by the SM calculations.
This property implies that, in  fits presented below, the EFT corrections modify only the shapes of
the HL-LHC distributions in the muon channel,  the right panels in  Figs.~\ref{fig:hllhc-nc} and~\ref{fig:hllhc-cc},
but not those of the electron pseudo-data.

\begin{figure}[t]
\begin{center}
  \includegraphics[width=0.49\textwidth]{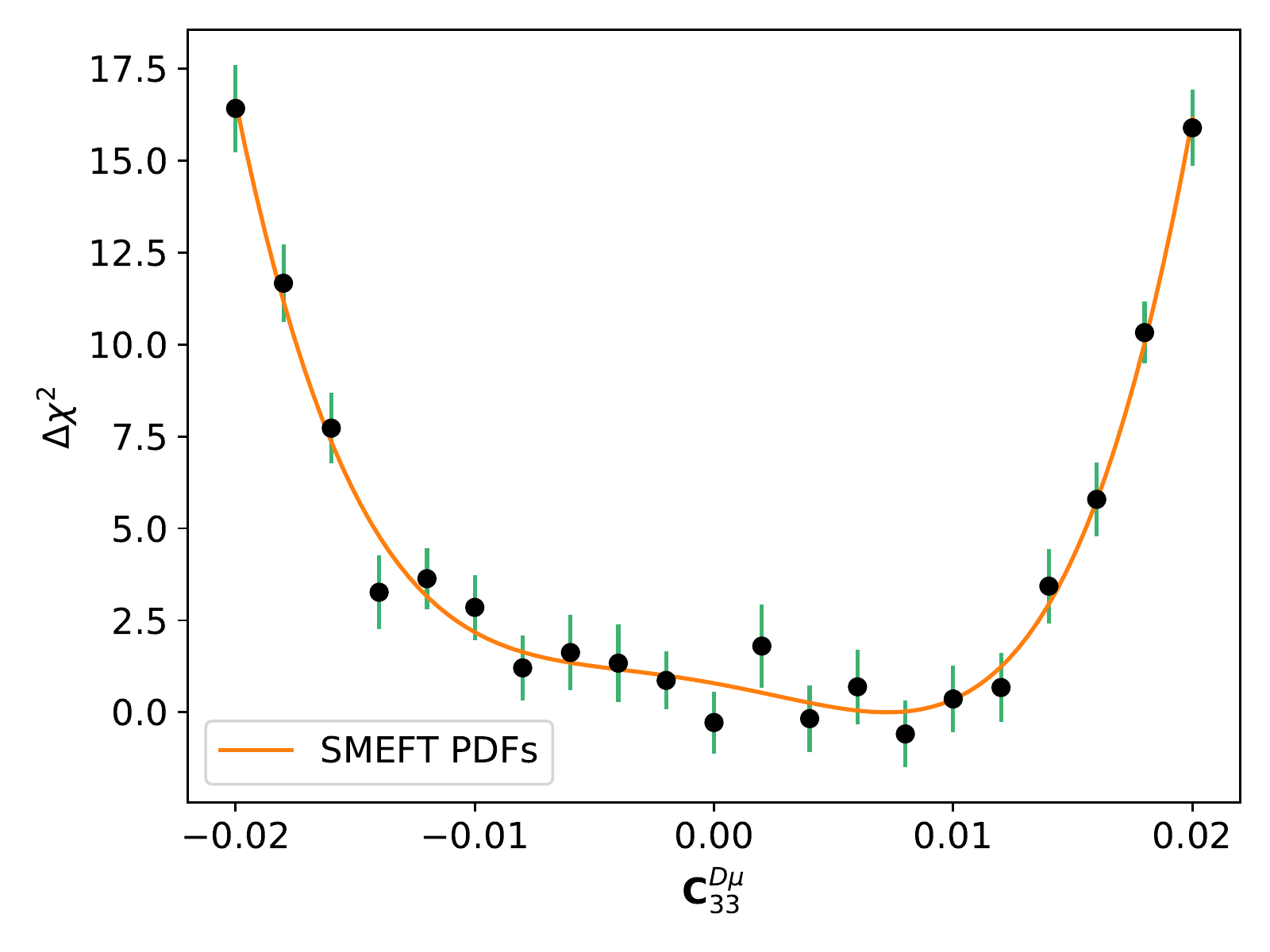}
  \includegraphics[width=0.49\textwidth]{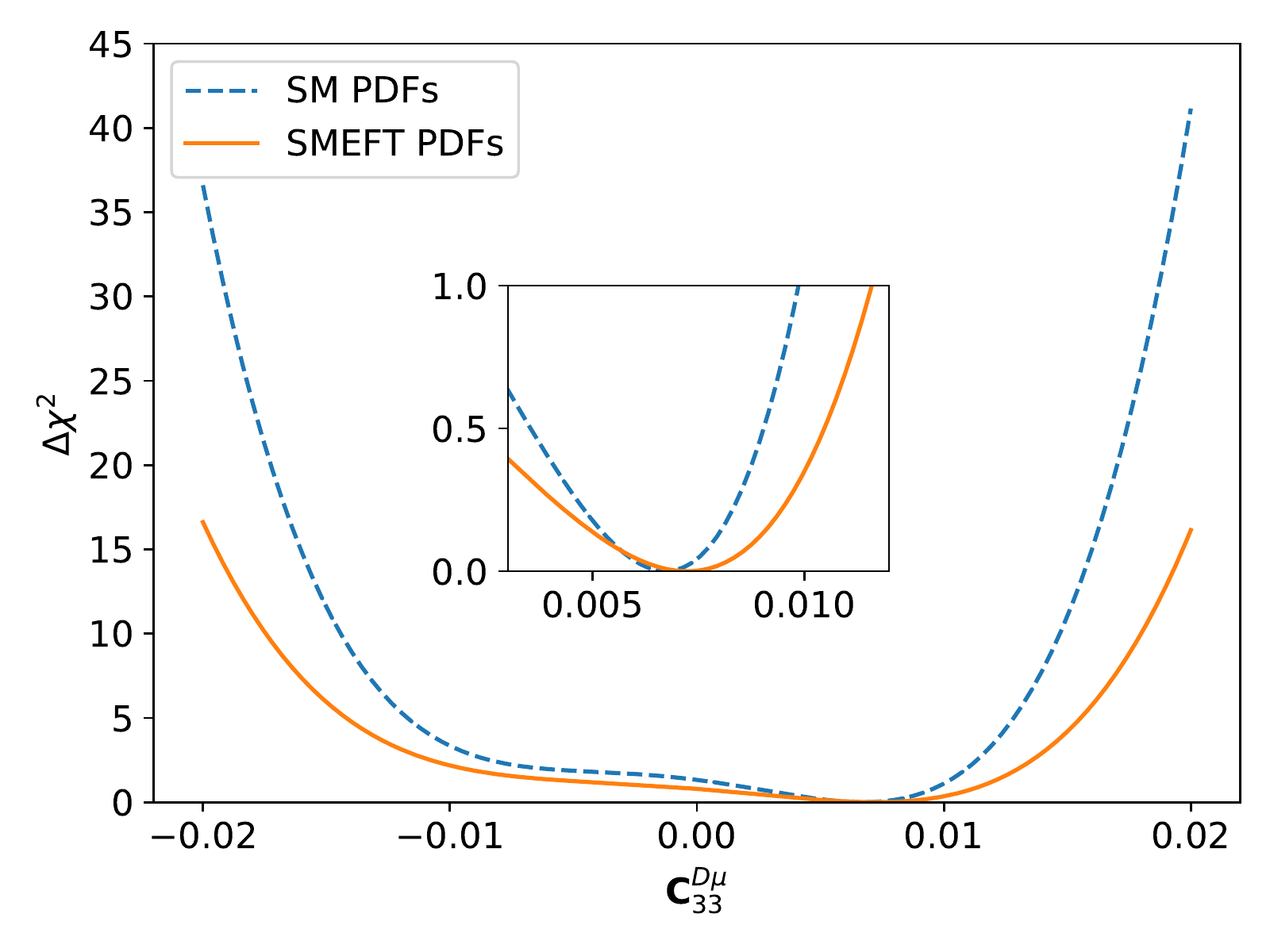}
  \caption{\label{fig:cb-bounds-hl} Left: the values of $\Delta \chi^2$
    obtained for the SMEFT PDFs 
    as a function of  ${\bf C}^{D\mu}_{33}$
    from the fits including the HL-LHC pseudo-data, together with the corresponding
      quartic polynomial fit.
       Right: comparison
      of the polynomial fit obtained with SMEFT PDFs and displayed in the left panel
      with its counterpart based on SM PDFs.}
\end{center}
\end{figure}

Fig.~\ref{fig:cb-bounds-hl} displays the values of $\Delta \chi^2$
obtained for the SMEFT PDFs 
as a function of the EFT parameter ${\bf C}^{D\mu}_{33}$ 
from the joint fits that include the HL-LHC pseudo-data.
The sampling is constituted by 21 points uniformly distributed in ${\bf C}^{D\mu}_{33} \in \lc -0.02,0.02\rc$.
As in Fig.~\ref{fig:parabolas1}, the error bars indicate the uncertainties
associated to the finite number of Monte Carlo replicas used for each value of ${\bf C}^{D\mu}_{33}$.
The profile in $\Delta\chi^2$ exhibits a double minimum structure (bimodal distribution), explained
by the fact that in this scenario it is the quadratic rather than the linear
terms in the EFT expansion that dominate.
The corresponding quartic polynomial fit using Eq.~(\ref{eq:parabolicfit}) can be seen
to successfully reproduce the $\Delta\chi^2$ values obtained in this joint analysis.
The right panel of Fig.~\ref{fig:cb-bounds-hl} then compares
the polynomial fit obtained with the SMEFT PDFs with the corresponding one
 when using instead fixed SM PDFs to determine the $\Delta\chi^2$ values,
with the inset focusing on the region close to $\Delta\chi^2\simeq 0$.
The associated 68\% and 95\% CL bounds are then reported in Table~\ref{tab:hlbounds2},
where we note that since the 68\% CL interval is disjoint we evaluate the
shift and broadening only for the 95\% CL bounds.

\begin{table}[t]
 \renewcommand{\arraystretch}{1.40}
  \centering
  \begin{tabular}{l|c|c|c|c}
    & SM PDFs & SMEFT PDFs  & best-fit shift  & broadening  \\
    \hline
    ${\bf C}^{D\mu}_{33}\times 10^2$  (68\% CL) &  $[-0.1, 1.1]$ &  $[-0.3, 1.2]$  & 0.06  & 25\% \\
    ${\bf C}^{D\mu}_{33}\times 10^2$ (95\% CL) & $[-1.0, 1.2]$ & $[-1.2, 1.4]$ & 0.06 & 18\%\\
   \hline
  \end{tabular}
  \caption{ \label{tab:hlbounds2}
    Same as Table~\ref{tab:hlbounds}, now for the ${\bf C}^{D\mu}_{33}$ 
    parameter from EFT benchmark scenario II.
 }
\end{table}

Inspection of Fig.~\ref{fig:cb-bounds-hl} and Table~\ref{tab:hlbounds2} indicates that,
even at the HL-LHC, the interplay between PDFs and EFT coefficients remains
 moderate in this particular scenario.
Indeed, in contrast with the marked effects in scenario I (Fig.~\ref{fig:hllhc-ellipse}),
where the bounds on the $\hat{W}$ and $\hat{Y}$ worsened by up to an order of
magnitude when the SMEFT PDFs were consistently used, in scenario
II the obtained bounds on ${\bf C}^{D\mu}_{33}$ would only loosen by around 30\%.
The origin of this rather different behaviour can be traced
back to the fact that in scenario II the electron channel data do not receive EFT corrections,
and hence all the information that they provide makes it possible to exclusively constrain the PDFs.
The muon channel distributions then determine the allowed range for ${\bf C}^{D\mu}_{33}$,
restricted by the well-constrained large-$x$ quarks and antiquark PDFs from the electron data.
This finding demonstrates how the availability of measurements in separate leptonic final states
is of utmost importance to test BSM scenarios that account for
violations of Lepton Flavour Universality.

\begin{figure}[t]
\begin{center}
  \includegraphics[width=0.49\textwidth]{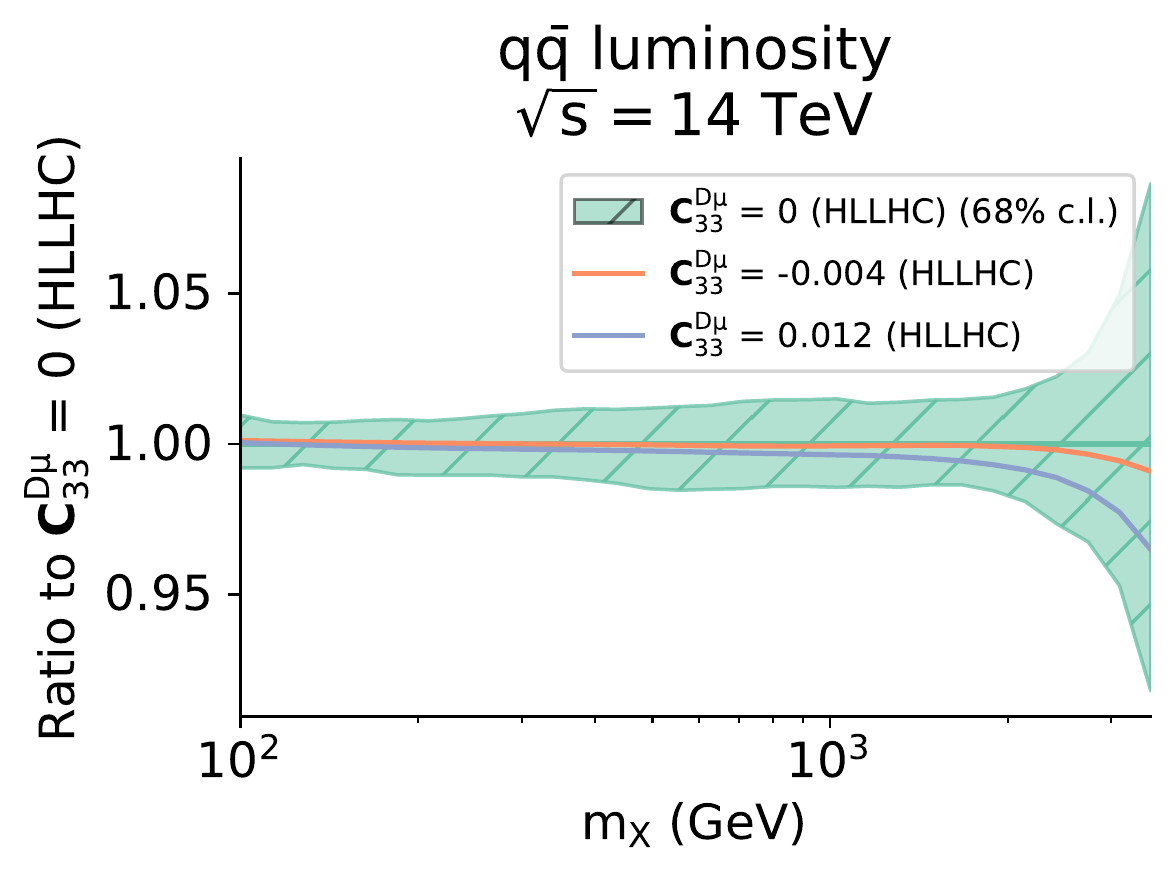}
  \includegraphics[width=0.49\textwidth]{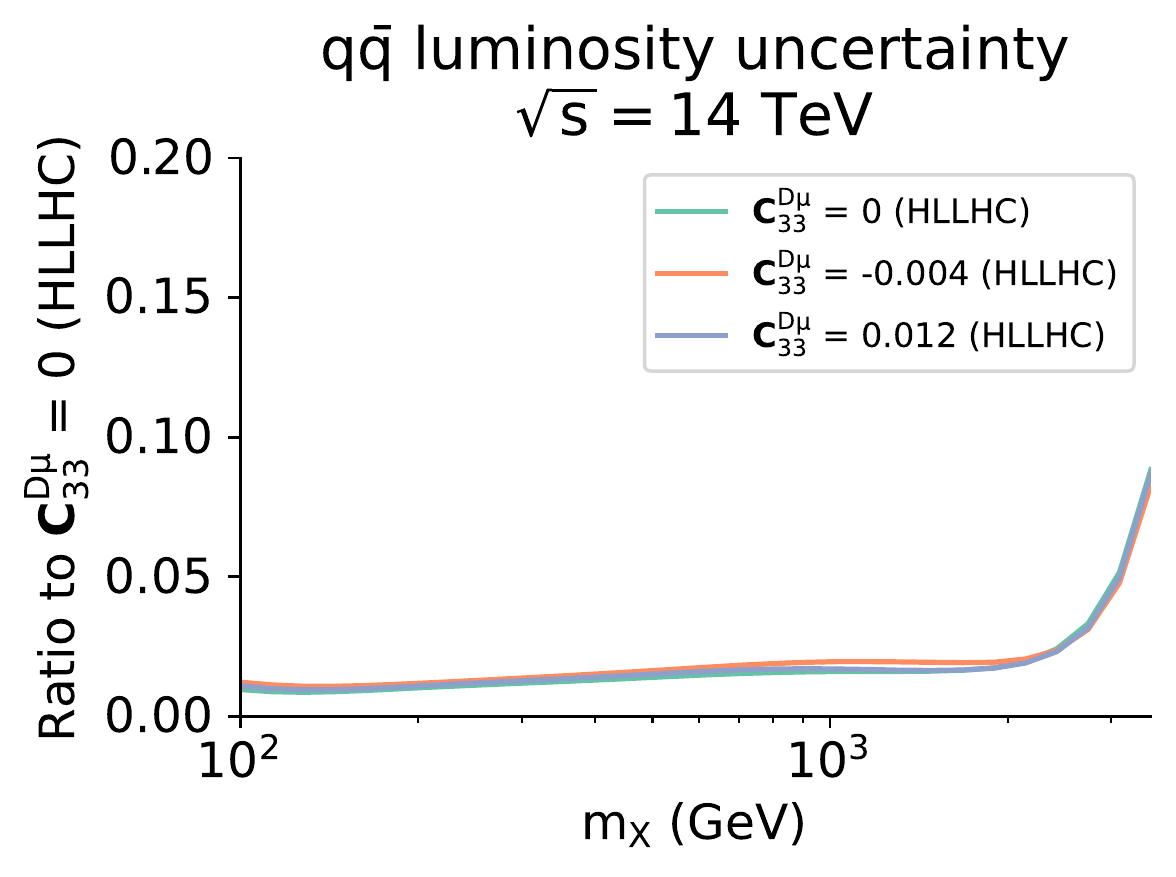}
  \caption{Same as Fig.~\ref{fig:HLlumiWY} now in the case of  the simultaneous fits of the PDFs
    and the ${\bf C}^{D\mu}_{33}$  EFT parameter taking into account HL-LHC pseudo-data.
    \label{fig:HLlumiCbCoef}
  }
\end{center}
\end{figure}

In the same manner as in Fig.~\ref{fig:HLlumiWY}, Fig.~\ref{fig:HLlumiCbCoef}
displays the comparison of the quark-antiquark luminosities at $\sqrt{s}=14$ TeV
in the fits with HL-LHC pseudo-data in the case of the SM PDFs and
for the SMEFT PDFs for representative values of ${\bf C}^{D\mu}_{33}$.
Specifically, we show ${\bf C}^{D\mu}_{33}=-0.004$ and $0.012$,
chosen to lie at the boundary of the 68\% CL interval
reported in Table~\ref{tab:hlbounds2}.
The result that the two values lead to the same effect on $\mathcal{L}_{q\bar{q}}$
follows from the dominance of the quadratic EFT terms in this scenario.
One finds that the shift in the central values of the quark-antiquark luminosity
induced by a non-zero value of ${\bf C}^{D\mu}_{33}$ is well within PDF uncertainties.
This is consistent with the result of Fig.~\ref{fig:cb-bounds-hl}
indicating that bounds on ${\bf C}^{D\mu}_{33}$ obtained with SM and with SMEFT PDFs are relatively
similar in this scenario even after accounting for the HL-LHC constraints.

\section{Conclusions and outlook} 
\label{sec:conclusions}

Indirect searches for new physics beyond the SM, such as those
carried out in the SMEFT framework, often aim at pinning down subtle distortions with respect
to the SM predictions, such as a few-percent deviation in the value of a given production
cross section or decay rate.
Exploiting the full potential of current and future precision measurements at the LHC
for these indirect BSM searches
requires the development of novel data interpretation frameworks that are able to account
for hitherto ignored effects that can no longer be neglected.
A pressing example of this is the interplay between PDF and EFT effects
in the high-energy tails of LHC distributions.
Indeed, the very same datasets
are being used both to determine the parton distributions (assuming SM cross sections)
and, independently, to constrain EFT coefficients (assuming SM PDFs).
Given that these LHC processes provide significant information for both PDF and EFT fits,
it is of paramount importance to ascertain the extent for which eventual EFT signals
can be reabsorbed into the PDFs, as well as how current bounds on the EFT coefficients are modified
within a consistent simultaneous determination together with the PDFs.

In this work, building upon our previous DIS-only study~\cite{Carrazza:2019sec}, we have presented
a first simultaneous determination of PDFs and EFT coefficients from high-energy LHC
data, specifically from high-mass Drell-Yan cross sections.
Our analysis has considered available unfolded measurements, detector-level
searches  based on the full Run II luminosity, and tailored HL-LHC projections.
The EFT interpretation of the Drell-Yan data is formulated
in terms of two benchmark scenarios, first a flavour
universal one leading to modifications of the  $\hat{W}$ and $\hat{Y}$ electroweak parameters~\cite{Farina:2016rws},
and second a flavour-specific scenario motivated by the recent evidence for lepton flavour universality violation in $B$-meson decays~\cite{Greljo:2017vvb}.

\begin{figure}[t]
\begin{center}
\includegraphics[width=0.99\textwidth]{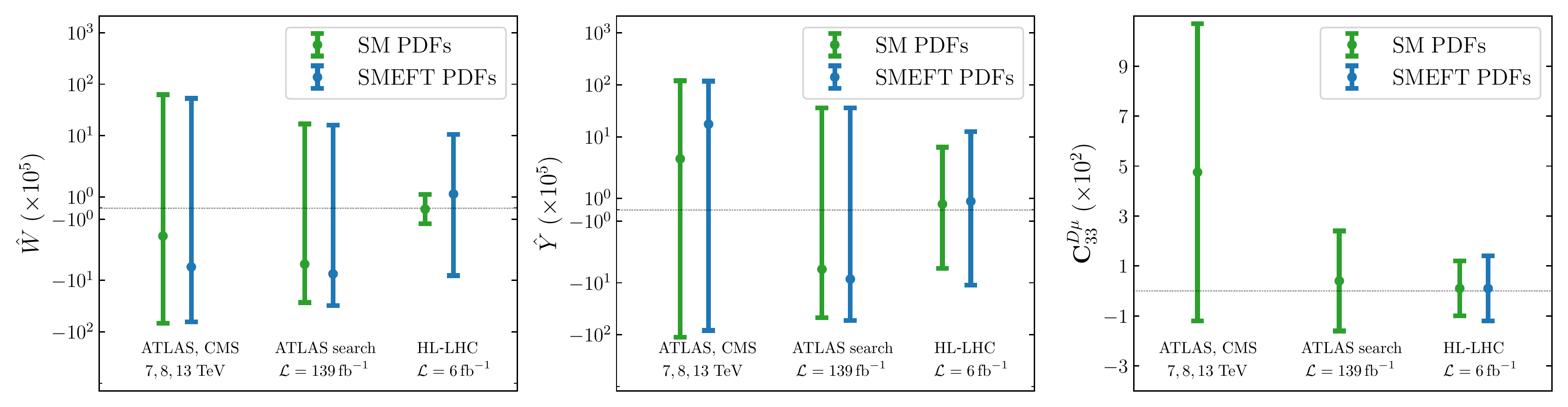}
\caption{\label{fig:plot_bounds_summary} Overview of the  95\% CL bounds
  obtained in this work
  the EFT parameters $\hat{W}$ (left), $\hat{Y}$
  (middle), and ${\bf C}^{D\mu}_{33}$ (right panel)
  based on either the SM PDFs of the SMEFT PDFs. Both PDF and
  methodological uncertainties are accounted in the bounds, when
  available. 
}
\end{center}
\end{figure}

The main findings of this work are summarised
in Fig.~\ref{fig:plot_bounds_summary}.
We demonstrate how, for the analysis of all available unfolded Drell-Yan data,
the consistent simultaneous extraction of the PDFs together with the EFT parameters leads to a modest increase in the uncertainties of the latter (up to 15\%, in the case of the $\hat{W}$ and $\hat{Y}$ parameters),
as well as to a shift in their central values by up to a third of a sigma.
Furthermore, while our results indicate that for current data the interplay between PDF and EFT effects
remains moderate, the impact of their cross-talk will become much larger
at the HL-LHC: using SM rather than SMEFT PDFs would lead to artificially precise
bounds, even mimicking new physics effects.
This result indicates that including high-energy data in PDF
fits should be done with care, as PDFs can actually absorbe the effects of
new physics. On the other hand, we have seen that in
this simple case, using a conservative set of PDFs that does not
include any of the high-mass Drell-Yan data and accounting for the
large contribution of the PDF uncertainty on the bounds inflates them 
and makes them of the same order of magnitude as those obtained in a
simultaneous fit of PDFs and SMEFT coefficients.

At the same time, once real data at HL-LHC are considered, neglecting the PDF interplay and simply
using conservative sets of PDFs might miss EFT manifestations of new
physics or misinterpret them.  One should also emphasise that
estimators such as those shown in Fig. 5.6 only become available in
the joint PDF+EFT fit, and cannot be defined in the ``conservative PDF"
approach. In particular, they provide information on the kinematic
dependence of any possible deviation between the data and the SM
predictions, and the extent to which this can be reabsorbed into the
PDFs. Hence, they represent a powerful diagnosic tool to separate QCD
effects from genuine BSM deviations.
A complementary strategy to disentangle QCD effects from BSM effects would
be to account for the constraints on the large-$x$ PDFs  arising from other processes
 for which EFT corrections can be neglected, such as forward $W,Z$
 production at LHCb. This way the uncertainties associated to the PDFs
 at large-$x$ would be reduced and the indirect signal for new physics
 could be more easily disentangled. A detailed study aimed at a
definition of conservative PDFs in a more general scenario is beyond the scope of this paper, and will be the topic of future work.

Furthermore, concerning the next steps in this program, it would be interesting to consider other more general EFT benchmark
scenarios, as well as accounting for new 13 TeV measurements based on increased
luminosity.
One could also envisage complementing the EFT analysis of inclusive DY cross sections
with related processes, such as dilepton production
in association with extra jets, which provides
sensitivity to different combinations of dimension-six operators.

In addition to increasing the dataset,
one would need to address a major challenge of the
current fitting methodology for the joint determination
of PDFs and EFT coefficients, namely that for each point in the EFT parameter space one needs
to carry out a full-fledged NNPDF fit, which requires intensive
computing resources. To tame the instability of the $\chi^2$, once all
datasets are kept into account, one needs to run a very large number
of replicas. Furthermore, in the simultaneous fit of EFT coefficients
and PDFs, we currently ignore the correlation between the two sets of
parameters. 
The kind of analysis presented in this work could then be streamlined and made
more efficient by adapting the
fitting methodology to exploit the relatively simple
dependence of hadronic cross sections on the EFT coefficients, which is either linear at $\mathcal{O}(\Lambda^{-2})$
or quadratic at $\mathcal{O}(\Lambda^{-4})$.
Such methodological developments would make it possible to simultaneously fit
the PDFs with a large number of EFT coefficients, something that
is currently infeasible but that is being developed. 

Finally, beyond single gauge boson production, it would also be important to ascertain the
interplay that arises between PDFs and EFT effects for the interpretation of gluon-dominated LHC processes,
such as top-quark pair production and inclusive jet and dijet (or even multijet) production.
The reason is that these two groups of processes have been shown to provide
crucial information for, on the one hand, pinning down
the gluon PDF over a broad range of $x$ values~\cite{AbdulKhalek:2020jut,Czakon:2016olj},
and on the other hand, constraining a large number of EFT dimension-six operators~\cite{Farina:2018lqo,Brivio:2019ius,Hartland:2019bjb,Alte:2017pme} which cannot be accessed by other probes.
Given that in modern global PDF analyses the gluon for $x\simeq 10^{-2}$ is almost entirely determined
by these and related high-energy LHC processes, it is conceivable that the PDF and EFT interplay there
could be more significant than for the inclusive DY processes studied in this work, even just accounting for
the already available measurements.

\section*{Acknowledgments}
We thank Emanuele Mereghetti, Tevong You and Celine Degrande for
insightful discussions about the project. We thank Claude Duhr and Bernhard Mistlberger for kindly
sending us the NNLO and N3LO QCD corrections for the Drell-Yan
invariant and transverse mass distributions. We thank Andrea Wulzer
and Lorenzo Ricci for benchmarking the charged current K-factors
and for suggesting to add the results obtained by using conservative
PDFs. 
M.~U. and Z.~K. are supported by the European Research Council under the 
European Union’s Horizon 2020 research and innovation Programme (grant agreement n.950246).
M.~U. and S.~I. are supported by the
Royal Society grant RGF/EA/180148.
The work of M.~U. is also funded by the Royal Society grant DH150088.
The work of J.~R. is partially supported by the Netherlands Science Council (NWO).
The work of A.~G. has received funding from the Swiss National Science Foundation (SNF) through the Eccellenza Professorial Fellowship ``Flavor Physics at the High Energy Frontier'' project number 186866, and is also partially supported by the European Research Council under the European Union’s Horizon 2020 research and innovation programme, grant agreement 833280 (FLAY).
The work of J.~M. is supported by the Sims Fund Studentship.
The work of M.~M. is supported by the University of Cambridge Schiff Foundation studentship.
C.~V. is supported by the STFC grant ST/R504671/1.
M.~U., S.~I., Z.~K., J.~M. and M.~M. are partially supported by STFC consolidated grants ST/P000681/1, ST/T000694/1.

\appendix
\section{Detailed SM PDF comparisons}
\label{app:pdfs}

\begin{figure}[t]
  \centering
  \includegraphics[width=0.32\textwidth]{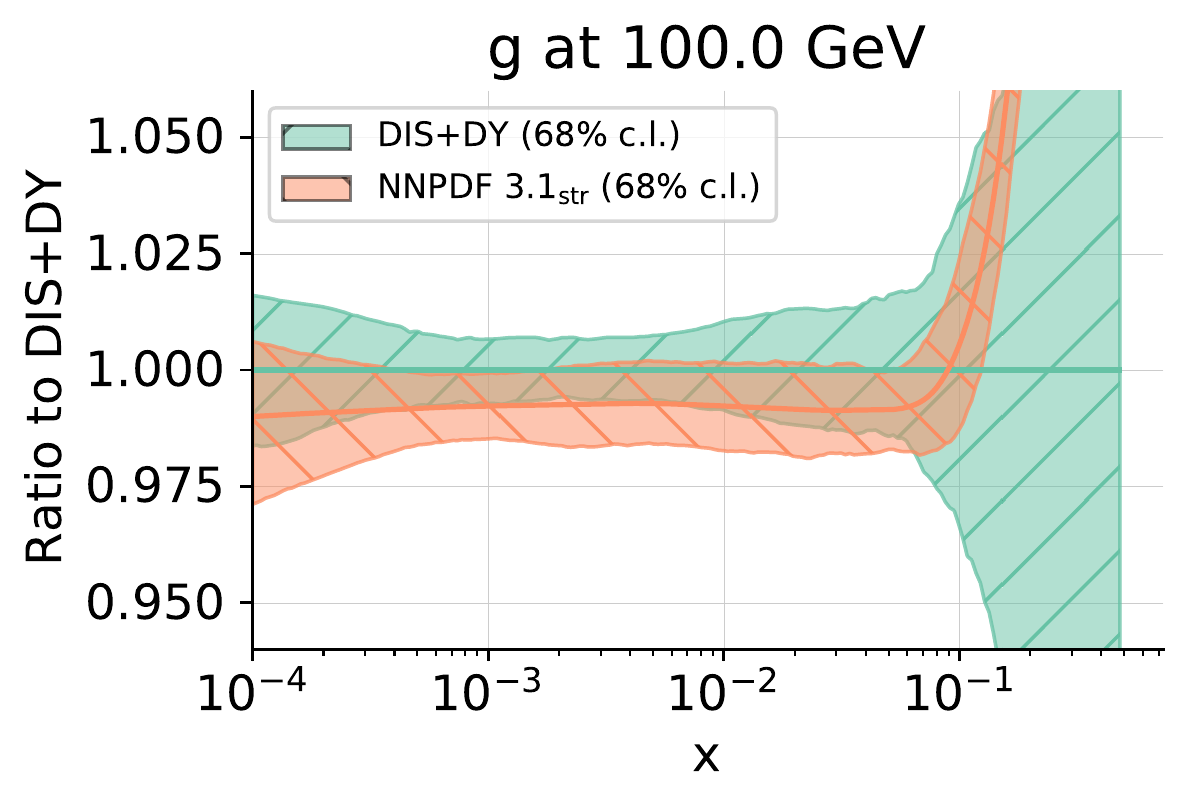}
  \includegraphics[width=0.32\textwidth]{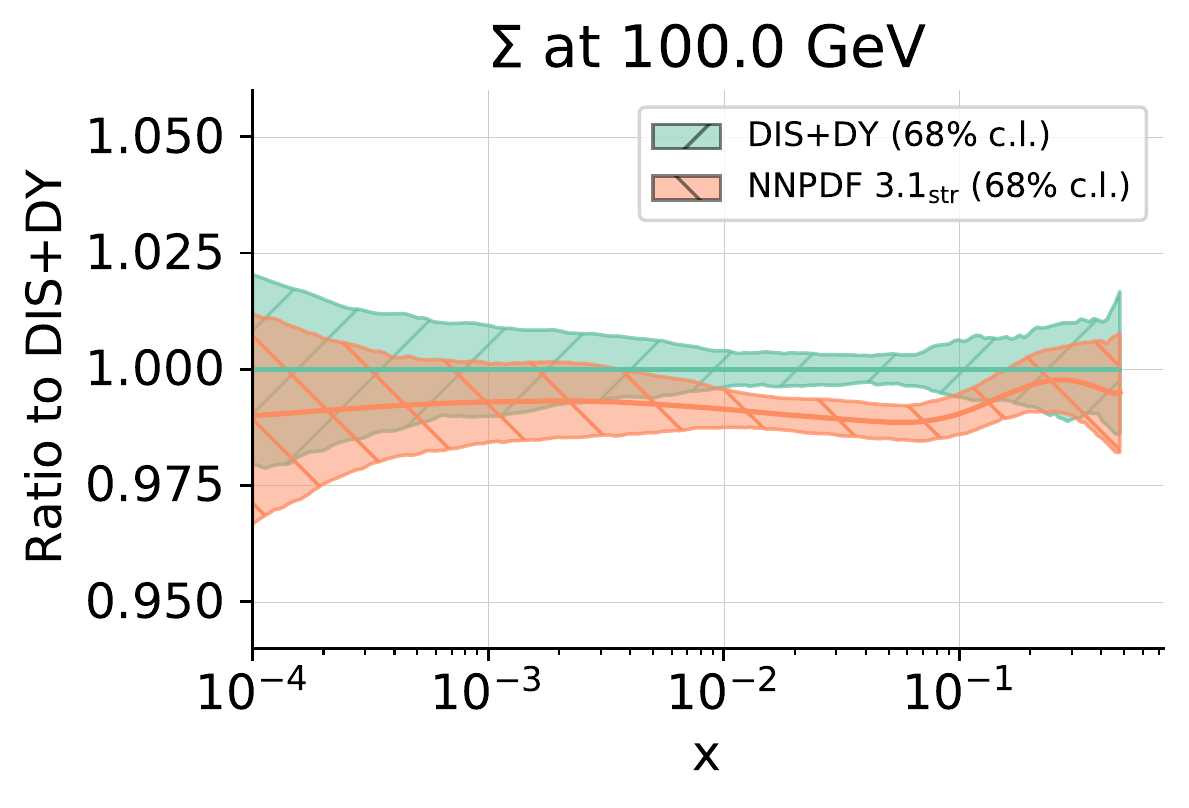}
  \includegraphics[width=0.32\textwidth]{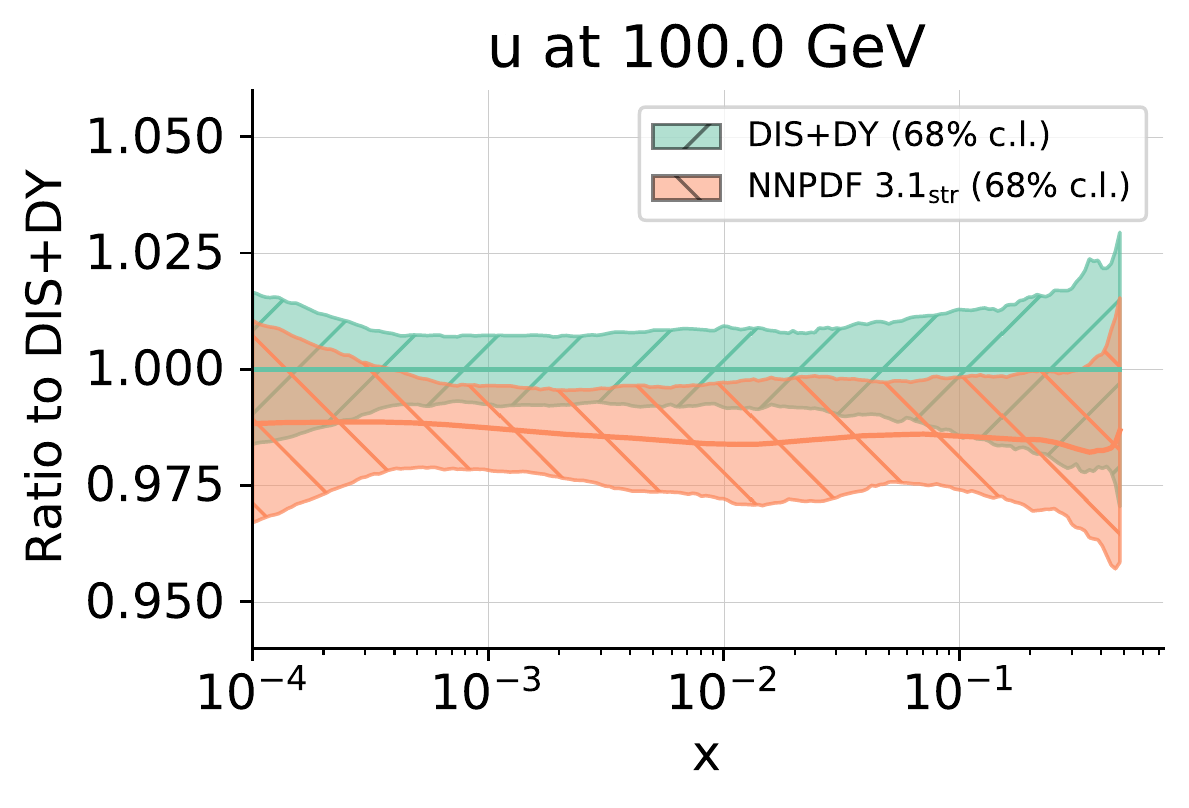}
  \includegraphics[width=0.32\textwidth]{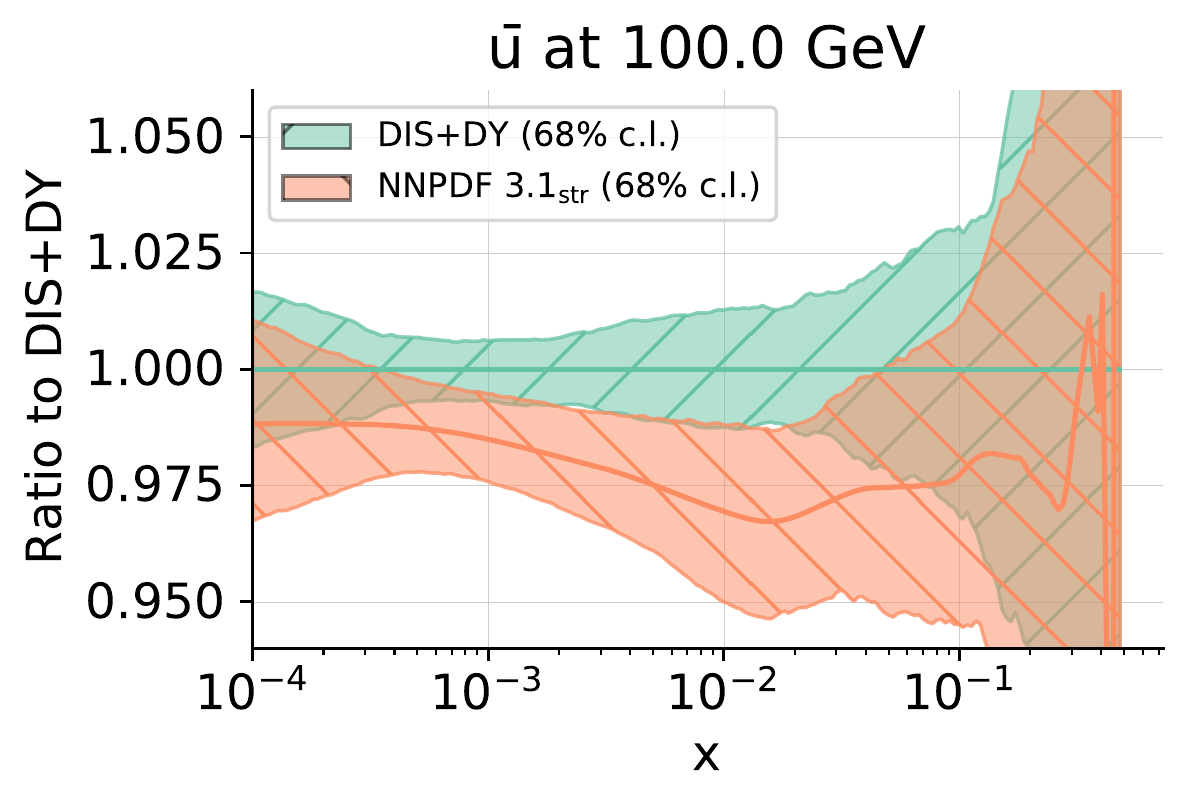}
  \includegraphics[width=0.32\textwidth]{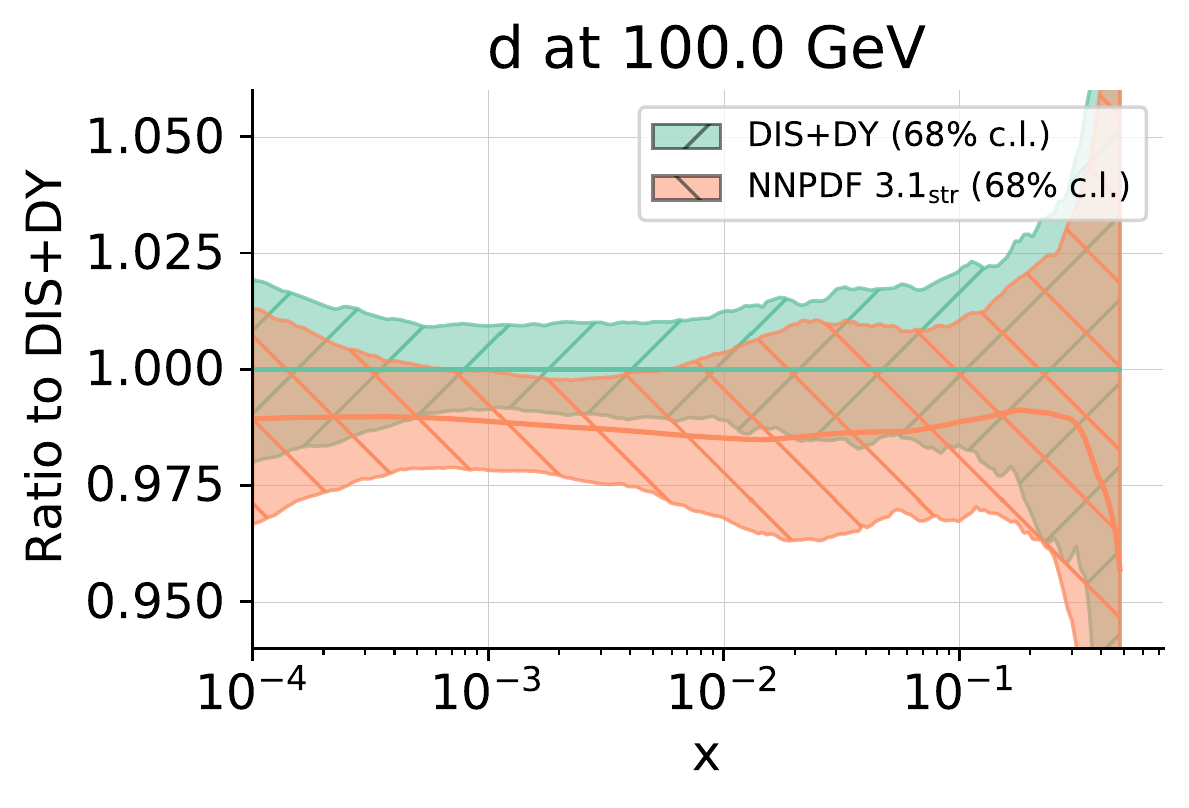}
  \includegraphics[width=0.32\textwidth]{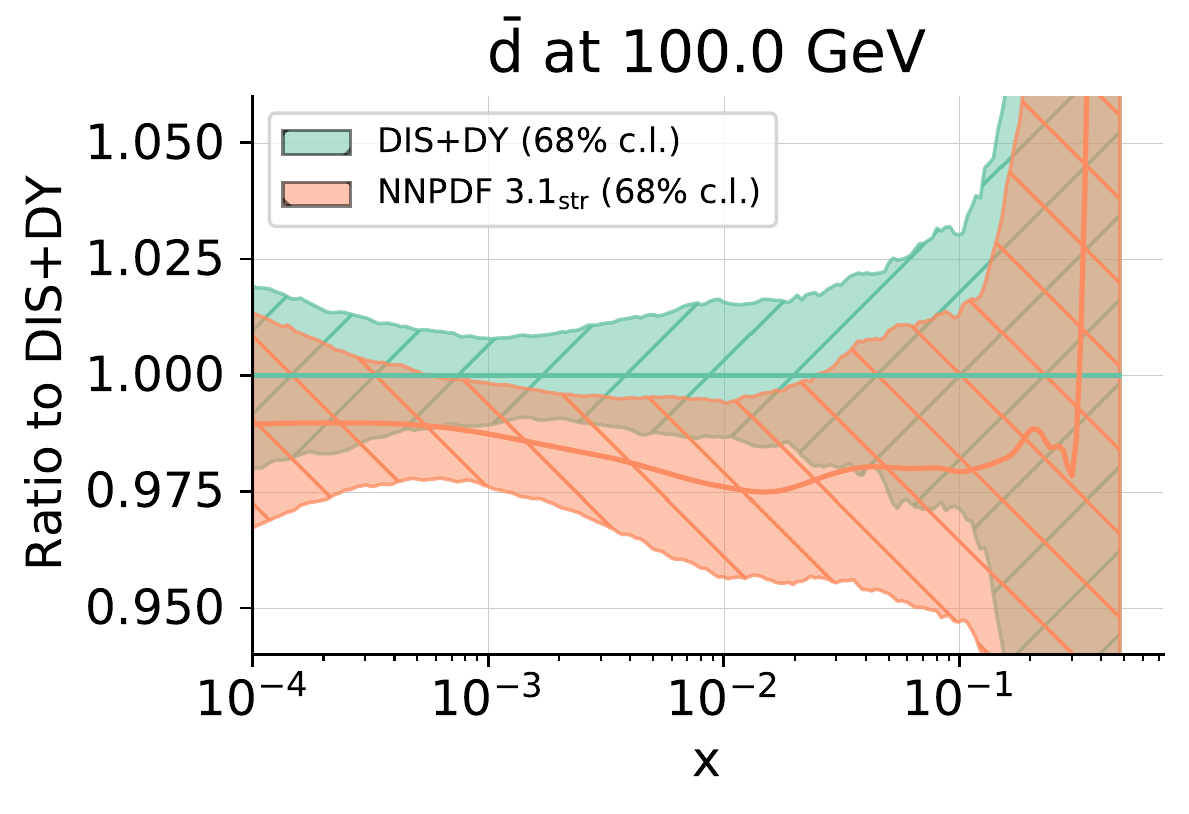}
  \includegraphics[width=0.32\textwidth]{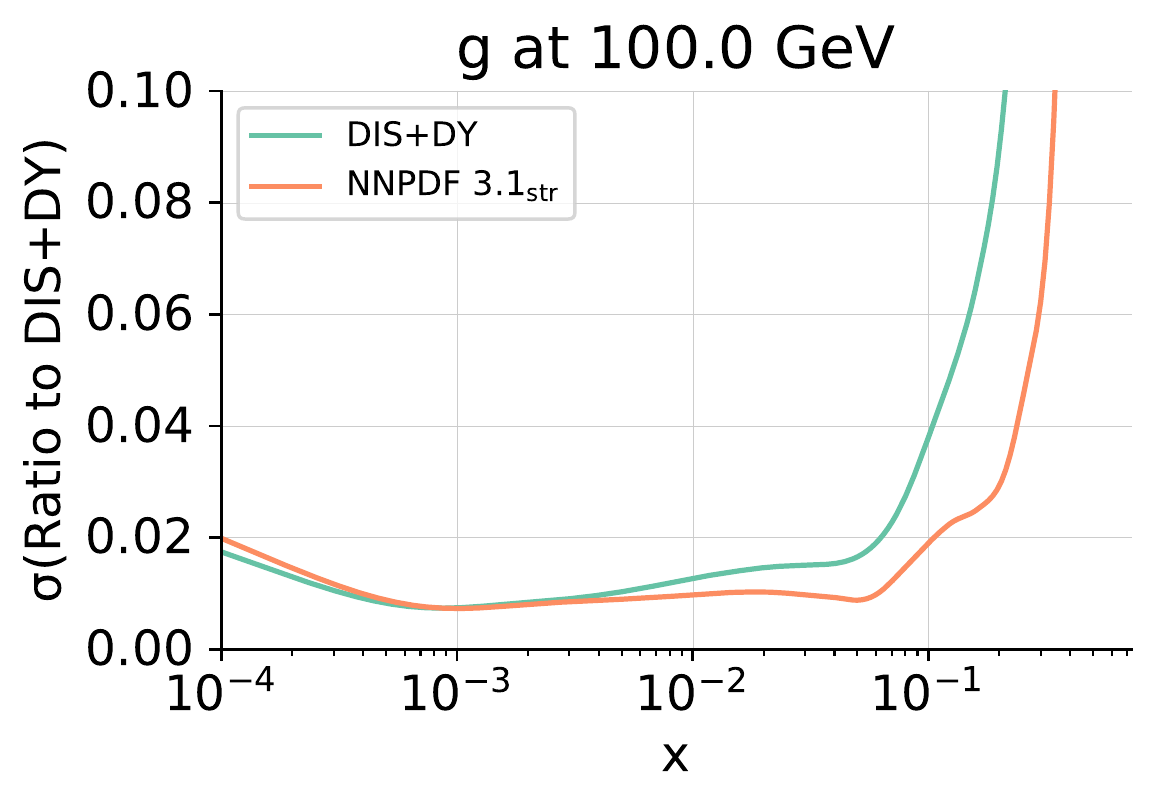}
  \includegraphics[width=0.32\textwidth]{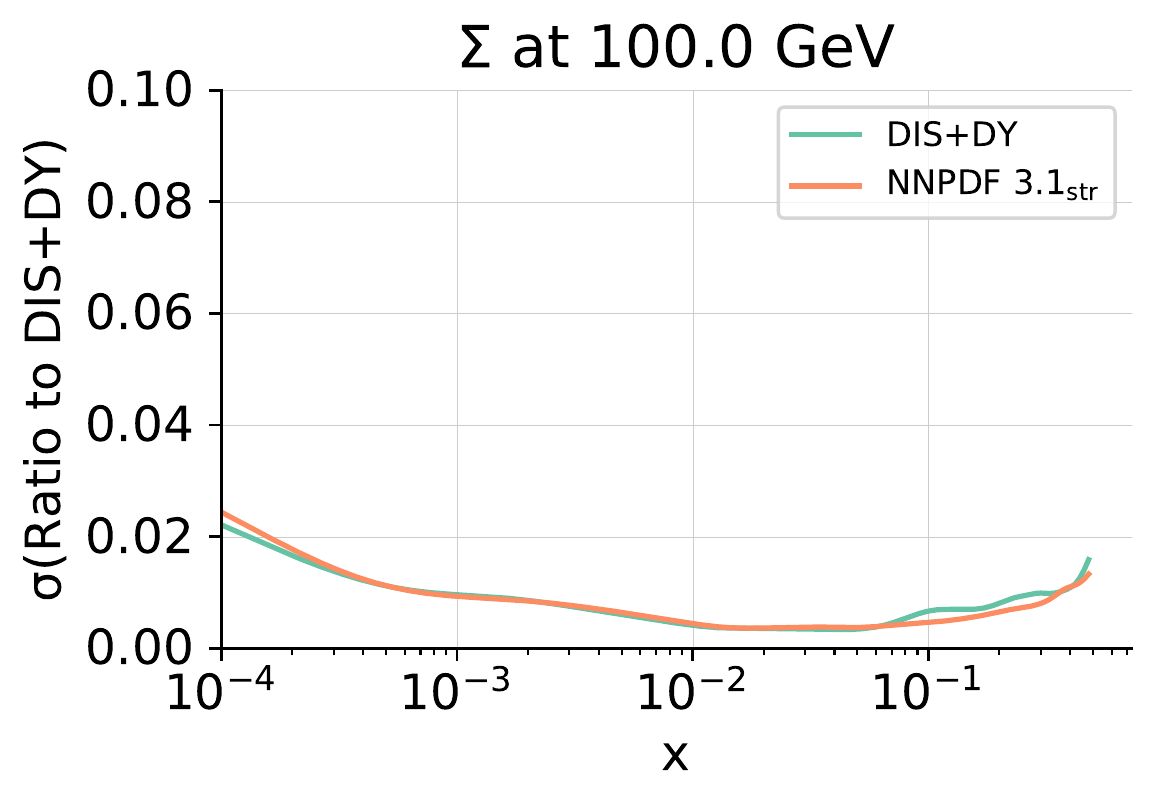}
  \includegraphics[width=0.32\textwidth]{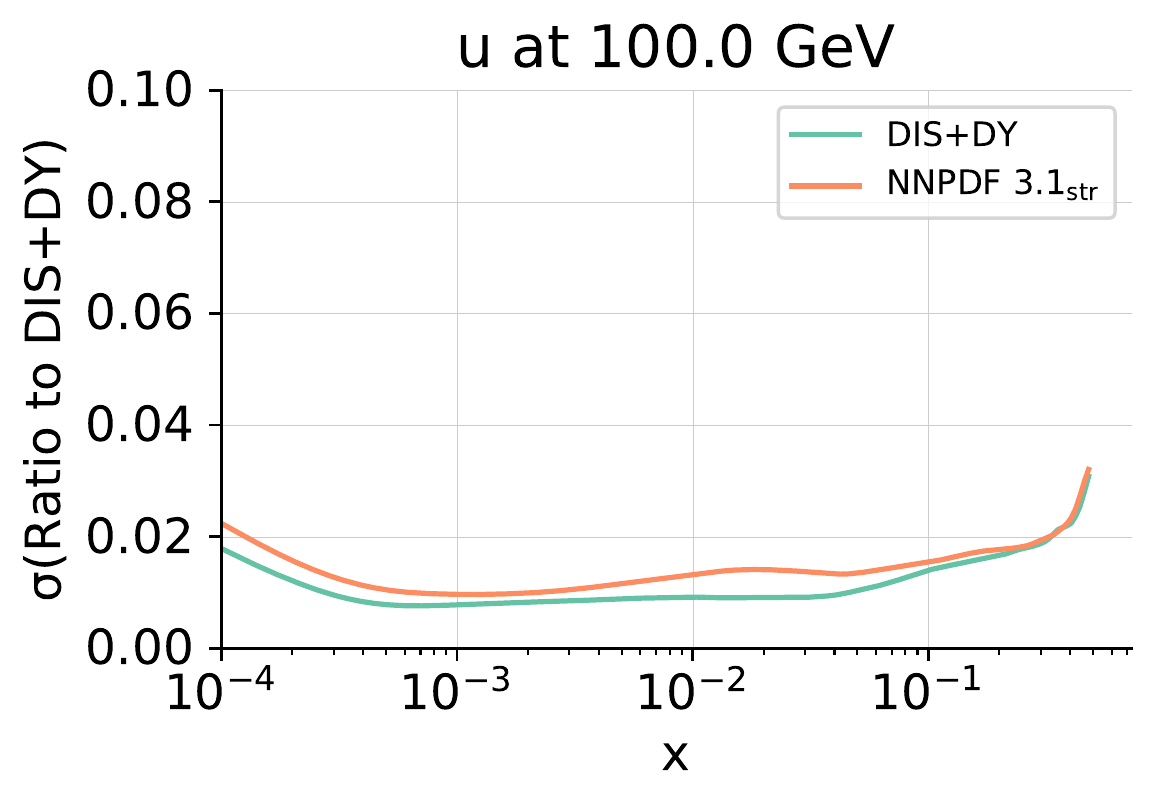}
  \includegraphics[width=0.32\textwidth]{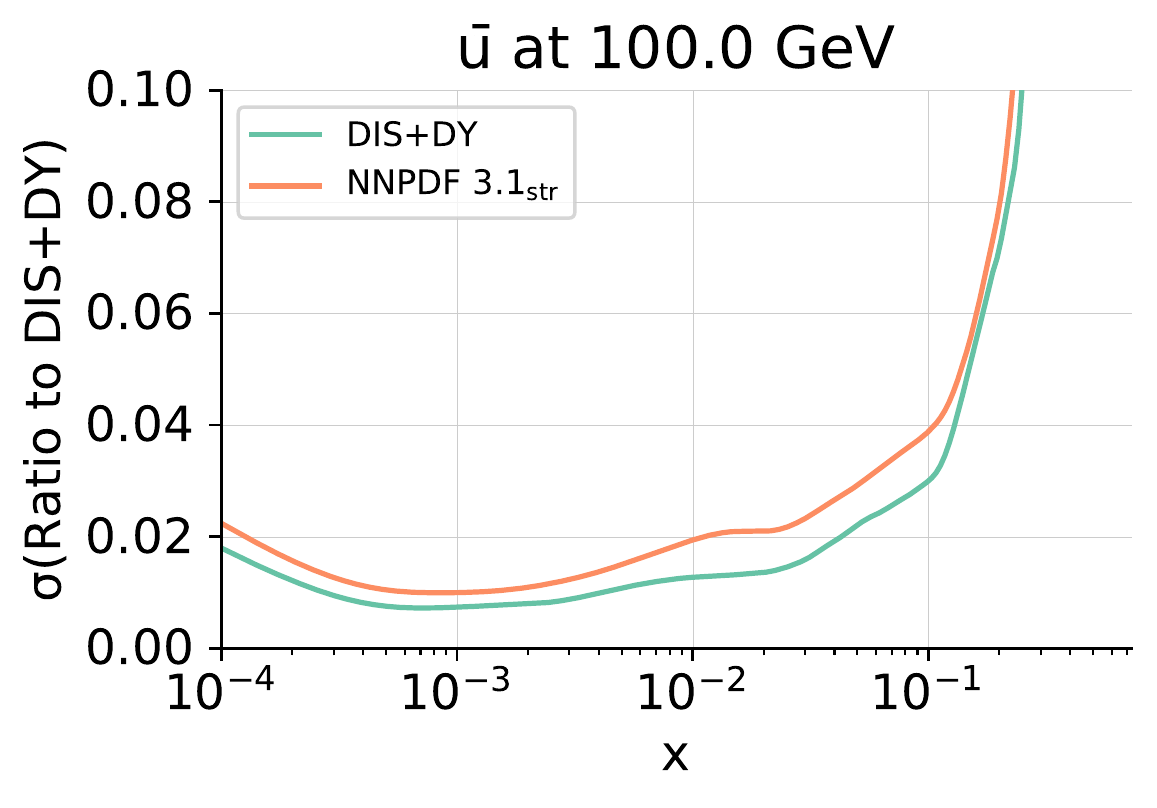}
  \includegraphics[width=0.32\textwidth]{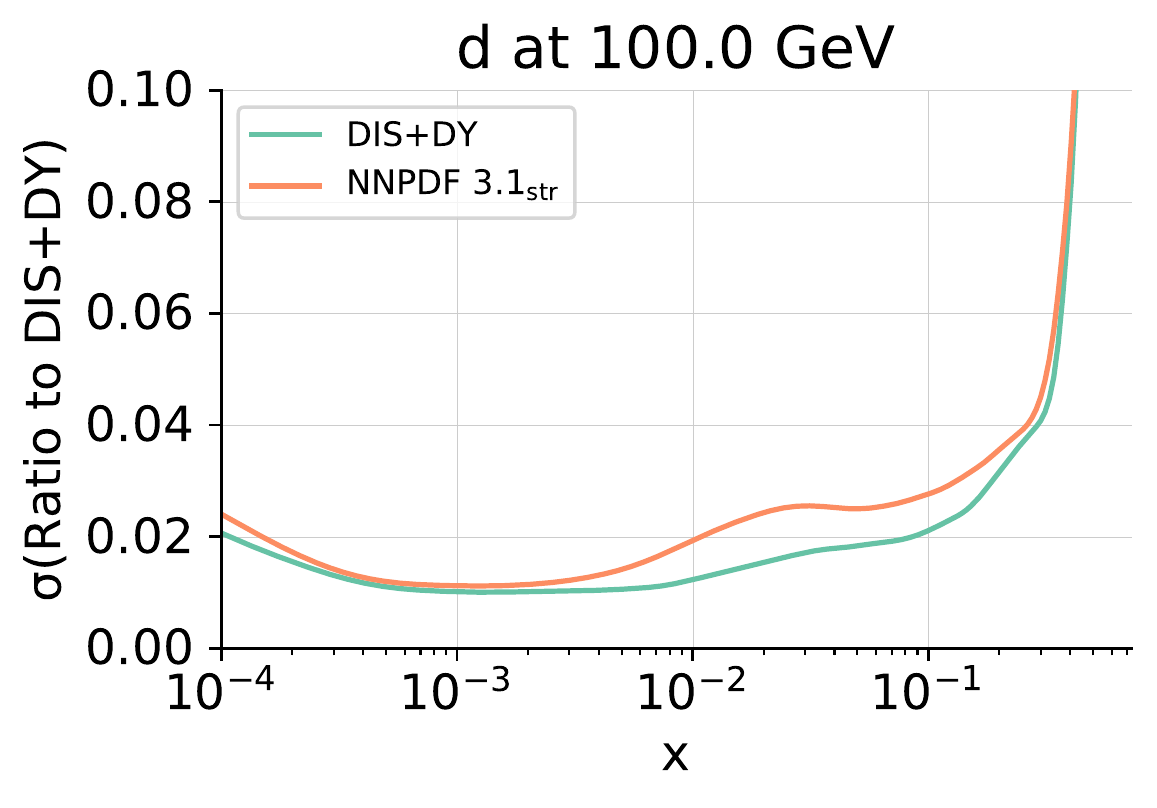}
  \includegraphics[width=0.32\textwidth]{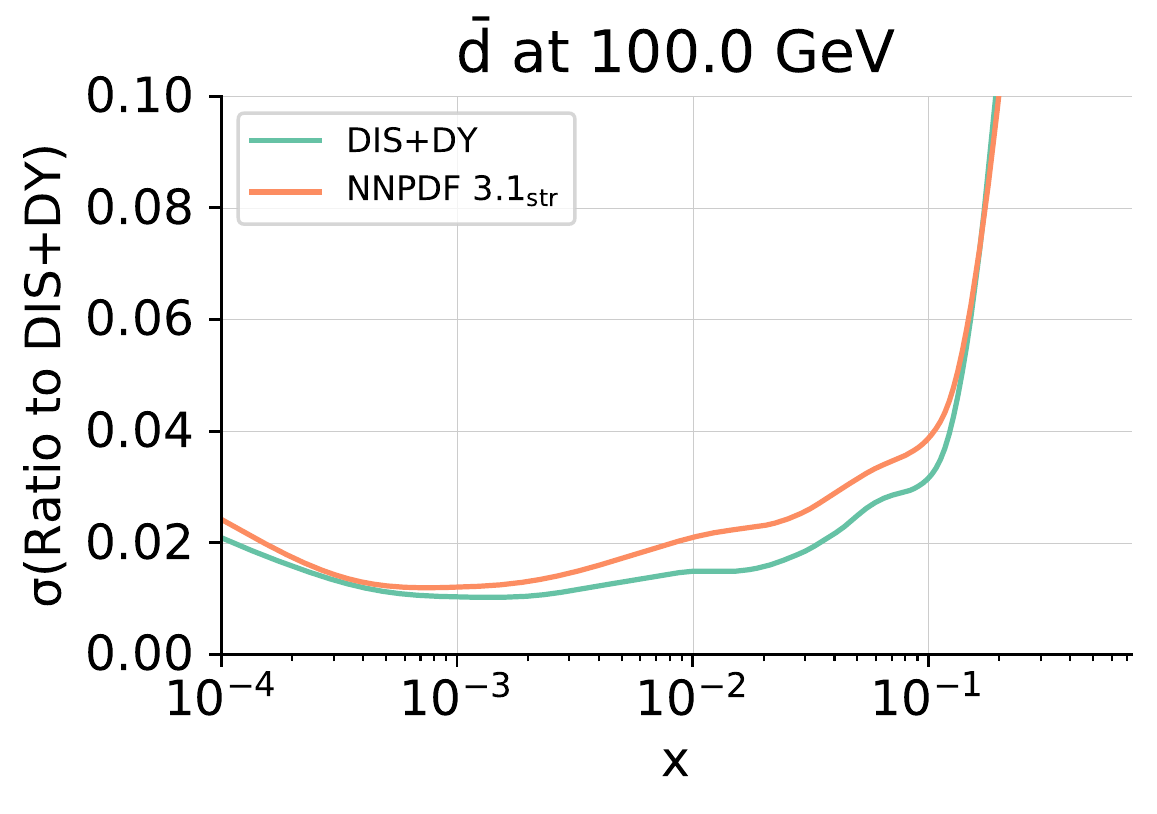}
  \caption{\small Same as Fig.~\ref{fig:pdfplot-impactHMDY} comparing the baseline SM PDF set used
    in this work (labelled ``DIS+DY'') with the global {\tt NNPDF3.1\_str} determination.
  \label{fig:pdfplot-rat-base_vs_global}}
\end{figure}

In this appendix we present detailed  comparisons
between different sets of SM PDFs to complement the discussions
in Sect.~\ref{sec:data}.
To begin with, we compare the baseline SM PDF of this work,
based on DIS+DY data, with the recent  global {\tt NNPDF3.1\_str} fit
obtained in the context of the proton strangeness study of~\cite{Faura:2020oom}.
Fig.~\ref{fig:pdfplot-rat-base_vs_global} is the counterpart of
Fig.~\ref{fig:pdfplot-impactHMDY}, now
displaying the gluon, singlet, up, anti-up, down, and anti-down quark
PDFs at $Q=100$ GeV
both for the baseline SM PDF
(labelled ``DIS+DY'') and for the global {\tt NNPDF3.1\_str} determination.

We observe an overall good compatibility between our DIS+DY baseline
and the {\tt NNPDF3.1\_str} global fit, with PDFs
in agreement at the one-sigma level in all cases except for the quark singlet $\Sigma$
in the region $0.01 \lsim x \lsim 0.1$.
As discussed in Sect.~\ref{sec:fitsettings}, the new  high-mass DY data included in this analysis
as compared to~\cite{Faura:2020oom}
are responsible for the bulk of the differences observed in Fig.~\ref{fig:pdfplot-rat-base_vs_global},
both in terms of central values and uncertainties, for the quark
and anti-quark PDFs.
Specifically, the upwards shift in the central values of the quark and anti-quark PDFs
in this $x$-region
for the DIS+DY baseline as compared to the {\tt NNPDF3.1\_str} determination
is consistent with the comparisons in Fig.~\ref{fig:pdfplot-impactHMDY}
illustrating the impact of the high-mass Drell-Yan data in the fit,
and the same applies for the associated reduction of the quark and anti-quark
PDF uncertainties.

As is well known,
the PDF uncertainties on the gluon become rather enlarged in the DIS+DY baseline due to the
lack of information from the top and jet cross sections.
However, this does not
impact the results of the present joint PDF and EFT interpretation, given that
gluon-induced contributions
to inclusive Drell-Yan processes enter only starting at NLO.
Furthermore, we also find somewhat larger uncertainties in the strangeness of
the DIS+DY baseline as compared to {\tt NNPDF3.1\_str} due to the
missing constraints from the NOMAD neutrino
dimuon cross sections.
All in all, with the exception of gluon-initiated processes, we can conclude
that the DIS+DY baseline to be used in this work is competitive with a full-fledged global
PDF determination.

\begin{figure}[t]
  \centering
  \includegraphics[width=0.32\textwidth]{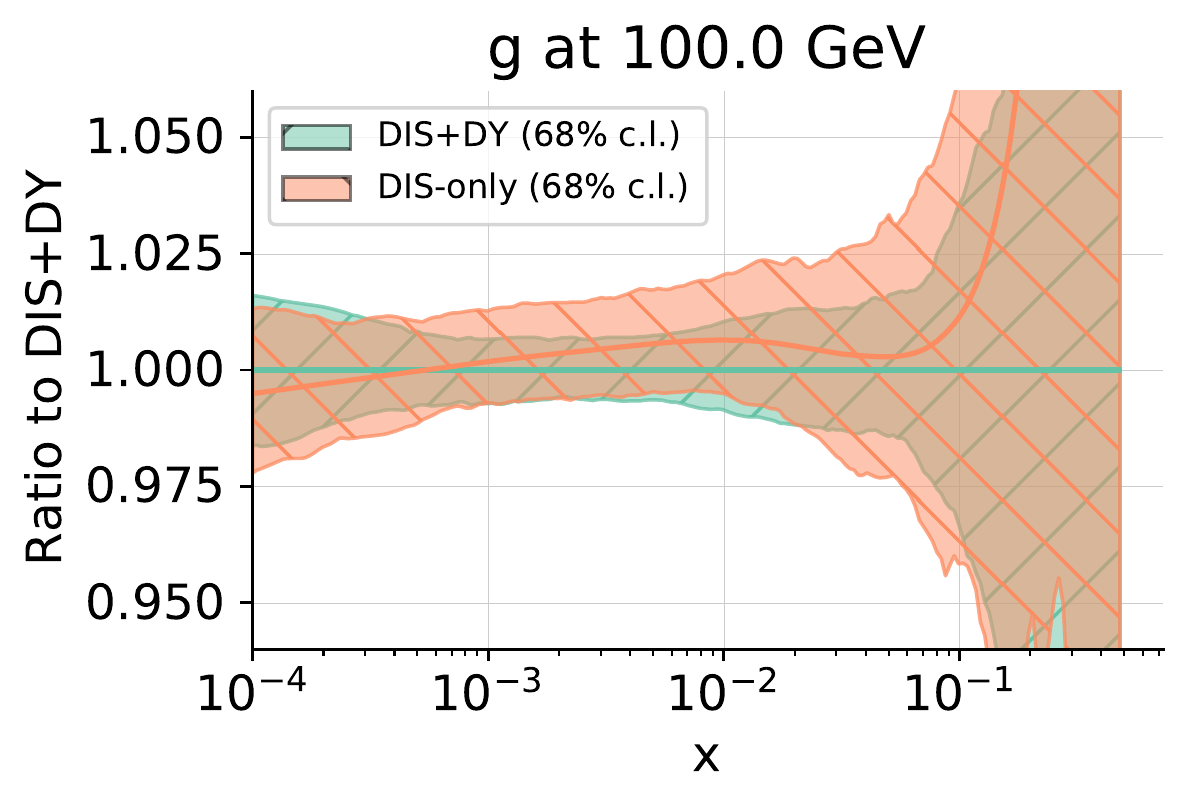}
  \includegraphics[width=0.32\textwidth]{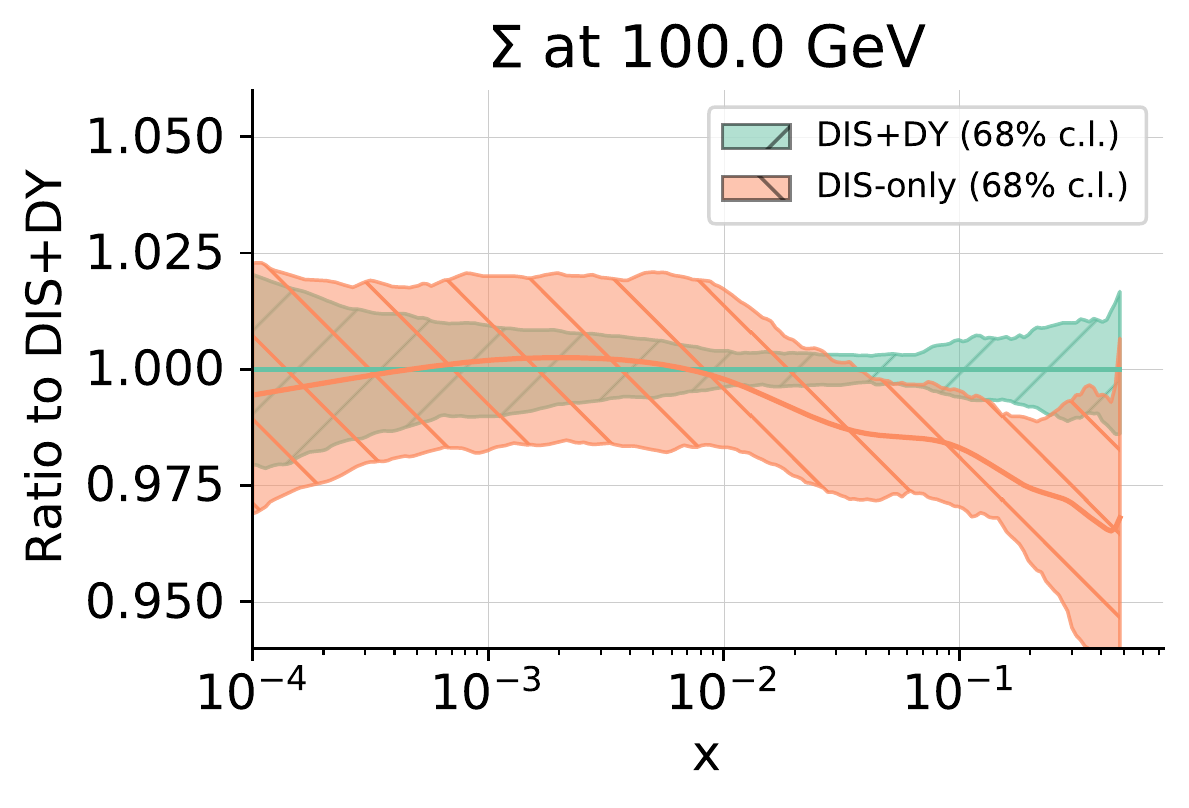}
  \includegraphics[width=0.32\textwidth]{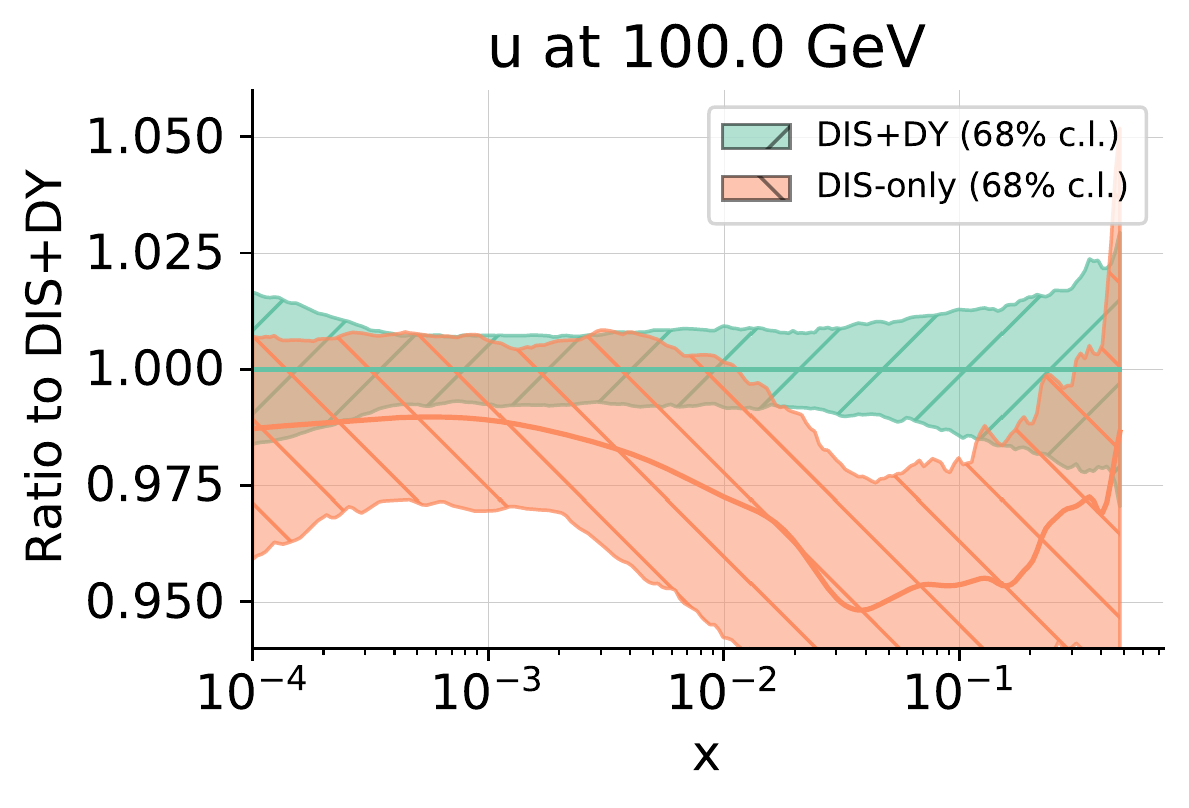}
  \includegraphics[width=0.32\textwidth]{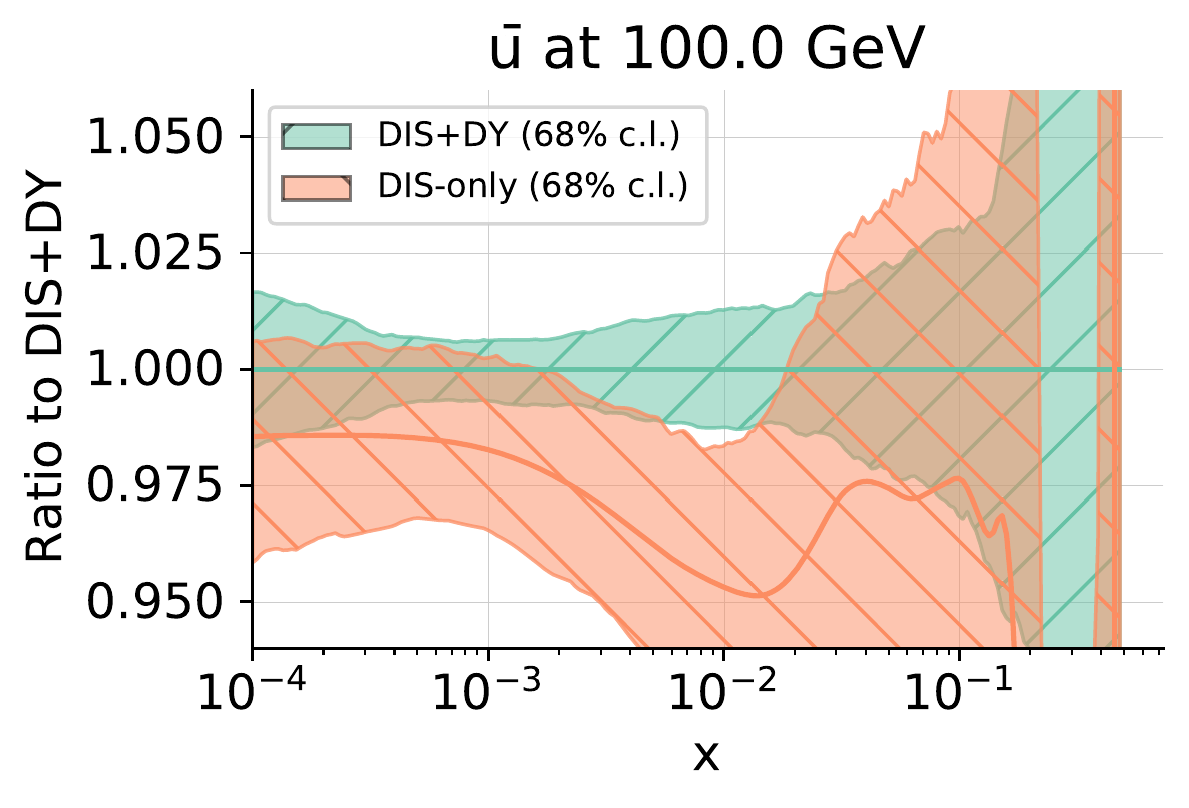}
  \includegraphics[width=0.32\textwidth]{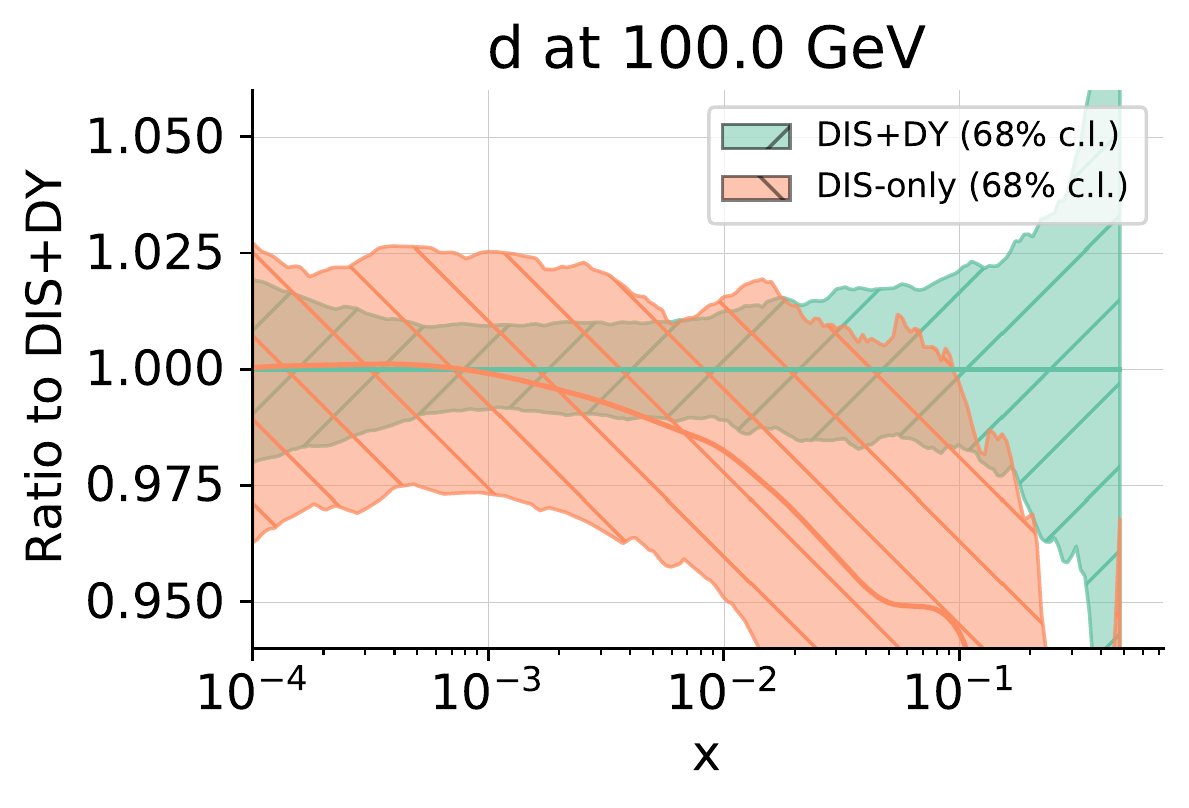}
  \includegraphics[width=0.32\textwidth]{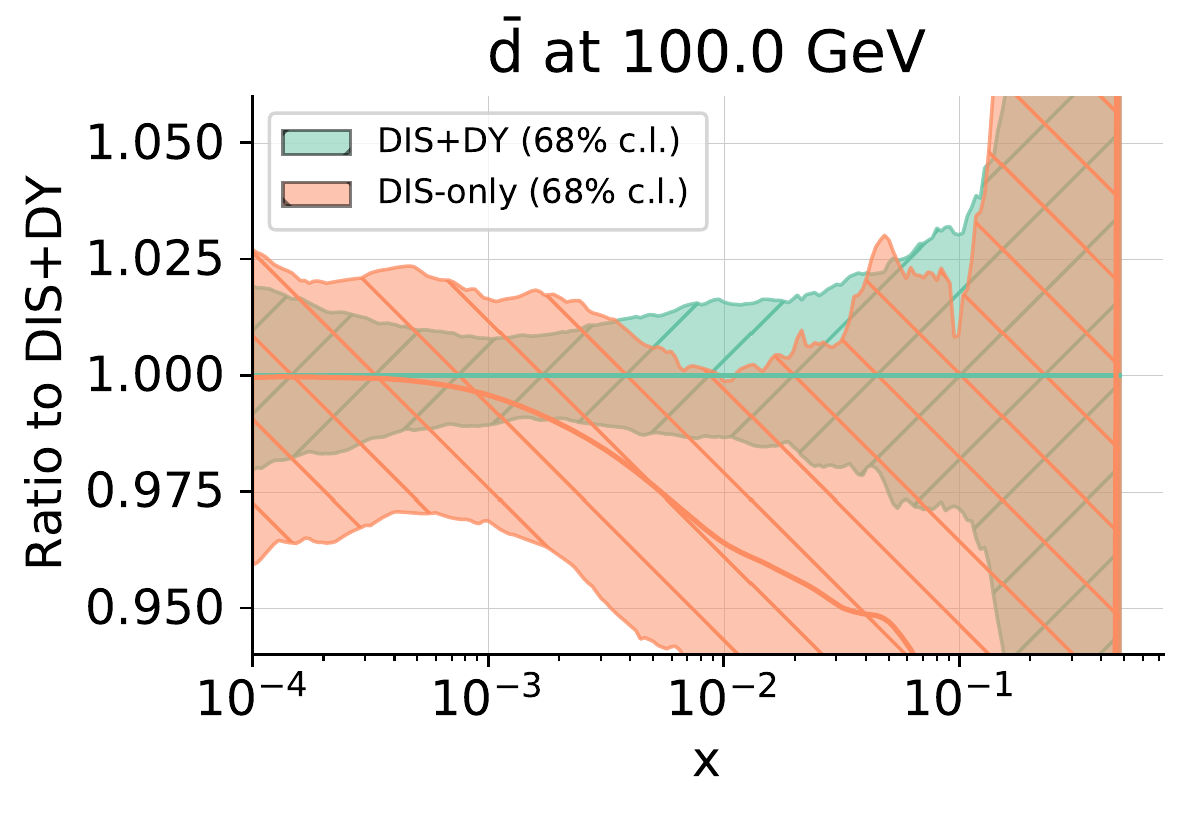}
  \caption{\small Same as the upper panels of Fig.~\ref{fig:pdfplot-impactHMDY} comparing now the
    baseline SM PDF set used
    in this work (labelled ``DIS+DY'') with the corresponding DIS-only fit.
    Note that the comparison between the PDF uncertainties in these two
    fits was already displayed in the lower panels of Fig.~\ref{fig:pdfplot-impactHMDY}.
  \label{fig:pdfplot-rat-dy_vs_dis}}
\end{figure}

Next, we display in Fig.~\ref{fig:pdfplot-rat-dy_vs_dis} the corresponding
comparison between the baseline SM PDF set based on DIS and DY data
(dubbed "DIS+DY'') with the same fit but only including DIS structure functions.
Note that the comparison between the PDF uncertainties in these two
fits was already displayed in the lower panels of Fig.~\ref{fig:pdfplot-impactHMDY}.
One can observe how in general there is excellent consistency between the two fits.
Indeed, PDFs are in agreement at the one-sigma level except for very specific cases,
such as the up quark PDF at $x\simeq 0.05$, but even there the differences are at most
at the 1.5$\sigma$ level.
The very marked reduction of PDF errors is also appreciable in the DIS+DY fit as compared to the DIS-only
fit, highlighting the importance of the DY data in the global PDF fit
to constrain the light quark and antiquark PDFs in a broad range of $x$.

\begin{figure}[t]
  \centering
  \includegraphics[width=0.32\textwidth]{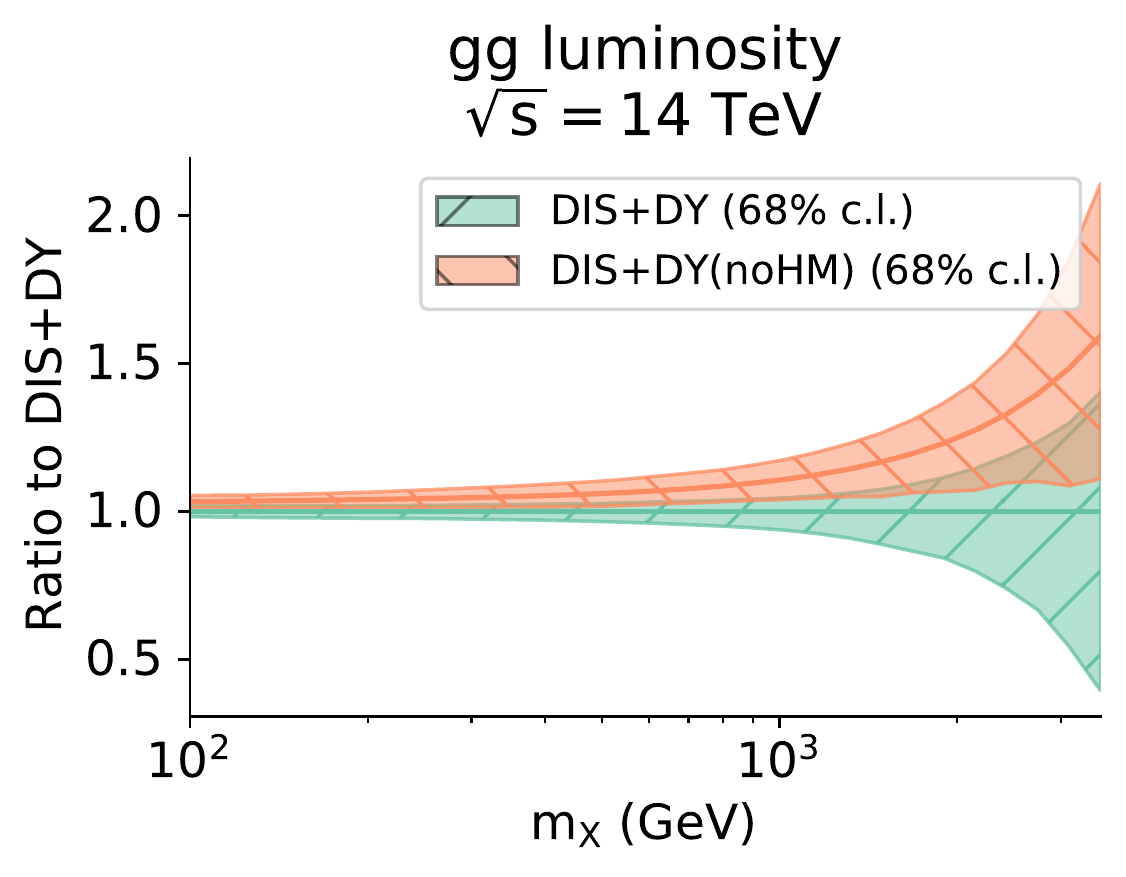}
  \includegraphics[width=0.32\textwidth]{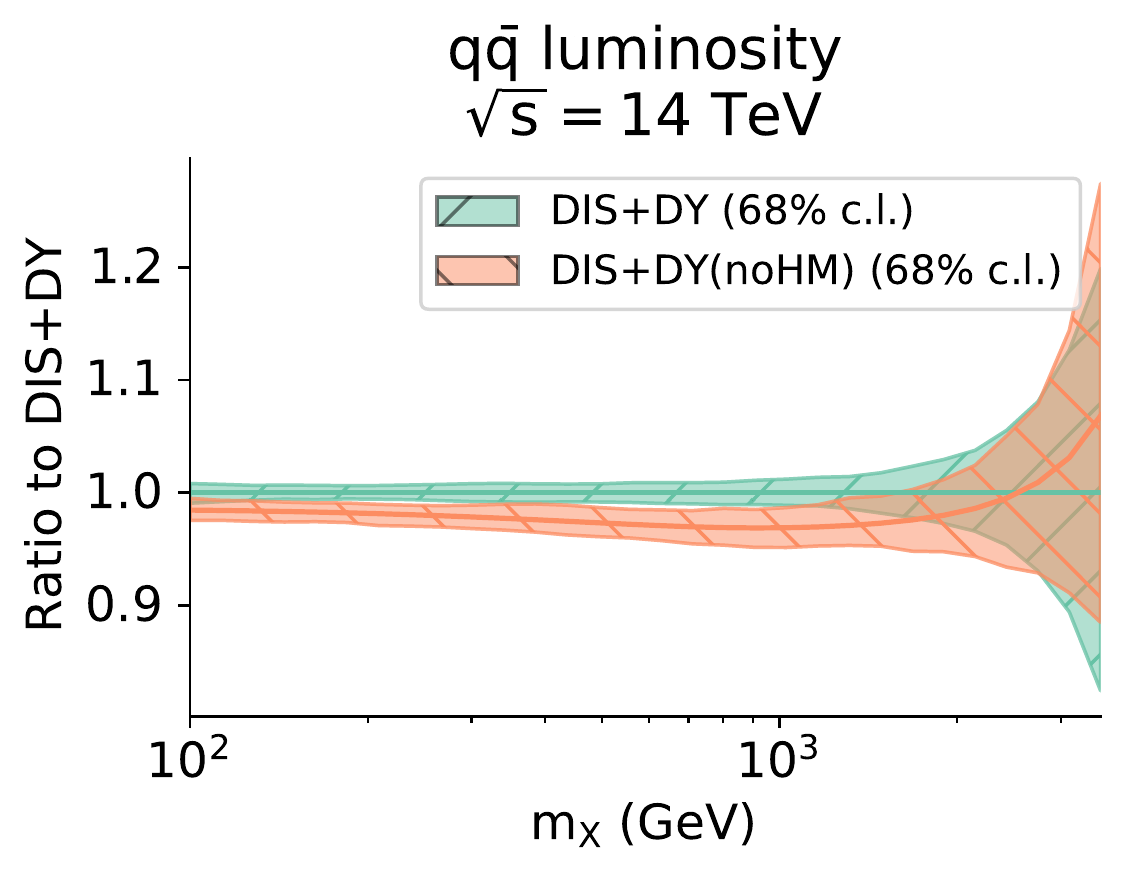}
  \includegraphics[width=0.32\textwidth]{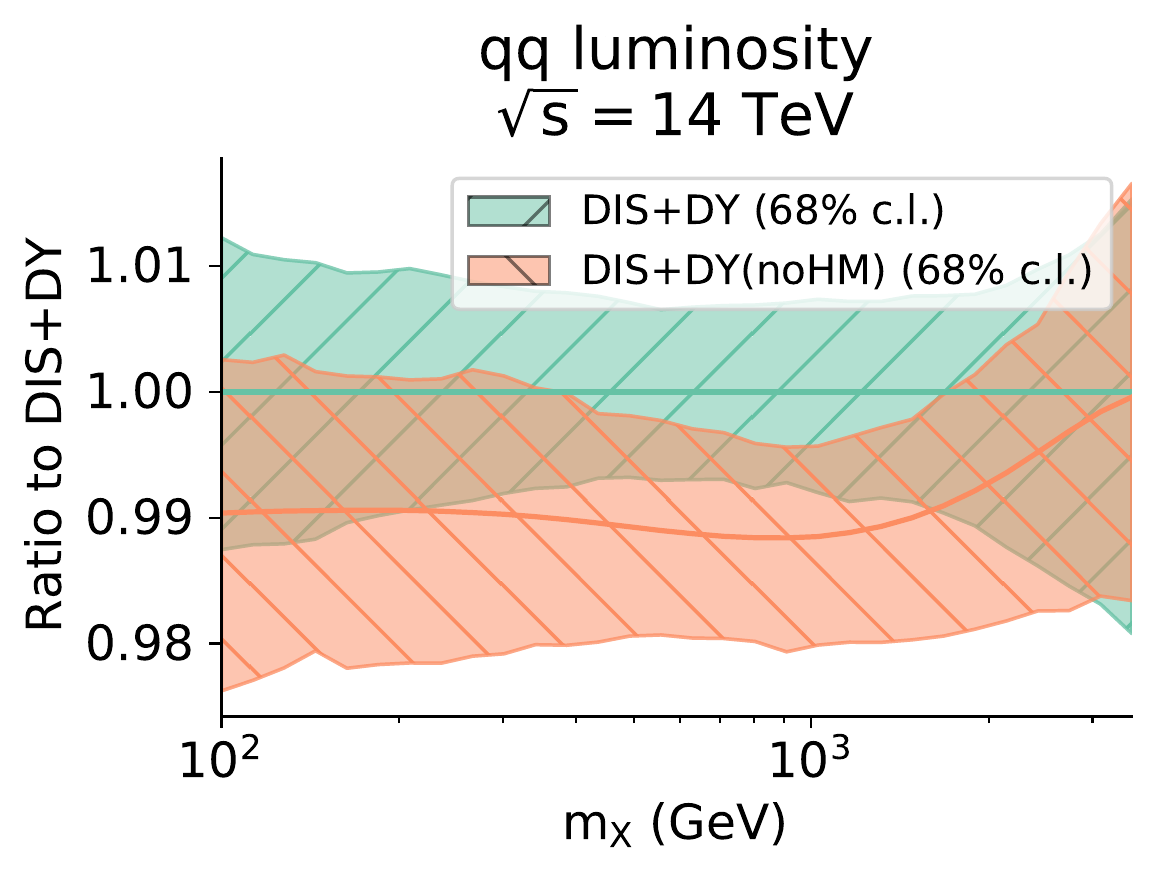}
  \includegraphics[width=0.32\textwidth]{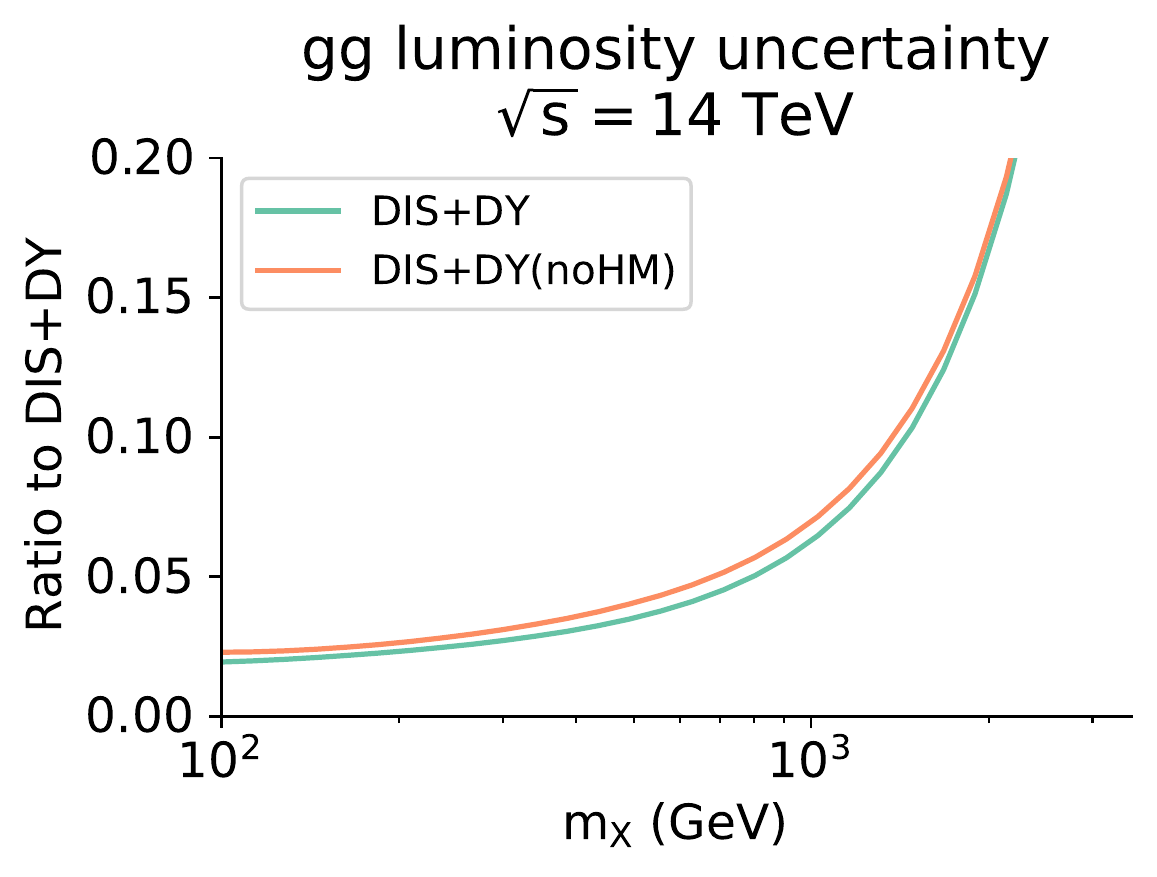}
  \includegraphics[width=0.32\textwidth]{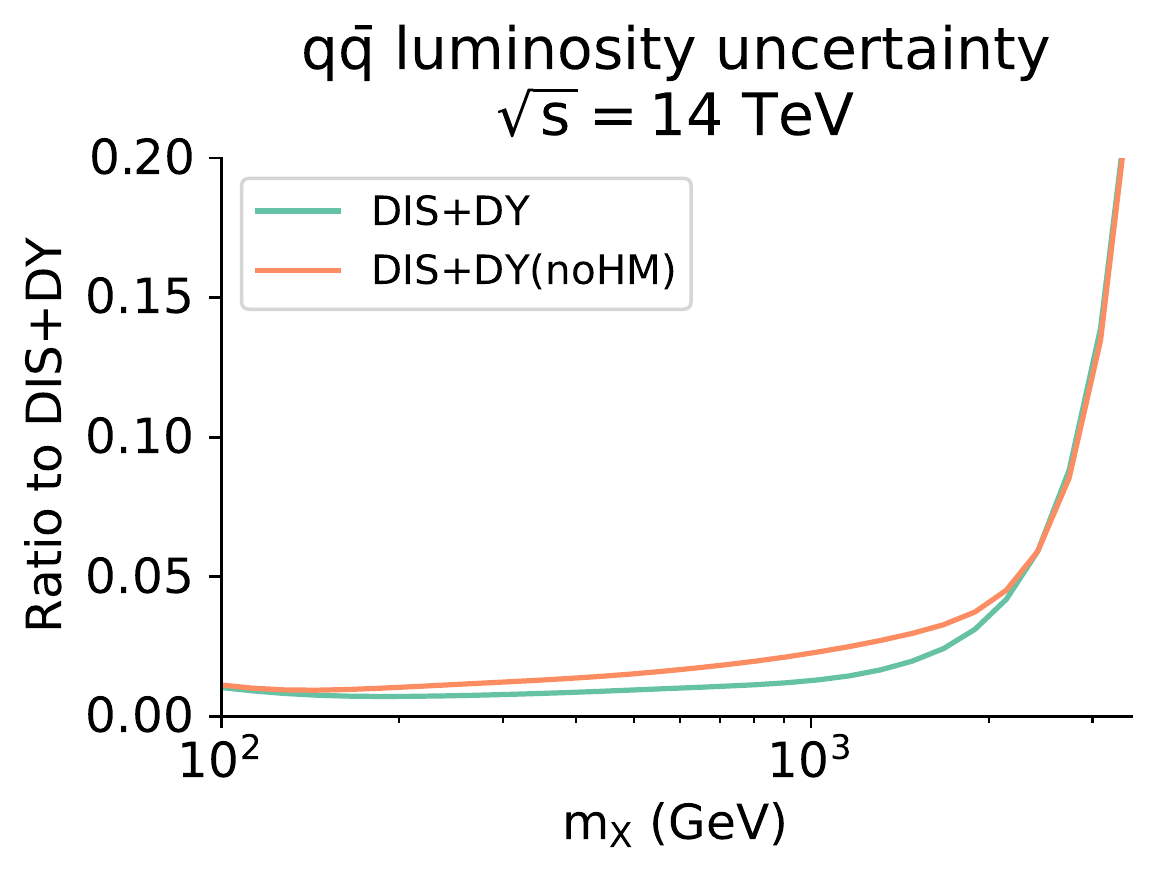}
  \includegraphics[width=0.32\textwidth]{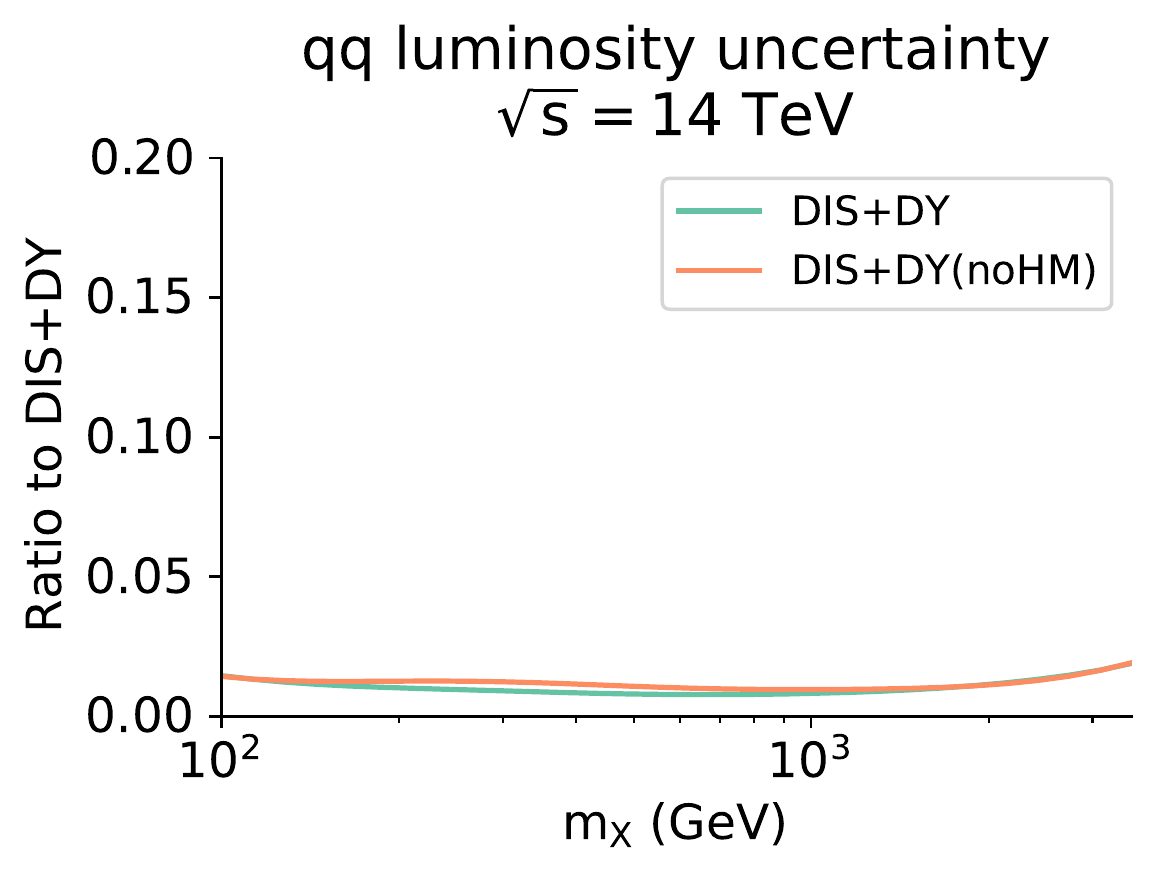}
  \caption{\small The gluon-gluon, quark-antiquark, and quark-quark luminosities
    at $\sqrt{s}=14$ TeV as a function of $m_X$ for the DIS+DY baseline
    and for the same fit excluding the all the datapoints in the high-mass
    DY experimental sets listed in Table.~\ref{tab:data-high-mass}.
    The top panels display the ratio of luminosities to the central value of the DIS+DY baseline,
    while the bottom panels compare the relative PDF uncertainties in each case.
  \label{fig:lumiplot-impactHMDY}}
\end{figure}

Finally, in Fig.~\ref{fig:lumiplot-impactHMDY} we compare the PDF luminosities in the DIS+DY baseline with those
from the same fit excluding all the data of the high-mass DY datasets
listed in Table~\ref{tab:data-high-mass}.
The corresponding comparisons at the PDF level was shown in Fig.~\ref{fig:pdfplot-impactHMDY}

We focus on the gluon-gluon, quark-antiquark, and quark-quark luminosities
at $\sqrt{s}=14$ TeV as a function of the invariant mass $m_X$ of the produced
final state, and display both the luminosity ratio to the reference
as well as the  relative PDF uncertainties in each case.

Again, one finds that the high-mass DY measurements constrain the luminosities
in the range $100~{\rm GeV}\lsim m_{X} \lsim 2$ TeV, consistent with the kinematic coverage
in $m_{\ell\ell}$ of the data used in the fit.
Their main effects are a reduction of the $q\bar{q}$ uncertainty for
$m_X$ between 500 GeV and 2 TeV and an upwards (downwards) shift in the central values
of the  $q\bar{q}$ ($gg$) luminosities within this $m_X$ region.
Th uncertainty of $\mathcal{L}_{qq}$ is barely changed and its central
value i shifted within $1\sigma$ PDF uncertainties once the high-mass
Drell-Yan datasets of Table.~\ref{tab:data-high-mass} are included in
the fit. 
This comparison further highlights how the high-mass DY data provide useful information
for constraining the PDF luminosities and in turn the
 high-$p_T$ processes relevant for both direct and indirect BSM searches at the LHC.

\section{Fit quality for SM and SMEFT PDFs}
\label{app:fit_quality}

In this appendix, we provide detailed information about the PDF fit quality as
quantified by the figure of merit used in the fits: the $\chi^2$ per data
point, defined in Eq.~(\ref{eq:chi2}) and evaluated with the $t_0$ prescription
described in Ref.~\cite{Ball:2009qv}.
We will do this both for the SM PDFs based on different datasets
and for the SMEFT PDFs from the fits with the baseline dataset
and for different values of the EFT parameters $\hat{W}$ and $\hat{Y}$.

\begin{table}[t]
   \renewcommand{\arraystretch}{1.40}
\begin{center}
\small
\begin{tabular}{ c c c  }
 \toprule
$\quad$ $\qquad$ Dataset $\qquad$ $\quad$ & NNPDF3.1  & This work \\
 \midrule
 ATLAS 7 TeV           & $m_{ee} \leq 200$~GeV & no cuts \\
 ATLAS 8 TeV              & not included &  $y_{\ell\ell} < 1.68$ (for $m_{\ell\ell} \ge 1000$~GeV) \\
 \midrule
 \multirow{2}{*}{ CMS  7 TeV}     & $m_{\mu\mu} \leq 200$~GeV  & $y_{\mu\mu} \leq 2.2$ (for all $m_{\mu\mu}$)  \\
    &  $y_{\mu\mu} \leq 2.2$ & $y_{\mu\mu} < 1.5$ (for $m_{\mu\mu} \ge 850$~GeV)  \\
 \midrule
 CMS 8 TeV        & not included & no cuts \\
 CMS 13 TeV     & not included & no cuts \\
 \bottomrule
\end{tabular}
\end{center}
\caption{\small The kinematic cuts applied to the high-mass Drell-Yan datasets listed
  in Table~\ref{tab:data-high-mass}, compared to those used in NNPDF3.1.}
\label{tab:cuts}
\end{table}

For completeness, we also provide here the $\chi^2$
values obtained when using {\tt NNPDF3.1\_str} as the input PDF set, with all other
settings such as the partonic
matrix elements unchanged.
Note that here the kinematical cuts are
slightly different as compared to~\cite{Faura:2020oom}, the differences being summarised in
Table~\ref{tab:cuts}.
The rationale behind having different cuts is that in this work we include electroweak corrections to the high-mass DY
 cross sections, thus the $m_{\ell\ell}\le 200$ GeV restriction applied in NNPDF3.1 is not necessary
 anymore.
 With the current cuts, essentially all high-mass DY data points can
 be included in the fits.
 The exception is a subset of points from the CMS 7 TeV and ATLAS 8 TeV datasets,
 where we restrict ourselves to the tree-level kinematic
 condition $ |y_{\ell\ell}| \leq \ln(\sqrt{s}/m_{\ell\ell})$.
 The reason is that our calculation of the EFT corrections is based on tree-level
 SM cross sections which must satisfy this requirement.
 Furthermore, for the CMS 7 TeV dataset
 the last rapidity bin is excluded for all $m_{\mu\mu}$ bins, since it is found to deviate
 from the SM predictions by a large amount suggesting the need to account for
 threshold resummation effects~\cite{Bonvini:2015ira}.
 Henceforth,
 here we evaluate the predictions based on {\tt NNPDF3.1\_str}
 for the same set of kinematical cuts as in this
 work.
 
Table~\ref{tab:chi2-baseline} summarises the values of the $\chi^2$
 for the baseline SM PDF fit labelled ``DIS+DY'' compared to the most
 recent NNPDF global fit  {\tt NNPDF3.1\_str}, as well as for the corresponding fits based
 on reduced datasets, namely the DIS-only fit and the fit excluding
 the high-mass DY data.
The entries in italic indicate the datasets that do not enter the corresponding fit. 
%
\begin{table}[htbp]
  \begin{center}
  \renewcommand{\arraystretch}{1.20}
\scriptsize
\begin{tabular}{ l| c| C{2.0cm}| C{2.2cm} |C{2.0cm} | c }
 \toprule
 \multirow{3}{*}{Dataset}    & \multirow{3}{*}{$n_{\rm dat}$}   &  \multicolumn{4}{c}{$\chi^2/n_{\rm dat}$ (SM PDFs)}  \\
&   & \multicolumn{3}{c|}{This work}  & Reference  \\
&   & DIS-only  & DIS+DY(noHM) & DIS+DY  & {\tt NNPDF3.1\_str} \\
 \midrule
 SLAC      & 67   & 1.032  &  0.807 & 0.780 & 0.772 \\
 BCDMS   & 581 & 1.150   &  1.222  & 1.230  & 1.229 \\
 NMC       & 325 & 1.320   &  1.347   & 1.378 & 1.346 \\
 CHORUS  & 832 & 1.058  &  1.188  & 1.228 & 1.191 \\
  NuTeV     & 76  & 0.796 &  0.642  & 0.684 & 0.703 \\
 HERA inclusive  & 1145 & 1.238   &  1.250  & 1.242 & 1.264 \\
 HERA charm  & 37 & 1.654 &  1.433  & 1.445 & 1.424 \\
HERA bottom  & 29 & 1.304 &  1.328  & 1.326 & 1.343 \\
 \midrule
     {\bf Total DIS}  & {\bf 3092} &  {\bf 1.172}  & {\bf 1.217}  & {\bf 1.230} & {\bf 1.225} \\
     \midrule
     \midrule
 E886    $\sigma^d_{\rm DY}/\sigma^p_{\rm DY}$ & 15  &  \textit{49.94}  &  0.484  & 0.484 & 0.509   \\
 E886    $\sigma^p_{\rm DY}$                               & 89  & \textit{1.306} &  1.061  &  1.094 & 1.064 \\
 E605   $\sigma^p_{\rm DY}$                                & 85  &  \textit{2.682}  &  0.972  & 0.982 & 1.006 \\
 \midrule
 CDF    $d\sigma_Z/dy_Z$                                    & 29 &  \textit{1.796}  &  1.443  & 1.460 & 1.459  \\
 D0     $d\sigma_Z/dy_Z$                                     & 28 &  \textit{0.650}  &  0.595  & 0.602 & 0.594  \\
 D0    	$W\to \mu \nu$ asy.                       & 9  &  \textit{6.729}  &  1.411  & 1.488 & 1.582 \\
 \midrule
 ATLAS  $W,Z$ 2010                                             & 30 &  \textit{1.353}  &  0.817  & 0.866  & 0.846  \\
 ATLAS low-mass $Z \rightarrow ee$                   & 6   & \textit{1.038}   &  0.985  & 0.949 & 0.995 \\
 ATLAS  $W,Z$ 2011                                             & 61  &\textit{6.077}&  1.704    &   1.681 & 1.760  \\
 ATLAS  	$W+c$ rapidity 		                      & 22 & \textit{0.497} &  0.469  &  0.468  & 0.487  \\
 ATLAS  $Z \, p_T$                                               & 92  &  \textit{1.110}  &  0.989  &  0.942  &  1.029 \\
 ATLAS  	$p_{T, W}$ in $W$+jets  	                    & 32  &  \textit{2.074}  &  1.574  &  1.690  & 1.567 \\
 \midrule
 CMS 	$W$ asy.				                     & 22 & \textit{5.362} & 1.291   &  1.287  & 1.292  \\
 CMS $\sigma_{W+c}$	7 TeV		              & 5  &  \textit{0.555}  &  0.495 &  0.478  & 0.505\\
 CMS $\sigma_{W^{+}+c}/\sigma_{W^{-}+c}$ 7 TeV & 5  &  \textit{2.526}  &  1.826  &  1.687  & 1.710\\
 CMS    $Z \, p_T$                                                  & 28  &  \textit{1.289}  & 1.336   &  1.296  & 1.354 \\
 CMS    	$W \rightarrow \mu\nu $ rapidity 	      & 22  &  \textit{5.022}  & 1.006   & 1.070 & 1.077 \\
 CMS 	$W+c$ rapidity   13 TeV 			      & 5  &  \textit{0.638}  & 0.661   &   0.658 & 0.671\\
 \midrule
 LHCb    $Z \rightarrow \mu\mu$                        & 9  &  \textit{2.440}  &  1.630  & 1.652 & 1.676 \\
 LHCb  $W,Z \rightarrow \mu$  7 TeV                 & 29  &  \textit{13.62}  &  2.032  & 2.209 & 2.136 \\
 LHCb    $Z \rightarrow ee$                                 & 17  &  \textit{1.273}  &  1.118  & 1.124 & 1.114 \\
 LHCb  $W,Z \rightarrow \mu$   8 TeV                & 30  & \textit{8.835}  &  1.496  & 1.769 & 1.475 \\
 \midrule
     {\bf Total DY (excl. HM)}                      & {\bf 670} &  \textit{\textbf{4.185}}  & \textbf{1.166}  &  \textbf{1.191}  & \textbf{1.193} \\
     \midrule
     \midrule
     ATLAS DY high-mass  7 TeV    & 13    &  \textit{2.261}  & \textit{2.014}   & 1.885 & 1.945 \\
 ATLAS DY high-mass 8 TeV          & 46   & \textit{1.393}   & \textit{1.227}    & 1.181 & \textit{1.215}  \\
 CMS DY high-mass 7 TeV             & 117  &  \textit{1.603}  & \textit{1.617}   &  1.589  & 1.584 \\
 CMS DY high-mass 8 TeV             &   41  &  \textit{0.796}  &  \textit{0.891}  &  0.805  &  \textit{0.838}\\
  CMS DY high-mass 13 TeV           & 43   & \textit{1.837}   &  \textit{1.981}  &  2.013  & \textit{1.952}\\
       \midrule
   {\bf Total DY (HM-only)}  & {\bf 260}  & \textit{\textbf{1.510}}   & \textit{\textbf{1.514}}   & \textbf{1.478}   &  \textbf{\textit{1.480}}  \\
     \midrule
     \midrule
     {\bf Total}   &  {\bf 4022} &  {\bf 1.733}  & {\bf 1.258}    &  {\bf 1.243}  &  {\bf 1.266}  \\
 \bottomrule
\end{tabular}
\end{center}
  \caption{\small \label{tab:chi2-baseline} The values of the $\chi^2$ per data point
    for the baseline SM PDF fit, labelled ``DIS+DY'', and for the corresponding fits based
    on reduced datasets.
    Here   Eq.~(\ref{eq:chi2}) is evaluated using the $t_0$ prescription.
    We also include the results obtained using  {\tt NNPDF3.1\_str} with the kinematic
    cuts used in this work and summarised in Table~\ref{tab:cuts}.
    Values in italics indicate datasets that do not enter the corresponding fit.
}
\end{table}

%
We observe that the quality of the description of the DIS data is 
similar across all fits considered.
As far as hadronic data are concerned, 
we observe that the fit quality of the LHCb data slightly deteriorates when the 
high-mass Drell-Yan data are included.
Also, the description of the CMS 13 TeV 
invariant mass distribution in the combined electron and muon 
channels is not optimal, even after including the data in the fit. 
However, the overall $\chi^2$ is statistically equivalent to the 
most recent NNPDF3.1 set.

Then in Table~\ref{tab:chi2-postfit} we list again the $\chi^2$ values in
the SM PDF fit (same as the ``DIS+DY'' column of Table~\ref{tab:chi2-baseline})
and compare them 
with those obtained with the SMEFT PDFs for the same representative values of
$\hat{W}$ and $\hat{Y}$ parameters
as used in Sect.~\ref{sec:scenarioIIresults}, see also the PDF-level comparisons
in Fig.~\ref{fig:SMEFT_PDFs}. Clearly the theoretical predictions are
computed consistently, namely the
partonic cross sections of the SM baseline "DIS+DY" are computed in
the SM, while the partonic cross section of the other columns are
augmented by the SMEFT contributions of the corresponding operators. 

First of all, we observe that as expected the addition of the EFT corrections
does not affect the description of the DIS structure functions.
Differences are also small for the low-mass and on-shell DY data, and slightly larger
for the HM measurements.
For instance, the $\chi^2$ to the high-mass Drell-Yan datasets
is 1.471  for $\hat{W}=0.0006$ to be compared with 1.478 for the SM PDFs.
In any case, the differences at the level of $\chi^2$ between the SM and SMEFT
PDFs are reasonably small, consistent with the finding that the best-fit values
of the $\hat{W}$ and $\hat{Y}$ parameters are close to the SM expectation.

\begin{table}[htbp]
  \begin{center}
 \renewcommand{\arraystretch}{1.20}
\scriptsize
\begin{tabular}{ l| c| C{1.3cm}| C{1.5cm} |C{1.7cm} |C{1.4cm} |C{1.6cm}  }
 \toprule
 \multirow{2}{*}{Dataset}    & \multirow{2}{*}{$n_{\rm dat}$}   &  \multicolumn{5}{c}{$\chi^2/n_{\rm dat}$}  \\
&   & SM  & $\hat{W}=0.0006$ & $\hat{W}=-0.0008$  &  $\hat{Y}=0.0012$ & $\hat{Y}=-0.0006$  \\
 \midrule
SLAC & 67   &          0.780 &            0.806 &             0.798 &             0.763 &             0.833 \\
BCDMS & 581  &           1.230 &             1.223 &              1.224 &              1.228 &              1.227 \\
NMC & 325  &           1.378 &             1.349 &              1.346 &              1.377 &              1.364 \\
CHORUS & 832  &           1.228 &             1.233 &              1.223 &              1.230 &              1.237 \\
 NuTeV & 76   &          0.684 &            0.698 &             0.651 &             0.706 &             0.655 \\
 HERA inclusive & 1145 &           1.242 &             1.247 &              1.244 &              1.243 &              1.245 \\
HERA charm & 37   &           1.445 &             1.440 &              1.431 &              1.433 &              1.416 \\
HERA bottom & 29   &           1.326 &             1.324 &              1.314 &              1.330 &              1.323 \\
 \midrule
     {\bf Total DIS}  & {\bf 3092}  & {\bf 1.230}   & {\bf 1.229}   & {\bf 1.224}   &  {\bf 1.230} & {\bf 1.231} \\
     \midrule
     \midrule
E886    $\sigma^d_{\rm DY}/\sigma^p_{\rm DY}$ &    15 &          0.484 &            0.454 &             0.475 &             0.513 &             0.458 \\
E886    $\sigma^p_{\rm DY}$   &    89 &           1.094 &             1.067 &              1.060 &              1.075 &              1.049 \\
E605   $\sigma^p_{\rm DY}$  &    85 &           0.982 &             1.025 &              1.013 &              1.005 &              1.032 \\
 \midrule
CDF    $d\sigma_Z/dy_Z$                       &    29 &           1.460 &             1.446 &              1.452 &              1.459 &              1.463 \\
 D0     $d\sigma_Z/dy_Z$    &    28 &          0.602 &            0.604 &             0.600 &             0.602 &             0.603 \\

D0    	$W\to \mu \nu$ asy.                     &     9 &           1.488 &             1.464 &              1.517 &              1.507 &              1.536 \\

 \midrule
 ATLAS  $W,Z$ 2010       &    30 &          0.866 &         0.875     &       0.869        &        0.868      &             0.863 \\
 ATLAS low-mass $Z \rightarrow ee$          &     6 &          0.949 &            0.960 &             0.964 &             0.949 &             0.952 \\
 ATLAS  $W,Z$ 2011  CC              &    46 &           1.837 &             1.921 &              1.899 &              1.893 &              1.864 \\
 ATLAS  $W,Z$ 2011  CF               &    15 &           1.254 &             1.264 &              1.269 &              1.260 &              1.253 \\
 ATLAS  	$W+c$ rapidity 					          & 22 &  0.468  & 0.470 & 0.445 & 0.478 & 0.460  \\
 ATLAS  $Z \, p_T$                            & 92  & 0.942   & 0.973   & 0.967 & 0.951 & 0.978 \\
    ATLAS  	$p_{T, W}$ in $W$+jets    & 32  & 1.690   & 1.675   &  1.737   & 1.640 & 1.697\\
 \midrule
 CMS 	$W$ $e$ asy.                &    11 &          0.838 &            0.829 &             0.827 &             0.856 &             0.825 \\
 CMS 	$W$ $\mu$ asy.                 &    11 &           1.736 &             1.747 &              1.746 &              1.760 &              1.735 \\
 CMS 	$\sigma_{W+c}$	7 TeV         &     5 &          0.478 &            0.483 &             0.487 &             0.471 &             0.501 \\
 CMS 		$\sigma_{W^{+}+c}/\sigma_{W^{-}+c}$ 7 TeV                 &     5 &           1.687 &             1.686 &              1.728 &              1.678 &              1.735 \\
 CMS    $Z \, p_T$    &    28 &           1.296 &             1.280 &              1.276 &              1.293 &              1.273 \\
 CMS    	$W \rightarrow \mu\nu $ rapidity                   &    22 &           1.070 &             1.054 &              1.088 &              1.031 &              1.048 \\
 CMS 	$W+c$ rapidity   13 TeV &     5 &          0.658 &            0.671 &             0.680 &             0.654 &             0.695 \\

 \midrule
LHCb    $Z \rightarrow \mu\mu$                  &     9 &           1.652 &             1.648 &              1.662 &              1.651 &              1.645 \\
 LHCb  $W,Z \rightarrow \mu$                 &    29 &           2.209 &             2.165 &              2.176 &              2.161 &              2.257 \\
LHCb    $Z \rightarrow ee$                   &    17 &           1.124 &             1.139 &              1.129 &              1.130 &              1.143 \\
LHCb  $W,Z \rightarrow \mu$                 &    30 &           1.769 &             1.773 &              1.748 &              1.741 &              1.823 \\
 \midrule
     {\bf Total DY (excl. HM)}                     & {\bf 670} & {\bf 1.191}   & {\bf 1.199}   &
     {\bf 1.198}  &  {\bf 1.192} & {\bf 1.202} \\
     \midrule
     \midrule
     ATLAS DY high-mass  7 TeV &    13 &           1.885 &             2.090 &              1.643 &              2.127 &              1.781 \\
 ATLAS DY high-mass 8 TeV    &    46 &           1.181 &             1.230 &              1.243 &              1.247 &              1.181 \\
 CMS DY high-mass 7 TeV                   &   117 &           1.589 &             1.585 &              1.553 &              1.598 &              1.577 \\
 CMS DY high-mass 8 TeV                    &    41 &          0.805 &            0.793 &             0.833 &             0.820 &             0.787 \\
  CMS DY high-mass 13 TeV               &    43 &           2.013 &             1.876 &              2.203 &              1.881 &              2.064 \\
     \midrule
    {\bf Total DY (HM-only)}   & {\bf 260}  & {\bf 1.478}   &  {\bf 1.471}   & {\bf 1.497}   &  {\bf 1.486} & {\bf 1.473} \\
     \midrule
     \midrule
     {\bf Total}   & {\bf 4022}  &  {\bf 1.243}   & {\bf 1.244}   & {\bf 1.242}   &  {\bf 1.245} & {\bf 1.247} \\
 \bottomrule
\end{tabular}
\end{center}
  \caption{\small Same as Table~\ref{tab:chi2-baseline}, now comparing the $\chi^2$ values (computed using the $t_0$ prescription,
  as above) 
    of the SM PDFs with those of the SMEFT PDFs for different values
    of the  $\hat{W}$ and $\hat{Y}$ parameters in benchmark scenario I,
    specifically those displayed in Figs.~\ref{fig:SMEFT_lumis} and~\ref{fig:SMEFT_PDFs}. \label{tab:chi2-postfit}
}
\end{table}


\afterpage{\FloatBarrier}

\section{Validation of the SMEFT $K$-factors}
\label{sec:benchmarking}

As described in Sect.~\ref{sec:theory}, in this work the effect of the dimension-six SMEFT
operators considered in the two benchmark scenarios is accounted for at the
level of cross sections via the $K$-factor approach, Eq.~(\ref{eq:theory_k_fac_app3}).
In this appendix, we provide further details about the calculation and validation
of these EFT $K$-factors.
Specifically, we compare the numerical values for these  $K$-factors, which have been obtained using
{\tt SMEFTsim}~\cite{Brivio:2017btx,Brivio:2020onw} interfaced with \amc, with the analytic calculation presented in~\cite{Greljo:2017vvb}.

\paragraph{DIS structure functions.} SMEFT corrections to the 
neutral-current deep-inelastic structure 
functions $F_2, F_3$ in the  benchmark
scenario I of Sect.~\ref{sec:scenarios} are obtained by means of a direct
calculation in perturbation theory\footnote{Scenario II is not relevant for DIS
  data, given that high-$Q^2$ structure functions only involve electrons.}.
In order to determine these corrections, we rewrite Eq.~(\ref{eq:WYWarsaw})
as the linear combination of four-fermion operators of the form
$\bar{q}_{\lambda} \gamma^{\mu} q_{\lambda} \bar{\ell}_{\lambda'} \gamma_{\mu} \ell_{\lambda'}$,
where $q_\lambda$ is a quark field of helicity $\lambda$ (with $\lambda = +1$ for a
right-handed field and $\lambda=-1$ for a left-handed field) and $\ell_{\lambda^\prime}$ is a
lepton field of helicity $\lambda'$.
The relevant operators for the $\hat{Y}$ parameter
are already of this form in Eq.~(\ref{eq:WYWarsaw}).
For the $\hat{W}$ parameter,
the associated operators can be expanded explicitly as:
\begin{align}
\label{eq:WYfourfermion}
\mathcal{L}_{\rm SMEFT} \supset   &-\frac{g^2 \hat{W}}{4 m_W^2}  \sum_{i=1}^{3} \Bigg(\bar{e}_L^i \gamma^{\mu} e_L^i\bar{u}_L^i \gamma_{\mu} u_L^i - \bar{e}_L^i \gamma^{\mu} e_L^i \bar{d}_L^i \gamma_{\mu} d_L^i \\&- \bar{\nu}_L^i \gamma^{\mu} \nu_L^i \bar{u}_L^i \gamma_{\mu} u_L^i + \bar{\nu}_L^i \gamma^{\mu} \nu_L^i \bar{d}_L^i \gamma_{\mu} d_L^i  \Bigg), \nonumber
\end{align}
where the index $i$ runs over generations, and the flavour-changing contributions have been dropped since they only contribute to low energy CC structure functions.

The EFT corrections to the DIS structure functions induced by a specific four-fermion operator of the form
\be
\mathcal{L}_{\rm SMEFT} \supset \frac{c_{\lambda\lambda'}^{q\ell}}{\Lambda^2} \bar{q}_{\lambda} \gamma^{\mu} q_{\lambda} \bar{\ell}_{\lambda'} \gamma_{\mu} \ell_{\lambda'}
\ee
can be shown to be given by
\begin{align}
\Delta F_2(x,Q^2) &=  \frac{c_{\lambda\lambda'}^{qe}}{\Lambda^2}  \frac{Q^2  }{2e^2} \left( e_q - K_Z \left(V^e - \lambda' A^e\right)\left( V^q - \lambda A^q \right)  \right) \left( xf_q(x,Q^2) + xf_{\bar{q}}(x,Q^2) \right), \nonumber\\[1.5ex]
\Delta F_3(x,Q^2) &=-  \frac{c_{\lambda\lambda'}^{qe}}{\Lambda^2} \frac{Q^2}{2e^2} \left( \lambda \lambda' e_q - K_Z \left( \lambda' V^e - A^e \right) \left( \lambda V^q - A^q \right)\right) \left( f_q(x,Q^2) - f_{\bar{q}}(x,Q^2) \right),\nonumber
\end{align}
where $e$ is the positron charge, $e_q$ is the charge on the quark $q$ in units of the positron charge, $\theta_W$ is the Weinberg angle, and $K_Z = Q^2/\sin^2(2\theta_W)(Q^2 + m_Z^2).$
The vector and axial couplings are given by $V^e = -\frac{1}{2} +
2\sin^2(\theta_W)$, $A^e = -\frac{1}{2}$, $V^q = I_3^q -
2\sin^2(\theta_W)e_q$ and $A^q = I^q_3$,
where $I_3^q$ is the third
component of the quarks' weak isospin. These formulae are the natural
generalisations of those derived in~\cite{Carrazza:2019sec}, where only right-handed four-fermion
operators were considered.
Taking combinations of these DIS structure-function corrections according to
Eq.~(\ref{eq:WYWarsaw}) for the $\hat{Y}$ parameter and to Eq.~(\ref{eq:WYfourfermion}) for the $\hat{W}$ parameter
yields the sought-for EFT corrections for DIS observables.

This calculation has been implemented in {\tt APFEL}~\cite{Bertone_2014} following the strategy 
presented in~\cite{Carrazza:2019sec}. FK tables are produced from {\tt
  APFEL}~\cite{Bertone:2016lga} and then used to evaluate the DIS $K$-factors
defined in Eq.~\eqref{eq:theory_k_fac_app3}. Furthermore, we have used {\tt APFEL} to
include the higher-order QCD corrections in the SMEFT sector, so that in fact Eq.~(\ref{eq:theory_k_fac_app3})
holds exactly for the DIS $K$-factors in our study.

\begin{figure}[t]
\centering
\includegraphics[width=0.49\textwidth]{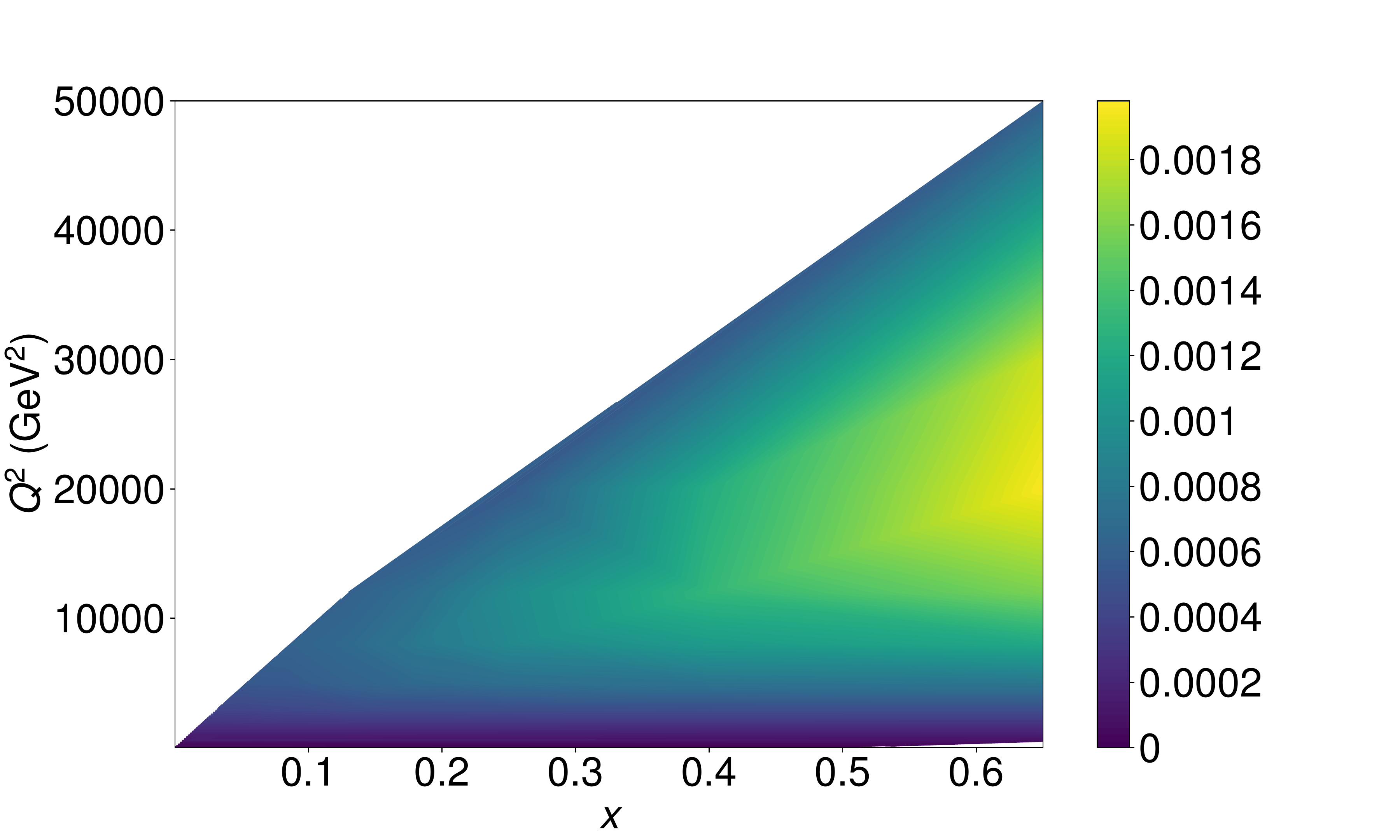}
\includegraphics[width=0.49\textwidth]{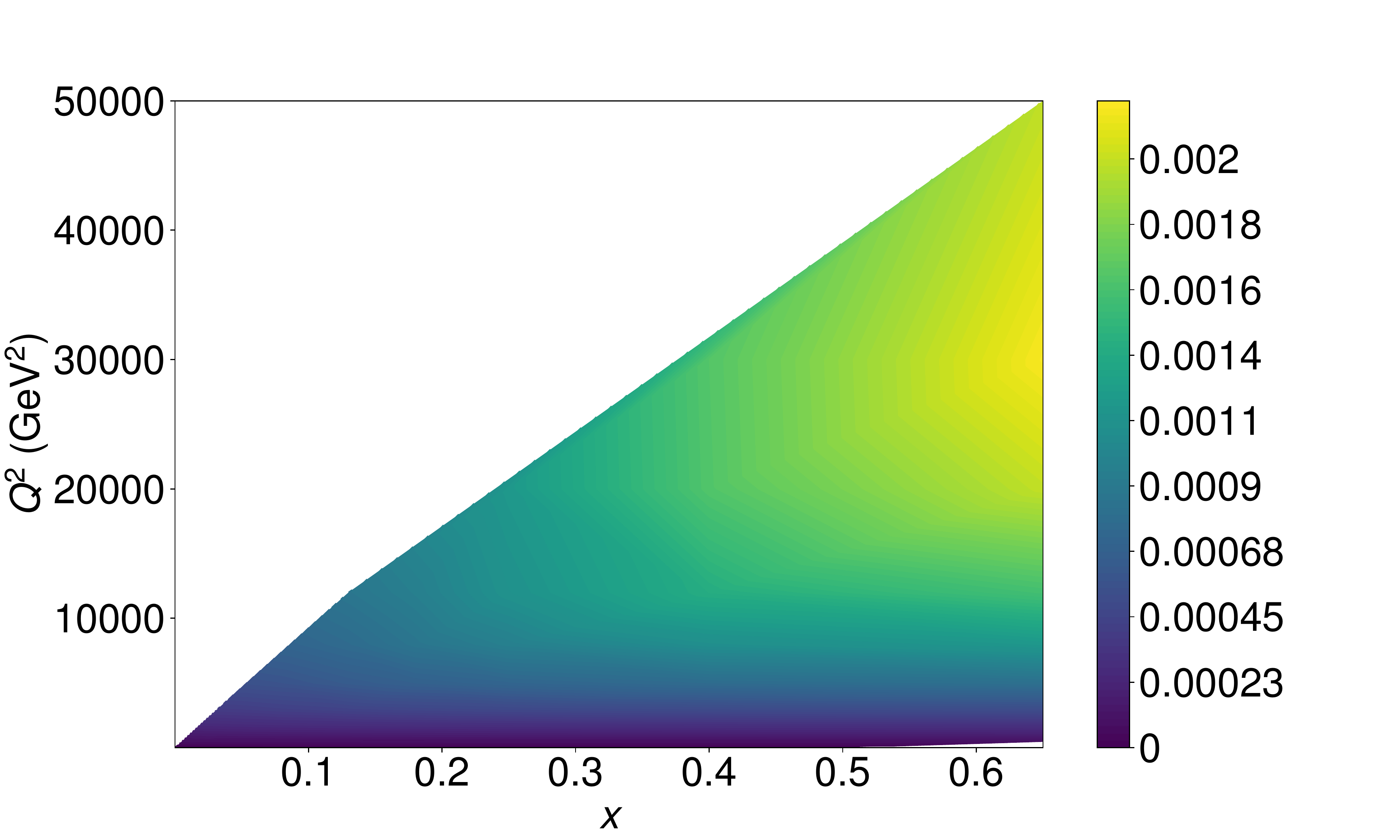}\\
\caption{\small Contour maps indicating the value of the EFT correction, $K_{\rm EFT}(\hat{W},\hat{Y})$$-$$1$ in Eq.~(\ref{eq:theory_k_fac}),
  for the DIS reduced cross sections as a function of $x$ and $Q^2$ for two
  representative values of the EFT parameters: $\hat{W}=-10^{-4}$ (left panel) and $\hat{Y}=-10^{-4}$
  (right panel).
  \label{fig:dissmeft}}
\end{figure}

Fig.~\ref{fig:dissmeft} displays contour maps indicating the EFT correction, $K_{\rm EFT}(\hat{W},\hat{Y})$$-$$1$ in Eq.~(\ref{eq:theory_k_fac}),
  for the DIS reduced cross sections (which include both $F_2$ and $xF_3$) as a function of $x$ and $Q^2$ for two
  representative values of the EFT parameters, $\hat{W}=-10^{-4}$ and $\hat{Y}=-10^{-4}$.
  These maps should be compared with Fig.~1 of~\cite{Carrazza:2019sec}, which considered
   different EFT scenarios.
  We find that the overall effect of
  non-zero $\hat{W}$ and $\hat{Y}$ parameters is rather small, well below the percent level
  even for the highest bins in $Q^2$ covered by the HERA data.
  This comparison highlights how, in this benchmark EFT scenario, the constraints
  on the $\hat{W}$ and $\hat{Y}$ parameters will be completely
  dominated by the high-mass Drell-Yan cross sections.

\paragraph{Drell-Yan distributions.} As demonstrated in Ref.~\cite{Greljo:2017vvb}, in
the case of the dilepton  $m_{\ell\ell}$ distribution in neutral-current
Drell-Yan, one can derive the following analytic expression 
\bea
  \label{eq:theory_k_fac_app4}
  \left(\frac{d\sigma}{dm_{\ell\ell}}\right)_{\rm SMEFT}& =&
  \left(\frac{d\sigma}{dm_{\ell\ell}}\right)_{\rm SM}
  \times
  \frac{\sum_{q,\ell} \mathcal{L}_{q\bar{q}}(m_{\ell\ell}^2/s, m_{\ell\ell}) |F_{q\ell}(m_{\ell\ell},\epsilon^{q\ell})|^2}{\sum_{q,\ell} \mathcal{L}_{q\bar{q}}(m_{\ell\ell}^2/s,m_{\ell\ell}) |F_{q\ell}(m_{\ell\ell},0)|^2 } \, ,
  \eea
where $F_{ql}(m_{ll},\epsilon^{q\ell})$ represents a form factor that depends
on the values of the SMEFT coefficients
\be
F_{q \ell}(m_{\ell \ell},\epsilon^{q\ell}) = \delta^{ij}\frac{e^2 Q_q Q_\ell}{m_{\ell \ell}^2} + \delta^{ij}\frac{ g_Z^q g_Z^\ell}{m_{\ell \ell}^2-m_Z^2 + i m_Z \Gamma_Z}+ \delta^{ij}{\frac{\epsilon^{q\ell}}{v^2}}~.
\ee
One can then match the Warsaw-basis
parametrisation of Eq.~(\ref{eq:Warsaw}) to the contact terms $\epsilon^{q \ell}_{i j k l} \equiv \epsilon^{q \ell} \delta_{i j} \delta_{k l}$ in the above equation, where $i, j$ ($k, l$) are the quark (lepton) flavour indices.
There are four combinations for quarks $q = u_L, u_R, d_L, d_R$ and two combinations for charged leptons $\ell = e_L, e_R$ relevant for the description of the neutral-current Drell-Yan process.
Specifically, the matching is
\begin{equation}
\begin{aligned}\label{eq:matching}
\epsilon^{u_L e_L} &= \hat W + \frac{1}{3} t^2_\theta \hat Y \,, 
&
\epsilon^{u_R e_L} &= \frac{4}{3} t^2_\theta \hat Y \,, 
&
\epsilon^{d_L e_L} &= - \hat W + \frac{1}{3} t^2_\theta \hat Y \,, 
&
\epsilon^{d_R e_L} &= - \frac{2}{3} t^2_\theta \hat Y \,, 
\\
\epsilon^{u_L e_R} &= \frac{2}{3} t^2_\theta \hat Y \,, 
&
\epsilon^{u_R e_R} &= \frac{8}{3} t^2_\theta \hat Y \,, 
&
\epsilon^{d_L e_R} &= \frac{2}{3} t^2_\theta \hat Y \,, 
&
\epsilon^{d_R e_R} &= - \frac{4}{3} t^2_\theta \hat Y \,,
\end{aligned}
\end{equation}
where $t_\theta$ is the tangent of the Weinberg angle. We confirm these results using the modified form of the Feynman propagator for $W^3$ and $B$ fields following directly from Eq.~\eqref{eq:WY}. In the unbroken phase, where $\hat s \gg m_W^2$, this simply amounts to the replacement $ 1/p^2 \to 1 / p^2 - \hat Y / m_W^2$ for $B$ and  $ 1/p^2 \to 1 / p^2 - \hat W / m_W^2$ for $W^3$ in the Standard Model
calculation of $p p \to \ell^+ \ell^-$.
Note that there are two diagrams at the leading order with $W^3$ and $B$ vector bosons propagating in the $s$-channel.

An analogous expression involving the transverse mass $m_T$ can be derived in the CC case. These $K$-factors can be matched to the experimental
data kinematics by integrating over the suitable ranges in
$m_{\ell\ell}$~($m_T$) and rapidity $y_{\ell\ell}$~($y_{\ell}$).

\paragraph{1D distributions.}
For the benchmarking between the analytical and numerical calculations,
we use the relation between the $\hat{W}$ coefficient and the $c_{lq}^{(3)}$
coefficient in the Warsaw basis, $c_{lq}^{(3)} = - \hat{W} \Lambda^2/v^2$,
where $\Lambda$ is the SMEFT cut-off and $v$ is the Higgs vev.
We set $\Lambda = 1$~TeV and determine $v$ using
the ($\alpha_{EW}, m_Z, G_F$) input scheme.
We set $m_Z$ and $G_F$
using their PDG values: $m_Z = 91.1876$~GeV and $G_F = 1.1663787\times 10^{-5}$~GeV$^{-2}$,
and set $\alpha_{EW} = 1/127.951$.
These values yield $v = 246.22$~GeV for the Higgs vev.

\begin{figure}[t]
  \begin{center}
    \includegraphics[scale=0.6]{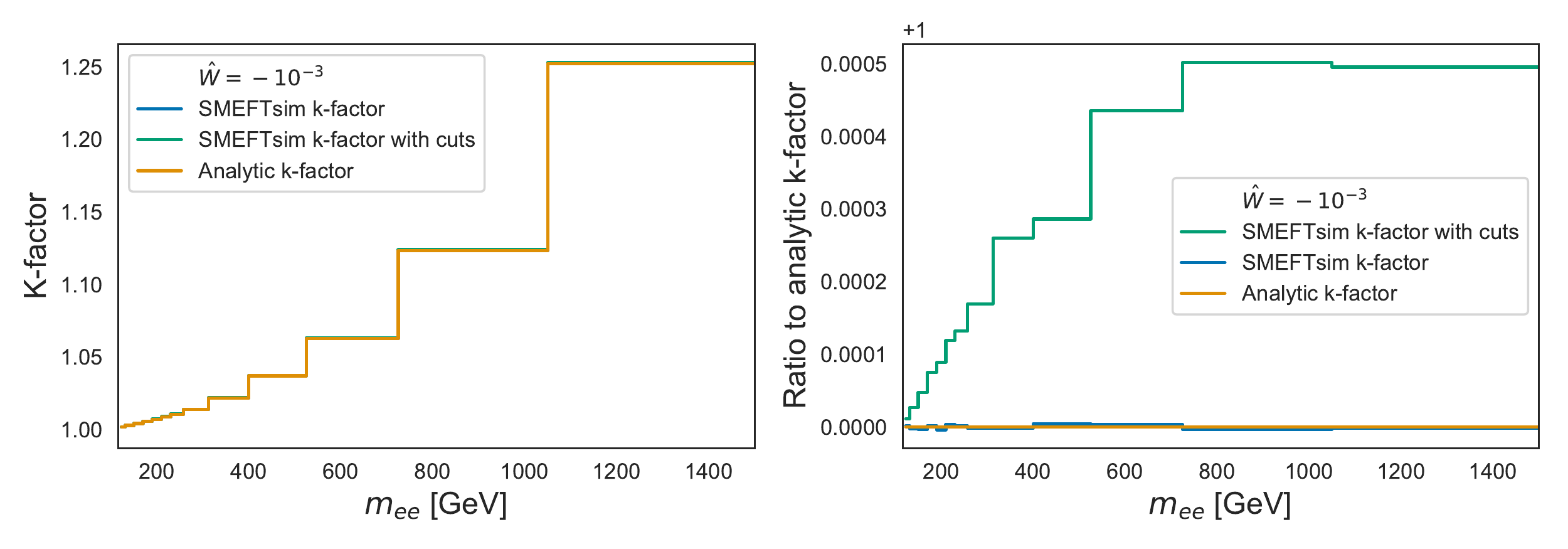}
  \end{center}
  \caption{Left: comparison of the SMEFT $K$-factors, Eq.~(\ref{eq:theory_k_fac_app3}),
    linearised in the EFT parameters,
    between the numerical and the analytical approaches for $\hat{W} = -10^{-3}$
    and the kinematics of the ATLAS 7 TeV DY data.
    The label ``cuts'' indicates that we impose acceptance
    requirements of $p_T^\ell \ge 25$~GeV
    and $|\eta_{\ell}|\le 2.5$
    in the numerical ({\tt SMEFTsim}) calculation.
    Right: relative difference in the $K$-factors shown in the left panel taking
    the analytical calculation as a reference.
    \label{fig:w_benchmark_allbins}
    }
\end{figure}

Fig.~\ref{fig:w_benchmark_allbins} displays the
comparison of the SMEFT $K$-factors, Eq.~(\ref{eq:theory_k_fac_app3}),
linearised in the EFT parameters,
between the numerical and the analytical approaches for a representative
value of $\hat{W} = -10^{-3}$
($c_{lq}^{(3)} =1.65 \times 10^{-2}$)
and the kinematics of the ATLAS 7 TeV DY data.
The label ``cuts'' indicates that we impose acceptance
requirements of $p_T^\ell \ge 25$~GeV
and $|\eta_{\ell}|\le 2.5$
in the numerical ({\tt SMEFTsim}) calculation on the final-state leptons;
these cuts cannot be applied in the analytical calculation.
The right panel shows the relative difference in these $K$-factors, with
the analytical calculation as a reference.

From this comparison we observe, first of all, the perfect agreement between
the analytical and numerical $K$-factors in the case of no acceptance cuts,
and second, that the acceptance cuts on the leptonic variances leave the $K$-factor value
essentially unchanged.
Furthermore, we have verified that the same level of agreement in the calculation
of the EFT $K$-factors  between the numerical and analytical approaches
is obtained in the case of the $\hat{Y}$
parameter, as well as once the quadratic EFT corrections are accounted for.
In addition, the calculation of the EFT $K$-factors for the neutral current DY pseudo-data
used in the HL-LHC projections of Sect.~\ref{sec:hllhc} has been validated
by ensuring that the analytical and {\tt SMEFTsim} calculations are
in perfect agreement.

As discussed in~\cite{Greljo:2017vvb}, far above the $Z$ peak
it is sufficient to include higher-order corrections
to 
the SM cross section in Eq.~(\ref{eq:theory_k_fac_app4}) to achieve good theoretical accuracy in the SMEFT,
due to the (approximate) factorisation of higher-order QCD corrections in this region.
We also note that renormalisation group evolution effects~\cite{Jenkins:2013wua,Alonso:2013hga}
are not required in this calculation, since for the operators considered in our benchmark scenarios
the corresponding anomalous dimensions are either Yukawa-suppressed
or suppressed by NLO electroweak contributions.

\begin{figure}[t]
  \begin{center}
     \includegraphics[scale=0.6]{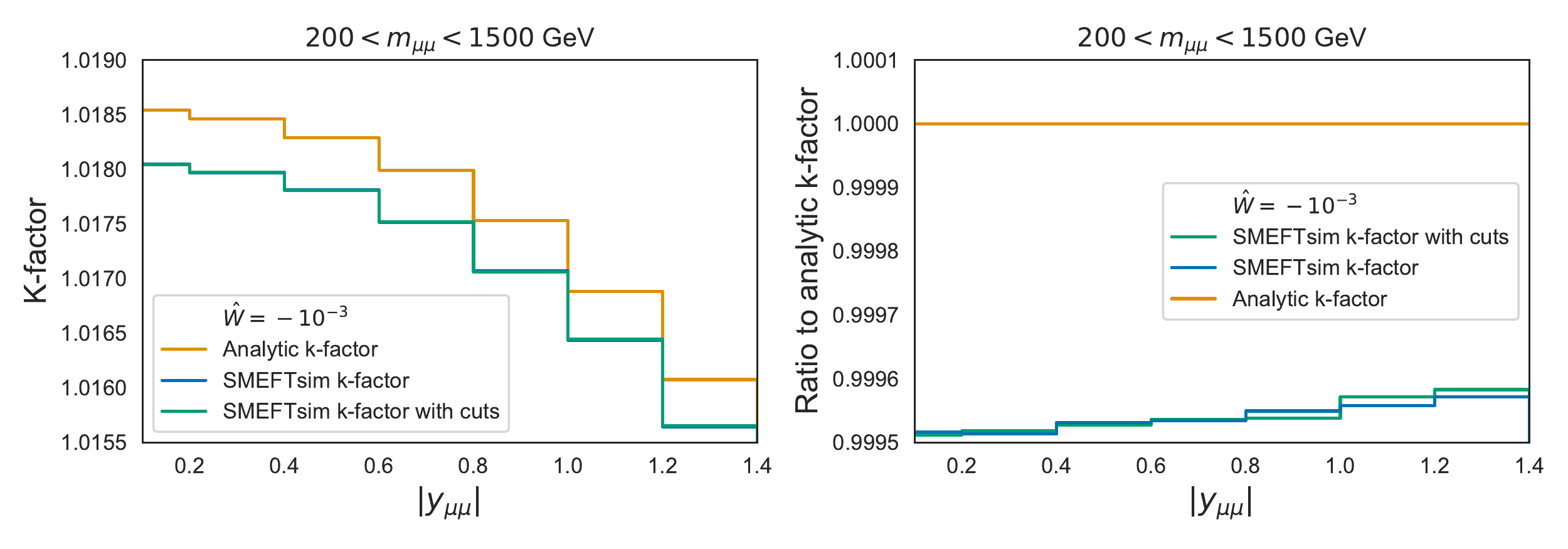}
  \end{center}
  \vspace{-0.7cm}
  \caption{Same as Fig.~\ref{fig:w_benchmark_allbins} (also for $\hat{W}=10^{-3}$) for
the EFT $K$-factors for the double-differential Drell-Yan cross sections in $m_{\ell\ell}$ and $|y_{\ell\ell}|$,
for the highest $m_{\ell\ell}$ bin of the CMS 7 TeV DY measurement.
We only compute  $K$-factors for $|y_{\ell\ell}|$  satisfying LO kinematics.
   }
  \label{fig:cms_7tev_120_200}
\end{figure}

\paragraph{2D distributions.}
In Drell-Yan datasets such as the CMS 7 TeV data of~\cite{Chatrchyan:2013tia}, the
measurement is presented as a double-differential distribution in the
dilepton invariant mass $m_{\ell\ell}$ (or equivalently $\tau=m_{\ell\ell}^2/s$)
and rapidity $|y_{\ell\ell}|$.
Also in this case, bin-by-bin EFT $K$-factors can be computed using
the prescription of Eq.~(\ref{eq:theory_k_fac_app3}).
Fig.~\ref{fig:cms_7tev_120_200} displays the same comparison as
in Fig.~\ref{fig:w_benchmark_allbins} (also for $\hat{W}=-10^{-3}$) now for
the EFT $K$-factors for the double-differential Drell-Yan cross sections in $m_{\ell\ell}$ and $|y_{\ell\ell}|$,
for the highest $m_{\ell\ell}$ bin of the ATLAS 7 TeV DY measurement.
Note that we only compute these $K$-factors for the $|y_{\ell\ell}|$ that satisfy the LO kinematics.
As in the case of the 1D distributions, we find good agreement between the analytical
and numerical calculations, with differences well below the absolute magnitude
of the $K$-factors, and also that the impact of the acceptance cuts in the leptonic
variables is negligible.

\begin{figure}[t]
\centering
\includegraphics[width=1\textwidth]{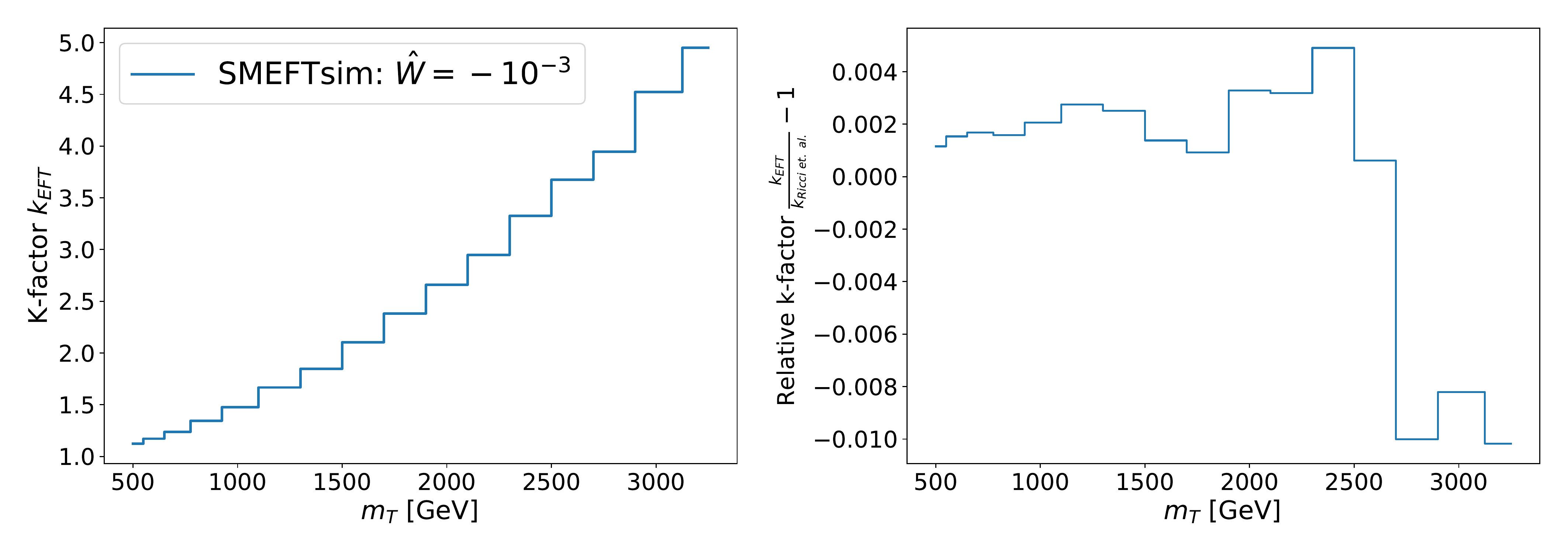}
\caption{\label{fig:benchHLcc} Left: the EFT $K$-factor
  for high-mass charged-current DY production at the HL-LHC
  as a function of $m_T$, the 
  transverse mass of the neutrino-lepton pair, for $\hat{W}=-10^{-3}$.
  Right: the same quantity, now compared to the corresponding k-factors
	provided by the authors of \cite{Ricci:2020xre}.}
\end{figure}

\paragraph{EFT $K$-factors for CC Drell-Yan at the HL-LHC.}
The left panel of Fig.~\ref{fig:benchHLcc} displays the EFT $K$-factor
  for high-mass charged-current DY production at the HL-LHC
  as a function of $m_T$, the 
  transverse mass of the neutrino-lepton pair, for a parameter value of $\hat{W}=-10^{-3}$.
  We find rather large EFT corrections, with $K$-factors as large as $K_{\rm EFT}\simeq 5$
  for the highest $m_T$ bin.
  To validate this {\tt SMEFTsim}-based calculation, we show a comparison between
  $k_{EFT}$ and the linear k-factor $k_{Ricci\ et.\ al}$ provided by the authors of
  \cite{Ricci:2020xre}.  The quadratic k-factors are also shown in the left panel.  
  In the right panel we plot the relative k-factor $\frac{k_{EFT}}{k_{Ricci\ et. et}} -1$,
  finding good agreement between the two calculations.
  In addition, one can observe that the quadratic EFT corrections
  become larger than the linear ones for $m_T\gsim 3$ TeV.
  Nevertheless, we point out that in our projections
  for the HL-LHC data from Sect.~\ref{sec:hllhc}
  we restrict ourselves to the linear approximation in benchmark scenario I.

\paragraph{Effect of varying PDFs in the computation of the SMEFT
  $K$-factors.}

The SMEFT $K$-factors in Eq.~(\ref{eq:theory_k_fac}) are
precomputed before the fit using a reference SM PDF set and then kept fixed.
Here, we quantitatively assess the effect of varying the input NNLO PDF in
Eqns.~(\ref{eq:mult_k_fac_app1}) and~(\ref{eq:mult_k_fac_app2}).

We will first spell out the approximation made when using
SMEFT K-factors, and then assess its impact.
For definiteness, consider corrections from a non-zero $\hat{W}$
coefficient.
Denote the partonic cross section by  $\hat{\sigma}_{\rm SMEFT} = \hat{\sigma}_{SM} + \hat{W} \hat{\sigma}_{\hat{W}}$ and
let $\mathcal{L}_{\rm SMEFT}$ denote the luminosity calculated with
SMEFT PDFs. The total cross section for $\hat{W}\neq 0$ is given by
 \begin{align}
 \sigma(\hat{W}) &= (\hat{\sigma}_{\rm SM} + \hat{W}
 \hat{\sigma}_{\hat{W}})_{ij} \otimes \mathcal{L}_{\rm SMEFT,ij}\label{eq:noapprox}\\
 &= \hat{\sigma}_{\rm SM,ij} \otimes \mathcal{L}_{\rm SMEFT,ij} +
 \hat{W} \hat{\sigma}_{\hat{W},ij} \otimes \mathcal{L}_{\rm SMEFT,ij},\notag
 \end{align}
 where we sum over the partons $i,j$.  By using the SMEFT K-factor
 approach we approximate Eq.~\eqref{eq:noapprox} as
 \begin{align}
  \sigma(\hat{W}) &\approx \hat{\sigma}_{\rm SM,ij} \otimes
  \mathcal{L}_{\rm SMEFT,ij} \Big(1 + \hat{W}
  \frac{\hat{\sigma}_{\hat{W},ij} \otimes \mathcal{L}_{\rm
      SM,ij}}{\hat{\sigma}_{\rm SM,ij} \otimes \mathcal{L}_{\rm SM,ij}}   \Big) \label{eq:approx}\\
  &= \hat{\sigma}_{\rm SM,ij} \otimes \mathcal{L}_{\rm SMEFT,ij} + W
  \hat{\sigma}_{\rm SM,ij} \otimes \mathcal{L}_{\rm SMEFT,ij}
  \left(\frac{\hat{\sigma}_{\hat{W},ij} \otimes \mathcal{L}_{\rm
      SM,ij}}{\hat{\sigma}_{\rm SM,ij} \otimes \mathcal{L}_{\rm SM,ij}}   \right).\notag
 \end{align}
 Note that the first term of Eqs.~\eqref{eq:noapprox} and
 \eqref{eq:approx} are equal, thus we can express the approximation as
 \begin{align}
   \label{eq:approx2}
\frac{\hat{\sigma}_{\hat{W},ij} \otimes \mathcal{L}_{\rm
    SMEFT,ij}}{\hat{\sigma}_{\rm SM,ij} \otimes \mathcal{L}_{\rm
	 SMEFT,ij}}  &\approx \frac{\hat{\sigma}_{\hat{W},ij} \otimes
  \mathcal{L}_{\rm SM,ij}}{\hat{\sigma}_{\rm SM,ij} \otimes
                       \mathcal{L}_{\rm SM,ij}},
                       \end{align}
                       or, equivalently, as
                       \begin{align}
                         R_{\hat{W}}({\rm SMEFT}) &\approx R_{\hat{W}}({\rm SM})
 \end{align}
 where $R$ is defined in Eqns.~(\ref{eq:mult_k_fac_app1})
 and~(\ref{eq:mult_k_fac_app2}) and is computed either with SMEFT PDFs
 or fixed SM PDFs.
In what follows we test whether $R(\rm SM)$ is a good approximation for
$R(\rm SMEFT)$, taking each of the coefficients of Scenario I and II at a time. 
 
\begin{figure}[htb]
\centering
\includegraphics[width=0.45\textwidth]{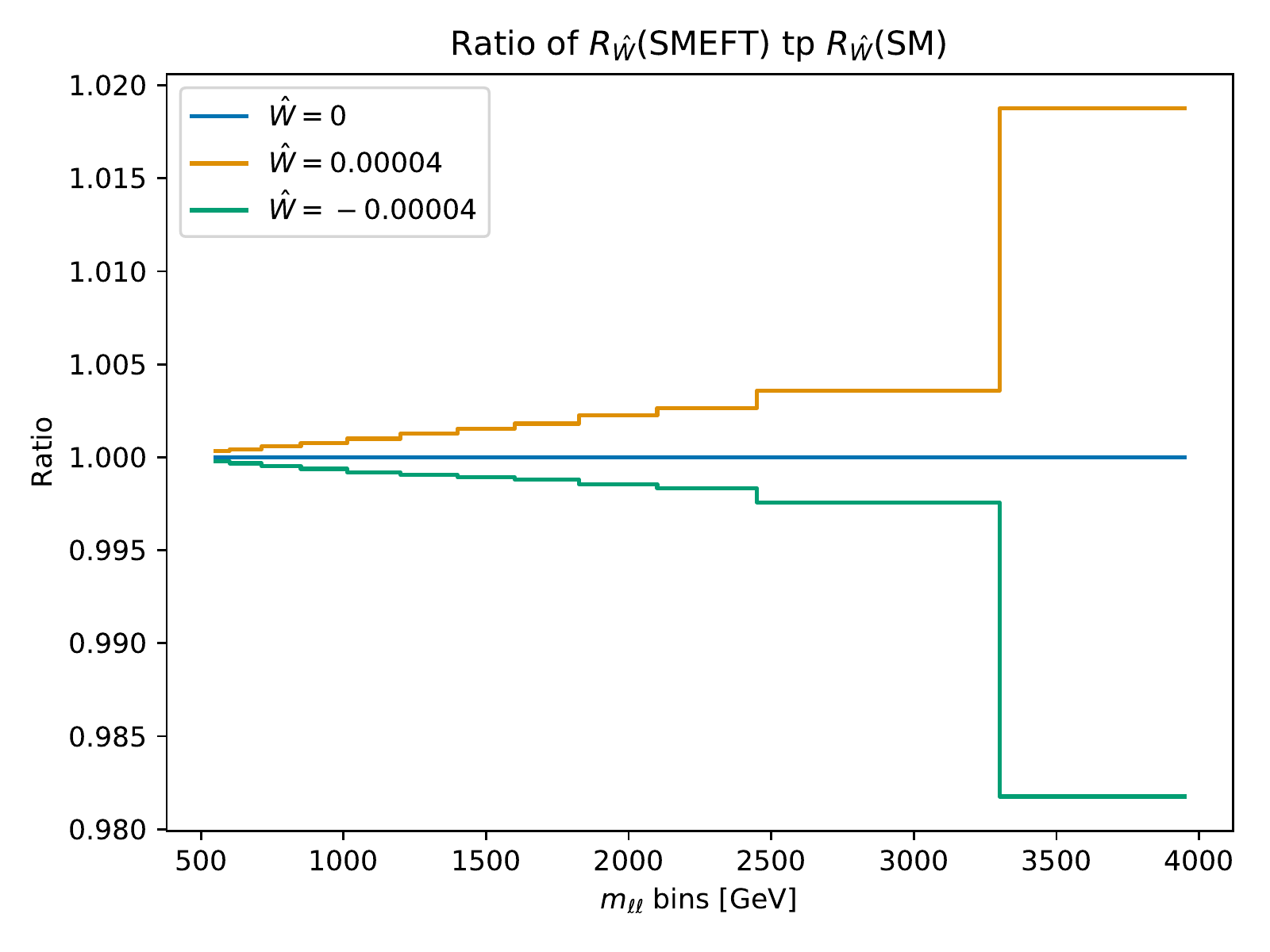}\\
\includegraphics[width=0.45\textwidth]{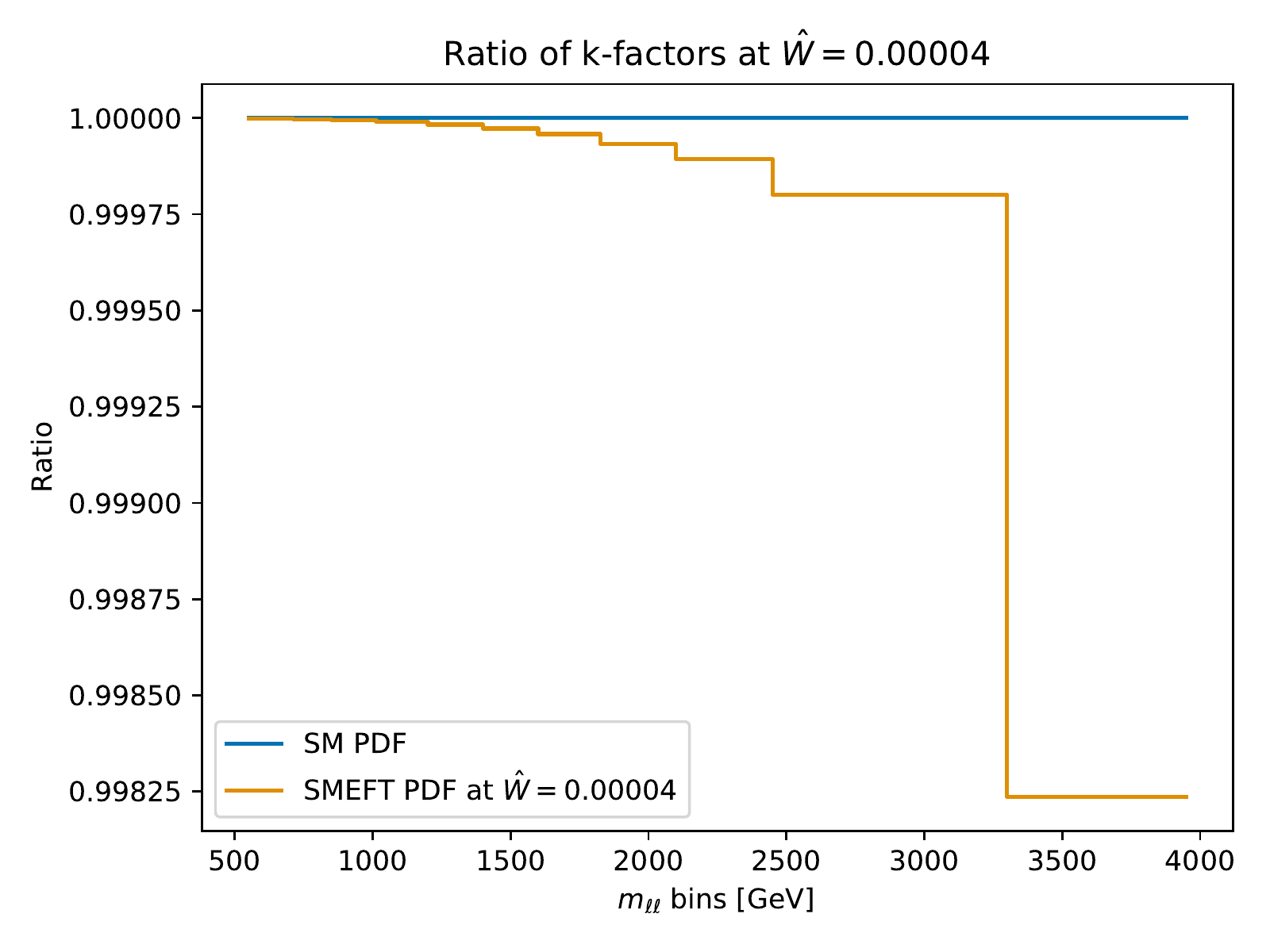}
\includegraphics[width=0.45\textwidth]{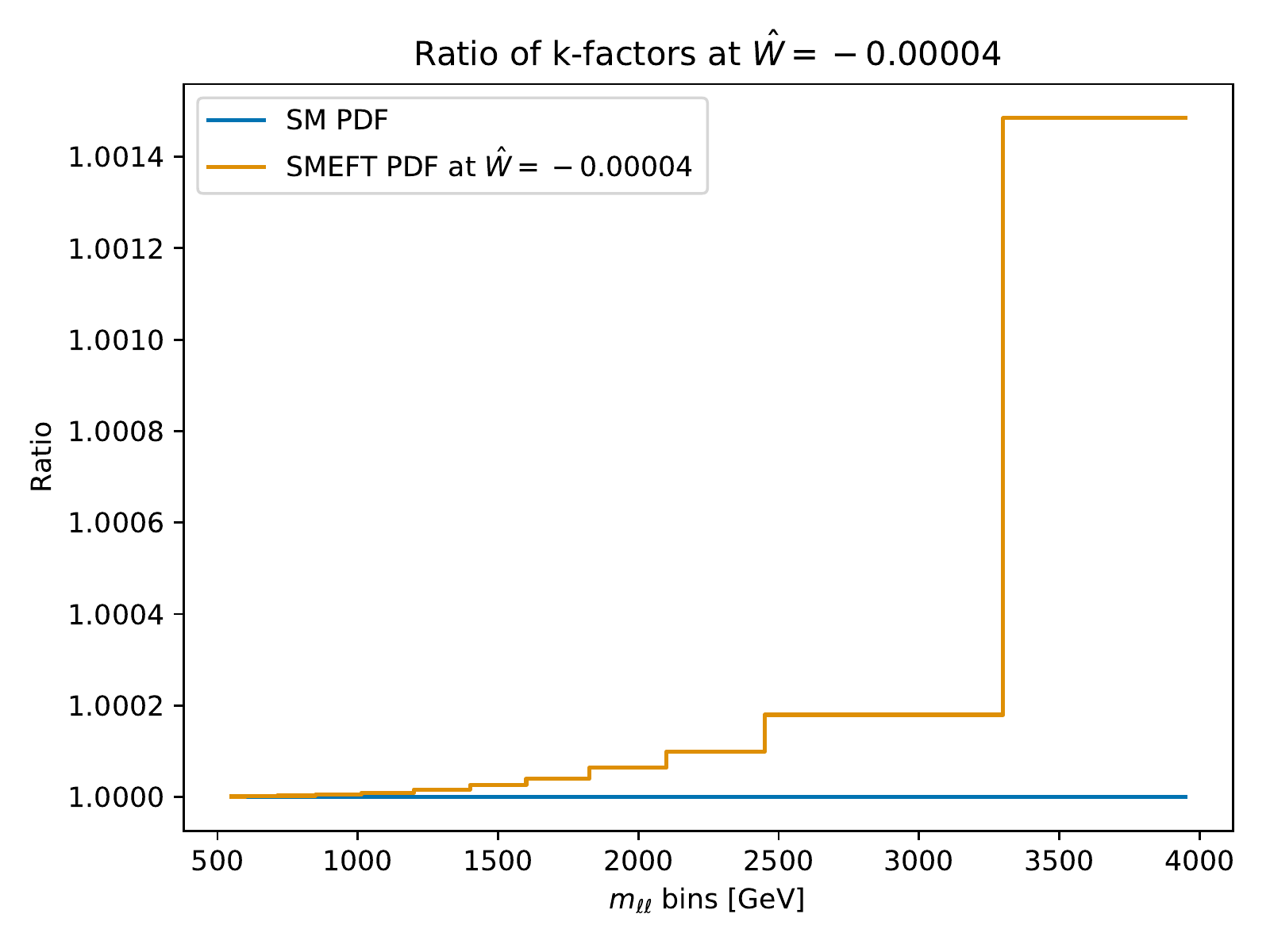}
	\caption{\label{fig:Wapproxim} The impact of the SMEFT PDFs on
          $R_{W}$ (above) and the SMEFT K-factor $K = 1 + \hat{W}
          R_{W}$ (below) calculated at the Scenario I benchmark points
          $\hat{W}=\pm 4\cdot 10^{-5}$. }
\end{figure}

We will first consider Scenario I and the $\hat{W}$ parameter. 
We use the PDFs including HL-LHC pseudodata and
calculate $R_{\hat{W}}$ for the HL-LHC NC Drell-Yan bins outlined in Section \ref{sec:hllhc}.
In Fig.~\ref{fig:Wapproxim} we observe that, using fixed SM PDFs in
the computation of $R_{\hat{W}}$ yields a 2\% deviation in
the highest invariant mass bin. In the same figure we assess the
impact of these differences on the SMEFT $K$-factors themselves, $K_{\hat{W}} = 1 + \hat{W} R_{\hat{W}}$,
calculated at each of the benchmark points $\hat{W}= \pm 4\cdot
10^{-5}$ and we observe that the difference at the level of the observable
is completely negligible, at the permil level. 

\begin{figure}[hbt]
\centering
\includegraphics[width=0.45\textwidth]{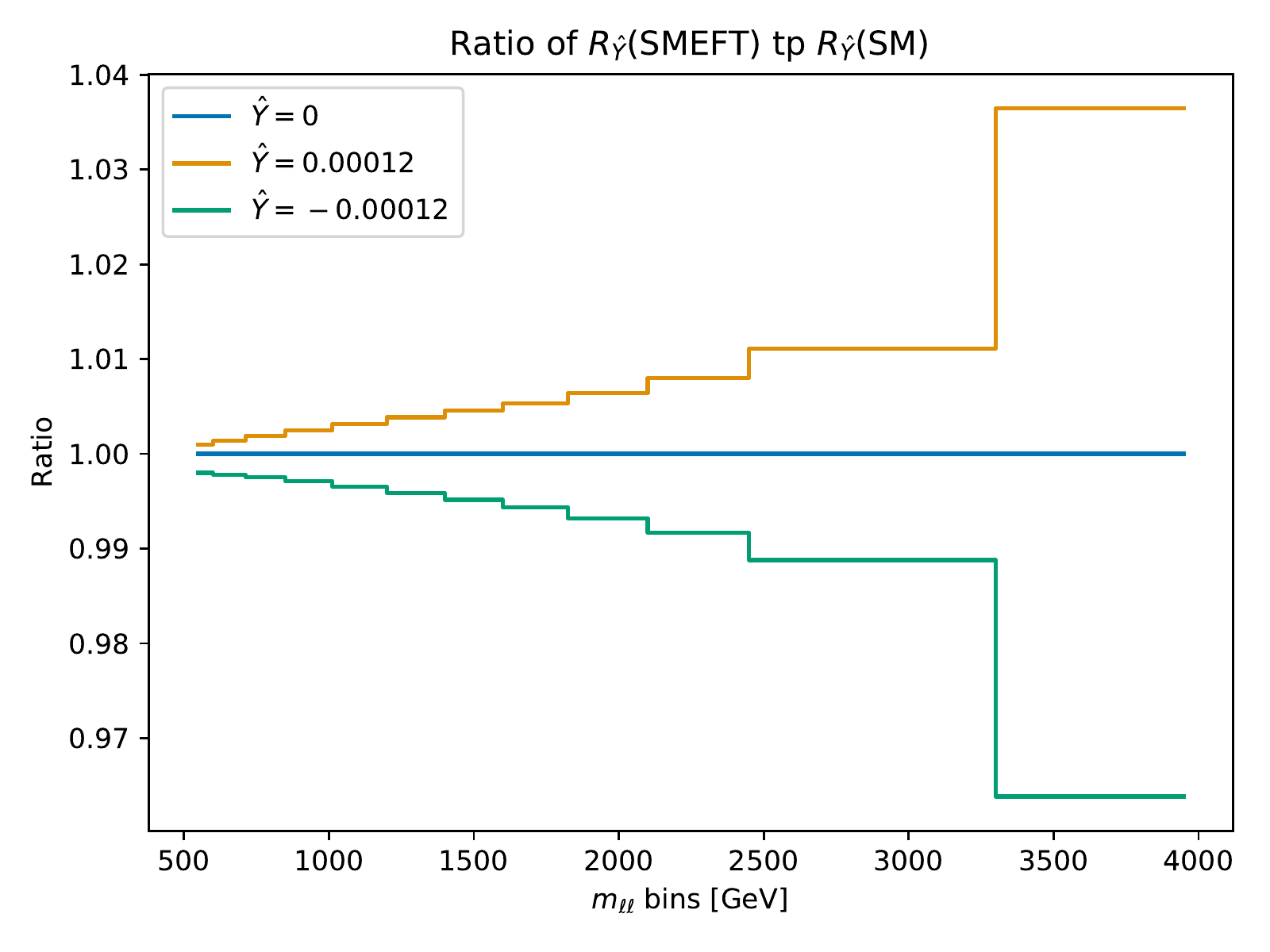}\\
\includegraphics[width=0.45\textwidth]{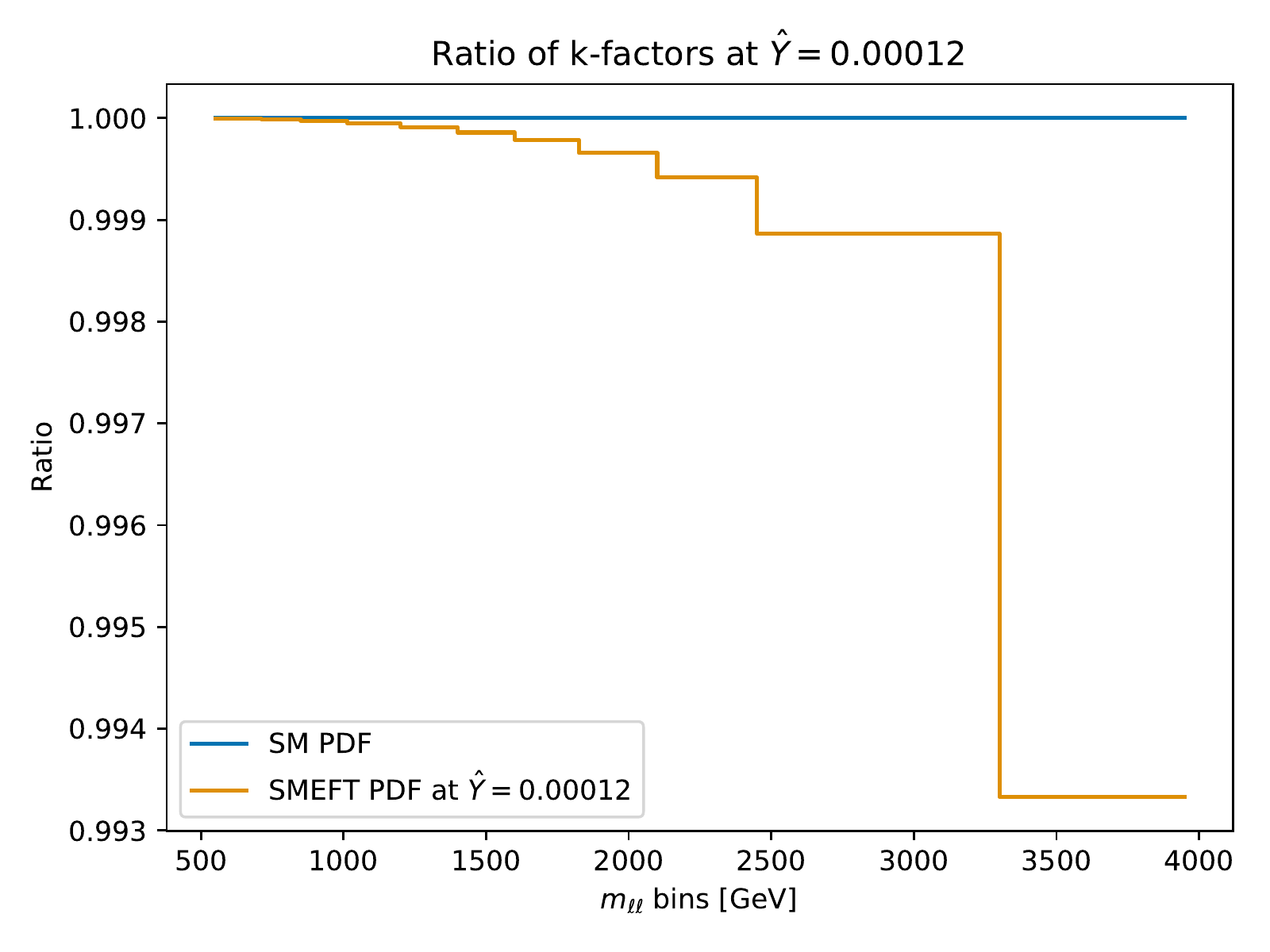}
\includegraphics[width=0.45\textwidth]{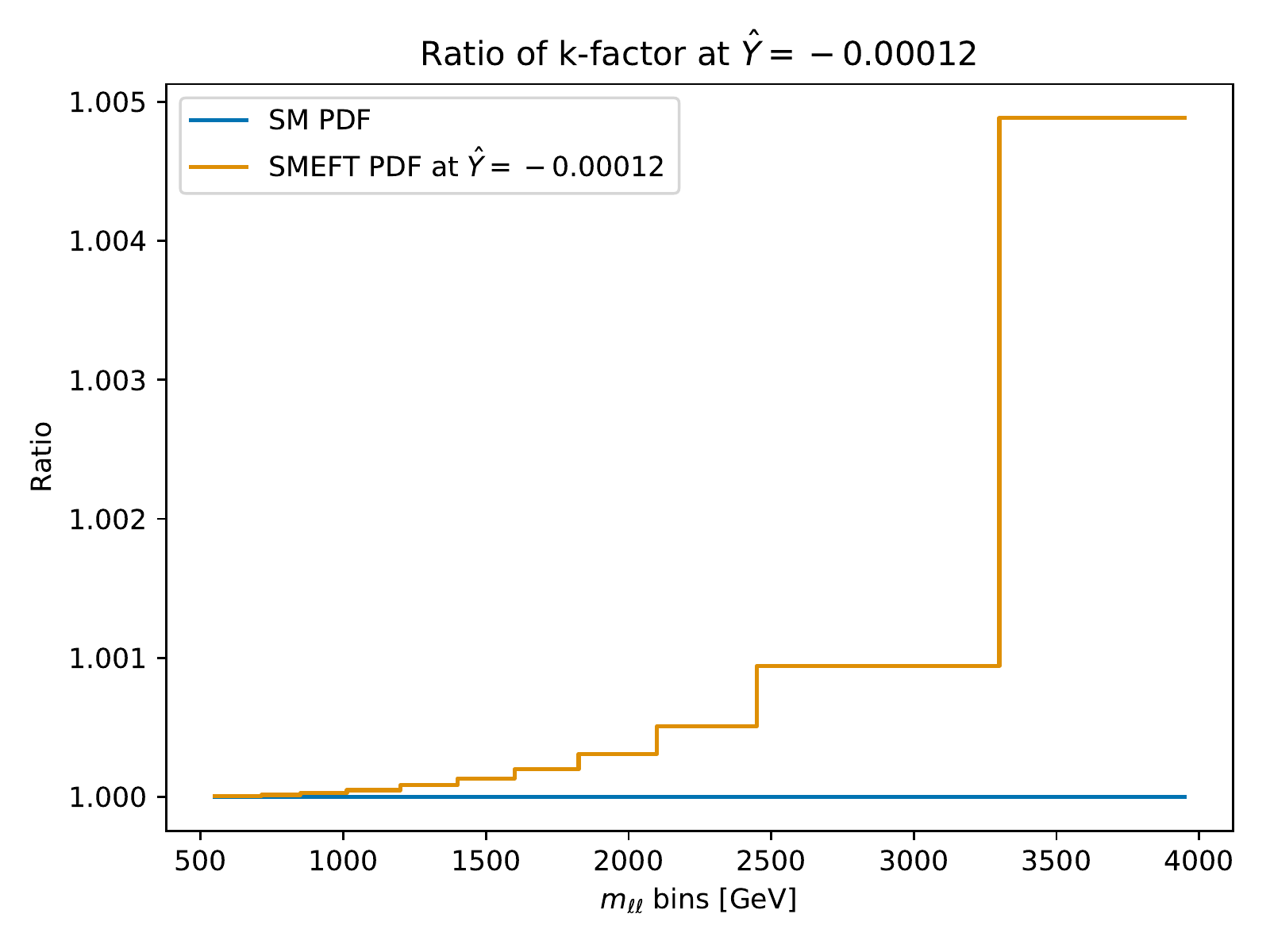}
\caption{\label{fig:Yapproxim} Same as Fig.~\ref{fig:Wapproxim}, in the
  case of $\hat{Y}$ parameter.}
\end{figure}

In the case of the $\hat{Y}$ parameter we observe
a rather larger deviation in the highest invariant mass
bin, which reaches to 4\%, as shown in
Fig~\ref{fig:Yapproxim}. However, as in the case of $\hat{W}$, the
impact of these discrepancies on the SMEFT $K$-factors, calculated at
each of the benchmark points $\hat{Y}= \pm 1.2 \cdot 10^{-4}$, thus on
the observable is below the percent level, which is still negligible
compared to the experimental and theoretical uncertainties associated
to the last bin if the invariant mass distribution.

\begin{figure}[hbt]
\centering
\includegraphics[width=0.45\textwidth]{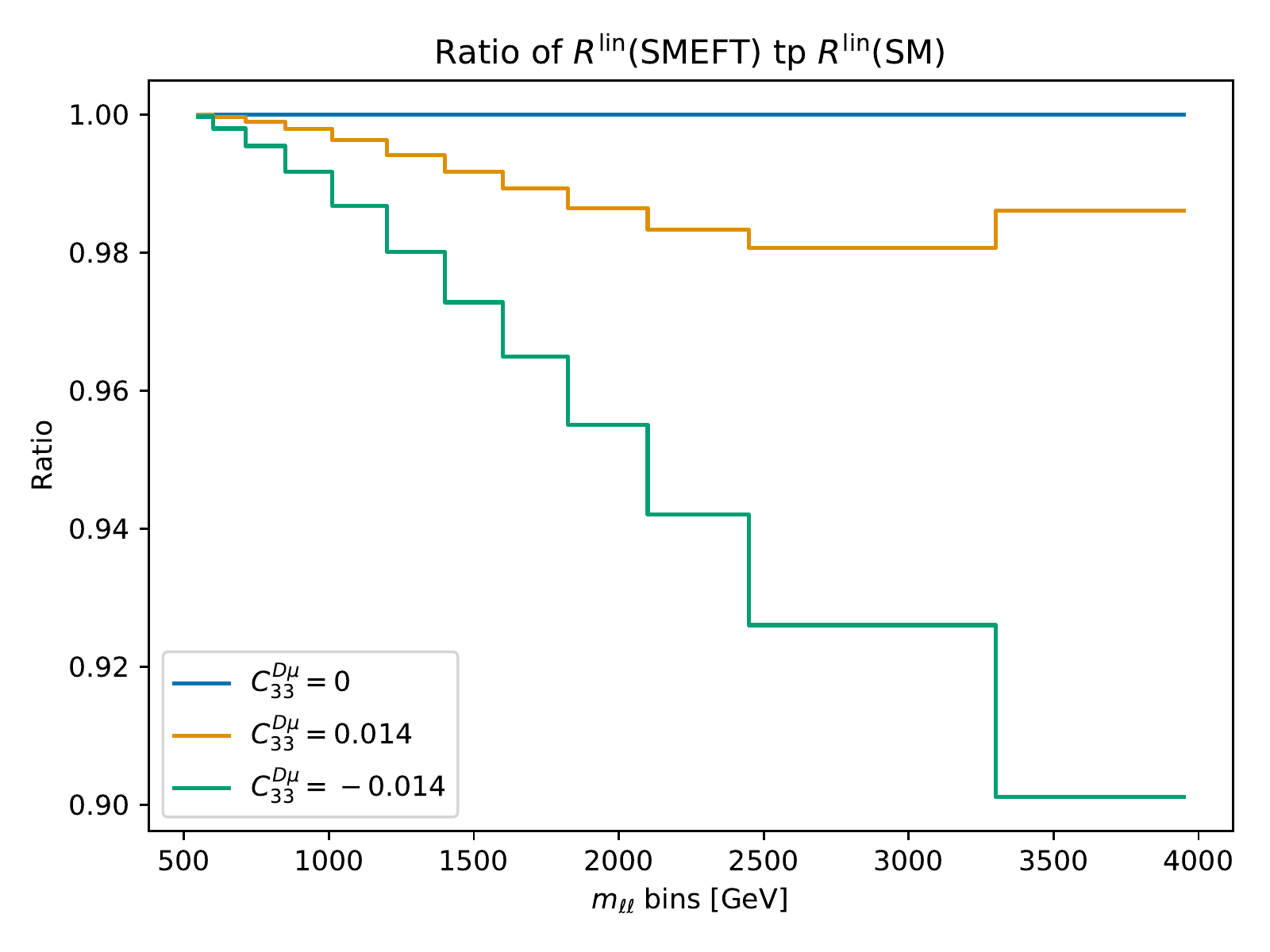}
\includegraphics[width=0.45\textwidth]{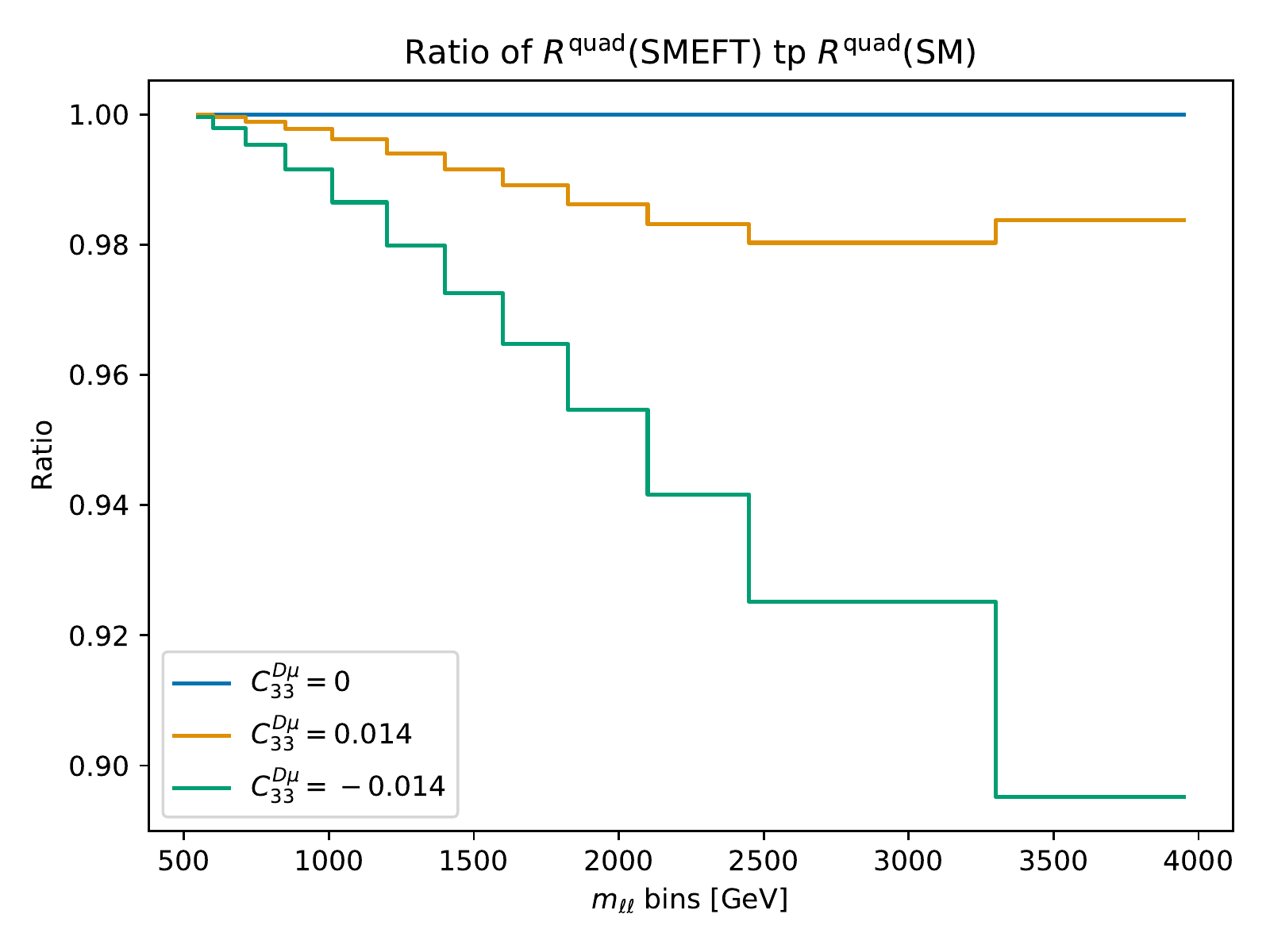}\\
\includegraphics[width=0.45\textwidth]{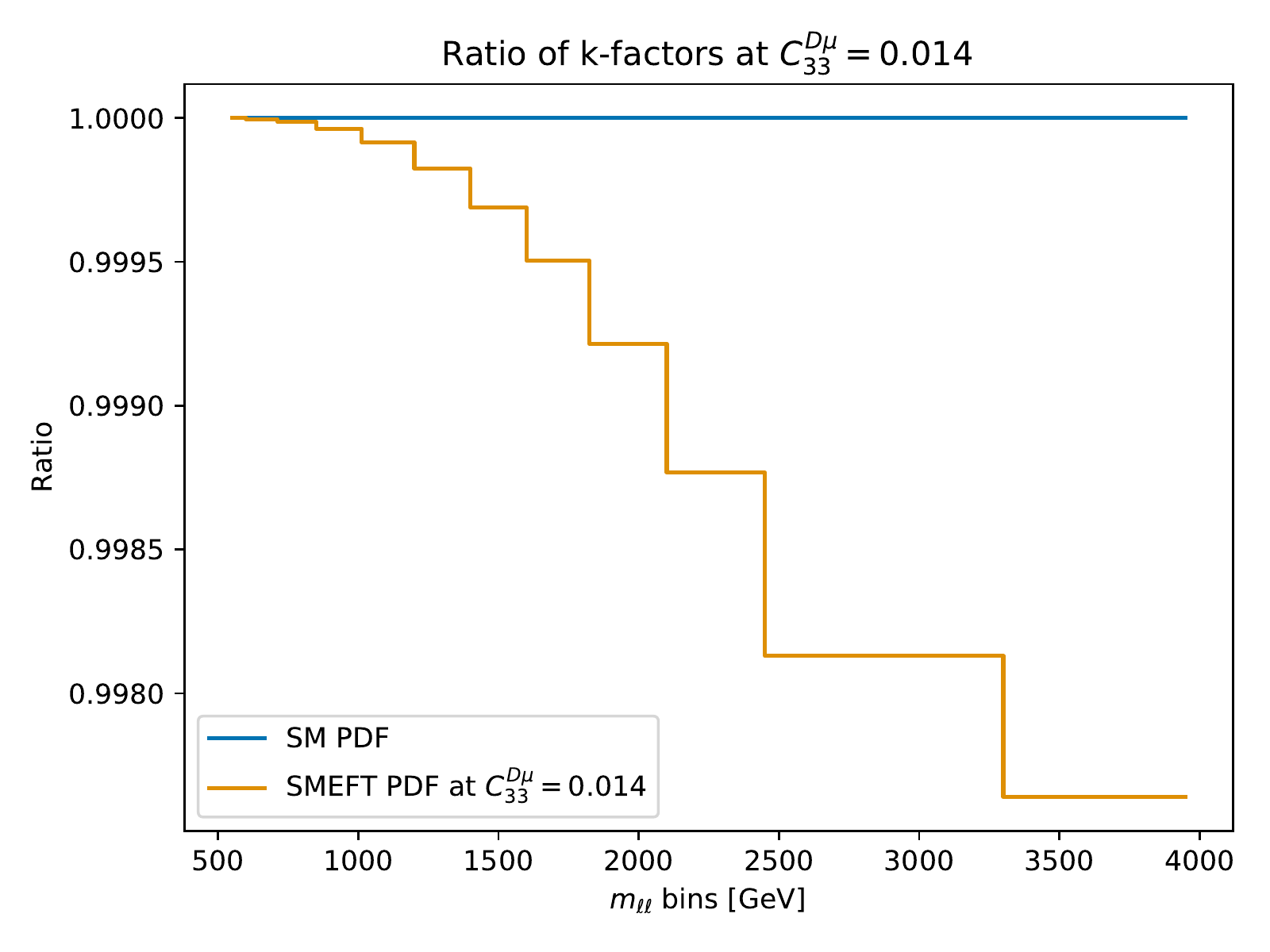}
\includegraphics[width=0.45\textwidth]{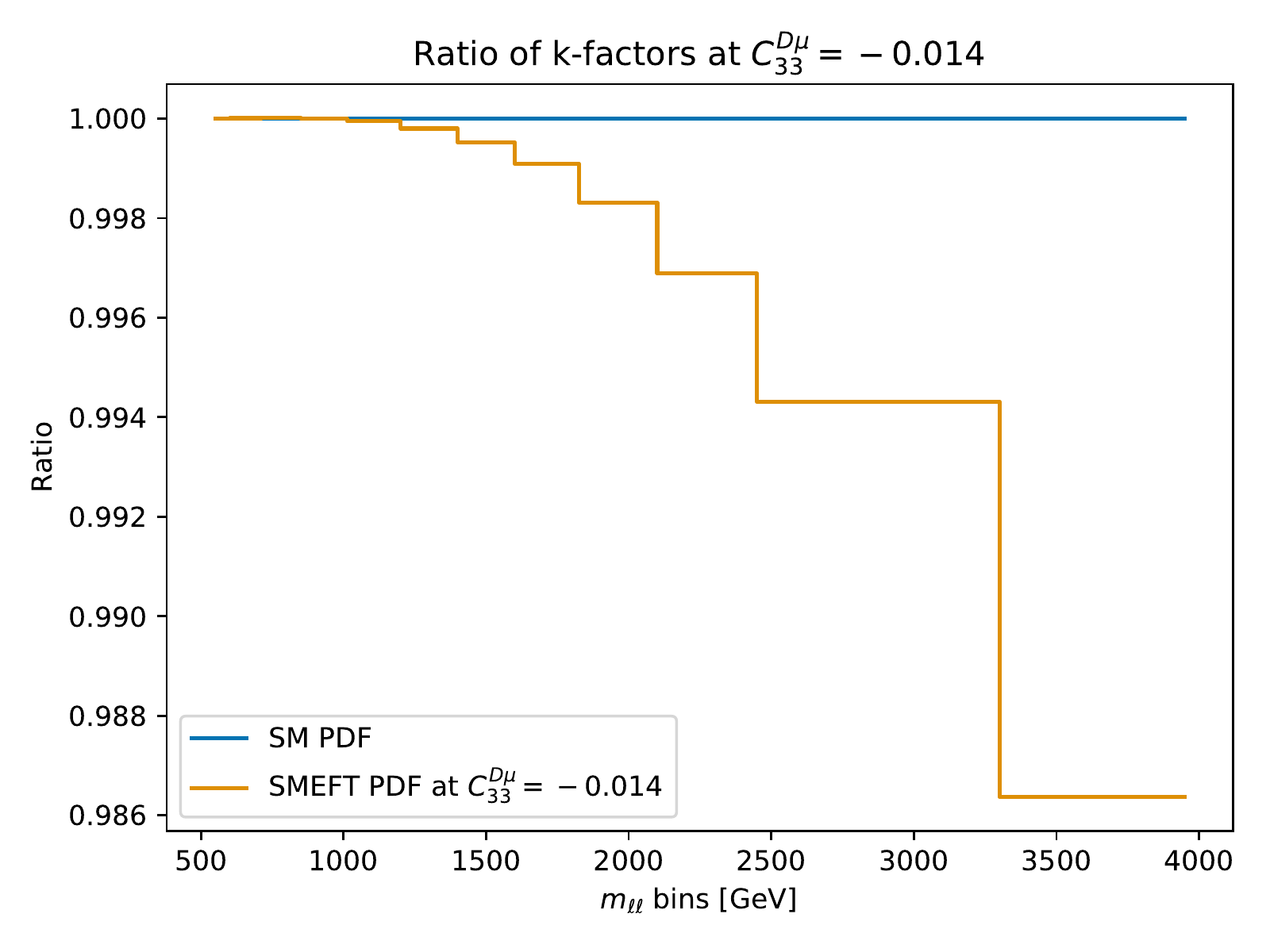}
\caption{\label{fig:Cbapproxim} The impact of the SMEFT PDFs on $R^{\rm lin}$, $R^{\rm quad}$ (above) and the SMEFT $K$-factor $K = 1 + C_{33}^{D\mu} R^{\rm lin} + (C_{33}^{D \mu})^{2}R^{ \rm quad}$ (below) calculated at the Scenario II benchmark points $C_{33}^{D \mu} = \pm 0.014$.}
\end{figure}

Finally, we turn to Scenario II, in which we expect to observe
the largest deviation between our approximation, based on using fixed
SM PDFs in the computation of the SMEFT K-factors, and the full calculation.  This
is expected because we include both the linear and quadratic terms in the EFT expansion,
and because of the flavour non-universal structure of Scenario II.  Here we will denote
the $K$-factor by $K = 1 + C_{33}^{D\mu} R^{\rm lin} + (C_{33}^{D \mu})^{2}R^{ \rm quad}$
and calculate the dependence of each $R^{\rm lin}$, $R^{\rm quad}$ on the SMEFT PDFs.
We observe a 10\% deviation in the computation of $R$  in the highest invariant
mass bin, as shown in Fig~\ref{fig:Cbapproxim}. However, the impact of
such differences on the actual SMEFT $K$-factors, thus on the
observables, calculated at each of the benchmark points $C_{33}^{D
  \mu} = \pm 0.014$, is still at the percent level, which is
acceptable compared to the experimental and theoretical
uncertainties. 

\section{Flavour dependence of the SMEFT PDFs}
\label{add:smeftpdfs}

\begin{figure}[t]
\begin{center}
\includegraphics[width=0.32\textwidth]{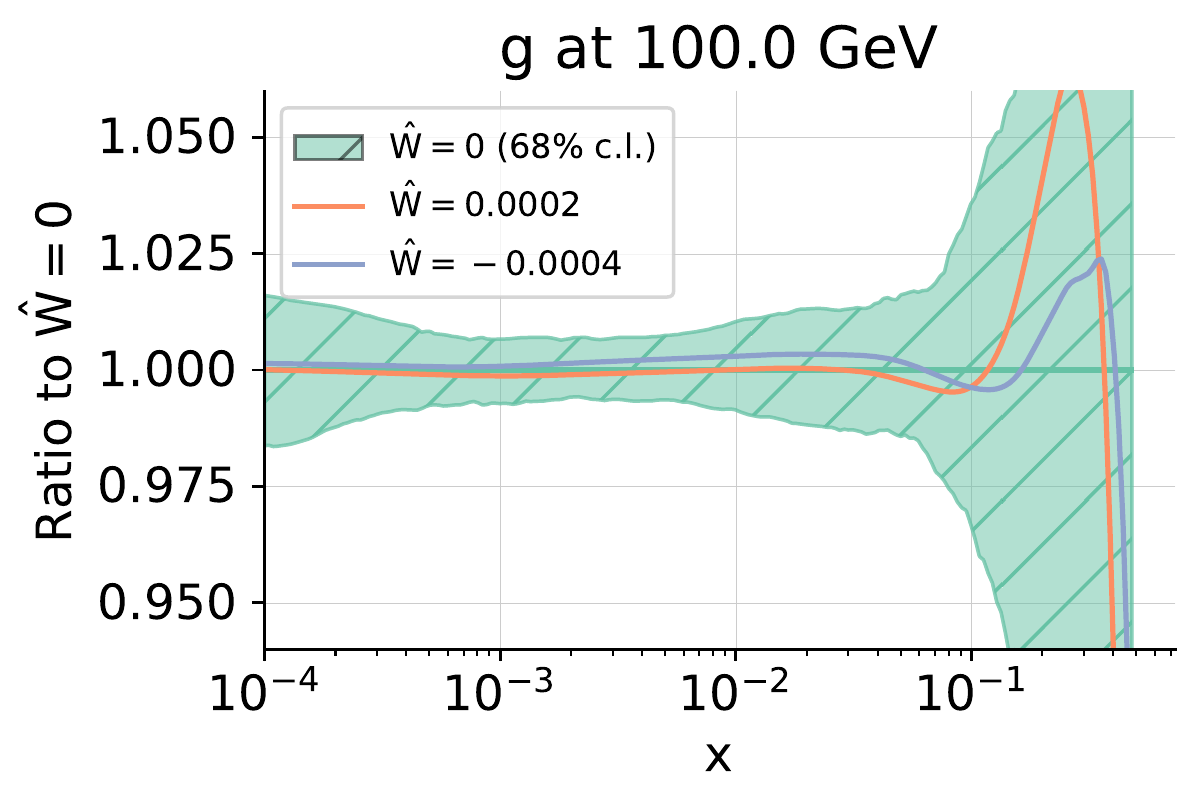}
\includegraphics[width=0.32\textwidth]{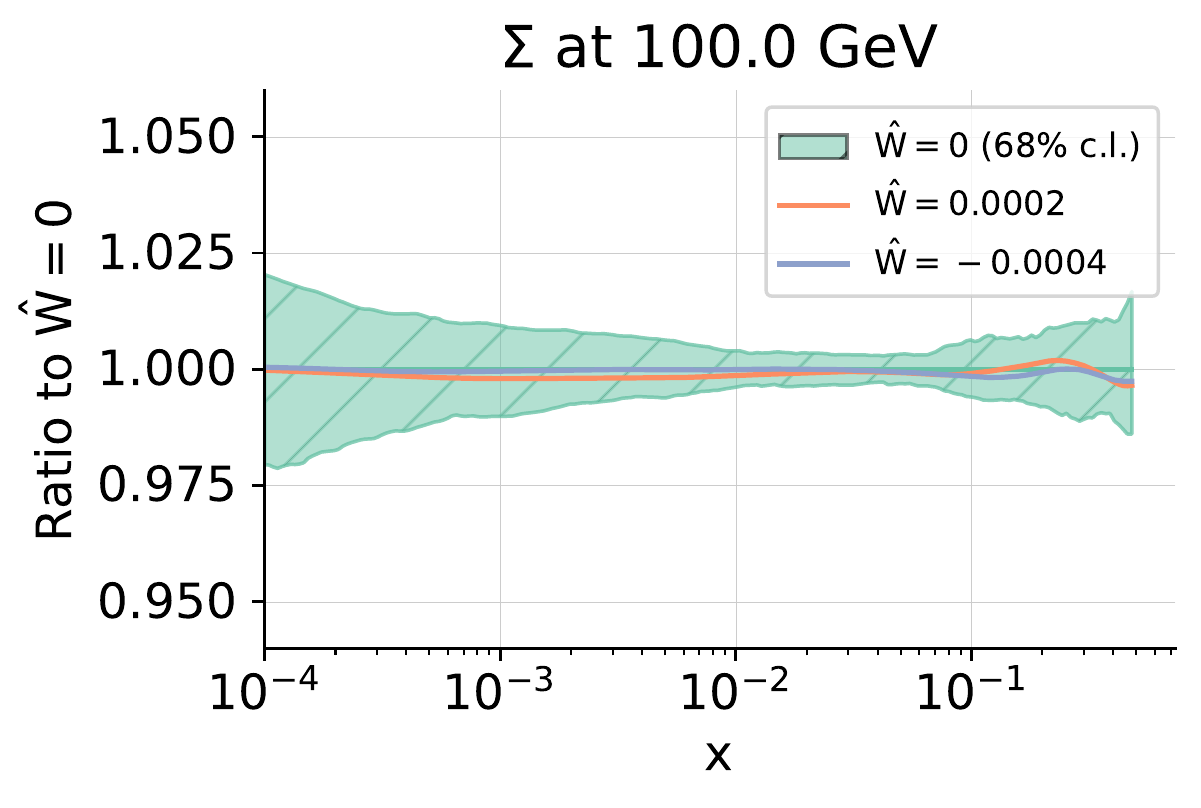}
\includegraphics[width=0.32\textwidth]{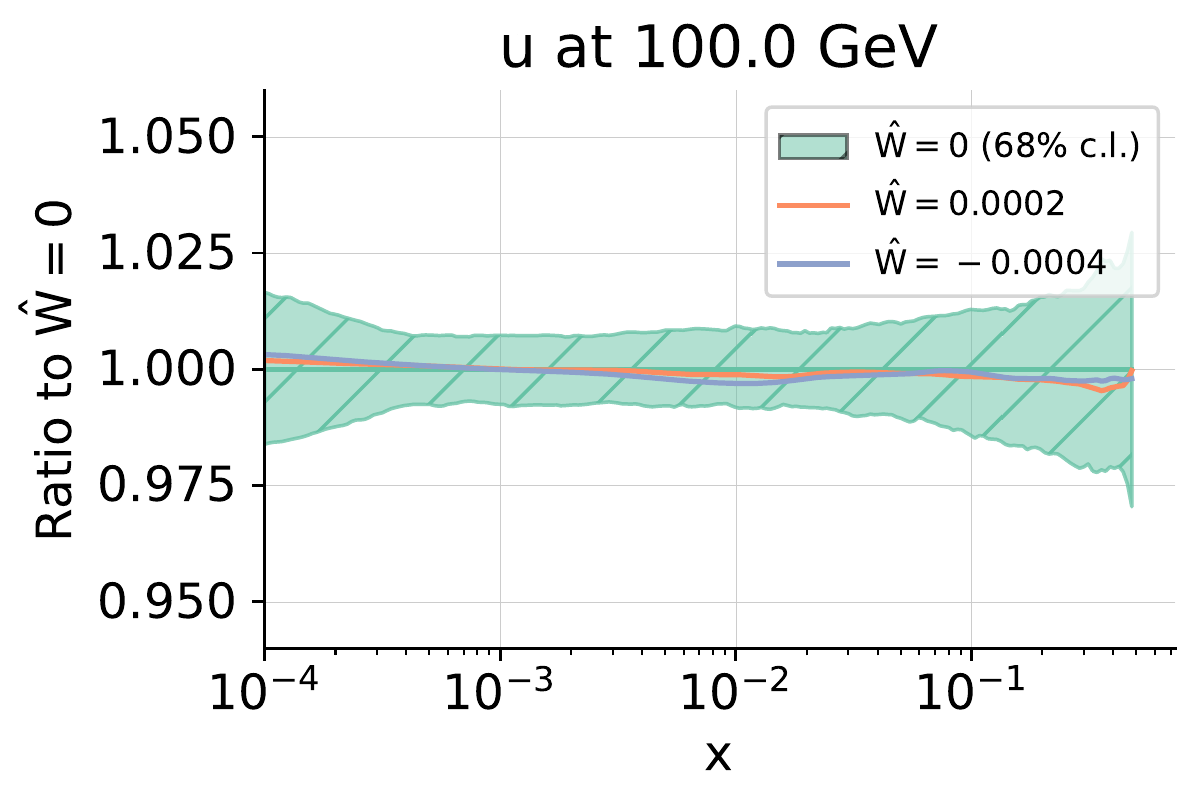}
\includegraphics[width=0.32\textwidth]{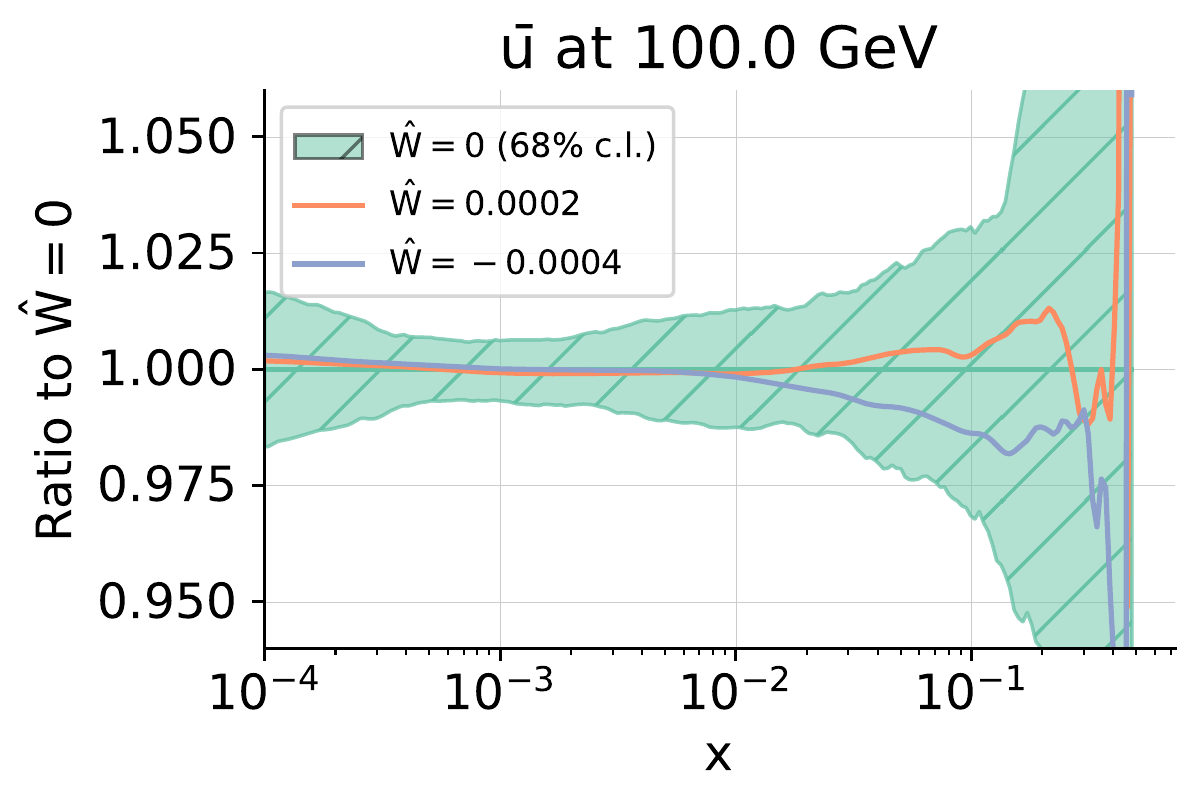}
\includegraphics[width=0.32\textwidth]{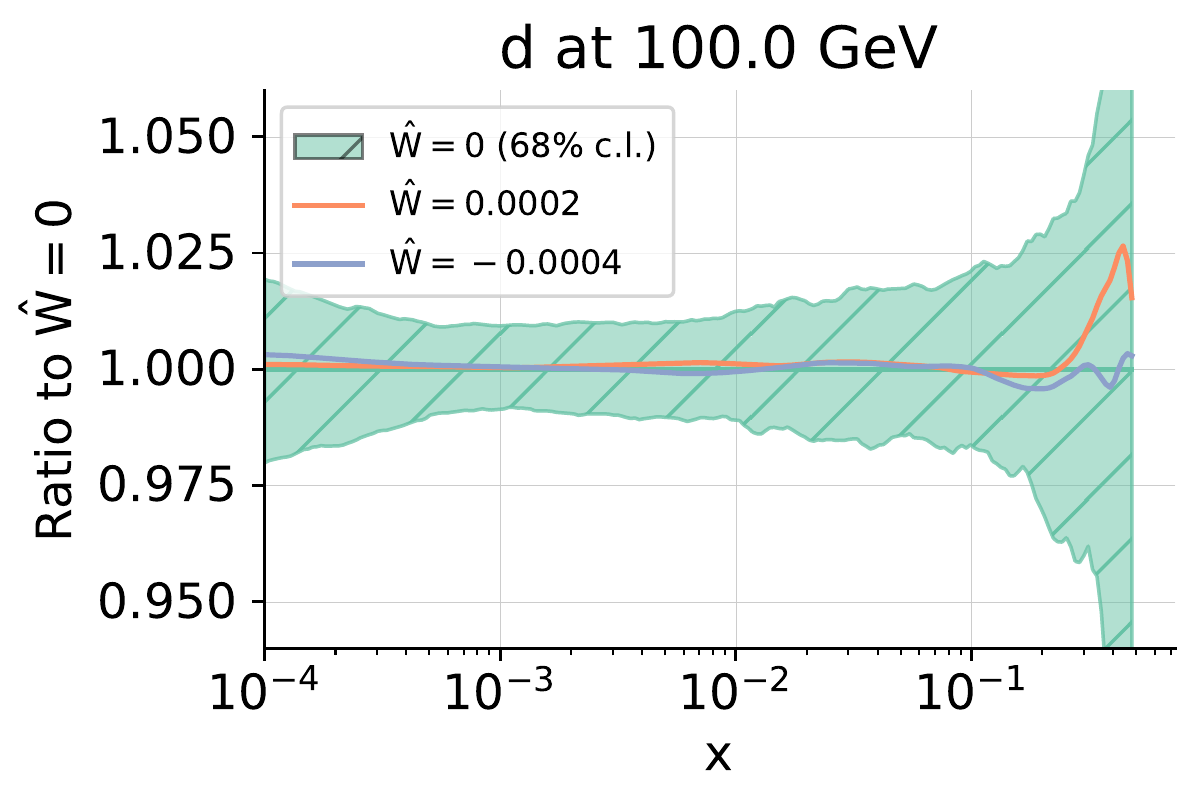}
\includegraphics[width=0.32\textwidth]{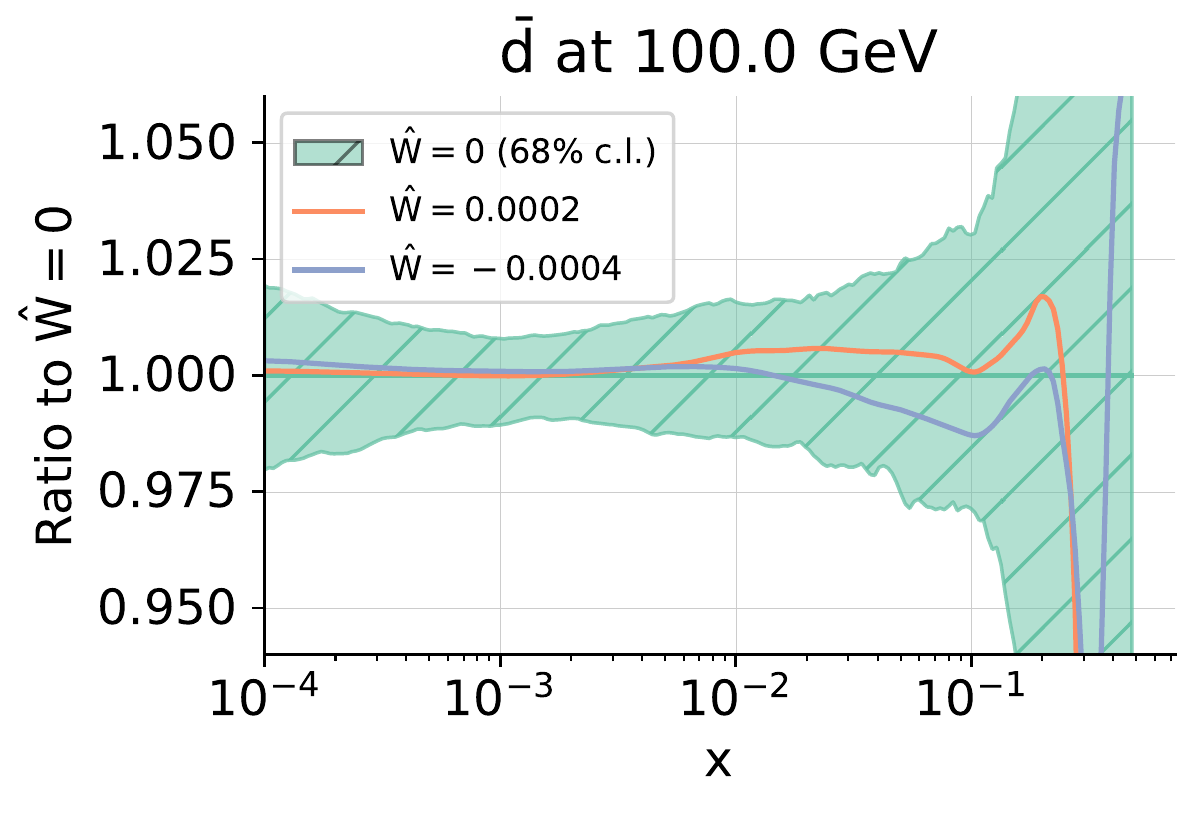}
\includegraphics[width=0.32\textwidth]{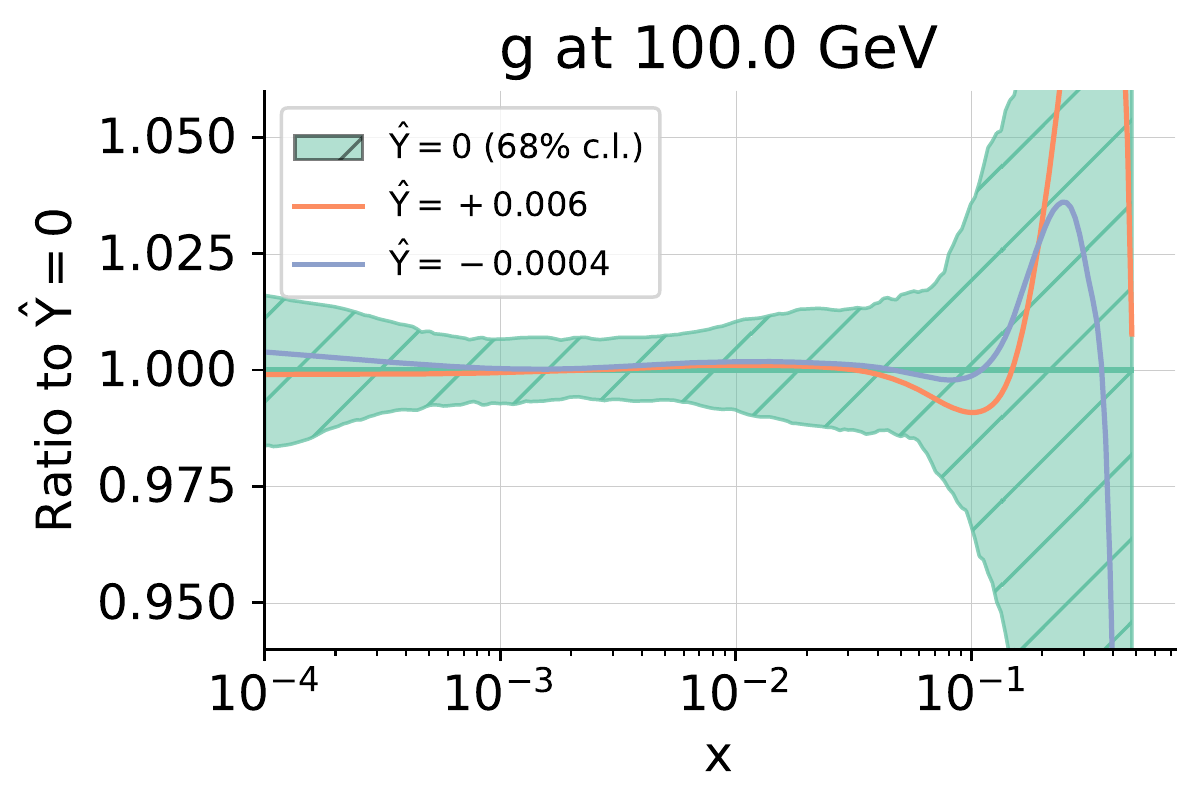}
\includegraphics[width=0.32\textwidth]{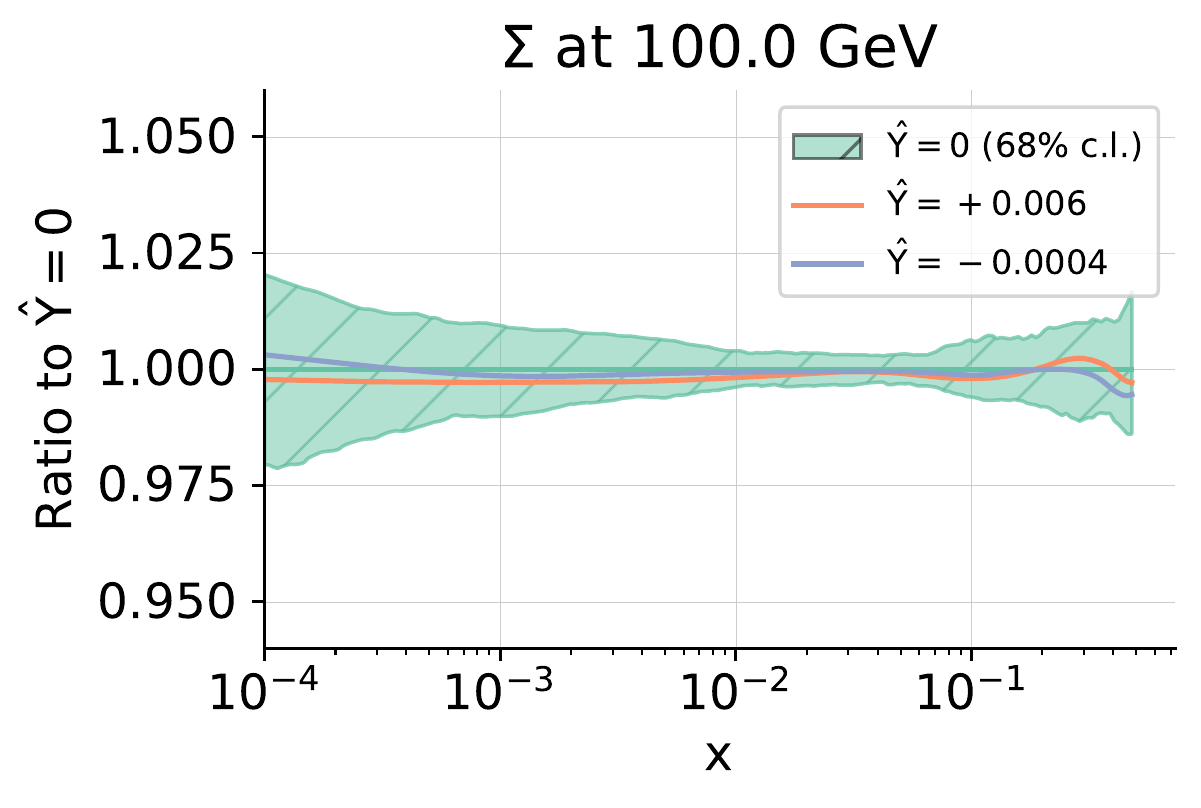}
\includegraphics[width=0.32\textwidth]{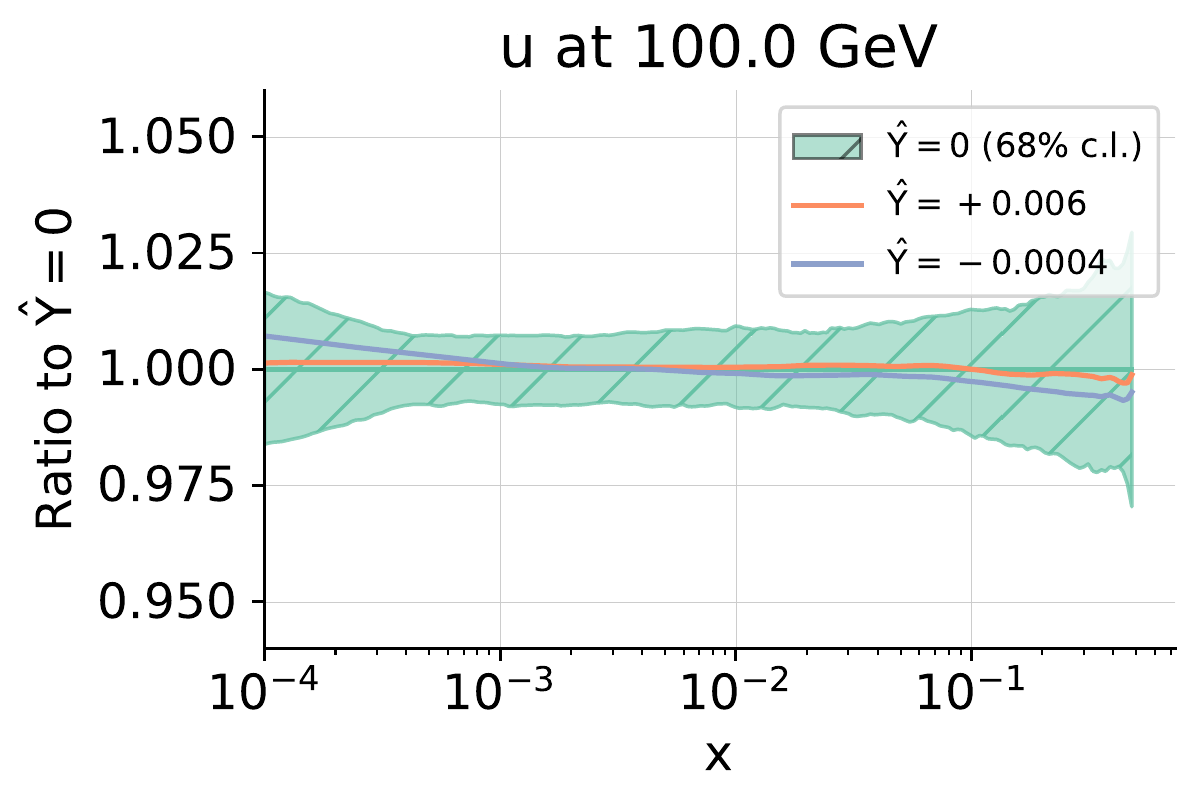}
\includegraphics[width=0.32\textwidth]{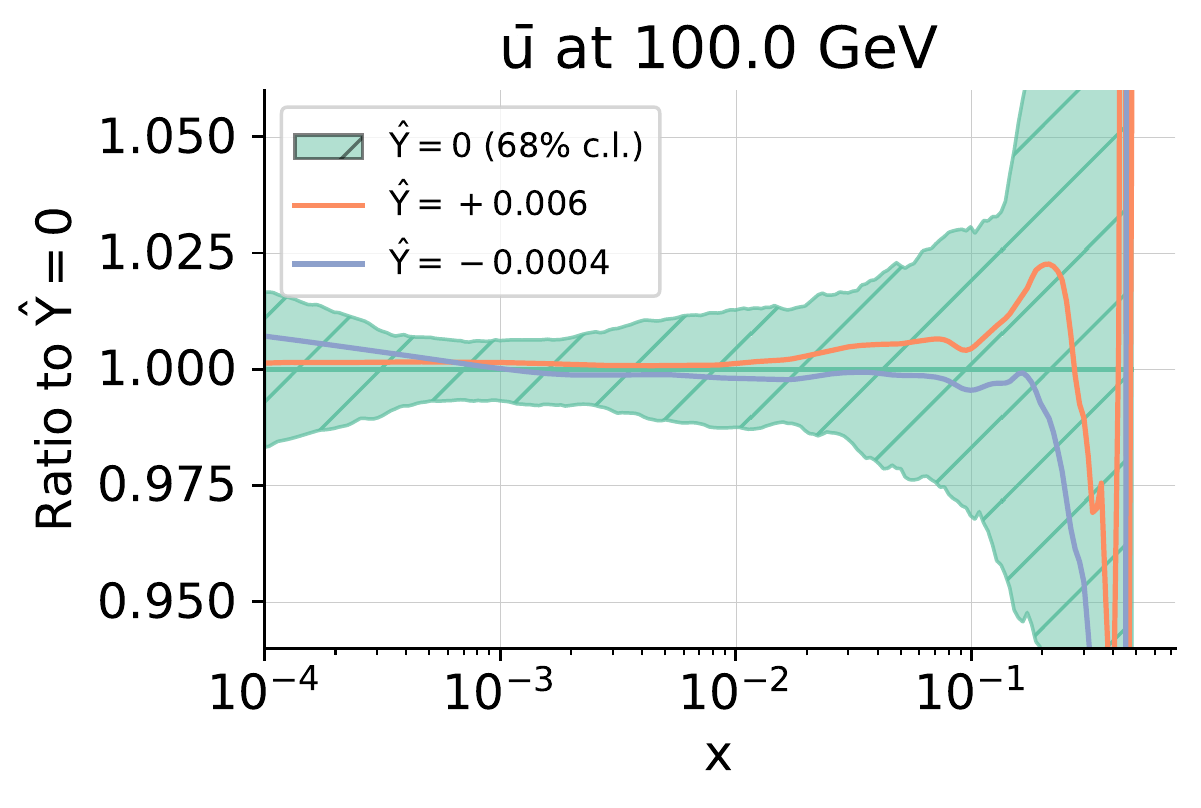}
\includegraphics[width=0.32\textwidth]{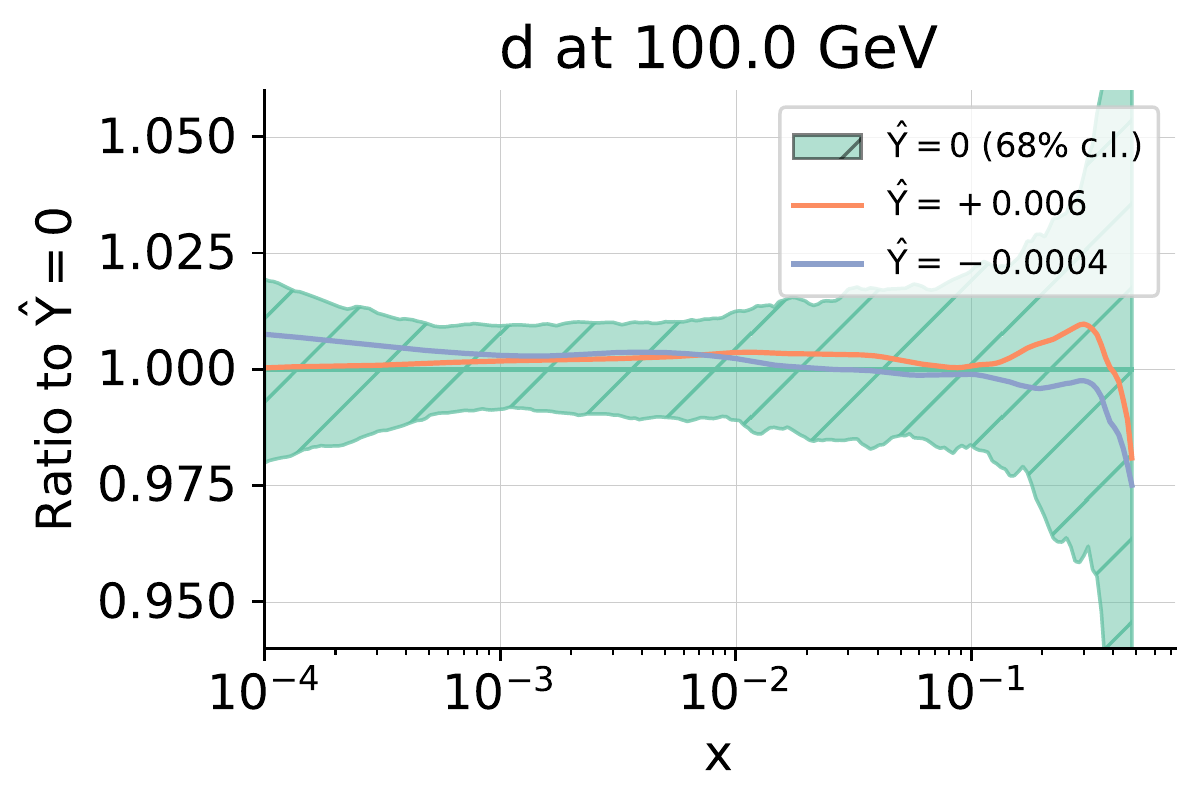}
\includegraphics[width=0.32\textwidth]{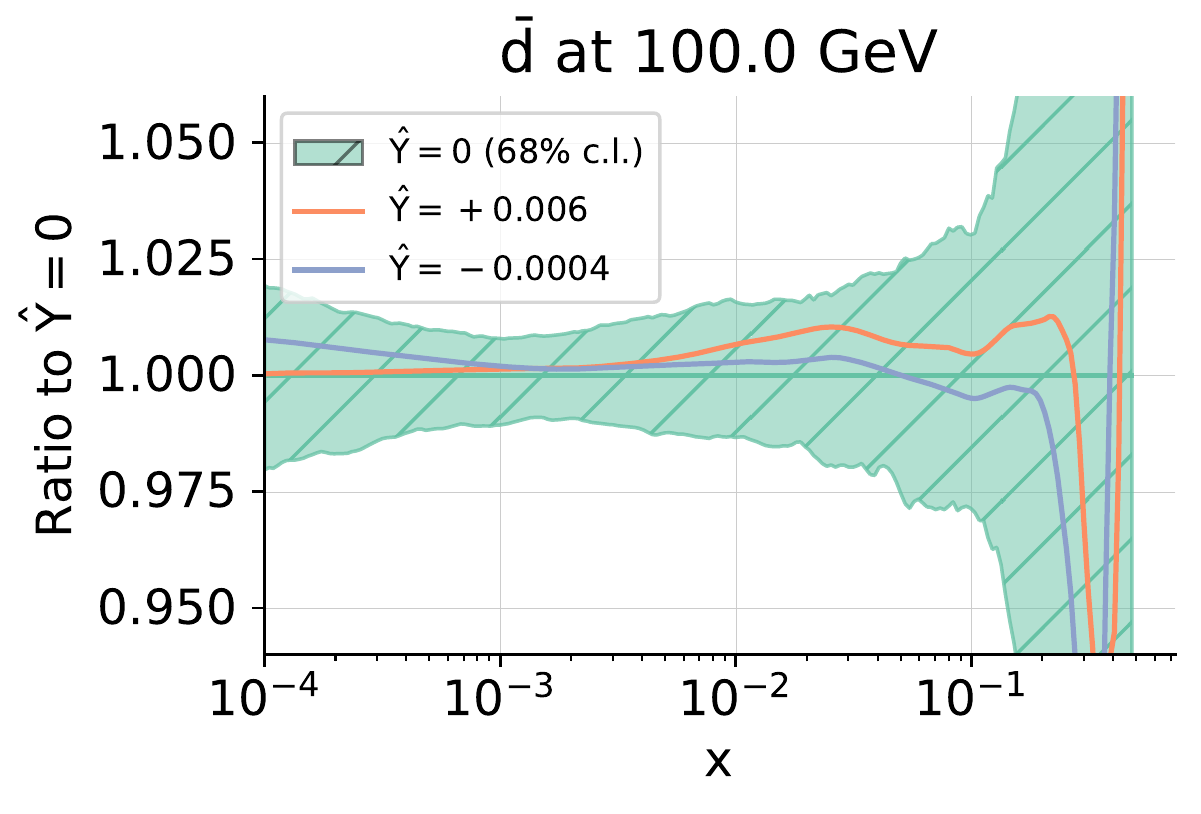}
\caption{\label{fig:SMEFT_PDFs} \small Comparison between the SM and the SMEFT PDFs
  at $Q=100$ GeV, displayed as ratios to the central value of the SM PDFs,
   for representative values of the $\hat{W}$ (upper) and of $\hat{Y}$ (lower panels) parameters.
  We show the gluon, the total quark singlet, the up quark and antiquark, and the down
  quark and antiquark PDFs.
  The values of $\hat{W}$ and $\hat{Y}$ are chosen to be close 
 to the upper and lower limits of the 68\% CL intervals reported in Table~\ref{tab:bound1w}.
}
\end{center}
\end{figure}

In Sect.~\ref{sec:res1}, when discussing the results of the joint
PDF and EFT fits to available Drell-Yan cross-section data, we presented the comparison between the
SM and SMEFT PDFs in benchmark scenario I for different values
of the $\hat{W}$ and $\hat{Y}$ parameters in terms of
the partonic luminosities, Fig.~\ref{fig:SMEFT_lumis}.
In this appendix we present the corresponding comparisons
between  SM and SMEFT PDFs at the level of individual PDF flavours.
Fig.~\ref{fig:SMEFT_PDFs} displays a
comparison between the SM and the SMEFT PDFs
  at $Q=100$ GeV
   for representative values of the $\hat{W}$ (upper) and of $\hat{Y}$ (lower panels) parameters.
  The values of $\hat{W}$ and $\hat{Y}$ are chosen to be close 
  to the upper and lower limits of the 68\% CL intervals reported in Table~\ref{tab:bound1w}.
  The error band in the SM PDFs corresponds to the 68\% CL PDF uncertainty, while for the SMEFT
  PDFs only the central values are shown.

In all cases, one finds that the EFT-induced shifts on the PDFs are smaller
than their uncertainties, though in some cases these shifts can represent up to one-third 
of a standard deviation.
In particular, the up and down antiquarks in the region $x\gsim 10^{-2}$ are the PDF flavours most affected
by the EFT effects.
This finding can be understood from the fact that the NC Drell-Yan cross section
is proportional to the $u\bar{u}$ and $d\bar{d}$ combinations at leading order,
but the up and down quark PDF are already well constrained by lower-energy DIS measurements.
Furthermore, we have verified
that the PDF uncertainties themselves are unchanged in the SMEFT fits.
The results of  Fig.~\ref{fig:SMEFT_PDFs} are consistent with those of Table~\ref{tab:bound1w} and demonstrate
that, with current data, the interplay between EFT effects and PDFs in the high-mass Drell-Yan tails
is appreciable but remains subdominant as compared to other sources of uncertainty.

\providecommand{\href}[2]{#2}\begingroup\raggedright\endgroup


\end{document}